\documentclass[a4paper,12pt]{article}
\pdfoutput=1
\usepackage{jheppub}
\usepackage[utf8]{inputenc}
\usepackage{url}
\usepackage{mathrsfs}  
\usepackage{pdflscape}
\usepackage{refcount}
\usepackage{booktabs}
\usepackage{todonotes}
\usepackage{enumitem}
\usepackage{adjustbox}
\usepackage{afterpage}
\usepackage{mathdots}

\usepackage{amsfonts}
\usepackage{bbm}

\usepackage{pdfpages}
\usepackage{caption}
\usepackage{subcaption}

\usepackage{tikz}
\usetikzlibrary{positioning}
\usetikzlibrary{arrows}
\usetikzlibrary{decorations.pathreplacing}
\usetikzlibrary{shapes.geometric}
\usetikzlibrary{calc}
\usetikzlibrary{decorations.pathmorphing}
\usetikzlibrary{shapes.misc}

\definecolor{goodgreen}{RGB}{55,169,49}

\definecolor{darkyellow}{RGB}{230,170,10}

\definecolor{brightyellow}{RGB}{255,240,190}

\tikzset{flavour/.style={draw=none,minimum size=0.3mm,fill=white, regular polygon,regular polygon sides=4,draw}}
\tikzset{gaugeBig/.style={inner sep=1mm,draw=none,fill=white,minimum size=2mm,circle, draw}}
\tikzset{bd/.style={circle, draw=black, inner sep=0pt, fill=black, minimum size=2mm}}
\tikzset{wd/.style={circle, draw=black, inner sep=0pt, fill=white, minimum size=2mm}}
\tikzset{Dynkin/.style={circle, draw=black, inner sep=0pt, fill=white, minimum size=2mm}}
\tikzstyle{ligne}=[draw, very thick] 
\tikzstyle{gridline}=[draw, gray] 
\tikzset{gauge/.style={circle, draw,inner sep=2.5pt}}
\tikzset{gaugeo/.style={circle, draw,inner sep=2.5pt,fill=orange}}
\tikzset{gaugec/.style={circle, draw,inner sep=2.5pt,fill=cyan}}
\tikzset{gauger/.style={circle, draw,inner sep=2.5pt,fill=red}}
\tikzset{gaugeb/.style={circle, draw,inner sep=2.5pt,fill=blue}}
\tikzset{gaugeg/.style={circle, draw,inner sep=2.5pt,fill=green}}
\tikzset{gaugem/.style={circle, draw,inner sep=2.5pt,fill=magenta}}
\tikzset{hasse/.style={circle, fill,inner sep=2pt}}
\tikzset{shrinky/.style={circle, fill,inner sep=1pt}}
\tikzset{sized/.style={circle, draw, inner sep=1.5pt}}
\tikzset{seven/.style={circle, draw,inner sep=3pt}}

\tikzset{dotto/.style={circle, orange, draw,inner sep=1.5pt,fill=orange}}
\tikzset{dottp/.style={circle, purple, draw,inner sep=1.5pt,fill=purple}}
\tikzset{dottc/.style={circle, cyan, draw,inner sep=1.5pt,fill=cyan}}
\tikzset{dottr/.style={circle, red, draw,inner sep=1.5pt,fill=red}}
\tikzset{dottb/.style={circle, blue, draw,inner sep=1.5pt,fill=blue}}
\tikzset{dottg/.style={circle, green, draw,inner sep=1.5pt,fill=green}}
\tikzset{dottm/.style={circle, magenta, draw,inner sep=1.5pt,fill=magenta}}

\tikzset{redgauge/.style={draw=none,minimum size=0.4cm,fill=red,circle, draw}}
\tikzset{gauge3/.style={draw=none,minimum size=0.4cm,fill=white,circle, draw}}
\tikzset{bluegauge/.style={draw=none,minimum size=0.4cm,fill=blue,circle, draw}}
\tikzset{redflavor/.style={draw=none,minimum size=0.6cm,fill=red, regular polygon,regular polygon sides=4,draw}}
\tikzset{blueflavor/.style={draw=none,minimum size=0.6cm,fill=blue, regular polygon,regular polygon sides=4,draw}}
\tikzset{flavour2/.style={draw=none,minimum size=0.6cm,fill=white, regular polygon,regular polygon sides=4,draw}}
\tikzset{rede/.style={line width=0.5mm,red}}
\tikzset{bluee/.style={line width=0.5mm,blue}}
\tikzset{pinkline/.style={line width=0.5mm,purple}}
\pgfdeclarelayer{edgelayer}
\pgfdeclarelayer{nodelayer}

\DeclareMathOperator{\U}{U}
\DeclareMathOperator{\SU}{SU}
\DeclareMathOperator{\SO}{SO}

\DeclareMathOperator{\USp}{USp}
\DeclareMathOperator{\tr}{Tr}

\usepackage{epsfig}
\usepackage{amsmath}
\usepackage{amssymb}
\usepackage{amsfonts}
\usepackage{amsxtra}
\usepackage{amsthm}
\usepackage{mathrsfs}
\usepackage{makeidx}
\usepackage{graphics}
\usepackage{dsfont}
\usepackage{mathtools}
\usepackage{graphicx}
\usepackage{placeins}
\usepackage{bm}
\usepackage[capitalise]{cleveref}
\usepackage{empheq}
\usepackage{colortbl}
\usepackage{xcolor}
\usepackage{titlesec}
\usepackage{longtable}
\usepackage{hhline}
\usepackage{float}
\usepackage{color}
\usepackage{tikz}
\usepackage{xfrac}
\usepackage{footnote}
\usepackage{rotating}
\usepackage{lscape}
\usepackage{makecell}
\usepackage{environ}
\usepackage{tabularx}
\usepackage{subfiles}
\usepackage{ytableau}
\usepackage{tikz-3dplot}
\usepackage{slashed}
\usepackage{pifont}
\usepackage{multirow}
\usepackage{mdframed}
\usepackage{bbm}
\usepackage[normalem]{ulem}
\usepackage{arydshln}
\usepackage{enumitem} 
\usepackage{fancybox}

\usetikzlibrary{positioning,trees,decorations.pathmorphing,decorations.markings,decorations.pathreplacing,calc,shapes,patterns,arrows,chains,arrows.meta,fit,fadings,decorations.markings,graphs,graphs.standard,quotes,plotmarks,calligraphy,decorations.pathreplacing}

\usetikzlibrary{positioning}
\usetikzlibrary{chains}
\usetikzlibrary{arrows, arrows.meta ,fit,decorations.pathreplacing}
\tikzstyle{every picture}+=[remember picture]
\tikzstyle{na} = [baseline]
\tikzstyle{ligne}=[draw, thick]

\usetikzlibrary{arrows, decorations.markings, calc, fadings, decorations.pathreplacing, patterns, decorations.pathmorphing, positioning}
\tikzset{>={Latex[width=1.5mm,length=1.5mm]}}
\tikzset{bd/.style={circle, draw=black, inner sep=0pt, fill=black, minimum size=1.2mm}}
\tikzset{bld/.style={circle, draw=blue, inner sep=0pt, fill=blue, minimum size=1.2mm}}
\tikzset{wd/.style={circle, draw=black, inner sep=0pt, fill=white, minimum size=1.2mm}}
\tikzset{rd/.style={circle, draw=red, inner sep=0pt, fill=red, minimum size=.9mm}}
\tikzset{wrd/.style={circle, draw=red, inner sep=0pt, fill=white, minimum size=.9mm}}
\usetikzlibrary{graphs,graphs.standard,quotes}

\def\node#1#2{\overset{#1}{\underset{#2}{{\color{gray} \bullet}}}}

\def\node#1#2{\overset{#1}{\underset{#2}{\circ}}}

\tikzstyle{every picture}+=[remember picture]
\tikzstyle{na} = [baseline=-.5ex]

\newcommand{\eg}{e.g. }

\newcommand{\ie}{i.e. }

\numberwithin{equation}{section}
\newcommand{\bes}[1]{\begin{equation} \begin{split} #1\end{split} \end{equation}}

\newcommand{\be}{\begin{equation}} \newcommand{\ee}{\end{equation}}
\newcommand{\bea}{\begin{equation} \begin{aligned}} \newcommand{\eea}{\end{aligned} \end{equation}}

\def\tilde{\widetilde}

\def\hat{\widehat}

\def\bar{\overline}

\def\rt2{\sqrt{2}}

\def\tr{\mathop{\rm tr}}

\def\CA{{\cal A}}

\def\CC{{\cal C}}

\def\CI{{\cal I}}

\def\CM{{\cal M}}
\def\CN{{\cal N}}

\def\CZ{{\cal Z}}

\def\1{{\ds 1}}

\newcommand{\cC}{\mathcal{C}}

\newcommand{\fz}{\mathfrak{z}}

\newcommand{\ID}{\mathds{1}}

\newcommand{\fc}{\mathfrak{c}}

\def\SO{\mathrm{SO}}
\def\PSO{\mathrm{PSO}}
\def\O{\mathrm{O}}
\def\SU{\mathrm{SU}}

\def\Spin{\mathrm{Spin}}
\def\Pin{\mathrm{Pin}}

\def\su{\mathfrak{su}}

\def\so{\mathfrak{so}}

\def\usp{\mathfrak{usp}}

\def\repa{\raise4pt\hbox{$\square$}\mkern-14mu\raise-4pt\hbox{$\square$}}
\def\repab{\overline{\raise4pt\hbox{$\square$}\mkern-14mu\raise-4pt\hbox{$\square$}\mkern-1mu}}

\def\smileface{\ensuremath{\hbox{\large$\bigcirc$}\mkern-15mu\raise-1pt\hbox{\scriptsize$\smallsmile$}%
\mkern-10mu\raise4pt\hbox{..}\mkern4mu}}
\def\frownface{\ensuremath{\hbox{\large$\bigcirc$}\mkern-15mu\raise-1pt\hbox{\scriptsize$\smallfrown$}%
\mkern-10mu\raise4pt\hbox{..}\mkern4mu}}

\newcommand{\ba}{\begin{array}}
\newcommand{\ea}{\end{array}}
\newcommand{\bi}{\begin{itemize}}
\newcommand{\ei}{\end{itemize}}
\def\vec#1{\bm{#1}}
\def\bea#1\eea{\allowdisplaybreaks \begin{align}#1\end{align}}
 \newcommand{\ben}{\begin{enumerate}}
\newcommand{\een}{\end{enumerate}}
\newcommand{\bean}{\begin{eqnarray*}}
\newcommand{\eean}{\end{eqnarray*}}
\newcommand{\eref}[1]{(\ref{#1})}

\newcommand{\PE}{\mathop{\rm PE}}

\newcommand{\BC}{\mathbb{C}}

\newcommand{\BZ}{\mathbb{Z}}

\newcommand{\diag}{\mathrm{diag}}

\definecolor{light-gray}{gray}{0.5}
\definecolor{new-green}{rgb}{0,0.7,0.3}
\definecolor{cerulean}{rgb}{0.0, 0.48, 0.65}
\definecolor{claret}{rgb}{0.50, 0.09, 0.20}
\definecolor{darkred}{rgb}{0.7, 0.11, 0.11}
\definecolor{scarlet}{rgb}{1.0, 0.13, 0.0}
\definecolor{orange-red}{rgb}{1.0, 0.27, 0.0}

\newcommand{\purple}{\color{purple}}
\newcommand{\brown}{\color{brown}}
\newcommand{\blue}{\color{blue}}
\newcommand{\gray}{\color{light-gray}}
\newcommand{\red}{\color{red}}
\newcommand{\green}{\color{new-green}}
\newcommand{\violet}{\color{violet}}

\newcommand{\claret}{\color{claret}}

\newcommand{\orangered}{\color{orange-red}}

\def\aup#1 {\overset{#1}{\uparrow} \, \overset{\tilde{#1}}{\downarrow}}

\tikzset{snake it/.style={decorate, decoration={snake, amplitude=.4mm, segment length=2mm,
                       post length=0mm,pre length=0mm}}}

\def\u{\mathfrak{u}}
\DeclareMathAlphabet{\mymathds}{U}{BOONDOX-ds}{m}{n}
\tikzstyle{double_border} = [draw, double, double distance=1pt]

\preprint{\hspace{1cm}}

\title{Orthosymplectic Quivers: Indices, Hilbert Series, and Generalised Symmetries}
\author[a,b]{William Harding,}
\author[b,c]{Noppadol Mekareeya,}
\author[d]{and Zhenghao Zhong}

\affiliation[a]{Dipartimento di Fisica, Universit\`a di Milano-Bicocca, Piazza della Scienza 3, I-20126 Milano, Italy}
\affiliation[b]{INFN, sezione di Milano-Bicocca,
Piazza della Scienza 3, I-20126 Milano, Italy}
\affiliation[c]{Department of Physics, Faculty of Science, Chulalongkorn University, Phayathai Road, Pathumwan, Bangkok 10330, Thailand}
\affiliation[d]{Mathematical Institute, University of Oxford,\\Andrew Wiles Building, Woodstock Road, Oxford, OX2 6GG, UK}

\emailAdd{w.harding@campus.unimib.it}
\emailAdd{n.mekareeya@gmail.com}
\emailAdd{zhenghao.zhong@maths.ox.ac.uk}

\abstract{We investigate generalised global symmetries in 3d $\mathcal{N}=4$ orthosymplectic quiver gauge theories. Using the superconformal index, we identify a $D_8$ categorical symmetry web in a class of theories featuring $\mathfrak{so}(2N) \times \mathfrak{usp}(2N)$ gauge algebra (at zero Chern-Simons levels) and $n$ bifundamental half-hypermultiplets, analogous to ABJ-type models. As a distinct contribution, we improve the prescription, previously studied in the literature, for computing Coulomb branch Hilbert series of $\mathrm{SO}(N)$ gauge theories with $N_f$ vector hypermultiplets. Our improved prescription extends these methods by incorporating fugacities for discrete zero-form symmetries - specifically charge conjugation and magnetic symmetries - and properly treating background magnetic fluxes for the flavour symmetry. This refinement enables calculations for various global forms ($\mathrm{O}(N)^\pm, \mathrm{Spin}(N), \mathrm{Pin}(N)$) and ensures consistency with the Coulomb branch limit of the superconformal index and known dualities. The proper treatment of fluxes is particularly essential for analysing orthosymplectic quivers where such a flavour symmetry is gauged. We verify our methods through several examples, including an analysis of the mapping of discrete symmetries under mirror symmetry for $T[\mathrm{SO}(N)]$ and $T[\mathrm{USp}(2N)]$ theories. The analysis also readily generalises to the $T_\rho[\mathrm{SO}(N)]$ and $T_\rho[\mathrm{USp}(2N)]$ theories associated with partition $\rho$.
}

\begin{document}
\maketitle

\section{Introduction and conclusion}
Three-dimensional gauge theories whose gauge group is based on the Lie algebra $\so(N)$ – including $\SO(N)$, $\Spin(N)$, $\O(N)^\pm$, and $\Pin(N)$\footnote{Throughout this paper, $\Pin(N)$ refers to the $\Pin^+(N)$ group, following the notation of \cite{Aharony:2013kma}.} – exhibit rich global symmetry structures. As a key example, consider the $\SO(N)$ gauge theory with an even Chern-Simons level $2k$, denoted $\SO(N)_{2k}$. This theory possesses a $\BZ^{[0]}_{2,\, \CC}$ zero-form charge conjugation symmetry and a $\BZ^{[0]}_{2,\, \CM}$ zero-form magnetic (or topological) symmetry. These zero-form symmetries can be coupled to background one-form gauge fields, denoted $A_1^\CC$ and $A_1^{\CM}$ respectively. Gauging these symmetries (promoting the background fields to dynamical ones) transforms the $\SO(N)$ theory into related theories with different global forms of the gauge group. In addition, the theory typically possesses a $\BZ_2^{[1]}$ one-form symmetry, which couples to a background two-form gauge field $A_2^B$. It was shown in \cite{Cordova:2017vab} that, for even $N=2n$, these global symmetries exhibit a mixed 't Hooft anomaly. This anomaly is described by the following inflow action on a four-manifold $M_4$ bounding the 3d spacetime:
\bes{ \label{anomaly_Cordova}
i \pi \int_{M_4} A_2^B \cup (n A_1^\CM \cup A_1^\CM + k A_1^\CC \cup A_1^\CC + A_1^\CC \cup A_1^\CM)~.
}
This mixed anomaly has far-reaching consequences. Following arguments in \cite{Tachikawa:2017gyf}, it was explicitly demonstrated in \cite{Bhardwaj:2022maz, Bartsch:2022ytj} that, for $k=0$, sequentially gauging these $\BZ_2$ symmetries generates a web of symmetries whose structure is controlled by the dihedral group $D_8$ of order 8. Specifically within this web, the $\mathrm{PSO}(2n) \cong \SO(2n)/\BZ_2$ gauge theory exhibits a finite non-Abelian $D_8$ zero-form global symmetry, while the $\Pin(2n)$ theory possesses non-invertible symmetries characterised by the representations of $D_8$. These arguments and structures generalise to other 3d theories with similar symmetries, such as variants of Aharony-Bergman-Jafferis (ABJ) theories \cite{Aharony:2008gk} involving $\so(2N)_{2k} \times \usp(2N)_{- k}$ gauge algebra, which generically possess $\CN=5$ supersymmetry, as explored in \cite{Mekareeya:2022spm, Bergman:2024its}.

In this paper, we investigate the structure of generalised global symmetries, including non-invertible symmetries, within 3d $\CN=4$ orthosymplectic quiver gauge theories.\footnote{See also \cite{Bhardwaj:2022dyt, Mekareeya:2022spm, Bhardwaj:2023zix, Nawata:2023rdx} for related work on 3d $\CN=4$ unitary quiver gauge theories.} Various aspects of these theories, notably the structure of their moduli spaces and techniques involving quiver manipulations, have been extensively studied in the literature (see, e.g., \cite{Hanany:2016gbz, Cabrera:2017njm, Cabrera:2017ucb, Cabrera:2018ldc, Hanany:2019tji, Bourget:2020xdz,Akhond:2021knl, Bourget:2021zyc, Bourget:2021xex, 
Sperling:2021fcf, Nawata:2021nse, Bennett:2024llh, Bennett:2025zor, Bennett:2025edk} for recent work). Using the superconformal index \cite{Bhattacharya:2008zy,Bhattacharya:2008bja, Kim:2009wb,Imamura:2011su, Kapustin:2011jm, Dimofte:2011py, Aharony:2013dha, Aharony:2013kma} as our primary analytic tool, we demonstrate the emergence of a $D_8$ categorical symmetry web in a class of 3d $\CN=4$ theories with $\so(2N) \times \usp(2N)$ gauge algebra (at zero Chern-Simons levels) coupled to $n$ bifundamental half-hypermultiplets, which share structural similarities with the ABJ-type models.

Furthermore, as a necessary tool for analysing the Coulomb branch moduli spaces of these orthosymplectic quiver theories, we revisit and refine the prescription for computing the associated Coulomb branch Hilbert series by focusing on $\SO(N)$ gauge theories with $N_f$ vector hypermultiplets, thereby extending the method presented in \cite{Cremonesi:2014uva}. Our refined prescription incorporates two crucial extensions: (1) the inclusion of the fugacity $\chi$ for the zero-form charge conjugation symmetry (in addition to the standard fugacity $\zeta$ for the zero-form magnetic symmetry), and (2) the proper treatment of background magnetic fluxes for the $\usp(2N_f)$ flavour symmetry. The refinement by both $\chi$ and $\zeta$ allows for the direct computation of Coulomb branch Hilbert series for theories with different global forms of the orthogonal gauge group, including $\O(N)^\pm$, $\Spin(N)$, and $\Pin(N)$. Simultaneously, the proper inclusion of background magnetic fluxes ensures agreement between the Hilbert series and the Coulomb branch limit of the superconformal index, and maintains consistency with known dualities. This accurate treatment of background fluxes is especially critical for the analysis of orthosymplectic quivers, where the original $\usp(2N_f)$ flavour symmetry is gauged, becoming an internal node of the quiver diagram. We demonstrate this refined prescription through several examples, verifying its consistency with superconformal index calculations and known dualities, including three-dimensional mirror symmetry \cite{Intriligator:1996ex}.

This paper is organised as follows.
Section \ref{sec:indexSONgaugetheory} reviews the computation of the superconformal index for 3d $\CN=4$ $\SO(N)$ gauge theories, adopting the notation and conventions of \cite{Aharony:2013kma}. We emphasise the subtle phase factor arising from the adjoint chiral field within the vector multiplet when the charge conjugation fugacity is $\chi=-1$. Section \ref{sec:dualities} discusses dualities between 3d $\CN=3$ and $\CN=4$ supersymmetric QCD theories with $\so(N_c)$-type gauge groups. These dualities, derived by combining results from \cite{Kapustin:2011gh} and \cite{Aharony:2013kma}, have been previously discussed in \cite[Section 4.1]{Mekareeya:2022spm} and utilised extensively in the study of generalised symmetries and anomalies in Chern-Simons-matter theories. Section \ref{sec:noninvertible} contains the analysis of variants of the theory with $\so(2N) \times \usp(2N)$ gauge algebra with $n$ bifundamental half-hypermultiplets that possess the $D_8$ categorical symmetry structure. Section \ref{sec:CBHS} presents our refined prescription for computing the Coulomb branch Hilbert series of $\SO(2N)$ and $\SO(2N+1)$ gauge theories with $N_f$ vector hypermultiplets. This prescription incorporates fugacities $\chi$ and $\zeta$ for charge conjugation and magnetic symmetries, respectively, as well as background magnetic fluxes for the $\usp(2N_f)$ flavour symmetry. We validate this prescription through numerous examples, checking consistency against corresponding index limits and the dualities discussed in Section \ref{sec:dualities}. We also analyse the mapping of these discrete symmetries under mirror symmetry. In Section \ref{sec:ccinlinearquivers}, we examine the actions of charge conjugation and magnetic symmetries associated with each orthogonal-type gauge group within the linear quiver theories $T[\SO(N)]$ and $T[\USp(2N)]$ \cite{Gaiotto:2008ak}, analysing how these actions map across mirror symmetry. Our results can be straightforwardly generalised to the $T_\rho[\SO(N)]$ and $T_\rho[\USp(2N)]$ theories. Explicit expressions for various superconformal indices used throughout the paper are collected in Appendix \ref{app:souspABJlike}. Appendix \ref{app:nilpotent} collects Coulomb branch Hilbert series for theories mirror dual to $T_{\rho}[\SO(2N+1)]$ and $T_{\rho}[\SO(2N)]$, computed for various partitions $\rho$ using the prescription developed in Section \ref{sec:CBHS}. Finally, Appendix \ref{app:affineD5} explores discrete gaugings of the affine $D_5$ quiver theory \eqref{affineD5}, and relates these results to the orthosymplectic theory \eqref{OSpSO10} discussed in Section \ref{sec:orthosymso10} and its mirror theory \eqref{USp2w5flv} described by the $\USp(2)$ gauge theory with five flavours.

\section{Index of the 3d $\CN=4$ $\SO(N)$ gauge theory} \label{sec:indexSONgaugetheory}
Let us consider the 3d $\CN=4$ $\SO(N)$ gauge theory with $N_f$ hypermultiplets in the vector representation, whose superconformal index can be computed as described in \cite{Aharony:2013kma}. We introduce the following fugacities and background magnetic fluxes associated with various global symmetries: the fugacity $a$ for the $\U(1)$ axial symmetry; the fugacities $f_i$ and the magnetic fluxes $n_i$ (with $i=1, \ldots, N_f$) for the $\usp(2N_f)$ flavour symmetry; the fugacity $\zeta=\pm 1$ for the $\BZ_{2, \, \CM}^{[0]}$ zero-form magnetic symmetry; and the fugacity $\chi=\pm 1$ for the $\BZ_{2, \, \CC}^{[0]}$ zero-form charge conjugation symmetry. In the following, we write $N = 2 r + \epsilon$, where $\epsilon \in \{0, 1\}$. The index for the case $\chi = +1$ is given by
\bes{ \label{indexSONwithNfchi1}
&\CI_{\text{$\SO(N)$, $N_f$}}(x;a; \vec f; \vec n; \zeta; \chi = +1)  \\
& = \frac{1}{r! 2^{r-1+\epsilon}}\sum_{(m_1, \ldots, m_r) \in \BZ^r} \zeta^{\sum_i m_i} \oint \left( \prod_{\ell=1}^r \frac{d z_\ell}{2\pi i z_\ell} \right) \CZ_{\text{vec}}^{\SO(N)}(x; \vec z; \vec m; \chi = +1)  \\
& \qquad \qquad \times \CZ^{\SO(N)}_{\mathcal{A}}(x; a; \vec z; \vec m; \chi=+1) \times \prod_{i=1}^{N_f} \prod_{s_f=\pm 1} \Bigg\{ \left[\CZ_{\text{chir}}^{1/2} (x;  a f_i^{s_f} ; s_f n_i)\right]^\epsilon  \\
& \qquad \qquad \times \prod_{\ell=1}^r \prod_{s =\pm 1} \CZ_{\text{chir}}^{1/2} (x; a z_\ell^{s} f_i^{s_f};s m_\ell +s_f n_i) \Bigg\}~,
}
where the contribution of a chiral multiplet with $R$-charge $R$ is
\bes{
\CZ_{\text{chir}}^{R} (x; z; m ) = (x^{1-R} z^{-1})^{|m|/2} \prod_{j=0}^\infty \frac{1-(-1)^m z^{-1} x^{|m|+2-R+2j}}{1-(-1)^m z x^{|m|+R+2j}}~,
}
the vector multiplet contribution is
\bes{
\scalebox{0.99}{$
\begin{split}
&\CZ_{\text{vec}}^{\SO(N)}(x; \vec z; \vec m; \chi = +1) = \Bigg[ \prod_{\ell=1}^r \prod_{s=\pm 1} x^{-| m_\ell|/2} (1-(-1)^{m_\ell} z^s_\ell x^{|m_\ell|} ) \Bigg ]^\epsilon \\ & \qquad \times \prod_{1 \leq a< b\leq r} \,\, \prod_{s_1, s_2 = \pm 1} x^{-|s_1 m_a +s_2 m_b|/2} \Big(1-(-1)^{|s_1 m_a +s_2 m_b|} z_a^{s_1} z_b^{s_2} x^{|s_1 m_a +s_2 m_b|} \Big)~,
\end{split}
$}
}
and the contribution of the adjoint chiral field in the $\SO(N)$ vector multiplet is
\bes{
&\CZ^{\SO(N)}_{\mathcal{A}}(x; a; \vec z; \vec m; \chi=+1) = \Bigg[ \prod_{\ell =1}^r \prod_{s = \pm 1} \CZ_{\text{chir}}^{R=1} (x; a^{-2} z_\ell^{s}; s m_\ell )  \Bigg]^\epsilon \\
& \qquad \times \Big[\CZ_{\text{chir}}^{R=1} (x; a^{-2}; 0)\Big]^r \times \prod_{1\leq i<j \leq r} \, \prod_{s_1, s_2 = \pm 1} \CZ_{\text{chir}}^{R=1} (x; a^{-2} z_i^{s_1} z_j^{s_2}; s_1 m_i +s_2 m_j ) ~.
}

For $\chi=-1$, the cases of odd and even $N$ must be treated separately. For $N=2r+1$, the $\O(2r+1)$ holonomy can be put in the form $(z_1, z_1^{-1}, \ldots, z_r, z_r^{-1},\chi= -1)$. The vector multiplet contribution is therefore:
\bes{ 
\scalebox{0.99}{$
\begin{split}
&\CZ_{\text{vec}}^{\SO(2r+1)}(x; \vec z; \vec m; \chi = -1) = \prod_{\ell=1}^r \prod_{s=\pm 1} x^{-| m_\ell|/2} (1-(-1)^{m_\ell} \chi z^s_\ell x^{|m_\ell|} ) \\ & \qquad \times \prod_{1 \leq a< b\leq r} \,\, \prod_{s_1, s_2 = \pm 1} x^{-|s_1 m_a +s_2 m_b|/2} \Big(1-(-1)^{|s_1 m_a +s_2 m_b|} z_a^{s_1} z_b^{s_2} x^{|s_1 m_a +s_2 m_b|} \Big)~.
\end{split}
$}
}
However, the contribution from the adjoint chiral field within the $\SO(N)$ vector multiplet acquires an additional phase factor (indicated in {\purple purple}), the origin of which will be discussed later:
\bes{ \label{adjchi-1odd}
&\CZ_{\CA}^{\SO(2r+1)}(x; \vec z; \vec m; \chi = -1) =  {\purple (-1)^{\sum_{j=1}^{r} m_j} } \times \prod_{\ell =1}^r \prod_{s = \pm 1} \CZ_{\text{chir}}^{R=1} (x; a^{-2} \chi z_\ell^{s}; s m_\ell ) \\
& \qquad \times \Big[\CZ_{\text{chir}}^{R=1} (x; a^{-2}; 0)\Big]^r \times \prod_{1\leq i<j \leq r} \, \prod_{s_1, s_2 = \pm 1} \CZ_{\text{chir}}^{R=1} (x; a^{-2} z_i^{s_1} z_j^{s_2}; s_1 m_i +s_2 m_j )~.
}
In summary, the index for the $\SO(2r+1)$ gauge theory with $\chi=-1$ is
\bes{ \label{indexSO2r+1withNfchi-1}
&\CI_{\text{$\SO(2r+1)$, $N_f$}}(x;a; \vec f; \vec n; \zeta; \chi = -1)  \\
& = \frac{1}{r! 2^{r-1+\epsilon}}\sum_{(m_1, \ldots, m_r) \in \BZ^r} \zeta^{\sum_i m_i} \oint \left( \prod_{\ell=1}^r \frac{d z_\ell}{2\pi i z_\ell} \right) \CZ_{\text{vec}}^{\SO(2r+1)}(x; \vec z; \vec m; \chi = -1)  \\
& \qquad \qquad \times \CZ^{\SO(2r+1)}_{\mathcal{A}}(x; a; \vec z; \vec m; \chi=-1) \times \prod_{i=1}^{N_f} \prod_{s_f=\pm 1} \Bigg[ \CZ_{\text{chir}}^{1/2} (x; a \chi f_i^{s_f} ; s_f n_i) \\
& \qquad \qquad \times \prod_{\ell=1}^r \prod_{s =\pm 1} \CZ_{\text{chir}}^{1/2} (x; a z_\ell^{s} f_i^{s_f};s m_\ell +s_f n_i) \Bigg]~.
}
For $N=2r$, the holonomy of $\O(2r)$ can be put in the form $(z_1, z_1^{-1}, \ldots, z_{r-1}, z_{r-1}^{-1},1,-1)$. Therefore, the vector multiplet contributes as 
\bes{ \label{vecchi-1}
&\CZ_{\text{vec}}^{\SO(2r)}(x; \vec z; \vec m; \chi = -1) = \Big[ \CZ_{\text{vec}}^{\SO(2r)}(x; \vec z; \vec m; \chi = +1) \Big]_{z_r = 1,\, z_r^{-1}=-1, \, m_r =0} \\
& = \prod_{1 \leq a< b\leq r-1} \,\, \prod_{s_1, s_2 = \pm 1} x^{-|s_1 m_a +s_2 m_b|/2}  \Big(1-(-1)^{|s_1 m_a +s_2 m_b|} z_a^{s_1} z_b^{s_2} x^{|s_1 m_a +s_2 m_b|} \Big) \\ & \quad \,\,\, \times \prod_{\ell=1}^{r-1} x^{-|2 m_\ell|} (1- (-1)^{2 m_\ell} z_\ell^{2} x^{|2 m_\ell|}) (1- (-1)^{2 m_\ell} z_l^{-2} x^{|2 m_\ell|})~.
}
The contribution of the adjoint chiral field in the $\SO(2r)$ vector multiplet also picks up an extra phase, highlighted in {\purple purple}: 
\bes{ \label{adjchi-1}
&\CZ_{\CA}^{\SO(2r)}(x; \vec z; \vec m; \chi = -1) \\&= \Big[\CZ_{\text{chir}}^{R=1} (x; a^{-2}; 0)\Big]^{r-1} \CZ_{\text{chir}}^{R=1} (x; -a^{-2}; 0) \times \Bigg[{\purple (-1)^{\sum_{j=1}^{r} m_j} } \\ & \qquad \times \prod_{1\leq i<j \leq r} \, \prod_{s_1, s_2 = \pm 1} \CZ_{\text{chir}}^{R=1} (x; a^{-2} z_i^{s_1} z_j^{s_2}; s_1 m_i +s_2 m_j ) \Bigg]_{z_r = 1,\, z_r^{-1}=-1,\, m_r =0}~.
}
In summary, the index for the $\SO(2r)$ gauge theory with $\chi=-1$ is
\bes{ \label{indexSO2rwithNfchi-1}
\scalebox{0.94}{$
\begin{split}
&\CI_{\text{$\SO(2r)$, $N_f$}}(x;a; \vec f; \vec n; \zeta; \chi = -1)  \\
& = \frac{1}{(r-1)! 2^{r-1}}\sum_{(m_1, \ldots, m_{r-1}) \in \BZ^{r-1}} \zeta^{\sum_i m_i} \oint \left( \prod_{\ell=1}^{r-1} \frac{d z_\ell}{2\pi i z_\ell} \right) \CZ_{\text{vec}}^{\SO(2r)}(x; \vec z; \vec m; \chi = -1)  \\
& \quad \times \CZ^{\SO(2r)}_{\mathcal{A}}(x; a; \vec z; \vec m; \chi=-1)  \times \prod_{i=1}^{N_f} \prod_{s_f=\pm 1}\Bigg[ \prod_{\ell=1}^{r-1} \prod_{s =\pm 1} \CZ_{\text{chir}}^{1/2} (x; a z_\ell^{s} f_i^{s_f};s m_\ell +s_f n_i) \\
& \quad \times \CZ_{\text{chir}}^{1/2} (x; a f_i^{s_f}; s_f n_i) \,\, \CZ_{\text{chir}}^{1/2} (x; a \chi f_i^{s_f}; s_f n_i) \Bigg]~.
\end{split}
$}
}
In any case, the index of the $\SO(N)$ gauge theory with $N_f$ flavours, refined with respect to both $\zeta$ and $\chi$, can be obtained as
\bes{
\CI_{\SO(2N), N_f}(x;a;\vec f; \vec n; \zeta; \chi)=\frac{1}{2} \sum_{s=\pm1} \CI_{\SO(2N), N_f}(x;a;\vec f; \vec n; \zeta; \chi=s) (1+s \chi)~.
}

Crucially, the contribution from the adjoint chiral field includes the phase factor ${\purple (-1)^{\sum_{j} m_j}}$ when $\chi = -1$.  This phase can be interpreted as the contribution of a mixed Chern-Simons term between the background gauge field for charge conjugation symmetry and the dynamical gauge field (see \cite[Page 39]{Aharony:2013kma}).  Physically, this phase ensures that the minimal monopole operator of $\SO(N)$ with flux $(1,0,\ldots,0)$ is even under charge conjugation, making it also the minimal monopole operator in the $\O(N)^+$ gauge theory (see also \cite[Section 2.4]{Aharony:2013kma}). This phase has been implicitly included in the literature when calculating indices for theories with orthogonal gauge groups at zero Chern-Simons level (e.g., \cite{Kim:2013cma, Hayashi:2022ldo}). This will also be used throughout this article. As an example, let us consider $N=4$ ($r=2, \epsilon=0$) and $(m_1, m_2) = (1,0)$.  In this case:
\bes{ \label{exampleN4flux10}
&\CZ^{\SO(4)}_{\CA}(x; a ; z_1,z_2;m_1=1,m_2=0; \chi=+1) = a^4\Big\{1 +2 \left(a^{-2}-a^2\right)x \\ & \qquad \qquad +\left[ 3a^{-4} + \left(\mathbf{[2,0]}+{\bf [0,2]} -2\right)\left(-a^{-2}+a^2\right) +a^4-4 \right]  x^2 +\ldots \Big\}~, \\ &
\CZ^{\SO(4)}_{\CA}(x; a ; z_1 ; m_1= 1; \chi=-1) = a^4\Big[1+ 0x+ \left(a^{-4}-a^4\right)x^2 + \ldots \Big]~,
}
where ${\bf [a,b]}$ denotes the character $\chi^{\so(4)}_{[a,b]} (z_1,z_2)$.  If the phase in \eqref{adjchi-1} \emph{were} omitted, the $\chi=-1$ sector contribution \emph{would} be:
\bes{
a^4\Big[-1+ 0x+ \left(-a^{-4}+a^4\right)x^2 + \ldots \Big]~,
}
where the term $-1$ indicates that the minimal monopole operator would be odd under charge conjugation, contradicting the expected behaviour.  

Furthermore, let $C_\pm(a^{- 2p} x^p)$ be the coefficients of the terms $a^{-2p} x^p$ in the $\chi=\pm1$ contribution given by \eqref{exampleN4flux10}, where the overall prefactor $a^4$ is removed. Then, $\sum_{p=0}^\infty C_+(a^{-2p} x^p) t^{2p} = \PE \left[2 t^2\right]$, which is the dressing factor in the Coulomb branch Hilbert series of $\SO(4)$ with flux $(1,0)$, as given in the third line of \cite[Table 6]{Cremonesi:2014uva} (see also \cite[Appendix A]{Cremonesi:2013lqa} for the definition of the dressing factor). If we gauge the charge conjugation symmetry by summing the contributions for $\chi = \pm 1$ from \eqref{exampleN4flux10} and dividing by two, then $\frac{1}{2} \sum_{p=0}^\infty \left[C_+(a^{-2p} x^p)+C_-(a^{-2p} x^p)\right] t^{2p} = \text{PE}\left[t^2 +t^4\right]$. This corresponds to the dressing factor for the $\O(4)$ gauge group given in the third line of \cite[Table 7]{Cremonesi:2014uva}, and it is also equal to the dressing factor for the $\SO(5)$ gauge group given in \cite[(A.8)--(A.9)]{Cremonesi:2013lqa}.  The presence of the phase in \eqref{adjchi-1} is essential to ensure that the Coulomb branch limit of the index\footnote{As pointed out in \cite{Razamat:2014pta}, the Coulomb branch limit of the index can be obtained as $\sum_{p=0}^\infty C(a^{-2p} x^p) t^{2p}$, whereas the Higgs branch limit of the index is given by $\sum_{p=0}^\infty C(a^{2p} x^p) t^{2p}$, where $C(a^{\pm2p} x^p)$ denote the coefficients of the terms $a^{\pm 2p} x^p$ in the series expansion of the index.} agrees with the Coulomb branch Hilbert series for theories with orthogonal gauge groups \cite[Appendix A]{Cremonesi:2014uva}.

 We can now mention how to adapt these results for the 3d $\CN=3$ $\SO(N)_k$ gauge theory with Chern-Simons level $k$ and $N_f$ flavours. The index is obtained from \eqref{indexSONwithNfchi1}, \eqref{indexSO2r+1withNfchi-1}, and \eqref{indexSO2rwithNfchi-1} via simple modifications. Specifically, the factor $\CZ^{\SO(N)}_{\CA}$ is omitted, as the adjoint chiral field in the vector multiplet becomes massive for $k \neq 0$. Additionally, we make the following replacement in the aforementioned expression of the index: $\left( \prod_{\ell} \frac{d z_\ell}{2\pi i z_\ell} \right) \,\, \rightarrow \,\, \left( \prod_{\ell} \frac{d z_\ell}{2\pi i z_\ell} z_\ell^{k m_\ell} \right)$.

Finally, we note that the indices for various variants of the orthogonal gauge group can be computed as follows \cite[(6.13)]{Aharony:2013kma}:
\bes{ \label{indvariants}
\scalebox{1}{$
\begin{split}
\CI_{\O(N)^+}(\zeta) &=\frac{1}{2} \left[ \CI_{\SO(N)}(\zeta; \chi=+1)+\CI_{\SO(N)}(\zeta; \chi=-1) \right]~, \\
\CI_{\Spin(N)}(\chi) &=\frac{1}{2} \left[ \CI_{\SO(N)}(\zeta=+1; \chi)+\CI_{\SO(N)}(\zeta=-1; \chi) \right]~, \\
\CI_{\O(N)^-}(\zeta) &=\frac{1}{2} \left[ \CI_{\SO(N)}(\zeta; \chi=1)+\CI_{\SO(N)}(-\zeta; \chi=-1) \right]~, \\
\CI_{\Pin(N)} &=\frac{1}{2} \left[ \CI_{\Spin(N)}(\chi=+1)+\CI_{\Spin(N)}(\chi=-1) \right]~,
\end{split}$}}
where, for conciseness, we omit the dependence on $N_f$, $x$, $a$, $\vec f$ and $\vec n$ in the above expressions. 

\section{Dualities in 3d $\CN=3$ and $\CN=4$ $\so(N)$ SQCD} \label{sec:dualities}
The following duality between 3d $\mathcal{N}=2$ gauge theories was established in \cite[Section 5.3]{Aharony:2013kma}: (a) the $\SO(n_c)_k$ (or $\O(n_c)^+_k$ and $\Spin(n_c)_k$) gauge theory with $n_f$ flavors of chiral multiplets $Q$ in the vector representation and zero superpotential, and (b) the $\SO(n_c')_{-k}$ (or $\O(n_c')^+_{-k}$ and $\O(n_c')^-_{-k}$) gauge theory, where $n'_c=  n_f+|k|-N_c+2$, with $n_f$ flavors of chiral multiplets $q$ in the vector representation, a collection of $n_f(n_f+1)/2$ gauge singlets $M$, and the superpotential $W=M q q $.  By employing the same argument as in \cite{Kapustin:2011gh}, we can obtain the following duality between 3d $\mathcal{N}=3$ gauge theories:
\bes{ \label{dualityN3}
\text{(1)} \,\ & \text{the $\SO(N)_k$ (or $\O(N)^+_k$ and $\Spin(N)_k$) gauge theory} \\ 
& \text{with $N_f$ hypermultiplets in the vector representation, and}\\
\text{(2)} \,\ & \text{the $\SO(N')_{-k}$ (or $\O(N')^+_{-k}$ and $\O(N'_c)^{-}_{-k}$) gauge theory,}  \\
& \text{where $N'= 2N_f +|k| -N +2$, with $N_f$ hypermultiplets} \\
& \text{in the vector representation.}
}
We remark that this duality was used extensively in \cite{Mekareeya:2022spm} to study various mixed 't Hooft anomalies in 3d $\mathcal{N}=3$ Chern-Simons-matter theories.

As discussed in \cite[Section 2]{Kapustin:2011gh}, the $\CN=3$ duality \eref{dualityN3} can be derived from the $\CN=2$ duality by identifying $n_c = N$, $n_f = 2N_f$ and perturbing the theory (a) in the latter with the quartic superpotential
\bes{ \label{perturba}
W_{\text{(a)}} = - \frac{2 \pi}{k} \tr(Q^t J Q)^2~,
}
where $J$ is the $2N_f \times 2N_f$ symplectic matrix, and by identifying the meson $QQ^t$ in theory (a) with the singlets $M$ in theory (b) as follows:
\bes{ \label{mapmesonstoM}
QQ^t = \lambda M~, \quad \text{with $\lambda = \pm \frac{k}{2\pi \sqrt{2}}$}~.
}
The perturbation \eqref{perturba} ensures that the theory flows to an $\mathcal{N}=3$ fixed point, and the specific value of $\lambda$ stated in \eqref{mapmesonstoM} ensures that the quartic superpotential in the dual theory (b) is
\bes{
W_{\text{(b)}} &= \sqrt{2} q M q^t -\frac{2\pi}{k} \lambda^2 \tr(J M)^2 = \sqrt{2} q M q^t -\frac{k}{4\pi} \tr(J M)^2 \\
&\xrightarrow{\text{integrate out $M$}} \quad \frac{2\pi}{k} \tr(qJq^t)^2,
}
which is consistent with the duality.

Although the above argument is presented for the case of $k \neq 0$, we note that, as can be verified explicitly using the index, the duality \eqref{dualityN3} also holds for $k=0$, where both theories (1) and (2) have $\mathcal{N}=4$ supersymmetry.  In this case, these theories are ``good" in the sense of Gaiotto-Witten \cite{Gaiotto:2008ak} if and only if $N_f \geq N-1$ and $N_f \geq N'-1 = 2N_f-N+1$, which is equivalent to $N_f =N-1$ and $N = N'$.  This implies that the gauge groups of theories (1) and (2) are both balanced.  In particular, this result implies the following $\mathcal{N}=4$ duality:
\bes{ \label{dualityN4}
\text{(I)} \,\ & \text{the $\SO(N)$ (or $\O(N)^+$ and $\Spin(N)$) gauge theory with} \\ 
& \text{$N_f= N-1$ hypermultiplets in the vector representation, and}\\
\text{(II)} \,\ & \text{the $\SO(N)$ (or $\O(N)^+$ and $\O(N)^-$) gauge theory with }  \\
& \text{$N_f=N-1$ hypermultiplets in the vector representation.}
}
Note that, for the $\SO(N)$ gauge group, even though theories (I) and (II) seem to be the same, the $\BZ_2$ charge conjugation symmetry of theory (II) is mapped under the duality to the diagonal subgroup of the $\BZ_2$ charge conjugation symmetry and the $\BZ_2$ magnetic symmetry of theory (I); see \eref{matchindexSOKapustin} below. 

The indices of theories (1) and (2) in \eqref{dualityN3} can be matched in a manner similar to that of \cite[(6.12)]{Aharony:2013kma} as follows:
\bes{
\CI_{\text{$\SO(N)_k$, $N_f$}}(x;a; \vec f; \vec n; \zeta; \chi)  = \left( \prod_{i=1}^{N_f}  {f_i}^{k n_i}   \right) \CI_{\text{$\SO(N')_{-k}$, $N_f$}}(x;a; \vec f; \vec n; \zeta; \zeta \chi)~.
}
We emphasise that the fugacity of the charge conjugation symmetry of the dual theory is identified with the product of the fugacities of the charge conjugation symmetry and magnetic symmetry of the original theory. Specialising this to the case of $k=0$ as discussed above, we have
\bes{ \label{matchindexSOKapustin}
\CI_{\text{$\SO(N)$, $N_f=N-1$}}(x;a; \vec f; \vec n; \zeta; \chi)  = \CI_{\text{$\SO(N)$, $N_f=N-1$}}(x;a; \vec f; \vec n; \zeta; \zeta \chi)~.
}
Summing over $\zeta=\pm 1$ on each side of this equation and dividing the corresponding result by two, we obtain the index for the $\Spin(N)$ gauge theory with $N_f=N-1$ on the left-hand side, and that for the $\O(N)^-$ gauge theory with $N_f = N-1$ on the right-hand side. This confirms the duality \eref{dualityN4}.

As an example, let us examine the $\SO(4)$ gauge theory with $N_f=3$ flavours. The index up to order $x^2$ reads
\bes{
&\CI_{\text{$\SO(4)$, $N_f=3$}}(x;a; \vec f; n_1 = n_2 = 0; \zeta; \chi)\\& =1+ \left( \zeta a^{-2} + \mathbf{21} a^2\right) x + \Big[ \left(2+\zeta+ \chi+ \zeta \chi \right) a^{-4} \\ & \qquad +\left( \mathbf{21} \zeta + \mathbf{14} \chi + \mathbf{14} \zeta  \chi \right) + {\bf 216} a^4 - {\blue \left(\mathbf{21} + \zeta + \mathbf{1}\right)}  \Big] x^2 + \ldots~,
}
where the {\blue blue} terms denote the contributions of the $\usp(6)  \oplus \u(1)_\CM \oplus \u(1)_A$ conserved currents, where the $\BZ^{[0]}_{2,\, \CM}$ magnetic symmetry gets enhanced to $\u(1)_\CM$ in the infrared because the $\SO(4)$ gauge group is balanced. The $\usp(6)$ representations that appear above are as follows:
\bes{
\mathbf{1}=[0,0,0]~, \quad \mathbf{14} = [0,1,0]~, \quad \mathbf{21} = [2,0,0]~, \quad \mathbf{216} = [2,0,1]~.
}
Let $C(a^{\pm 2p} x^p)$ be the coefficients of the terms $a^{\pm 2p} x^p$ in the above index. The Higgs branch limit $\sum_{p=0}^\infty C(a^{2p} x^p) t^{2p}$ gives the Hilbert series of the closure of the orbit $[2^3]$ of $\usp(6)$ (see \eg~ \cite[Table 12, Page 37]{Hanany:2016gbz}), where upon setting $f_1=f_2=1$ gives
\bes{
\left[\sum_{p=0}^\infty C(a^{2p} x^p) t^{2p}\right]_{f_1=f_2=1} = \frac{(1 + 7 t^2 + 15 t^4 + 7 t^6 + t^8)(1+t^2)^2}{(1-t^2)^{12}}~.
}
The Coulomb branch limit $\sum_{p=0}^\infty C(a^{-2p} x^p) t^{2p}$, on the other hand, admits a simple closed form, since the Coulomb branch is a complete intersection \cite[Section 5.4]{Cremonesi:2013lqa}:
\bes{ \label{CBSO4w3flv}
\sum_{p=0}^\infty C(a^{-2p} x^p) t^{2p} = \PE \left[\zeta t^2 +(1 +\zeta +\chi +\zeta  \chi)t^4 +\zeta t^6 -t^8-t^{12}  \right]~.
}
The Coulomb branch generators are listed explicitly in \cite[(5.23)]{Cremonesi:2013lqa}. We now discuss these generators and identify their contributions within the refined Coulomb branch Hilbert series \eref{CBSO4w3flv}. The two Casimir operators of $\SO(4)$, built from the adjoint scalar $\varphi$ in the vector multiplet, are the Pfaffian $\mathrm{Pf}(\varphi) \sim \epsilon^{i_1 i_2 i_3 i_4} \varphi_{i_1 i_2} \varphi_{i_3 i_4}$ and the trace-squared $\mathrm{tr}(\varphi^2) \sim \delta^{i_2 i_3} \delta^{i_1 i_4} \varphi_{i_1 i_2} \varphi_{i_3 i_4}$. These correspond to the terms $\chi t^4$ and $t^4$ respectively in the plethystic exponent of \eref{CBSO4w3flv}. The Pfaffian operator $\mathrm{Pf}(\varphi)$, involving the epsilon tensor, is odd under charge conjugation and therefore carries the fugacity $\chi$. The remaining generators involve the minimal monopole operator $X_{(1,0)}$, which carries magnetic flux $(1,0)$, dressed with appropriate scalar fields. Recall that, in the presence of the monopole $X_{(1,0)}$, the gauge symmetry is broken to a residual $\U(1) \times \SO(2)$. Let $\varphi_1$ and $\varphi_2$ denote the scalar components associated with the residual $\U(1)$ and $\SO(2)$ factors, respectively. Their contributions to the Hilbert series \eref{CBSO4w3flv} are identified as follows: the bare monopole $X_{(1,0)}$ contributes $\zeta t^2$; the dressed operators $X_{(1,0)} \varphi_1$, $X_{(1,0)} \epsilon^{i_1 i_2} (\varphi_2)_{i_1 i_2}$, and $X_{(1,0)} \varphi_2^2$ contribute $\zeta t^4$, $\zeta \chi t^4$, and $\zeta t^6$, respectively. These operators account for all six generators contributing to the Coulomb branch Hilbert series at these orders. Under the duality map \eref{matchindexSOKapustin}, which transforms the fugacities as $(\zeta, \chi) \rightarrow (\zeta, \zeta \chi)$, the operator $\mathrm{Pf}(\varphi)$ (contributing $\chi t^4$) is exchanged with the dressed monopole $X_{(1,0)} \epsilon^{i_1 i_2} (\varphi_2)_{i_1 i_2}$ (contributing $\zeta \chi t^4$). We reiterate that the phase factor discussed in \eqref{adjchi-1odd} and \eqref{adjchi-1} is vital for this consistency. Without this phase, the minimal monopole operator $X_{(1,0)}$ would incorrectly appear to be odd under charge conjugation, contradicting the operator mapping required by the duality \eref{dualityN4}.

It is also instructive to consider the effect of turning on a background magnetic flux for the flavour symmetry, for example, $\vec n = (1,0,0)$. In this case, the index reads
\bes{\label{indSO4wNf3flux100}
&\CI_{\text{$\SO(4)$, $N_f=3$}}(x; a;  f_1 = f_2 =1; n_1 = 1, n_2=n_3= 0; \zeta; \chi) \\
&= \left(\chi+ \zeta \chi\right) a^{-4} x^2 +\Big[\left(\zeta +\chi+\zeta  \chi\right)a^{-6}+\left(5 \zeta +10 \chi+10 \zeta  \chi \right)a^{-2} \Big] x^3 + \ldots~.
}
The first term $\chi a^{-4} x^2$ corresponds to the monopole operator $V_{(0,0;1,0,0)}$ with $\SO(4)$ gauge magnetic flux $\vec m =(0,0)$ and $\usp(6)$ background flavour magnetic flux $\vec n = (1,0,0)$. Similarly, the term $\zeta \chi a^{-4} x^2$ corresponds to the monopole operator $V_{(1,0;1,0,0)}$, carrying gauge flux $\vec m = (1,0)$ and flavour flux $\vec n = (1,0,0)$.\footnote{Both $V_{(0,0;1,0,0)}$ and $V_{(1,0;1,0,0)}$ have $R$-charge 2, consistent with the $x^2$ scaling.} We observe explicitly that these two leading-order monopole operators, distinguished by their gauge flux and charge conjugation properties (encoded in the $\chi$ and $\zeta\chi$ factors), are exchanged under the duality \eqref{dualityN4}, consistently with its action on the fugacities $(\zeta, \chi) \rightarrow (\zeta, \zeta \chi)$.

\section{Non-invertible symmetry in 3d $\CN=4$ gauge theories} \label{sec:noninvertible}
Recent studies \cite{Mekareeya:2022spm, Bergman:2024its} have shown that certain variants of the ABJ theory, specifically those described by the $\so(2N)_{2k} \times \usp(2N)_{-k}$ gauge theory with two bifundamental half-hypermultiplets, possess non-invertible symmetries. In particular, the authors of \cite{Bergman:2024its} demonstrated that the $\left[\SO(2N)_{2k} \times \USp(2N)_{-k} \right] /\mathbb{Z}_2$ theory has a dihedral group of order $8$, \ie $D_8$, global symmetry, and that the $D_8$ symmetry web \cite{Bhardwaj:2022maz, Bartsch:2022ytj} can be reproduced by sequentially gauging various $\BZ_2$ global symmetries of such a theory.  

In this section, we generalise this result to a theory with $\CN=4$ supersymmetry, namely the $\so(2N) \times \usp(2N)$ gauge theory, with zero Chern-Simons level and $n$  bifundamental half-hypermultiplets. For the theory to be ``good'' in the sense of \cite{Gaiotto:2008ak}, the $\so(2N)$ gauge group must effectively couple to at least $2N-1$ vector flavours, and the $\usp(2N)$ gauge group to at least $2N+1$ fundamental flavours. Since the $n$ bifundamental half-hypermultiplets provide flavours for both factors, we will take $n \geq 3$ in the following analysis.\footnote{For $n=2$, the monopole operator that carries the minimal magnetic fluxes under both gauge factors has zero $R$-charge, rendering the theory ``bad''.}

The continuous flavour symmetry of such a theory is $\so(n)$, which rotates the $n$ bifundamental half-hypermultiplets. As discussed in \cite{Mekareeya:2022spm, Bergman:2024its} (see also \cite{Tachikawa:2019dvq,Bergman:2020ifi}), the $\SO(2N) \times \USp(2N)$ gauge theory possesses discrete zero-form symmetries originating from the $\SO(2N)$ factor: a magnetic symmetry $\BZ^{[0]}_{2, \, \CM}$ and a charge conjugation symmetry $\BZ^{[0]}_{2,\, \CC}$. Additionally, there is a $\BZ_2^{[1]}$ one-form symmetry, identified as the diagonal subgroup of the $\BZ_2 \times \BZ_2$ centre symmetry corresponding to the $\SO(2N)$ and $\USp(2N)$ gauge factors. The reduction from $\BZ_2 \times \BZ_2$ to a single diagonal $\BZ_2$ one-form symmetry is due to screening by the bifundamental fields. As pointed out in \cite{Cordova:2017vab}, the 't Hooft anomaly involving these finite symmetries is captured by the following anomaly theory supported on a spin four-manifold $M_4$:
\bes{ \label{anomABJlike}
i \pi \int_{M_4} \left( N A_2^B \cup A_1^\CM \cup A_1^\CM  + A_2^B \cup A_1^\CM \cup A_1^\CC \right)~,
}
where $A_2^B$ is the two-form background field for the $\BZ_2^{[1]}$ one-form symmetry, and $A_1^\CM$ and $A_1^\CC$ are the one-form background fields for the $\BZ^{[0]}_{2, \, \CM}$ and $\BZ^{[0]}_{2, \, \CC}$ symmetries respectively. Note that the $\BZ_2^{[1]}$ one-form symmetry itself is free of 't Hooft anomalies \cite{Cordova:2017vab} and can therefore be gauged. Gauging this $\BZ_2^{[1]}$ one-form symmetry leads to the $\left[\SO(2N) \times \USp(2N)\right] /\BZ_2$ theory. The manifest global symmetry of the gauged theory is $\so(n) \times \BZ_{2, \, B}^{[0]} \times \BZ_{2, \, \CM}^{[0]} \times \BZ_{2, \, \CC}^{[0]}$, where $\BZ_{2, \, B}^{[0]}$ is the zero-form symmetry dual to the $\BZ_2^{[1]}$ one-form symmetry of the parent theory. Based on the arguments in \cite{Bergman:2024its}, it is expected that the Abelian group of finite zero-form symmetries $\BZ_{2, \, B}^{[0]} \times \BZ_{2, \, \CM}^{[0]} \times \BZ_{2, \, \CC}^{[0]}$ enhances to the non-Abelian dihedral group $D_8$. We will show this using the index.

\subsubsection*{The case of $N=2$ and $n=3$}
For convenience in studying such a theory with the index, we consider the case of $N=2$ and $n=3$. For such values of $N$ and $n$, the first term of \eref{anomABJlike} vanishes modulo $2 \pi i$, leaving only the mixed anomaly $i \pi \int_{M_4} A_2^B \cup A_1^\CM \cup A_1^\CC$. We thus consider the $\left[\SO(4) \times \USp(4)\right]/\BZ_2$ gauge theory. Its index, refined by the fugacities for the manifest global symmetry $\SO(3) \times \BZ^{[0]}_{2,\, B} \times \BZ^{[0]}_{2,\, \CM} \times \BZ^{[0]}_{2,\, \CC}$, is detailed in Appendix \ref{app:souspABJlike}. The explicit result is given by
\bes{ \label{indD2C2modZ2}
&\CI\left\{\left[\SO(4) \times \USp(4)\right]/\BZ_2\right\} \\
&= 1+ {\bf [2]} x + \Bigg\{ \frac{1}{2} \Big[4 \left(1+\chi\right) + g \left(1+ \zeta + \chi + \zeta \chi\right)\Big] a^{-4} +{\bf [4]} \left(1 + \chi\right)\\ & \quad \,\,\,\ + \Big[ \left( {\bf [8]}+2{\bf [4]}+2\right) + \left({\bf [8]+[4]}+1\right)\chi \Big] a^4 -{\blue \left({\bf [2]}+1\right)} \Bigg\} x^2 + \ldots~,
}
where $a$ is the fugacity for the $\U(1)_A$ axial symmetry; $g$, $\zeta$, and $\chi$ are the fugacities for $\BZ^{[0]}_{2,\, B}$, $\BZ^{[0]}_{2,\, \CM}$, and $\BZ^{[0]}_{2,\, \CC}$ respectively; and ${\bf [x]}$ denotes the character of the irreducible representation $[x]$ of $\SO(3)$. The terms highlighted in {\blue blue} denote the contribution of the $\U(1)_A \times \SO(3)$ conserved currents. For reference, specialising to unit $\SO(3)$ fugacity and expanding up to order $x^3$, the index takes the form
\bes{ \label{fixformindex}
&1+ 3 a^2 x + \Big[C(a^{-4} x^2) a^{-4} + C(a^{0} x^2)   +  C(a^{4} x^2) a^{4} \Big] x^2 \\
& \,\ + \Big[ C(a^{-6} x^3) a^{-6}  +  C(a^{-2} x^3) a^{-2} +  C(a^{2} x^3) a^{2} +  C(a^{6} x^3) a^{6} \Big] x^3 \\
& \,\ +\Big[ C(a^{-8} x^4) a^{-8} + \ldots \Big] x^4 + \ldots~,
}
where 
\bes{ \label{indSO4USp4modZ2}
&\scalebox{0.85}{$
\begin{array}{ll} 
C(a^{-4} x^2) = 2 \left(1+\chi\right)+ \frac{1}{2} g \sum_{i,j=0}^1 \zeta^i \chi^j~,  &\,\, C(a^{0} x^2) = 1+ 5 \chi~,\\ 
C(a^{4} x^2)= 21+ 15 \chi~, & \,\, 
C(a^{-6} x^3) = \chi \left(1+\zeta \right) + g \sum_{i,j=0}^1 \zeta^i \chi^j ~, \\
C(a^{-2} x^3)= (3-\chi) + \frac{7}{2} g \sum_{i,j=0}^1 \zeta^i \chi^j~,   &\,\,  C(a^{2} x^3)=  10\chi-1~, \\  
C(a^{6} x^3) = 97 + 75 \chi ~, &\,\, 
C(a^{-8} x^4) = 8+3 \zeta +6 \chi +2 \zeta  \chi + 3 g \sum_{i,j=0}^1 \zeta^i \chi^j~.
\end{array}
$}
}

Let us examine the coefficient $C(a^{-4} x^2)$ of the term $a^{-4} x^2$. The contributions of various operators to this coefficient are listed below:
\bes{ \label{termsata-4x2}
\begin{tabular}{c|c|c}
\hline
Term in $C(a^{-4} x^2)$  & Operator &  $\SO(4) \times \USp(4)$ magnetic flux \\
\hline
$2$ &  $\tr(\phi_D^2)$, $\tr(\phi_C^2)$  &  $(0,0;0,0)$, $(0,0;0,0)$ \\
$2\chi$ &  $\mathrm{Pf}(\phi_D)$, $V_{(0,0;1,0)}$ & $(0,0; 0,0)$, $(0,0; 1,0)$ \\
$\frac{1}{2} g (1+\chi)$ &  $V_{(\frac{1}{2}, -\frac{1}{2}; \frac{1}{2}, \frac{1}{2})}$ &  $\left(\frac{1}{2}, -\frac{1}{2} ; \frac{1}{2}, \frac{1}{2}\right)$ \\
$\frac{1}{2} g (\zeta+\zeta \chi)$ & $V_{(\frac{1}{2}, \frac{1}{2}; \frac{1}{2}, \frac{1}{2})}$ & $\left( \frac{1}{2}, \frac{1}{2}; \frac{1}{2}, \frac{1}{2} \right)$ \\
\hline
\end{tabular}
}
where $\phi_D$ and $\phi_C$ denote the adjoint scalar fields in the $\SO(4)$ and $\USp(4)$ vector multiplets, respectively. The notation $V_{(a_1, a_2; b_1, b_2)}$ denotes the monopole operator carrying $\SO(4) \times \USp(4)$ magnetic flux $(a_1, a_2; b_1, b_2)$. The charge-conjugation oddness ($\chi$-dependence) of the monopole operator $V_{(0,0; 1,0)}$ is consistent with the discussion surrounding \eref{indSO4wNf3flux100}. Let us now focus on the last two rows of the table \eref{termsata-4x2} involving fractional fluxes. The structure relates to how charge conjugation acts on monopole operators. As explained in \cite[Pages 8-9]{Aharony:2013kma} for the $\SO(4)$ gauge theory, the $\BZ_{2,\, \CC}^{[0]}$ symmetry exchanges monopole operators with fluxes $(1,1)$ and $(1,-1)$, denoted $Y^{(1)}$ and $Y^{(2)}$ in that reference. Neither monopole operator ($Y^{(1)}$ or $Y^{(2)}$) has a definite charge conjugation parity. Instead, the linear combinations $Y_\pm \equiv Y^{(1)}\pm Y^{(2)}$ are the monopole operators with definite (even/odd) parity under $\BZ_{2,\, \CC}^{[0]}$. Applying similar logic to the fractional flux monopoles appearing in the $\left[\SO(4) \times \USp(4)\right]/\BZ_2$ theory, the factor $\frac{1}{2}(1+\chi)$ associated with $V_{(\frac{1}{2}, -\frac{1}{2}; \frac{1}{2}, \frac{1}{2})}$ (and similarly $\frac{1}{2}(\zeta+\zeta \chi)$ for $V_{(\frac{1}{2}, \frac{1}{2}; \frac{1}{2}, \frac{1}{2})}$) reflects that these specific monopole operators in the quotient theory do not have a definite parity under the original $\BZ_{2,\, \CC}^{[0]}$ symmetry alone, but are projected in a way captured by these fugacity combinations.

The prefactor $\frac{1}{2}$ in the term $\frac{1}{2} g \sum_{i,j=0}^1 \zeta^i \chi^j$ in the coefficient of $a^{-4} x^2$ indicates that the index is invalid upon turning on the fugacities $g$, $\zeta$ and $\chi$ simultaneously.\footnote{In fact, the terms $\frac{1}{2} g (1+ \zeta + \chi+ \zeta \chi)$ correspond to the two-dimensional irreducible representation of $D_8$. Each individual term $1$, $\zeta$, $\chi$ and $\zeta \chi$ corresponds to the four one-dimensional irreducible representations. The presence of the factor of $\frac{1}{2}$ indicates that we cannot refine $g$, $\zeta$ and $\chi$ simultaneously, since the associated global symmetries do not commute. Setting $g=1$ does not yield the consistent terms in the index, since the product group of $\BZ_2$ generated by $\zeta$ and $\BZ_2$ generated by $\chi$ is not a subgroup of $D_8$. We thank the JHEP referee for pointing out this interpretation.} Note, however, that among these fugacity, if we either set $\chi = 1$ or $\zeta=1$, or $\zeta=\chi=1$, then the index becomes well-defined. As pointed out in \cite{Mekareeya:2022spm}, setting a fugacity to one amounts to turning off the background gauge field for the corresponding global symmetry. Therefore, this discussion is consistent with the presence of the 't Hooft anomaly $i \pi \int_{M_4} A_2^B \cup A_1^\CM \cup A_1^\CC$ in the $\SO(4) \times \USp(4)$ gauge theory. Another way to see this anomaly is to gauge all of the symmetries associated with $A_2^B$, $A_1^\CM$ and $A_1^\CC$ in the the $\SO(4) \times \USp(4)$ gauge theory and see what goes wrong in the resulting $\left[\mathrm{Pin}(4) \times \USp(4)\right]/\BZ_2$ gauge theory. From the perspective of the index \eref{indD2C2modZ2}, this amounts to summing over $\zeta=\pm 1$ and $\chi = \pm 1$ and then dividing by four. The result is
\bes{ \label{indPin4USp4modZ2}
\scalebox{0.99}{$
\begin{split}
&\CI\left\{\left[\mathrm{Pin}(4) \times \USp(4)\right]/\BZ_2\right\} \\ &= 1+ {\bf [2]} a^2 x + \Bigg[\frac{1}{2} (4+g)  a^{-4} +{\bf [4]} + \left( {\bf [8]}+2{\bf [4]}+2\right) a^4 -{\blue \left({\bf [2]}+1\right)} \Bigg] x^2 + \ldots~.
\end{split}
$}
}
This index is ill-defined even if we set $g=1$ due to the non-integral coefficient of the term $a^{-4} x^2$.  This confirms the obstruction to gauging the full $\BZ_{2, \, B}^{[0]} \times \BZ_{2, \, \CM}^{[0]} \times \BZ_{2, \, \CC}^{[0]}$ symmetry simultaneously, signalling the 't Hooft anomaly.

We now examine the superconformal indices for various global forms of the gauge group. For presentational simplicity, we set the $\SO(3)$ flavour fugacities to unity, but turn on the fugacities $g$, $\zeta$, and $\chi$ for the discrete zero-form symmetries. The index retains the general structure shown in \eref{fixformindex}, but the specific coefficients $C(a^{\pm 2p} x^p)$ differ for each global form, as reported below.
\bi 
\item $\left[\O(4)^+ \times \USp(4)\right]/\BZ_2:$
\bes{ \label{indO4pUSp4modZ2}
\begin{array}{ll}
C(a^{-4} x^2) =2+ \frac{1}{2} g (1+\zeta)~,  &\quad C(a^{0} x^2) = 1~,\\ 
C(a^{4} x^2)= 21~, & \quad C(a^{-6} x^3) = g (1+\zeta) ~, \\
C(a^{-2} x^3)= 3 + \frac{7}{2} g (1+\zeta)~,   &\quad  C(a^{2} x^3)=  -1~, \\  
C(a^{6} x^3) =97~,  &\quad C(a^{-8} x^4) = 8 + 3 g (1+\zeta)~.
\end{array}
}
Note that this index yields integer coefficients if the fugacity $\zeta$ is set to 1 (i.e., the $\BZ_{2,\CM}^{[0]}$ background field is turned off). The presence of terms involving $\frac{1}{2}(1+\zeta)$ signals an obstruction to gauging the $\BZ^{[0]}_{2,\, \CM}$ symmetry (which would involve summing over $\zeta=\pm 1$ and dividing by 2), as this procedure would lead to the ill-defined index previously identified for the $\left[\Pin(4) \times \USp(4)\right]/\BZ_2$ theory \eref{indPin4USp4modZ2}.
\item $\left[\Spin(4) \times \USp(4)\right]/\BZ_2:$
\bes{ \label{indSpin4USp4modZ2}
\begin{array}{ll}
C(a^{-4} x^2) =2 (1+\chi)+ \frac{1}{2} g (1+\chi)~,  &\, C(a^{0} x^2) = 1+5 \chi~,\\ 
C(a^{4} x^2)= 21+15 \chi~, & \, C(a^{-6} x^3) = \chi + g (1+\chi) ~, \\
C(a^{-2} x^3)= (3-\chi) + \frac{7}{2} g (1+\chi)~,   &\,  C(a^{2} x^3)=  10\chi-1~, \\  
C(a^{6} x^3) =97+75 \chi~, &\, C(a^{-8} x^4) =  8+6 \chi + 3 g (1+\chi)~.
\end{array}
}
Similarly, this index is well-defined if $\chi=1$. The presence of terms involving $\frac{1}{2}(1+\chi)$ indicates an obstruction to gauging the $\BZ^{[0]}_{2,\, \CC}$ symmetry (summing over $\chi=\pm 1$), which would again result in the ill-defined $\left[\Pin(4) \times \USp(4)\right]/\BZ_2$ index.
\item $\left[\O(4)^- \times \USp(4)\right]/\BZ_2:$
\bes{
\begin{array}{ll}
C(a^{-4} x^2) =2+ \frac{1}{2} g (1+\zeta)~,  &\quad C(a^{0} x^2) = 1~,\\ 
C(a^{4} x^2)= 21~, & \quad C(a^{-6} x^3) = \zeta + g (1+\zeta) ~, \\
C(a^{-2} x^3)= 3 + \frac{7}{2} g (1+\zeta)~,   &\quad  C(a^{2} x^3)=  -1~, \\  
C(a^{6} x^3) =97~, &\quad C(a^{-8} x^4) = 8+ 2 \zeta  + 3 g (1+\zeta)~.
\end{array}
}
This index yields integer coefficients if $\zeta = 1$. The presence of terms involving $\frac{1}{2}(1+\zeta)$ signals an obstruction to gauging the leftover $\BZ_{2,\, \CM}^{[0]}$ symmetry, which is still surviving after the diagonal subgroup of $\BZ_{2,\, \CM}^{[0]} \times \BZ_{2,\, \CC}^{[0]}$, which we denote by $\BZ_{2,\, \CM \cC}^{[0]}$, has been gauged.\footnote{As manifest from \eref{indvariants}, the index of the $\O(2N)^-$ gauge theory still depends on the fugacity $\zeta$ associated with the $\BZ_{2,\, \CM}^{[0]}$ symmetry, whose gauging leads to the $\Pin(2N)$ theory (see \cite[Figure 3]{Cordova:2017vab}).} Again, a valid transition to the $\left[\Pin(4) \times \USp(4)\right]/\BZ_2$ theory via this route is prevented.
\ei
We now perform the gauging of the $\BZ^{[0]}_{2,\, B}$ symmetry for each of the theories considered above. This corresponds to summing the respective index expressions over $g =\pm 1$ and dividing by two.
\bi
\item $\SO(4) \times \USp(4):$
\bes{ \label{indSO4USp4}
\begin{array}{ll}
C(a^{-4} x^2) =2 \left(1+\chi\right)~,  &\quad C(a^{0} x^2) = 1+5 \chi~,\\ 
C(a^{4} x^2)= 21+15 \chi~, & \quad C(a^{-6} x^3) = \chi \left(1+\zeta \right)  ~, \\
C(a^{-2} x^3)= 3-\chi~,   &\quad  C(a^{2} x^3)=  10\chi-1~, \\  
C(a^{6} x^3) =97+75 \chi~, &\quad C(a^{-8} x^4)= 8+3 \zeta +6 \chi+2 \zeta  \chi ~.
\end{array}
}
\item $\O(4)^+ \times \USp(4):$
\bes{ \label{indO4plusUSp4}
\begin{array}{ll}
C(a^{-4} x^2) =2~,  &\quad C(a^{0} x^2) = 1~,\\ 
C(a^{4} x^2)= 21~, & \quad C(a^{-6} x^3) = 0  ~, \\
C(a^{-2} x^3)= 3~,   &\quad  C(a^{2} x^3)=  -1~, \\  
C(a^{6} x^3) =97~, &\quad C(a^{-8} x^4)= 8+3 \zeta~.
\end{array}
}
\item $\Spin(4) \times \USp(4):$
\bes{ \label{indSpin4USp4}
\begin{array}{ll}
C(a^{-4} x^2) =2 \left(1+\chi\right)~,  &\quad C(a^{0} x^2) = 1+5 \chi~,\\ 
C(a^{4} x^2)= 21+15 \chi~, & \quad C(a^{-6} x^3) = \chi  ~, \\
C(a^{-2} x^3)= 3-\chi~,   &\quad  C(a^{2} x^3)=  10\chi-1~, \\  
C(a^{6} x^3) =97+75 \chi~, &\quad C(a^{-8} x^4)=8+ 6 \chi ~.
\end{array}
}
\item $\O(4)^- \times \USp(4):$
\bes{ \label{indO4mUSp4}
\begin{array}{ll}
C(a^{-4} x^2) =2~,  &\qquad C(a^{0} x^2) = 1~,\\ 
C(a^{4} x^2)= 21~, & \qquad C(a^{-6} x^3) = \zeta ~, \\
C(a^{-2} x^3)= 3~,   &\qquad  C(a^{2} x^3)=  -1~, \\  
C(a^{6} x^3) =97~, &\qquad C(a^{-8} x^4)= 8+2 \zeta ~.
\end{array}
}
\item $\Pin(4) \times \USp(4):$
\bes{ \label{indPin4USp4}
\begin{array}{ll}
C(a^{-4} x^2) =2~,  &\qquad \quad C(a^{0} x^2) = 1~,\\ 
C(a^{4} x^2)= 21~, & \qquad \quad C(a^{-6} x^3) = 0  ~, \\
C(a^{-2} x^3)= 3~,   &\qquad \quad C(a^{2} x^3)=  -1~, \\  
C(a^{6} x^3) =97~, &\qquad \quad C(a^{-8} x^4)= 8~.
\end{array}
}
\ei
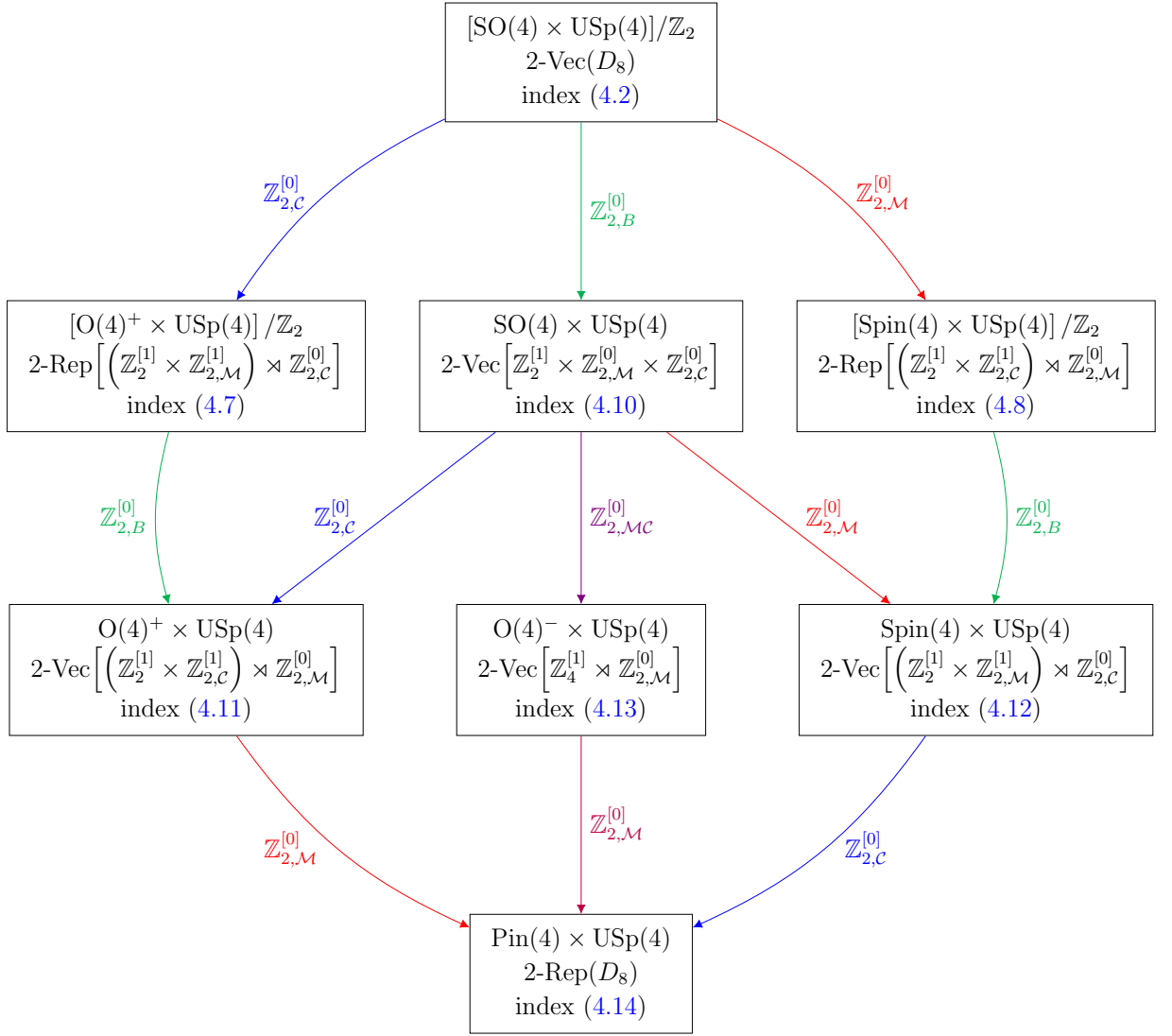
\begin{figure}
\centering
\scalebox{0.85}{
\begin{tikzpicture} 
			\node[draw] (Pin4) at (0,-7.5) {\begin{tabular}{c}
			$\Pin(4) \times \USp(4)$ \\ 2-Rep$(D_8)$ \\ index \eref{indPin4USp4} \end{tabular} }; 
                \node[draw] (O4m) at (0,-2.5) {\begin{tabular}{c}
			$\O(4)^- \times \USp(4)$ \\ 2-Vec$\left[\BZ_{4}^{[1]} \rtimes \BZ_{2, \CM}^{[0]}\right]$ \\ index \eref{indO4mUSp4} \end{tabular} };
                \node[draw] (O4p) at (-6.5,-2.5) {\begin{tabular}{c}
			$\O(4)^+ \times \USp(4)$ \\ 2-Vec$\left[\left(\BZ_{2}^{[1]} \times \BZ_{2, \cC}^{[1]}\right) \rtimes \BZ_{2, \CM}^{[0]}\right]$ \\ index \eref{indO4plusUSp4} \end{tabular} };
                \node[draw] (Spin4) at (6.5,-2.5) {\begin{tabular}{c}
			$\Spin(4) \times \USp(4)$ \\ 2-Vec$\left[\left(\BZ_{2}^{[1]} \times \BZ_{2, \CM}^{[1]}\right) \rtimes \BZ_{2, \cC}^{[0]}\right]$ \\ index \eref{indSpin4USp4} \end{tabular} };
                \node[draw] (SO4) at (0,2.5) {\begin{tabular}{c}
			$\SO(4) \times \USp(4)$ \\ 2-Vec$\left[\BZ_{2}^{[1]} \times \BZ_{2, \CM}^{[0]} \times \BZ_{2, \cC}^{[0]}\right]$ \\ index \eref{indSO4USp4} \end{tabular} };
                \node[draw] (O4pmodZ2) at (-6.5,2.5) {\begin{tabular}{c}
			$\left[\O(4)^+ \times \USp(4)\right]/\BZ_2$ \\ 2-Rep$\left[\left(\BZ_{2}^{[1]} \times \BZ_{2, \CM}^{[1]}\right) \rtimes \BZ_{2, \cC}^{[0]}\right]$ \\ index \eref{indO4pUSp4modZ2} \end{tabular} };
                \node[draw] (Spin4modZ2) at (6.5,2.5) {\begin{tabular}{c}
			$\left[\Spin(4) \times \USp(4)\right]/\BZ_2$ \\ 2-Rep$\left[\left(\BZ_{2}^{[1]} \times \BZ_{2, \cC}^{[1]}\right) \rtimes \BZ_{2, \CM}^{[0]}\right]$ \\ index \eref{indSpin4USp4modZ2} \end{tabular} };
                \node[draw] (SO4modZ2) at (0,7.5) {\begin{tabular}{c}
			$[\SO(4) \times \USp(4)]/\BZ_2$ \\ 2-Vec$(D_8)$ \\ index \eref{indD2C2modZ2} \end{tabular} }; 
            \draw[->,blue] (SO4modZ2) to [bend right=15] node[midway, left=0.2] {\blue $\BZ_{2,\cC}^{[0]}$} (O4pmodZ2);
            \draw[->,new-green] (SO4modZ2) to node[midway,right] {\green $\BZ_{2,B}^{[0]}$} (SO4);
            \draw[->,red] (SO4modZ2) to [bend left=15] node[midway, right=0.2] {\red $\BZ_{2,\CM}^{[0]}$} (Spin4modZ2);
            \draw[->,new-green] (O4pmodZ2) to [bend right=15] node[midway, left] {\green $\BZ_{2,B}^{[0]}$} (O4p);
            \draw[->,violet] (SO4) to node[midway,right] {\violet $\BZ_{2,\CM \cC}^{[0]}$} (O4m);
            \draw[->,new-green] (Spin4modZ2) to [bend left=15] node[midway, right] {\green $\BZ_{2,B}^{[0]}$} (Spin4);
            \draw[->,blue] (SO4) to node[midway,left=0.3] {\blue $\BZ_{2,\cC}^{[0]}$} (O4p);
            \draw[->,red] (SO4) to node[midway,right=0.3] {\red $\BZ_{2,\CM}^{[0]}$} (Spin4);
            \draw[->,red] (O4p) to [bend right=15] node[midway, left=0.2] {\red $\BZ_{2,\CM}^{[0]}$} (Pin4);
            \draw[->,purple] (O4m) to node[midway,right] {\purple $\BZ_{2,\CM}^{[0]}$} (Pin4);
            \draw[->,blue] (Spin4) to [bend left=15] node[midway, right=0.2] {\blue $\BZ_{2,\cC}^{[0]}$} (Pin4);
\end{tikzpicture}
}
    \caption[Dih8 web]{The $D_8$ symmetry web for variants of the $\so(4) \times \usp(4)$ gauge theory with three bifundamental half-hypermultiplets. Each arrow labelled by $\BZ^{[0]}_{2,x}$ connecting two boxes denotes the gauging of the zero-form symmetry $\BZ^{[0]}_{2,x}$. In each box, which is associated with a specific non-anomalous global form of the theory, we report the corresponding symmetry category and the index. Furthermore, observe that, from the anomaly theory \eref{anomABJlike}, this diagram retains the same structure for the $\so(4m) \times \usp(4m)$ gauge theory with $n$ bifundamental half-hypermultiplets.} \label{figDih8web}
\end{figure}
The pattern observed – namely the inconsistency encountered when attempting to simultaneously gauge combinations corresponding to the $\left[\Pin(4) \times \USp(4)\right]/\BZ_2$ theory, as seen in \eref{indPin4USp4modZ2}, contrasted with the well-defined indices obtained for the $\left[\O(4)^+ \times \USp(4)\right]/\BZ_2$ and $\left[\Spin(4) \times \USp(4)\right]/\BZ_2$ forms – is the hallmark of the underlying non-Abelian $D_8$ global zero-form symmetry structure present in the initial $\left[\SO(4) \times \USp(4)\right]/\BZ_2$ theory. This allows us to place these different global forms of the gauge group within the $D_8$ symmetry web, as depicted in Figure \ref{figDih8web}.

It is worth noting that the $\left[\O(4)^- \times \USp(4)\right]/\BZ_2$ gauge theory is not contained in this $D_8$ symmetry web (neither in \cite[Figure 3]{Bergman:2024its}, \cite[Figure 1]{Bhardwaj:2022maz}, nor \cite[Figure 34]{Bartsch:2022ytj}). Nevertheless, our index calculation demonstrates that this theory is distinct from the $\left[\O(4)^+ \times \USp(4)\right]/\BZ_2$ theory, necessitating their differentiation. Our analysis confirms that the $\left[\O(4)^- \times \USp(4)\right]/\BZ_2$ theory possesses the expected $\SO(3)$ continuous flavour symmetry. Furthermore, its Coulomb branch limit yields the unrefined Hilbert series presented in \eref{HSO4minusUSp4modZ2}, which exhibits properties consistent with $\CN=4$ supersymmetry, such as a palindromic numerator and the expected pole order at $t=1$. We will also provide a further check using the volume of the base of the Coulomb branch, viewed as a hyperK\"ahler cone, in \eref{ratiovolume}.
\subsubsection*{The case of $N=3$ and $n=3$}
Let us now briefly discuss the $\so(6) \times \usp(6)$ gauge theory with three bifundamental half-hypermultiplets. When the global form of the gauge group is $\SO(6) \times \USp(6)$, the mixed anomaly \eref{anomABJlike} takes the form 
\bes{ \label{anomN3n3}
i \pi \int_{M_4} A_2^B \cup \left[\mathrm{Bock} \left(A_1^\CM\right) + A_1^\CM \cup A_1^\CC \right]~,
}
where $\mathrm{Bock}$ is the obstruction to lifting $A^{\CM}_{1}$ from a $\BZ_2$ gauge field to a $\BZ_4$ gauge field. As discussed in \cite[Page 21]{Cordova:2017vab}, if we gauge $\BZ^{[0]}_{2,\CM}$, the mixed anomaly forces the resulting one-form global symmetry to be extended to $\BZ^{[1]}_4$. Similarly, if we gauge the $\BZ^{[1]}_{2}$ one-form symmetry, the resulting zero-form symmetry is extended to $\BZ^{[0]}_4$. 

The index of the theory, whose expression can be derived as described in Appendix \ref{app:souspABJlike}, can be employed as a tool to detect the aforementioned mixed anomaly. Explicitly, let us consider the  $[\SO(6) \times \USp(6)]/\BZ_2$ theory with a discrete $\BZ^{[0]}_{2,B}$ zero-form symmetry arising from gauging the $\BZ^{[1]}_{2}$ one-form symmetry, along with the $\BZ^{[0]}_{2,\CM}$ and $\BZ^{[0]}_{2,\CC}$ symmetries. Let us denote the associated fugacities as $g$, $\zeta$ and $\chi$, respectively. It follows that, if we denote with $V_{(a_1, a_2,a_3; b_1, b_2,b_3)}$ the monopole operator carrying $\SO(6) \times \USp(6)$ magnetic flux $(a_1, a_2,a_3; b_1, b_2,b_3)$, then the operators $V_+ \equiv V_{(\frac{1}{2},\frac{1}{2},\frac{1}{2};\frac{1}{2},\frac{1}{2},\frac{1}{2})}$ and $V_- \equiv V_{(\frac{1}{2},\frac{1}{2},-\frac{1}{2};\frac{1}{2},\frac{1}{2},\frac{1}{2})}$ contribute $\frac{1}{2} g \left(1+\chi\right) \zeta^{\frac{3}{2}} a^{-9}$ and $\frac{1}{2} g \left(1+\chi\right) \zeta^{\frac{1}{2}} a^{-9}$ to the index at order $x^{\frac{9}{2}}$, respectively.\footnote{Note that such monopole operators do not carry a definite $\BZ^{[0]}_{2,\CC}$ charge, but the linear combinations $V_+ \pm V_-$ do.}

A clear manifestation of the anomaly comes from the prefactor $\frac{1}{2}$, which signals that we cannot refine the index with respect to the fugacities $g$, $\zeta$ and $\chi$ simultaneously, whereas the index is well-defined upon setting $\zeta=1$ and/or $\chi=1$.
\begin{figure}
\centering
\scalebox{0.8}{
\begin{tikzpicture} 
			\node[draw] (Pin6) at (0,-7.5) {\begin{tabular}{c}
			$\Pin(6) \times \USp(6)$ \\ 2-Rep$(D_8)$ \end{tabular} }; 
                \node[draw] (Spin6) at (0,-2.5) {\begin{tabular}{c}
			$\Spin(6) \times \USp(6)$ \\ 2-Vec$\left[\BZ_{4}^{[1]} \rtimes \BZ_{2,\cC}^{[0]}\right]$ \end{tabular} };
                \node[draw] (O6p) at (-6.5,-2.5) {\begin{tabular}{c}
			$\O(6)^+ \times \USp(6)$ \\ 2-Vec$\left[\left(\BZ_{2}^{[1]} \times \BZ_{2, \cC}^{[1]}\right) \rtimes \BZ_{2, \CM}^{[0]}\right]$ \end{tabular} };
                \node[draw] (O6m) at (6.5,-2.5) {\begin{tabular}{c}
			$\O(6)^- \times \USp(6)$ \\ 2-Vec$\left[\left(\BZ_{2}^{[1]} \times \BZ_{2, \CM \cC}^{[1]}\right) \rtimes \BZ_{2, \CM}^{[0]}\right]$ \end{tabular} };
                \node[draw] (SO6) at (0,2.5) {\begin{tabular}{c}
			$\SO(6) \times \USp(6)$ \\ 2-Vec$\left[\BZ_{2}^{[1]} \times \BZ_{2, \CM}^{[0]} \times \BZ_{2, \cC}^{[0]}\right]$ \end{tabular} };
                \node[draw] (O6pmodZ2) at (-6.5,2.5) {\begin{tabular}{c}
			$\left[\O(6)^+ \times \USp(6)\right]/\BZ_2$ \\ 2-Rep$\left[\left(\BZ_{2}^{[1]} \times \BZ_{2, \CM}^{[1]}\right) \rtimes \BZ_{2, \cC}^{[0]}\right]$ \end{tabular} };
                \node[draw] (O6mmodZ2) at (6.5,2.5) {\begin{tabular}{c}
			$\left[\O(6)^- \times \USp(6)\right]/\BZ_2$ \\ 2-Rep$\left[\left(\BZ_{2}^{[1]} \times \BZ_{2, \CM}^{[1]}\right) \rtimes \BZ_{2, \CM \cC}^{[0]}\right]$ \end{tabular} };
                \node[draw] (SO6modZ2) at (0,7.5) {\begin{tabular}{c}
			$[\SO(6) \times \USp(6)]\BZ_2$ \\ 2-Vec$(D_8)$  \end{tabular} }; 
            \draw[->,blue] (SO6modZ2) to [bend right=15] node[midway, left=0.2] {\blue $\BZ_{2,\cC}^{[0]}$} (O6pmodZ2);
            \draw[->,new-green] (SO6modZ2) to node[midway,right] {\green $\BZ_{2,B}^{[0]}$} (SO6);
            \draw[->,violet] (SO6modZ2) to [bend left=15] node[midway, right=0.2] {\violet $\BZ_{2,\CM \cC}^{[0]}$} (O6mmodZ2);
            \draw[->,new-green] (O6pmodZ2) to [bend right=15] node[midway, left] {\green $\BZ_{2,B}^{[0]}$} (O6p);
            \draw[->,red] (SO6) to node[midway,right] {\red $\BZ_{2,\CM}^{[0]}$} (Spin6);
            \draw[->,new-green] (O6mmodZ2) to [bend left=15] node[midway, right] {\green $\BZ_{2,B}^{[0]}$} (O6m);
            \draw[->,blue] (SO6) to node[midway,left=0.3] {\blue $\BZ_{2,\cC}^{[0]}$} (O6p);
            \draw[->,violet] (SO6) to node[midway,right=0.3] {\violet $\BZ_{2,\CM \cC}^{[0]}$} (O6m);
            \draw[->,red] (O6p) to [bend right=15] node[midway, left=0.2] {\red $\BZ_{2,\CM}^{[0]}$} (Pin6);
            \draw[->,blue] (Spin6) to node[midway,right] {\blue $\BZ_{2,\cC}^{[0]}$} (Pin6);
            \draw[->,purple] (O6m) to [bend left=15] node[midway, right=0.2] {\purple $\BZ_{2,\CM}^{[0]}$} (Pin6);
\end{tikzpicture}
} 
    \caption[Dih8 web SO6]{The $D_8$ symmetry web for variants of the $\so(6) \times \usp(6)$ gauge theory with three bifundamental half-hypermultiplets, where the notation for the arrows is as explained in Figure \ref{figDih8web}. Each box contains the specific global form of the gauge group, as well as the corresponding symmetry category. Observe that this $D_8$ symmetry web holds in general for the $\so(4m+2) \times \usp(4m+2)$ gauge theory with $n$ bifundamental half-hypermultiplets.} \label{figDih8webSO6}
\end{figure}
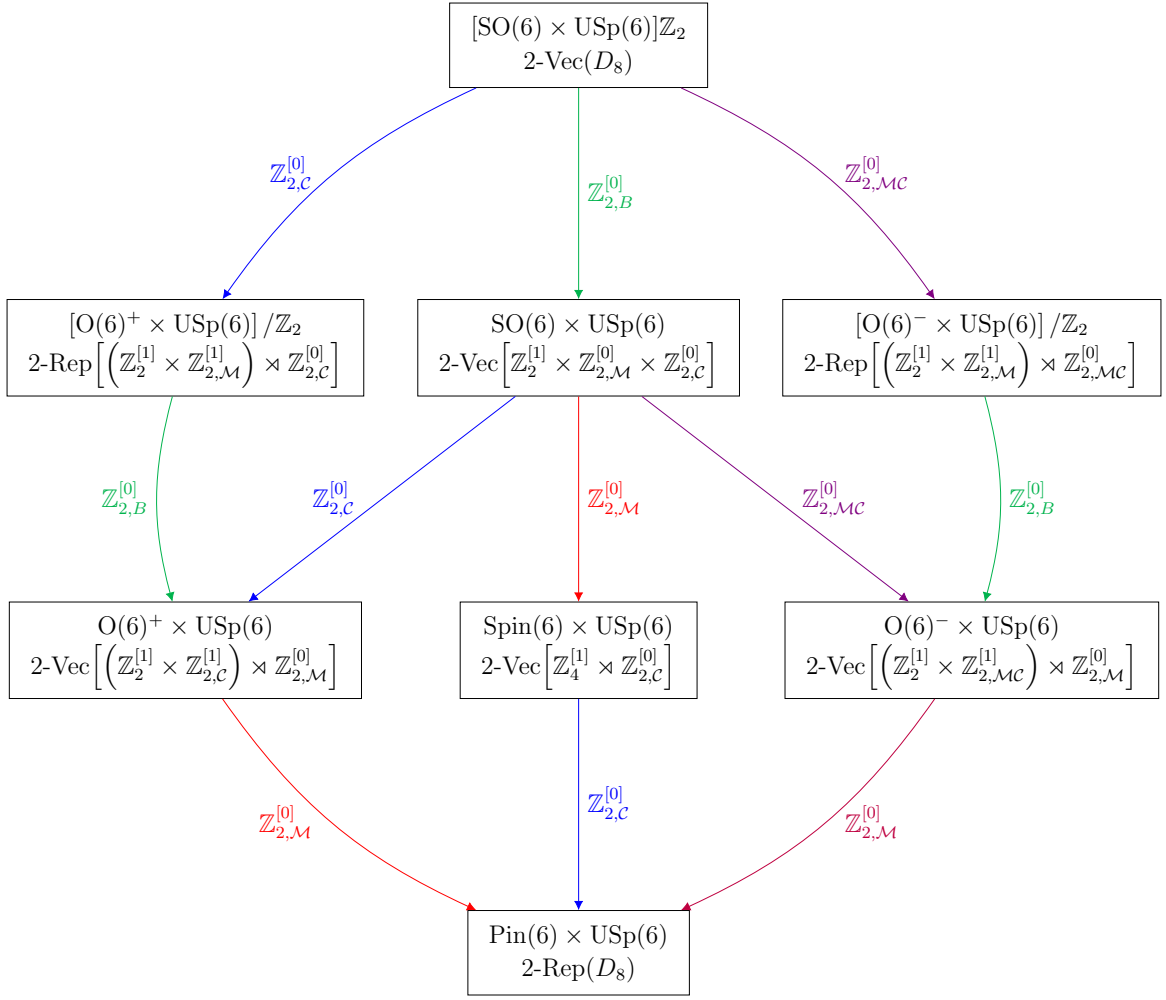

Furthermore, notice that half-integer powers of the fugacity $\zeta$ appear together with the fugacity $g$. A first observation is that, upon gauging the $\BZ^{[0]}_{2,B}$ symmetry to land on the $\SO(6) \times \USp(6)$ variant, whose index  can be obtained from that of the $[\SO(6) \times \USp(6)]/\BZ_2$ theory by summing over $g=\pm 1$ and dividing by two, such terms are projected out. However, this also signals that we cannot gauge both $\BZ^{[1]}_{2}$ and $\BZ^{[0]}_{2,\CM}$ in the $\SO(6) \times \USp(6)$ theory to reach the $[\Spin(6) \times \USp(6)]/\BZ_2$ variant, due to the mixed anomaly between such $\BZ_2$ symmetries. As a consequence, the $\BZ^{[0]}_{2,B}$ and $\BZ^{[0]}_{2,\CM}$ symmetries of the $[\SO(6) \times \USp(6)]/\BZ_2$ theory are extended to a $\BZ^{[0]}_4$ symmetry. This can be detected using the index by defining a $\BZ^{[0]}_4$ fugacity $d = g \zeta^{\frac{1}{2}}$, with $d^4 = 1$, where the contributions coming from the monopole operators $V_+$ and $V_-$ now read $\frac{1}{2} d^3 \left(1+\chi\right) a^{-9}$ and $\frac{1}{2} d \left(1+\chi\right) a^{-9}$ at order $ x^{\frac{9}{2}}$, respectively. Upon gauging the $\BZ^{[0]}_4$ symmetry of the $[\SO(6) \times \USp(6)]/\BZ_2$ theory, we land on the $\Spin(6) \times \USp(6)$ variant, whose index can be obtained from that of the latter theory by summing the fugacity $d$ over the four fourth roots of unity, and dividing by four. Finally, by further gauging the $\BZ^{[0]}_{2, \CC}$ symmetry, we reach the $\Pin(6) \times \USp(6)$ theory, which possesses a non-invertible 2-Rep$(D_8)$ symmetry.\footnote{Note that, due to the anomaly \eref{anomN3n3}, the $[\Pin(6) \times \USp(6)]/\BZ_2$ variant is anomalous. From the perspective of the index, this can be understood as follows. First, if the $\BZ^{[0]}_{2, \CC}$ symmetry of the $[\SO(6) \times \USp(6)]/\BZ_2$ theory is gauged by summing over $\chi=\pm 1$ and dividing by two, the total contribution to the index at order $x^{\frac{9}{2}}$ coming from $V_+$ and $V_-$ is $\frac{1}{2} g \left(\zeta^{\frac{1}{2}}+\zeta^{\frac{3}{2}}\right) a^{-9} = \frac{1}{2} \left(d+d^3\right) a^{-9}$. Next, we would gauge the $\BZ^{[0]}_{2, \CM}$ symmetry, but this is forbidden by the half-integer powers of $\zeta$ appearing together with $g$.} The various global forms of the gauge group, giving rise to the $D_8$ symmetry web, are summarised in Figure \ref{figDih8webSO6}.

Let us conclude with a final remark. As pointed out in \cite[(3.63)]{Grimminger:2024mks} and \cite[Footnote 1]{Zafrir:2025xca}, when thinking of $\BZ^{[0]}_4$ as a non-trivial extension between the $\BZ^{[0]}_{2,\mathfrak{n}}$ normal subgroup and the $\BZ^{[0]}_{2,\mathfrak{q}}$ quotient group, it is useful to parametrise the fugacity $d$ as 
\bes{ \label{dqn}
d = \exp \left[\frac{2 \pi i}{4} \left(\mathfrak{q} + 2 \mathfrak{n}\right)\right]~, \quad \text{with} \quad \mathfrak{q}=0,1~, \quad \mathfrak{n}=0,1~,
}
where the variables $\mathfrak{n}$ and $\mathfrak{q}$ are associated with $\BZ^{[0]}_{2,\mathfrak{n}}$ and $\BZ^{[0]}_{2,\mathfrak{q}}$, respectively. The gauging of $\BZ^{[0]}_{2,\mathfrak{n}}$ (resp. $\BZ^{[0]}_{2,\mathfrak{q}}$) can be implemented using the index by summing over $\mathfrak{n}=0,1$ (resp. $\mathfrak{q}=0,1$), and dividing by two. From \eref{dqn}, we see that the combination $d+d^3$, hence the contribution coming from $V_+$ and $V_-$ at order $x^{\frac{9}{2}}$, is projected out from the index of the $[\SO(6) \times \USp(6)]/\BZ_2$ theory upon gauging $\BZ^{[0]}_{2,\mathfrak{n}}$. As described above, this is what happens when the $\BZ^{[0]}_{2,B}$ symmetry is gauged to reach the $\SO(6) \times \USp(6)$ variant, leading to the identification of $\BZ^{[0]}_{2,B}$ with $\BZ^{[0]}_{2,\mathfrak{n}}$. Since, by further gauging $\BZ^{[0]}_{2,\mathfrak{q}}$, we reach the theory with the whole $\BZ^{[0]}_{4}$ symmetry gauged, whose gauge group is $\Spin(6) \times \USp(6)$, we conclude that we can also identify $\BZ^{[0]}_{2,\mathfrak{q}}$ with the $\BZ^{[0]}_{2,\CM}$ symmetry of the $\SO(6) \times \USp(6)$ theory.

\section{Coulomb branch Hilbert series: an improved prescription} \label{sec:CBHS}
This section focuses on the Coulomb branch Hilbert series for quiver gauge theories involving orthogonal gauge groups. We propose an improved prescription for computing the Coulomb branch Hilbert series of the $\SO(N)$ gauge theory with $N_f$ vector hypermultiplets, incorporating background magnetic fluxes for the $\usp(2N_f)$ flavour symmetry. This extends the methods presented in \cite[Appendix A]{Cremonesi:2014uva}. Our improvements include:
\begin{enumerate}
\item A method for incorporating the fugacity $\chi$ corresponding to the $\BZ_{2,\, \CC}^{[0]}$ charge conjugation symmetry of the $\SO(N)$ gauge group. \label{turnonchi}
\item The inclusion of a crucial phase factor when background magnetic fluxes for the $\usp(2N_f)$ flavour symmetry are non-zero. \label{phaseHS} This phase factor ensures consistency between the computed Hilbert series and the Coulomb branch limit of the superconformal index in the presence of these background fluxes.
\end{enumerate}
The first improvement (Point \ref{turnonchi}) is essential for directly calculating the Coulomb branch Hilbert series for related gauge groups like $\O(N)^+$, $\O(N)^-$, and $\Pin(N)$, bypassing the need to first compute the full superconformal index and subsequently take its Coulomb branch limit. The second improvement (Point \ref{phaseHS}) is particularly crucial in scenarios where the $\usp(2N_f)$ flavour symmetry is gauged, such as in the orthosymplectic quiver gauge theories discussed in subsequent sections.

\subsection{Prescription for the $\so(2N)$ gauge algebra}
A method for computing the Hilbert series for the $\O(2N)^+$ gauge theory with $N_f$ vector hypermultiplets, with vanishing background magnetic fluxes, was proposed in \cite[Appendix A]{Cremonesi:2014uva}. We begin by reviewing this prescription before presenting our generalisations.

Consider first the $\SO(2N)$ gauge theory with $N_f$ vector hypermultiplets. The Coulomb branch Hilbert series of this theory was discussed in detail in \cite[Section 5.4]{Cremonesi:2013lqa} and \cite{Cremonesi:2014kwa}:
\bes{ \label{HSSO2NN_f}
H_{\SO(2N), N_f}(t; \zeta; \vec n) = \sum_{\overset{m_1 \geq m_2 \geq \cdots \geq |m_N|}{m_N \in \BZ}} \zeta^{\sum_{i=1}^N m_i} \,\, t^{2\Delta(\vec m; \vec n)} P_{\SO(2N)} (t; \vec m)~.
}
Here, $\zeta$ is the fugacity for the $\BZ^{[0]}_{2,\, \CM}$ magnetic symmetry, $\vec m = (m_1, \ldots, m_N)$ denotes the magnetic fluxes for the $\SO(2N)$ gauge group, $\vec n = (n_1, \ldots, n_{N_f})$ represents the background magnetic fluxes for the $\usp(2N_f)$ flavour symmetry, and $\Delta(\vec m; \vec n)$ is the conformal dimension of the monopole operator characterised by these fluxes:
\bes{ \label{Deltamn}
\Delta(\vec m; \vec n) &= \frac{1}{2} \sum_{j=1}^{N_f} \sum_{i=1}^N \big(|m_i+ n_j|+|m_i- n_j|\big) \\&- \sum_{1\leq i<j \leq N} \big(|m_i+m_j| + |m_i-m_j|\big) ~.
}
The dressing factor $P_{\SO(2N)} (t; \vec m)$ (also known as the $P$-factor) represents the Hilbert series counting of gauge-invariant operators constructed from the adjoint scalar field $\varphi$ in the residual gauge group preserved by the magnetic flux $\vec m$. It is given by \cite[(A.10)]{Cremonesi:2013lqa}:
\bes{
P_{\SO(2N)}(t; \vec m) = Z^{D}_{\#_0(\vec m)} \prod_{j \geq 1} Z^U_{\#_j(\vec m)}~,
}
where $\#_x(\vec m)$ counts the number of times that $x$ appears in $(m_1, m_2, \ldots, m_{N-1}, |m_N|)$, with $m_1 \geq m_2 \geq \cdots \geq |m_N|$, $m_N \in \BZ$, and
\bes{
Z^{D}_{k} = \begin{cases} \frac{1}{1-t^{2k}} \prod_{i=1}^{k-1} \frac{1}{1-t^{4i}}  &\,\, k\geq 1\\
1 &\,\, k=0
\end{cases}~, \qquad
Z^{U}_{k} = \begin{cases} 
\prod_{i=1}^k \frac{1}{1-t^{2i}} & \,\, k \geq 1
\\
1 &\,\, k=0
\end{cases}~.
}

To compute the Hilbert series for the $\O(2N)^+$ gauge theory with $\vec{n}=\vec0$, the authors of \cite[Appendix A]{Cremonesi:2014uva} proposed to modify the summation range in \eref{HSSO2NN_f}. The magnetic lattice is restricted to $m_1 \geq m_2 \geq \cdots \geq m_N \geq 0$. This restriction arises because the charge conjugation symmetry $\BZ^{[0]}_{2, \CC}$, which is gauged in $\O(2N)^+$, identifies monopole fluxes $(m_1, m_2, \dots, m_N)$ and $(m_1, m_2, \dots, -m_N)$ when $m_N \neq 0$. Moreover, the Casimir invariant $\mathrm{Pf}(\varphi) \sim \epsilon_{i_1 i_2 \ldots i_{2N}} \varphi^{i_1 i_2} \cdots \varphi^{i_{2N-1} i_{2N}}$ exists for $\SO(2N)$, but not for $\O(2N)^+$. Consequently, the dressing factor $P_{\SO(2N)}$ in \eref{HSSO2NN_f} must be replaced. The appropriate factor for $\O(2N)^+$ corresponds to that of $\SO(2N+1)$, which coincides with the dressing factor for $\USp(2N)$. In summary, the authors of \cite[Appendix A]{Cremonesi:2014uva} proposed the following formula for $\vec{n}=\vec 0$:
\bes{ \label{HSO2NN_f}
H_{\O(2N)^+, N_f}(t; \zeta; \vec n=\vec 0) = \sum_{m_1 \geq m_2 \geq \cdots \geq m_N \geq 0} \zeta^{\sum_{i=1}^N m_i} \,\, t^{2\Delta(\vec m; \vec n=\vec 0)} P_{\USp(2N)} (t; \vec m)~,
}
where
\bes{ \label{PUSp2N}
P_{\USp(2N)}(t; \vec m) = Z^{C}_{\#_0(\vec m)} \prod_{j \geq 1} Z^U_{\#_j(\vec m)}~,
}
where $\#_x(\vec m)$ is now the multiplicity of $x$ in the set $\{m_1, m_2, \ldots, m_N\}$ with the restricted ordering $m_1 \geq m_2 \geq \cdots \geq m_N \geq 0$, and $Z^U_k$ is as defined before, while $Z^C_k$ is:
\bes{
Z^{C}_{k} = \begin{cases} \prod_{i=1}^{k} \frac{1}{1-t^{4i}}  &\,\, k\geq 1\\
1 &\,\, k=0
\end{cases}~.
}
We have verified that the result from \eref{HSO2NN_f} accurately matches the Coulomb branch limit of the superconformal index (with $\chi$ averaged) in numerous examples where $\vec{n}=\vec 0$. We now describe our proposals for incorporating the charge conjugation fugacity $\chi$ into \eref{HSSO2NN_f} and for correctly including non-zero background fluxes $\vec n$ in the formula for $\O(2N)^+$ and related groups, generalising \eref{HSO2NN_f}.

The formula \eref{HSSO2NN_f} correctly computes the $\chi=+1$ sector of the $\SO(2N)$ Hilbert series:
\bes{ \label{HSSO2NN_fchi1}
H_{\SO(2N), N_f}(t; \zeta; \chi=+1; \vec n) = \sum_{\overset{m_1 \geq m_2 \geq \cdots \geq |m_N|}{m_N \in \BZ}} \zeta^{\sum_{i=1}^N m_i} \,\, t^{2\Delta(\vec m; \vec n)} P_{\SO(2N)} (t; \vec m)~.
}
For $\chi=-1$, we propose that
\bes{ \label{HSSO2NN_fchi-1}
H_{\SO(2N), N_f}(t; \zeta; \chi=-1; \vec n) &= \sum_{m_1 \geq m_2 \geq \cdots \geq m_{N-1} \geq 0} \zeta^{\sum_{i=1}^{N-1} m_i} \,  {\blue \chi^{\sum_{j=1}^{N_f} n_j} } 
 \, t^{2\Delta(\tilde{\vec m}; \vec n)} \\
& \qquad \times \Big[ 2P_{\USp(2N)} (t; \tilde{\vec m}) - P_{\SO(2N)} (t; \tilde{\vec m}) \Big]  ~,
}
where $\tilde{\vec m} = (m_1, m_2, \dots, m_{N-1} ,0)$, which is simply $\vec m$ with the last entry set to zero. We emphasise that, in the $\chi = -1$ sector, $m_N$ is always set to zero. This restriction stems from the fact that, for $\SO(2N)$, any gauge holonomy corresponding to $\chi = -1$ can be brought to the form $\diag( z_1, z_1^{-1}, \dots, z_{N-1}, z_{N-1}^{-1}, 1, -1)$, effectively setting the associated magnetic flux component $m_N$ to zero; see \cite[(6.9)]{Aharony:2013kma}. The phase factor ${\blue \chi^{\sum_{j=1}^{N_f} n_j}}$ (written using $\chi=-1$) highlighted in {\blue blue} is crucial to ensure that the $\BZ_{2,\, \CC}^{[0]}$ charge assignments for operators match those obtained from the superconformal index when the background magnetic fluxes $\vec{n}$ are activated. We will discuss this in detail shortly. The full Hilbert series for the $\SO(2N)$ gauge group, refined by both $\zeta$ and $\chi$ and including background fluxes $\vec n$, is obtained by combining the $\chi=\pm 1$ sectors:
\bes{ \label{fullCBHSSO2NwN_f}
&H_{\SO(2N), N_f}(t; \zeta; \chi; \vec n) \\
&= \frac{1}{2} \left[ H_{\SO(2N), N_f}(t; \zeta; \chi=+1; \vec n) + H_{\SO(2N), N_f}(t; \zeta; \chi=-1; \vec n)\right] \\
& + \frac{1}{2} \left[ H_{\SO(2N), N_f}(t; \zeta; \chi=+1; \vec n) - H_{\SO(2N), N_f}(t; \zeta; \chi=-1; \vec n)\right] \chi \\
&=  \frac{1}{2} \sum_{s=\pm1} H_{\SO(2N), N_f}(t; \zeta; \chi=s; \vec n) (1+s \chi)~.
}
Using this Hilbert series for $\SO(2N)$, we can readily obtain the Hilbert series for other global forms of the gauge group via appropriate projections, described in \eqref{indvariants}:
\bes{
\scalebox{0.9}{$
\begin{split}
H_{\O(2N)^+, N_f}(t; \zeta; \vec n) &=\frac{1}{2} \left[ H_{\SO(2N), N_f}(t; \zeta; \chi=+1; \vec n)+H_{\SO(2N), N_f}(t; \zeta; \chi=-1; \vec n) \right]~, \\
H_{\Spin(2N), N_f}(t; \chi; \vec n) &=\frac{1}{2} \left[ H_{\SO(2N), N_f}(t; \zeta=+1; \chi; \vec n)+H_{\SO(2N), N_f}(t; \zeta=-1; \chi; \vec n) \right]~, \\
H_{\O(2N)^-, N_f}(t; \zeta; \vec n) &=\frac{1}{2} \left[ H_{\SO(2N), N_f}(t; \zeta; \chi=1; \vec n)+H_{\SO(2N), N_f}(t; -\zeta; \chi=-1; \vec n) \right]~, \\
H_{\Pin(2N), N_f}(t; \vec n) &=\frac{1}{2} \left[ H_{\Spin(2N), N_f}(t; \chi=+1; \vec n)+H_{\Spin(2N), N_f}(t; \chi=-1; \vec n) \right]~.
\end{split}$}
\label{variants}
}
Furthermore, the Coulomb branch Hilbert series computed using these prescriptions, including non-zero background fluxes $\vec n$, satisfies the $\CN=4$ duality relation \eqref{dualityN4} in exactly the same manner as the superconformal index identity \eref{matchindexSOKapustin}:
\bes{ \label{dualityHS}
H_{\text{$\SO(N_c)$, $N_f=N_c-1$}}(t; \zeta, \chi; \vec n)  = H_{\text{$\SO(N_c)$, $N_f=N_c-1$}}(t; \zeta, \zeta \chi; \vec n)~.
}

\subsection{Examples}
We now illustrate the proposed prescription with several examples.

\subsubsection{$\SO(4)$ gauge theory with three flavours}
We first test our prescription against known results. Consider the $\SO(4)$ gauge theory ($N=2$) with $N_f=3$ vector hypermultiplets. With vanishing background fluxes, $\vec n = \vec 0$, formula \eref{fullCBHSSO2NwN_f} precisely reproduces the known result \eref{CBSO4w3flv}, previously obtained as a limit of the index. This serves as a successful check on the incorporation of the charge conjugation fugacity $\chi$ via \eref{HSSO2NN_fchi-1} and \eref{fullCBHSSO2NwN_f}.

Next, let us turn on the background magnetic flux $\vec n = (1,0,0)$. Applying \eref{fullCBHSSO2NwN_f} with this flux, the Coulomb branch Hilbert series up to order $t^6$ is found to be:
\bes{ \label{CBHSSO4w3flv100}
\scalebox{0.98}{$
H_{\SO(4), N_f=3}(t; \zeta; \chi; n_1=1, n_2 = n_3 = 0) = (\chi+ \zeta  \chi  ) t^4 +  ( \zeta +\chi +\zeta  \chi) t^6+\ldots  ~.
$}
}
The coefficients of $t^{2p}$ in this expansion perfectly match the coefficients of $a^{-2p} x^p$ in the Coulomb branch limit of the superconformal index, previously given in \eref{indSO4wNf3flux100}.

Crucially, omitting the phase factor ${\blue (-1)^{\sum_{j=1}^{N_f} n_j}}$ highlighted in \eref{HSSO2NN_fchi-1}, this would yield
\bes{ \label{indexaaa}
(1 + \zeta)t^4 +  (1 + \zeta +  \zeta\chi ) t^6 +\ldots~.
} 
This result contradicts the Coulomb branch limit of the index \eref{indSO4wNf3flux100}. Furthermore, expression \eref{indexaaa} is not invariant under the transformation $\chi \rightarrow \zeta \chi$, violating the duality relation \eref{dualityHS} which is a direct consequence of \eref{dualityN4}. This confirms the necessity of the phase factor included in our prescription.

\subsubsection{$\so(4) \times \usp(4)$ gauge theory with $n$ bifundamental half-hypers}
Let us start by considering the $[\SO(4) \times \USp(4)]/\BZ_2$ gauge group. The Coulomb branch Hilbert series can be computed similarly to the index, detailed in Appendix \ref{app:souspABJlike}. The Hilbert series for the $\chi=+1$ sector is given by
\bes{
&H_{\left[\SO(4) \times \USp(4)\right]/\BZ_2, \, n}(t; g; \zeta; \chi=+1) \\
&= \sum_{e=0}^1 \,g^e  \sum_{\overset{m_1 \geq |m_2|}{m_i \in \BZ+\frac{e}{2}}} \,\, \sum_{\overset{n_1 \geq n_2 \geq 0}{n_i \in \BZ+\frac{e}{2}}} \zeta^{m_1+m_2} \,\, t^{2\Delta(\vec m; \vec n)}  P_{\SO(4)} (t;m_1,m_2) P_{\USp(4)} (t; n_1,n_2)~,
}
where
\bes{
\scalebox{0.99}{$
\begin{split}
\Delta(\vec m; \vec n) &= \frac{n}{2} \sum_{i, j=1}^2 \big(|m_i+n_j| + |m_i -n_j|\big) - \sum_{1 \leq i< j\leq 2} \big(|m_i+m_j|+|m_i-m_j|\big) \\
& \,\ - \sum_{1 \leq i< j\leq 2} \big(|n_i+n_j|+|n_i-n_j|\big) - 2\sum_{i=1}^2 |n_i|~.
\end{split}
$}
}
On the other hand, the Hilbert series for the $\chi=-1$ sector is given by
\bes{ \label{SO4minusUSp4}
&H_{\left[\SO(4) \times \USp(4)\right]/\BZ_2, \, n}(t; \zeta; \chi=-1)  \\
&= \sum_{m_1 \in \BZ_{\geq 0}} \,\, \sum_{\overset{n_1 \geq n_2 \geq 0}{n_i \in \BZ}} \zeta^{m_1}\, (-1)^{n(n_1+n_2)} \,\, t^{2\Delta(\vec m; \vec n)}  P_{\USp(4)} (t; n_1,n_2) \\
& \qquad \qquad \qquad \times \left[ 2P_{\USp(4)} (t; m_1,0)- P_{\SO(4)} (t; m_1,0) \right]  ~.
}
Note that, in the $\chi=-1$ sector, the last flux $m_2$ of the $\SO(4)$ gauge group is set to zero. As a consequence, there is no summation over half-integral fluxes, since otherwise these would violate the Dirac quantisation condition, which imposes $m_i - n_j \in \BZ$ for all $i,j$. (The fact that $m_2=0$ requires $n_j$ to be integers, which in turn requires $m_1$ to be an integer.) Moreover, we emphasise that, for $n$ odd, the phase $(-1)^{n(n_1+n_2)}$ is crucial, while it becomes trivial for even $n$.

The Hilbert series with the fugacities $g$, $\zeta$ and $\chi$ turned on can then be computed as in \eref{fullCBHSSO2NwN_f}: 
\bes{
&H_{\left[\SO(4) \times \USp(4)\right]/\BZ_2, \, n}(t; g; \zeta; \chi) \\
&= \frac{1}{2} \sum_{s = \pm 1} H_{[\SO(4) \times \USp(4)]/\BZ_2, \, n}(t; g; \zeta; \chi=s) \left(1+ s \chi \right)~.
}
This result perfectly matches the Coulomb branch limit $\sum_{p} C(a^{-2p} x^p) t^{2p}$ obtained from the coefficients of the superconformal index previously calculated in \eref{fixformindex} and \eref{indSO4USp4modZ2}. 

These formulae provide an efficient method for computing closed-form expressions for the Coulomb branch Hilbert series. Such closed forms are valuable for verifying expected geometric properties of the Coulomb branch moduli space compatible with $\CN=4$ supersymmetry. As a specific example, the Hilbert series for the $[\O(4)^- \times \USp(4)]/\BZ_2$ theory with $n=3$, setting fugacities to unity is:
\bes{ \label{HSO4minusUSp4modZ2}
&H_{\left[\O(4)^- \times \USp(4)\right]/\BZ_2, \, n=3}(t; g=1; \zeta=1) \\
&= \frac{1}{(1-t)^8 (1+t)^8 (1+t^2)^4 (1-t+t^2) (1+t+t^2) (1+t^4)^2} \\
& \times \,\, \frac{1}{(1-t+t^2-t^3+t^4) (1+t+t^2+t^3+t^4) (1+t^8)} \\
& \times \Big( 1 - 2t^2 + 2t^4 + 7t^8 - 3t^{10} + 12t^{12} + 20t^{16} - t^{18} + 24t^{20} + 4t^{22} \\
& \qquad + 24t^{24} - t^{26} + 20t^{28} + 12t^{32} - 3t^{34} + 7t^{36} + 2t^{40} - 2t^{42} + t^{44} \Big) \\
&= 1 + 3 t^4 + 3 t^6 + 16 t^8 + 23 t^{10} + 65 t^{12} +\ldots~.
}
The numerator is indeed palindromic and the Hilbert series has a pole at $t=1$ of order eight. This is as expected for a normal hyperK\"ahler variety of quaternionic dimension four. Let us also provide further examples for other gauge groups:
\begin{subequations}
\begin{align}
\begin{split} 
&H_{[\SO(4) \times \USp(4)]/\BZ_2, \, n=3}(t; g = 1; \zeta=1; \chi=1) \\
&= \frac{1}{(1-t)^8 (1+t)^8 (1+t^2)^4 (1-t+t^2) (1+t+t^2) (1+t^4)} \\
& \times \frac{1}{(1-t+t^2-t^3+t^4) (1+t+t^2+t^3+t^4)} \times \Big( 1 - 2t^2 + 4t^4 - t^6 \\
&\quad  +8t^8 - t^{10} + 12t^{12} + 2t^{14} + 16t^{16} + 2t^{18} +\text{palindrome} + t^{32} \Big) \\
& = 1 + 6 t^4 + 6 t^6 + 31 t^8 + 46 t^{10} + 130 t^{12}+\ldots~, 
\end{split} \\
\begin{split} 
&H_{\SO(4) \times \USp(4), \, n=3}(t; \zeta=1; \chi=1) \\
&=\frac{1}{(1-t)^8 (1+t)^8 (1+t^2)^4 (1-t+t^2) (1+t+t^2) (1+t^4)^3 (1-t^2+t^4)} \\
&\times  \frac{1}{(1-t+t^2-t^3+t^4) (1+t+t^2+t^3+t^4)} \times \Big( 1 - 3t^2 + 7t^4 - 11t^6 \\
&\quad +  22t^8 - 25t^{10} + 42t^{12} - 38t^{14} + 63t^{16} - 48t^{18}  \\
& \quad + 77t^{20}- 50t^{22} + 77t^{24} + \text{palindrome} + t^{44} \Big)\\
&= 1 + 4 t^4 + 2 t^6 + 19 t^8 + 22 t^{10} + 72 t^{12} +\ldots~.
\end{split}
\end{align}
\end{subequations}
As a further consistency check, we can compute the relative volume of the base of the Coulomb branch cone using the formula from \cite[(1.17)]{Martelli:2006yb} (see also \cite{Hanany:2018dvd} for applications involving discrete gauging). For a Hilbert series $H(t)$ corresponding to a cone $Y$ of complex dimension $d$, this relative volume $\mathrm{Vol}(Y)$ is proportional to $\lim_{t \rightarrow 1} (1-t)^d H(t)$. Applying this to our results (where $d=8$):
\bes{ \label{ratiovolume}
\lim_{t \rightarrow 1} \frac{H_{[\SO(4) \times \USp(4)]/\BZ_2, \, n=3}(t; g=1; \zeta=1; \chi=1)}{H_{\SO(4) \times \USp(4), \, n=3}(t; \zeta=1; \chi=1)} &= 2~, \\
\lim_{t \rightarrow 1} \frac{H_{[\O(4)^- \times \USp(4)]/\BZ_2, \, n=3}(t; g=1; \zeta=1)}{H_{\SO(4) \times \USp(4), \, n=3}(t; \zeta=1; \chi=1)} &= 1~.
}
These ratios ($2$ for gauging $\BZ_2^{[1]}$ starting from $\SO(4) \times \USp(4)$, $1$ for the $\O(4)^-$ case relative to $\SO(4) \times \USp(4)$ after gauging) match the expected values based on the discrete gauging procedure.

Finally, let us revisit the significance of the phase factor ${\blue (-1)^{n \sum n_j}}$ in the $\chi=-1$ prescription \eref{HSSO2NN_fchi-1}. If this phase were omitted, the calculation for the $[\SO(4) \times \USp(4)]/\BZ_2$ theory ($n=3$) would yield the following leading terms in the Hilbert series:
\bes{ \label{HSwithoutphase}
1+ \left[3+\chi + \frac{1}{2} g (1+ \chi)+\frac{1}{2} g (\zeta+\zeta \chi) \right] t^4+ \ldots~.
}
It is instructive to compare this with the operator contributions listed in Table \eref{termsata-4x2}. The terms proportional to $g$ in \eref{HSwithoutphase} originate from the operators listed in the last two rows of Table \eref{termsata-4x2}. Omitting the phase effectively amounts to assigning charge conjugation parity in a different way. For example, it would imply that the operator $V_{(0,0;1,0)}$ is even under charge conjugation (independent of $\chi$). As a result, the term $3$ in \eref{HSwithoutphase} would correspond to $\tr(\phi_D^2)$, $\tr(\phi_C^2)$ and $V_{(0,0;1,0)}$, and the term $\chi$ in \eref{HSwithoutphase} would correspond to $\mathrm{Pf}(\phi_D)$. Moreover, in this case, the Coulomb branch Hilbert series for the $\O(4)^+ \times \USp(4)$ gauge theory would be as follows:
\bes{
&\sum_{m_1 \geq m_2 \geq 0} \,\, \sum_{n_1 \geq n_2 \geq 0} t^{2 \Delta(\vec m; \vec n)} P_{\USp(4)} (t; m_1, m_2) P_{\USp(4)} (t; n_1, n_2) \\
&= 1 + 3 t^4 + 2 t^6 + 14 t^8 + 15 t^{10} + 47 t^{12}+\ldots~.
}
This result corresponds to a naive application of the prescription from \cite[Appendix A]{Cremonesi:2014uva} without the necessary phase correction needed when background fluxes (or dynamical fields transforming under flavour symmetries that are gauged, like in the quiver) are present. As this formula disagrees with the correct Coulomb branch limit obtained from the index \eref{indO4plusUSp4}, it underscores the necessity of the proposed phase factor correction. We will not pursue this further.

\subsubsection{Quivers related to $\bar{\mathrm{min}\, \so(6)}$ and $\bar{\mathrm{n.min}\, \so(5)}$}
Let us consider the 3d $\CN=4$ $\USp(2)$ gauge theory with $N_f=3$ flavours of fundamental hypermultiplets. The Higgs branch of this theory is known to be the closure of the minimal nilpotent orbit of the Lie algebra $\mathfrak{so}(6)$, denoted $\overline{\text{min}\,\mathfrak{so}(6)}$, while its Coulomb branch is the orbifold $\BC^2/\BZ_4$.  This theory can be represented by the following quiver diagrams:
\bes{ \label{USp2SO6}
\begin{tikzpicture}[baseline, font=\footnotesize]
\node[gauge,label=below:{$\USp(2)$}] (gg) at (0,0) {};
\node[flavour,label=below:{$\SO(6)$}] (ff) at (2,0) {};
\draw[-,bend right=0] (gg) to (ff);
\end{tikzpicture} 
\quad \text{or equivalently}  \quad
\begin{tikzpicture}[baseline, font=\footnotesize]
\node[flavour,label=below:{$\underset{{\chi_1}}{\SO(1)}$}] (ff1) at (-2,0) {};
\node[gauge,label=below:{$\USp(2)$}] (gg) at (0,0) {};
\node[flavour,label=below:{$\underset{{\chi_2}}{\SO(5)}$}] (ff2) at (2,0) {};
\draw[-,bend right=0] (ff1)--(gg)--(ff2);
\end{tikzpicture} 
}
where $\chi_1, \chi_2$ represent fugacities for the charge conjugation symmetries associated with the $\so(1)$ and $\so(5)$ flavour symmetry factors, respectively. Henceforth, we denote these zero-form symmetries as $\BZ^{[0]}_{2, \chi_1}$ and $\BZ^{[0]}_{2, \chi_2}$.

The index of this theory is given by (see \cite[(4.11)]{Giacomelli:2024sex}):
\bes{\label{indexUSp2w3flv}
\scalebox{0.96}{$
\begin{split}
&\CI_{\eref{USp2SO6}}(x; y_1, y_2; \chi_1, \chi_2; \delta) \\
&= \frac{1}{2}  \sum_{l \in \mathbb{Z}} {\brown (\chi_1 \chi_2)^{l \delta}} \oint \frac{dz}{2\pi i z} \CZ_{\text{vec}}^{\text{USp}(2)}(x;z; l) \prod_{s=0, \pm 1} \CZ_{\text{chir}}^{1}(x;a^{-2} z^{2s} ; 2sl) \\
&\quad \times \prod_{s = \pm 1} \CZ_{\text{chir}}^{1/2}(x;a \chi_{1} z^{s} ; sl) \CZ_{\text{chir}}^{1/2}(x;a \chi_{2} z^{s}; sl) \times {\claret \prod_{i=1}^{2} \prod_{s_1, s_2 = \pm 1} \CZ_{\text{chir}}^{1/2}(x;a z^{s_1} y_i^{s_2}; s_1 l)}~.
\end{split}
$}
}
The phase factor ${\brown (\chi_1 \chi_2)^{l \delta}}$, highlighted in brown, represents a potential modification to the standard index formula which, to our knowledge, has not been explicitly considered previously in the literature. We introduce a parameter $\delta \in \{0, 1\}$ to control the inclusion ($\delta=1$) or exclusion ($\delta=0$) of this phase. The case $\delta = 0$ corresponds to the index studied in \cite{Giacomelli:2024sex, Grimminger:2024mks}. We will show that, while this phase does not alter the Higgs branch limit of the index, it significantly impacts the Coulomb branch limit.

Expanding the index up to order $x^2$ (and setting the $\so(5)$ Cartan fugacities $y_i=1$ for simplicity), we find:
\bes{ \label{SO1USp2SO5delta0}
\scalebox{0.99}{$
\begin{split}
&\CI_{\eref{USp2SO6}}(x; y_i=1; \chi_1, \chi_2; \delta = 0)\\
& = 1+ \left[\chi_1 \chi_2 a^{-2} + (6+4\chi_1+4\chi_2+\chi_1 \chi_2) a^2  \right] x + \Big[ (2+ \chi_1 \chi_2 )a^{-4} \\
& \quad \,\,\,\  + (29 + 20 \chi_1  + 20 \chi_2 + 15 \chi_1 \chi_2 ) a^4 - \left(7 + 4 \chi_1 + 4 \chi_2+2\chi_1 \chi_2\right)\Big] x^2 + \ldots~,
\end{split}
$}
}
\bes{ \label{SO1USp2SO5delta1}
&\CI_{\eref{USp2SO6}}(x; y_i=1; \chi_1, \chi_2; \delta = 1)\\
& = 1 +  \left[ a^{-2} + (6 + 4\chi_1 + 4\chi_2 + \chi_1\chi_2)a^2  \right] x + \Big[ 3 a^{-4}  \\
& \quad \,\,\,\ + (29 + 20\chi_1 + 20\chi_2 + 15\chi_1\chi_2)a^4 -\left(8 + 4\chi_1 + 4\chi_2+ \chi_1\chi_2 \right) \Big] x^2+ \ldots~.
}
Observe the coefficient of the $a^{-2}x$ term, corresponding to the contribution from the minimal monopole operator of the $\USp(2)$ gauge group. Its dependence on $\chi_1, \chi_2$ differs between the $\delta=0$ and $\delta=1$ cases. Specifically, for $\delta=1$, this operator contributes $a^{-2}x$, independent of the charge conjugation fugacities $\chi_1, \chi_2$ in this formula (\ie it is even under $\BZ^{[0]}_{2, \chi_1}$ and $\BZ^{[0]}_{2, \chi_2}$). In the subsequent analysis, we will investigate the consequences of gauging various discrete symmetries, considering three primary scenarios:
\ben
\item Gauging $\BZ^{[0]}_{2, \chi_1}$ (equivalent to gauging $\BZ^{[0]}_{2, \chi_2}$).
\item Gauging the diagonal subgroup $\diag(\BZ^{[0]}_{2, \chi_1} \times \BZ^{[0]}_{2, \chi_2})$ of $\BZ^{[0]}_{2, \chi_1} \times \BZ^{[0]}_{2, \chi_2}$.
\item Gauging $\diag(\BZ^{[0]}_{2, \chi_1} \times \BZ^{[0]}_{2, \chi_2})$ and another $\BZ_2^{[0]}$ symmetry, which will be specified later.
\een

\subsubsection*{Option 1: Gauging $\BZ^{[0]}_{2, \chi_1}$}
We now consider gauging the $\BZ^{[0]}_{2, \chi_1}$ charge conjugation symmetry associated with the $\so(1)$ flavour factor in \eref{USp2SO6}. This procedure effectively promotes the $\so(1)$ flavour node to an $\O(1)$ gauge node, resulting in the following gauge theory:
\bes{ \label{O1USp2SO5}
\begin{tikzpicture}[baseline, font=\footnotesize]
\node[gauge,label=below:{$\O(1)$}] (ff1) at (-2,0) {};
\node[gauge,label=below:{$\USp(2)$}] (gg) at (0,0) {};
\node[flavour,label=below:{$\SO(5)$}] (ff2) at (2,0) {};
\draw[-,bend right=0] (ff1)--(gg)--(ff2);
\end{tikzpicture} 
}
At the level of the superconformal index, this gauging corresponds to averaging the index \eref{indexUSp2w3flv} over $\chi_1 = \pm 1$. With $y_i=1$, the results for $\delta=0$ and $\delta=1$ are:
\bes{
&\CI_{\eref{O1USp2SO5}}(x; y_i=1; \chi_2; \delta = 0)\\
& = 1+ \left(6+4\chi_2\right) a^2 x + \Big[ 2 a^{-4} + (29 + 20 \chi_2 ) a^4 - \left(7+ 4 \chi_2 \right)\Big] x^2 + \ldots~,
}
\bes{
\scalebox{0.99}{$
\begin{split}
&\CI_{\eref{O1USp2SO5}}(x; y_i=1; \chi_2; \delta = 1)\\
& = 1 +  \left[ a^{-2} + (6 + 4\chi_2 )a^2  \right] x + \Big[ 3 a^{-4} + (29 + 20\chi_2 )a^4 -\left(8 + 4\chi_2\right) \Big] x^2+ \ldots~.
\end{split}$}
}
Setting the remaining charge conjugation fugacity $\chi_2=1$ yields the unrefined indices:
\bes{ \label{unrefindexO1USp2SO5}
\scalebox{0.99}{$
\begin{split}
\CI_{\eref{O1USp2SO5}}(x; y_i=1; \chi_2=1; \delta = 0)&=1 + 10 a^2 x + (2 a^{-4}  + 49 a^4 - 11) x^2\\& \quad \,\,\,\ + (a^{-6} - a^{-2} - 93 a^2 + 165 a^6) x^3 +\ldots~, \\
\CI_{\eref{O1USp2SO5}}(x; y_i=1; \chi_2=1; \delta = 1)&= 1 + (a^{-2} + 10 a^2) x + (3 a^{-4} + 49 a^4- 12) x^2 \\
&\quad \,\,\,\ + (3 a^{-6} - 2 a^{-2}  - 93 a^2 + 165 a^6) x^3  + \ldots~.
\end{split}$}
}
The Higgs branch limit in both cases is the Hilbert series of the closure of the next to minimal orbit of $\so(5)$, donoted $\overline{\text{n.min}\,\mathfrak{so}(5)}$. The Coulomb branch of this theory, on the other hand, depends on the value of $\delta$:
\begin{equation} \label{HSO1USp2SO5}
\text{Coulomb branch of \eqref{O1USp2SO5}} =
\begin{cases}
\BC^2/\hat{D}_4~, &\quad \delta =0 \\
\BC^2/\BZ_4~, &\quad \delta =1
\end{cases}~.
\end{equation}
Note that, for the choice of $\delta=0$, we have the following web of dualities:
\bes{
\scalebox{0.89}{$
\begin{split}
\delta=0: \qquad &\begin{tikzpicture}[baseline, font=\footnotesize]
\node[gauge,label=below:{$\O(1)$}] (ff1) at (-2,0) {};
\node[gauge,label=below:{$\USp(2)$}] (gg) at (0,0) {};
\node[flavour,label=below:{$\SO(5)$}] (ff2) at (2,0) {};
\draw[-,bend right=0] (ff1)--(gg)--(ff2);
\end{tikzpicture}  \quad \overset{\text{dual}}{\longleftrightarrow} \quad \begin{tikzpicture}[baseline, font=\footnotesize]
\node[gauge,label=below:{$\O(2)$}] (gg) at (0,0) {};
\node[flavour,label=below:{$\USp(4)$}] (ff) at (2,0) {};
\draw[-,bend right=0] (gg)--(ff);
\end{tikzpicture} \\
&\overset{\text{mirror}}\longleftrightarrow  \hspace{-2cm}
\begin{tikzpicture}[baseline]
\node[flavour,label=below:{$1$}] (ffl) at (-4,0) {};
\node[gauge,label=below:{$1$}] (gg1) at (-2,0) {};
\node[gauge,label={below,xshift=0.2cm}:{$1$}] (gg2) at (0,0) {};
\node[gauge,label=below:{$1$}] (gg3) at (2,0) {};
\node[flavour,label=below:{$1$}] (ffr) at (4,0) {};
\draw[-,bend right=0] (ffl)--(gg1)--(gg2)--(gg3)--(ffr);
\draw[blue, dashed] (0,1.5)--(0,-1.5);
\draw [<->,red] (ffl) to [out=-150,in=-30,looseness=1] (ffr);
\end{tikzpicture}
\end{split}
$}
}
where the double-headed {\red red} arrow denotes wreathing \cite{Bourget:2020bxh} (see also \cite{Arias-Tamargo:2021ppf,Giacomelli:2024sex,Grimminger:2024mks,Lawrie:2025exx}) with respect to the vertical axis of symmetry (indicated by the {\blue blue} dashed line), the index computation for which was detailed in \cite{Grimminger:2024mks}. As pointed out in Section 4.1 of this reference, the two theories in the first line (related by duality) share the same index for $\delta=0$. Furthermore, this index matches that of the unitary wreathed quiver shown in the second line upon exchanging $a \leftrightarrow a^{-1}$, which is consistent with the expected mirror symmetry.

More generally, we can consider the theory
\bes{ \label{O1USp2SO2n+1}
\begin{tikzpicture}[baseline, font=\footnotesize]
\node[gauge,label=below:{$\O(1)$}] (ff1) at (-2,0) {};
\node[gauge,label=below:{$\USp(2)$}] (gg) at (0,0) {};
\node[flavour,label=below:{$\SO(2n+1)$}] (ff2) at (2,0) {};
\draw[-,bend right=0] (ff1)--(gg)--(ff2);
\end{tikzpicture} 
}
The Higgs branch is isomorphic to $\overline{\text{n.min}\,\mathfrak{so}(2n+1)}$, and the Coulomb branch depends on $\delta$: 
\begin{equation} 
\text{Coulomb branch of \eqref{O1USp2SO2n+1}} =
\begin{cases}
\BC^2/\hat{D}_{2n}~, &\quad \delta =0 \\
\BC^2/\hat{D}_{n+1}~, &\quad \delta =1
\end{cases}~.
\end{equation}
The reasoning for the differing Coulomb branches is as follows (see also the argument above \cite[(4.25)]{Giacomelli:2024sex}). Consider first the $\USp(2)$ gauge theory with $n+1$ flavours. For $\delta=1$, the minimal monopole operator $\mathfrak{M}$ of the $\USp(2)$ gauge group is neutral under both $\BZ^{[0]}_{2, \chi_1}$ and $\BZ^{[0]}_{2, \chi_2}$. Gauging $\BZ^{[0]}_{2, \chi_1}$, which creates the $\O(1)$ node in \eqref{O1USp2SO2n+1}, does not affect the Coulomb branch moduli space; therefore, it remains $\BC^2/\hat{D}_{n+1}$. The three generators of $\BC^2/\hat{D}_{n+1}$ are as follows: (1) the Casimir operator $\tr(\varphi^2)$ (order $4$ in the convention of \cite{Benvenuti:2010pq}), where $\varphi$ is the scalar in the $\USp(2)$ vector multiplet; (2) the monopole operator $\mathfrak{M}$ (order $2n-2$); and (3) the dressed monopole operator $G_{2n} = \mathfrak{M} \varphi$ (order $2n$). They satisfy the following defining equation of $\BC^2/\hat{D}_{n+1}$: 
\bes{
G_{2n}^2 +\mathfrak{M}^2 \tr(\varphi^2) = [\tr(\varphi^2)]^{n}~.
}
In contrast, for $\delta=0$, the contribution of $\mathfrak{M}$ in the index carries the fugacity $\chi_1 \chi_2$. Consequently, when gauging $\BZ_{2, \chi_1}^{[0]}$, the operators $\mathfrak{M}$ and $\mathfrak{M} \varphi$ are projected out as they are not invariant (they transform non-trivially under $\chi_1$). However, $\mathfrak{M}^2$ and $\mathfrak{M}^2 \varphi$ are invariant. Upon gauging $\BZ_{2, \chi_1}^{[0]}$, the effective generators become $\tr(\varphi^2)$ and $\mathfrak{M}^2$, along with a dressed operator $G_{4n-2} \equiv \mathfrak{M}^2 \varphi$. The defining relation transforms into:
\bes{
G_{4n-2}^2 +\mathfrak{M}^4 \tr(\varphi^2) = [\tr(\varphi^2)]^{2n-1}~, \quad \text{with $G_{4n-2} = \mathfrak{M}^2 \varphi$}~.
}
This is indeed the defining relation for $\BC^2/\hat{D}_{2n}$.

\subsubsection*{Option 2: Gauging the diagonal subgroup $\diag(\BZ^{[0]}_{2, \chi_1} \times \BZ^{[0]}_{2, \chi_2})$}
We now gauge the diagonal subgroup of $\BZ^{[0]}_{2, \chi_1} \times \BZ^{[0]}_{2, \chi_2}$.  
The corresponding index can be obtained as
\bes{ \label{gaugediagchi1}
& \frac{1}{2} \left[ \CI_{\eref{USp2SO6}}(x; y_i=1; \chi_1= \chi_2 =1 ; \delta)+ \CI_{\eref{USp2SO6}}(x; y_i=1; \chi_1 =\chi_2 =-1 ; \delta) \right] \\
&=1 + (a^{-2} + 7 a^2) x + (3 a^{-4} + 44 a^4-9) x^2 \\& \quad \,\,\,\,\, + (3 a^{-6} - 2 a^{-2}- 85 a^2 + 148 a^6) x^3 +\ldots~.
}
Note that the result is independent of $\delta$. In fact, \eref{gaugediagchi1} is the index of the following gauge theory:
\begin{equation} \label{O1=USp2SO4}
\begin{tikzpicture}[baseline, font=\footnotesize]
\node[gauge,label=below:{$\USp(2)$}] (C1) at (0,0) {};
\node[gauge,label=below:{$\O(1)$}] (O1) at (-2,0) {};
\node[flavour,label=below:{$\SO(4)$}] (SO4) at (2,0) {};
\draw[-,bend left=20] (O1) to (C1);
\draw[-,bend right=20] (O1) to (C1);
\draw[-,bend right=0] (C1) to (SO4);
\end{tikzpicture}
\end{equation}
This theory has an $\so(2) \oplus \so(4)$ continuous flavour symmetry, along with an $\so(2)$ magnetic (topological) symmetry. The latter is present due to the fact that the $\USp(2)$ gauge group has $3$ flavours of hypermultiplets in the fundamental representation and is therefore balanced \cite[Section 5.1]{Gaiotto:2008ak}. The Coulomb branch moduli space is $\BC^2/\BZ_4$. Note that the independence of $\delta$ of \eref{gaugediagchi1} is due to the presence of two bifundamental half-hypermultiplets under $\O(1) \times \USp(2)$ in \eref{O1=USp2SO4}.


\subsubsection*{Option 3: Gauging $\diag(\BZ^{[0]}_{2, \chi_1} \times \BZ^{[0]}_{2, \chi_2})$ and another $\BZ_2^{[0]}$ symmetry}
Theory \eref{O1=USp2SO4} can also be rewritten as
\begin{equation} \label{O1=USp2SO2SO2}
\begin{tikzpicture}[baseline, font=\footnotesize]
\node[gauge,label=below:{$\USp(2)$}] (C1) at (0,0) {};
\node[gauge,label=below:{$\O(1)$}] (O1) at (-2,0) {};
\node[flavour,label=above:{$\SO(2)$}] (SO2) at (0,1) {};
\node[flavour,label=right:{$\underset{{\chi_3}}{\SO(1)}$}] (SO1u) at (2,1) {};
\node[flavour,label=right:{$\underset{{\chi_4}}{\SO(1)}$}] (SO1d) at (2,-1) {};
\draw[-,bend left=20] (O1) to (C1);
\draw[-,bend right=20] (O1) to (C1);
\draw[-,bend right=0] (C1) to (SO2);
\draw[-,bend right=0] (C1) to (SO1u);
\draw[-,bend right=0] (C1) to (SO1d);
\end{tikzpicture}
\end{equation}
where we split the $\so(4)$ flavour symmetry into an $\so(2)$ and two $\so(1)$ sets, and we turn on the charge conjugation fugacities $\chi_3$ and $\chi_4$ associated with the two $\so(1)$ flavour nodes. The corresponding index can be derived by replacing the term in {\claret claret} in \eref{indexUSp2w3flv} with
\bes{
\prod_{i=3}^4\prod_{s = \pm 1} \CZ_{\text{chir}}^{1/2}(x;a \chi_i z^{s}; s l) \times \prod_{s_1, s_2 = \pm 1} \CZ_{\text{chir}}^{1/2}(x;a z^{s_1} y_1^{s_2}; s_1 l)~.
}
We can now gauge the diagonal subgroup of the $\BZ^{[0]}_{2, \chi_3} \times \BZ^{[0]}_{2, \chi_4}$ symmetry. At the level of the index, this is implemented by setting $\chi_3 = \chi_4 = s$, summing the index of theory \eref{O1=USp2SO2SO2} over $s=\pm 1$, and then dividing by two. Upon setting $y_1 = 1$, this yields
\bes{ \label{gaugewholechi1chi2}
\scalebox{0.92}{$
1 + (a^{-2} + 3 a^2) x + (3 a^{-4}  + 24 a^4 - 5) x^2 + (3 a^{-6} - 2 a^{-2} - 41 a^2 + 72 a^6) x^3+\ldots~,
$}
}
where again this result is independent of $\delta$. This index is equal to that of the following gauge theory:
\begin{equation} \label{O1=USp2=O1}
\begin{tikzpicture}[baseline, font=\footnotesize]
\node[gauge,label=below:{$\USp(2)$}] (C1) at (0,0) {};
\node[gauge,label=below:{$\O(1)$}] (O1l) at (-2,0) {};
\node[gauge,label=below:{$\O(1)$}] (O1r) at (2,0) {};
\node[flavour,label=right:{$\SO(2)$}] (D1u) at (0,1) {};
\draw[-,bend left=20] (O1l) to (C1);
\draw[-,bend right=20] (O1l) to (C1);
\draw[-,bend left=20] (O1r) to (C1);
\draw[-,bend right=20] (O1r) to (C1);
\draw[-,bend right=0] (C1) to (D1u);
\end{tikzpicture}
\end{equation}
This theory has an $\so(2) \oplus \so(2)\oplus \so(2)$ continuous flavour symmetry, along with an $\so(2)$ magnetic (topological) symmetry. The Coulomb branch moduli space is $\BC^2/\BZ_4$.

\subsubsection*{Mirror theory}
Let us now discuss a mirror theory of \eqref{USp2SO6} described by the following quiver diagram (see \eg~\cite{Feng:2000eq}, \cite[Table 2]{Cremonesi:2014uva}):
\bes{ \label{mirrorquiv}
\begin{tikzpicture}[baseline, font = \small]
\node[gauge,label=below:{$\USp(2)$}] (C1) at (0,0) {};
\node[gauge,label=below:{$\underset{\zeta_2, \, c_2}{\SO(2)}$}] (D1r) at (2,0) {};
\node[gauge,label=below:{$\underset{\zeta_1, \, c_1}{\SO(2)}$}] (D1l) at (-2,0) {};
\node[flavour,label=right:{$\SO(2)$}] (D1u) at (0,1) {};
\draw (D1l)--(C1)--(D1r);
\draw (C1)--(D1u);
\end{tikzpicture}
}
Here $(\zeta_1, c_1)$ and $(\zeta_2, c_2)$ denote the fugacities for the magnetic and charge conjugation symmetries associated with the left and right $\SO(2)$ gauge groups, respectively. Initially setting the charge conjugation fugacities $c_1=c_2=1$, the index of \eref{mirrorquiv} is
\bes{ \label{indexmirrorquivc2+1}
&\CI_{\eref{mirrorquiv}}(x; f; m_f; \zeta_1, \zeta_2;c_1 = c_2 =1) \\
&= \frac{1}{2} \sum_{m_u, m_z, m_v \in \BZ} \oint \frac{d u}{2\pi i u}\,\, \frac{d z}{2\pi i z} \,\, \frac{d v}{2\pi i v} \zeta_1^{m_u} \zeta_2^{m_v} \CZ_{\text{vec}}^{\USp(2)}(x;z; m_z) \\
&\quad\,\, \times \CZ_{\text{chir}}^{1}(x;a^{-2};0) \left[\prod_{\ell=-1}^1 \CZ_{\text{chir}}^{1}(x;a^{-2}z^{2 \ell};2 \ell m_z)  \right] \CZ_{\text{chir}}^{1}(x;a^{-2};0) \\
& \quad \,\, \times \prod_{s_1, s_2 = \pm 1}\Big[ \CZ_{\text{chir}}^{1/2}(x; a u^{s_1} z^{s_2}; s_1 m_u + s_2 m_z) \,\, \CZ_{\text{chir}}^{1/2}(x; a v^{s_1} z^{s_2}; s_1 m_v + s_2 m_z) \\
& \hspace{2.1cm} \times \CZ_{\text{chir}}^{1/2}(x; a f^{s_1} z^{s_2}; s_1 m_f + s_2 m_z) \Big]~,
}
whose series expansion, setting $f=1$ and $m_f=0$, reads
\bes{
\scalebox{0.99}{$
\begin{split}
& \CI_{\eref{mirrorquiv}}(x; f=1; m_f=0; \zeta_1, \zeta_2;c_1 = c_2 =1) \\&=  1 + \left[ (3 + 4 \zeta_1 + 4 \zeta_2 + 4 \zeta_1 \zeta_2) a^{-2} +a ^2\right] x + \left[ (24 + 20 \zeta_1 + 20 \zeta_2 + 20 \zeta_1 \zeta_2) a^{-4} \right. \\
& \quad \,\,\,\ \left. + 3 a^4  - (5 + 4 \zeta_1 + 4 \zeta_2 + 4 \zeta_1 \zeta_2) \right] x^2+ \ldots~.
\end{split}
$}
}
On the other hand, the index for \eref{mirrorquiv} when $c_1=1$ and $c_2=-1$ is
\bes{ \label{indexmirrorquivc2-1}
&\CI_{\eref{mirrorquiv}}(x; f; m_f; \zeta_1, \zeta_2;c_1 = 1, c_2 =-1) \\
&= \frac{1}{2} \sum_{\substack{m_u, m_z \in \BZ \\ m_v=0}}\,\, \oint \frac{d u}{2\pi i u}\,\, \frac{d z}{2\pi i z} \,\, \frac{d v}{2\pi i v} \zeta_1^{m_u} \zeta_2^{m_v} \CZ_{\text{vec}}^{\USp(2)}(x;z; m_z) \\
&\quad\,\, \times \CZ_{\text{chir}}^{1}(x;a^{-2};0) \left[\prod_{\ell=-1}^1 \CZ_{\text{chir}}^{1}(x; a^{-2} z^{2 \ell} ;2 \ell m_z)  \right] \CZ_{\text{chir}}^{1}(x;-a^{-2};0) \\
& \quad \,\, \times \prod_{s_1, s_2 = \pm 1}\Big[ \CZ_{\text{chir}}^{1/2}(x; a u^{s_1} z^{s_2}; s_1 m_u + s_2 m_z) \,\, \CZ_{\text{chir}}^{1/2}(x; a v^{s_1} z^{s_2}; s_1 m_v + s_2 m_z) \\
& \hspace{2.1cm} \times \CZ_{\text{chir}}^{1/2}(x; a f^{s_1} z^{s_2}; s_1 m_f + s_2 m_z) \Big]_{v =1, \, v^{-1} = -1 \, m_v =0}~.
}
The calculation can be straightforwardly adapted for the cases $(c_1, c_2)=(-1, 1)$ (by setting $m_u=0$ and evaluating at $u=1, u^{-1}=-1$) and $(c_1, c_2)=(-1, -1)$ (by setting $m_u=m_v=0$ and evaluating at $u=v=1, u^{-1}=v^{-1}=-1$).\footnote{Note that, in the case $(c_1, c_2)=(-1, -1)$, both contributions coming from the adjoint chiral fields in the two $\SO(2)$ vector multiplets are given by $\CZ_{\text{chir}}^{1}(x;-a^{-2};0)$.} Setting $f=1, m_f=0$, the expansion for any case where at least one $c_i=-1$ yields
\bes{
& \CI_{\eref{mirrorquiv}}(x; f=1; m_f=0; \zeta_1, \zeta_2;c_1,c_2) \\&= 1 + \left(-a^{-2} + a^2\right) x + \left(4a^{-4} + 3a^4-1\right) x^2 \\& \quad \,\,\,\ + \left(-4a^{-6} + 3a^{-2} - 2a^2 + 3a^6\right)x^3+\ldots~, 
\quad \text{if $c_1 = -1$ and/or $c_2 = -1$}~.
}
The index, refined with the charge conjugation fugacities $c_1, c_2$, is obtained by summing over the four sectors as follows:
\bes{ \label{fullindexmirrorquiv}
& \CI_{\eref{mirrorquiv}}(x; f=1; m_f=0; \zeta_1, \zeta_2; c_1,c_2)  \\
&= \frac{1}{4} \sum_{s_1, s_2 = \pm 1} \Big[\CI_{\eref{mirrorquiv}}(x; f=1; m_f =0; \zeta_1, \zeta_2; c_1=s_1, c_2=s_2) \\& \qquad \qquad \qquad \times \left(1+s_1 c_1 + s_2 c_2 + s_1 s_2 c_1 c_2\right) \Big] \\
&=1+\left[C(a^{-2} x) a^{-2} + a^2 \right] x +\left[C(a^{-4} x^2) a^{-4} + C(a^0 x^2) + 3a^4 \right] x^2 \\
&\quad \,\,\,\ +\left[C(a^{-6} x^3) a^{-6} + C(a^{-2} x^3) a^{-2} -2a^2 + 3a^6 \right] x^3 +\ldots~. 
}
The Higgs branch limit, given by the coefficients of the term $a^{2p} x^p$, implies that the Higgs branch of \eref{mirrorquiv} is $\BC^2/\BZ_4$. Note that this is independent of $\zeta_i$ or $c_i$, and so does not change upon gauging any charge conjugation or magnetic symmetry. Instead, the other coefficients in \eref{fullindexmirrorquiv} are given by
\bes{ \label{coefffullindexmirrorquiv}
C(a^{-2} x) &= \zeta _1+\zeta _2+\zeta _1 \zeta _2 +\left(c_1+c_2+c_1c_2\right) \left(1+\zeta _1+\zeta _2+\zeta _1 \zeta _2\right)~, \\
C(a^0 x^2) &= -\left[2+C(a^{-2} x) \right]~, \\
C(a^{-4} x^2) &= 9+ 5 C(a^{-2} x)~, \\
C(a^{-6} x^3) &= 15+ 19 C(a^{-2} x)~, \\
C(a^{-2} x^3) &= -\left[8 + 11 C(a^{-2} x)\right]~. \\
}
Let us now examine the gauging of the charge conjugation symmetries in this mirror theory.

\paragraph{Gauging $\BZ^{[0]}_{2, c_1} \times \BZ^{[0]}_{2, c_2}$.} Let us gauge the whole $\BZ^{[0]}_{2, c_1} \times \BZ^{[0]}_{2, c_2}$. The resulting index is
\bes{
&\frac{1}{4} \sum_{c_1, c_2 =\pm 1} \CI_{\eref{mirrorquiv}}(x; f=1; m_f=0; \zeta_1, \zeta_2; c_1,c_2)~.
}
The index is given by \eref{fullindexmirrorquiv}, where now the coefficient $C(a^{-2} x)$ in \eref{coefffullindexmirrorquiv} is changed to $\zeta_1 + \zeta_2 + \zeta_1 \zeta_2$, while the other coefficients take the same form.
Setting $\zeta_i=1$, we obtain
\bes{
\scalebox{0.94}{$
1 + (3a^{-2} + a^2 )x + (24a^{-4}  + 3a^4 - 5 )x^2 + (72a^{-6} - 41a^{-2}  - 2a^2 + 3a^6 )x^3+\ldots~,
$}
}
which is equal to \eref{gaugewholechi1chi2} upon exchanging $a \leftrightarrow a^{-1}$. We thus conclude that
\bes{
\begin{tikzpicture}[baseline, font = \small]
\node[gauge,label=below:{$\USp(2)$}] (C1) at (0,0) {};
\node[gauge,label=below:{${\O(2)}$}] (D1r) at (2,0) {};
\node[gauge,label=below:{${\O(2)}$}] (D1l) at (-2,0) {};
\node[flavour,label=right:{$\SO(2)$}] (D1u) at (0,1) {};
\draw (D1l)--(C1)--(D1r);
\draw (C1)--(D1u);
\end{tikzpicture}
\quad \overset{\text{mirror}}{\longleftrightarrow} \quad
\begin{tikzpicture}[baseline, font=\footnotesize]
\node[gauge,label=below:{$\underset{\eref{O1=USp2=O1}}{\USp(2)}$}] (C1) at (0,0) {};
\node[gauge,label=below:{$\O(1)$}] (O1l) at (-2,0) {};
\node[gauge,label=below:{$\O(1)$}] (O1r) at (2,0) {};
\node[flavour,label=right:{$\SO(2)$}] (D1u) at (0,1) {};
\draw[-,bend left=20] (O1l) to (C1);
\draw[-,bend right=20] (O1l) to (C1);
\draw[-,bend left=20] (O1r) to (C1);
\draw[-,bend right=20] (O1r) to (C1);
\draw[-,bend right=0] (C1) to (D1u);
\end{tikzpicture}
}

\paragraph{Gauging $\BZ^{[0]}_{2, c_1}$ or gauging the diagonal subgroup of $\BZ^{[0]}_{2, c_1} \times \BZ^{[0]}_{2, c_2}$.}
Let us now consider gauging only $\BZ^{[0]}_{2, c_1}$ in \eref{mirrorquiv}. The resulting index is
\bes{
&\frac{1}{2} \sum_{c_1 =\pm 1} \CI_{\eref{mirrorquiv}}(x; f=1; m_f=0; \zeta_1,\zeta_2; c_1,c_2)~.
}
The index is still given by \eref{fullindexmirrorquiv}, with the coefficient $C(a^{-2} x)$ in \eref{coefffullindexmirrorquiv} changed to $\zeta _1+\zeta _2+\zeta _1 \zeta _2 +c_2 \left(1+\zeta _1+\zeta _2+\zeta _1 \zeta _2\right)$, while the other coefficients take the same form. Similarly, gauging the diagonal subgroup of $\BZ^{[0]}_{2, c_1} \times \BZ^{[0]}_{2, c_2}$ leads to the index
\bes{
&\frac{1}{2} \sum_{s =\pm 1} \CI_{\eref{mirrorquiv}}(x; f=1; m_f=0; \zeta_1, \zeta_2; c_1=c_2=s)~,
}
which can also be expressed as \eref{fullindexmirrorquiv}, with the coefficient $C(a^{-2} x)$ in \eref{coefffullindexmirrorquiv} changed to $1+ 2(\zeta _1+\zeta _2+\zeta _1 \zeta _2)$, while the other coefficients take the same form.  In both cases, if we set all the fugacities to unity, we obtain
\bes{
\scalebox{0.93}{$
1 + (7a^{-2} + a^2)x + (44a^{-4} + 3a^4 - 9)x^2 + (148a^{-6} - 85a^{-2} - 2a^2 + 3a^6)x^3+\ldots~.
$}
}
This agrees with \eref{gaugediagchi1} upon exchanging $a \rightarrow a^{-1}$. We thus conclude that
\bes{
&\begin{tikzpicture}[baseline, font = \small]
\node[gauge,label=below:{$\USp(2)$}] (C1) at (0,0) {};
\node[gauge,label=below:{${\SO(2)}$}] (D1r) at (2,0) {};
\node[gauge,label=below:{${\O(2)}$}] (D1l) at (-2,0) {};
\node[flavour,label=right:{$\SO(2)$}] (D1u) at (0,1) {};
\draw (D1l)--(C1)--(D1r);
\draw (C1)--(D1u);
\end{tikzpicture}
\qquad \overset{\text{dual}}{\longleftrightarrow} \qquad
\begin{tikzpicture}[baseline, font = \small]
\node[gauge,label=below:{$\USp(2)$}] (C1) at (0,0) {};
\node[gauge,label=below:{${\blue \mathrm{O}(2)}$]}] (D1r) at (2,0) {};
\node[gauge,label=below:{\blue [${\mathrm{O}(2)}$}] (D1l) at (-2,0) {};
\node[flavour,label=right:{$\SO(2)$}] (D1u) at (0,1) {};
\draw (D1l)--(C1)--(D1r);
\draw (C1)--(D1u);
\end{tikzpicture} \\
& \hspace{2cm} \overset{\text{mirror}}{\longleftrightarrow} \qquad
\begin{tikzpicture}[baseline, font=\footnotesize]
\node[gauge,label=below:{$\underset{\eref{O1=USp2SO4}}{\USp(2)}$}] (C1) at (0,0) {};
\node[gauge,label=below:{$\O(1)$}] (O1) at (-2,0) {};
\node[flavour,label=below:{$\SO(4)$}] (SO4) at (2,0) {};
\draw[-,bend left=20] (O1) to (C1);
\draw[-,bend right=20] (O1) to (C1);
\draw[-,bend right=0] (C1) to (SO4);
\end{tikzpicture}
}
where the notation {\blue $[\mathrm{O}(2)$}, {\blue $\mathrm{O}(2)]$} denotes the theory that results from gauging the diagonal subgroup of $\BZ^{[0]}_{2, c_1} \times \BZ^{[0]}_{2, c_2}$.

\paragraph{What about a mirror theory of \eref{O1USp2SO5}?}
Let us return to the index \eref{unrefindexO1USp2SO5} calculated for theory \eref{O1USp2SO5}. A candidate mirror dual to this theory must reproduce the index \eref{unrefindexO1USp2SO5} under the standard mirror map exchange $a \leftrightarrow a^{-1}$. However, the structure of the coefficient $C(a^{-2}x)$ in the mirror quiver index \eref{fullindexmirrorquiv} (before gauging $c_i$) reveals that its dependence on $c_1, c_2$ is such that gauging any combination of these $\BZ_2$ symmetries cannot yield the coefficient $10$ required to match the $a^2 x$ term in \eref{unrefindexO1USp2SO5} (which maps to $10 a^{-2} x$ under $a \leftrightarrow a^{-1}$), since the equation $x+y(1+x)=10$ does not admit any integer solutions such that $0\leq x \leq 3$ and $0 \leq y \leq 3$.

Intriguingly, if one were to perform a calculation omitting the crucial phase factors discussed earlier (analogous to the issue highlighted for the Hilbert series prescription \eref{HSSO2NN_fchi-1}), one might hypothetically calculate the Coulomb branch Hilbert series for quiver
\bes{\label{mirrSO2USp2O2}
\begin{tikzpicture}[baseline, font = \small]
\node[gauge,label=below:{$\USp(2)$}] (C1) at (0,0) {};
\node[gauge,label=below:{${\O(2)}$}] (D1r) at (2,0) {};
\node[gauge,label=below:{${\SO(2)}$}] (D1l) at (-2,0) {};
\node[flavour,label=right:{$\SO(2)$}] (D1u) at (0,1) {};
\draw (D1l)--(C1)--(D1r);
\draw (C1)--(D1u);
\end{tikzpicture}
}
as follows:\footnote{This calculation resembles a naive application of methods from \cite[Appendix A]{Cremonesi:2014uva} without necessary phase corrections, particularly relevant when dealing with background fluxes or gauged flavour symmetries.}
\bes{ \label{indexnaive}
&\sum_{m_u \in\BZ}\, \sum_{m_z \in \BZ_{\geq 0}} \, \sum_{m_v \in\BZ_{\geq 0}} \,\, \zeta_1^{m_u} \zeta_2^{m_v}\,\, t^{\sum_{s = \pm 1} \left( |s m_z|+|m_u+sm_z|+|m_v+sm_z| \right)-2|2m_z|} \\
& \qquad \qquad \times  \frac{1}{1-t^2} P_{\USp(2)}(t; m_z) P_{\USp(2)}(t; m_v)  \\
&= 1 + 10 t^2 + 49 t^4 + 165 t^6 + 440 t^8 + 1001 t^{10}+\ldots~, \,\, \text{for $\zeta_1=\zeta_2=1$}~.
}
This resulting series remarkably matches the Hilbert series of $\overline{\text{n.min}\,\mathfrak{so}(5)}$, which is indeed the Higgs branch of theory \eref{O1USp2SO5}. However, as previously established, omitting necessary phase factors leads to results inconsistent with the superconformal index and known dualities, rendering such calculations unreliable for establishing mirror symmetry.

This prompts a different question: can we {\it ad hoc} modify the index computation for the mirror quiver \eref{mirrorquiv} itself, by inserting a phase factor analogous to our proposed correction, to achieve a better match? Specifically, let us tentatively modify the expression for the $c_2=-1$ sector, based on \eref{indexmirrorquivc2-1}, by inserting a phase ${\blue (-1)^{m_z}}$:
\bes{ \label{indexmirrorquivc2-1xxxx}
&\hat{\CI}_{\eref{mirrorquiv}}(x; f; m_f; \zeta_1, \zeta_2;c_1 = 1, c_2 =-1) \\
&= \frac{1}{2} \sum_{\substack{m_u, m_z \in \BZ \\ m_v=0}}\,\, {\blue (-1)^{m_z}} \,\, \oint \frac{d u}{2\pi i u}\,\, \frac{d z}{2\pi i z} \,\, \frac{d v}{2\pi i v} \zeta_1^{m_u} \zeta_2^{m_v} \CZ_{\text{vec}}^{\USp(2)}(x; z; m_z) \\
&\quad\,\, \times \CZ_{\text{chir}}^{1}(x;a^{-2};0) \left[\prod_{\ell=-1}^1 \CZ_{\text{chir}}^{1}(x;a^{-2} z^{2 \ell};2 \ell m_z)  \right] \CZ_{\text{chir}}^{1}(x;-a^{-2};0) \\
& \quad \,\, \times \prod_{s_1, s_2 = \pm 1}\Big[ \CZ_{\text{chir}}^{1/2}(x; a u^{s_1} z^{s_2}; s_1 m_u + s_2 m_z) \,\, \CZ_{\text{chir}}^{1/2}(x; a v^{s_1} z^{s_2}; s_1 m_v + s_2 m_z) \\
& \hspace{2.1cm} \times \CZ_{\text{chir}}^{1/2}(x; a f^{s_1} z^{s_2}; s_1 m_f + s_2 m_z) \Big]_{v =1, \, v^{-1} = -1, \, m_v =0}~,
}
while keeping the calculation for the $c_1=c_2=1$ sector \eref{indexmirrorquivc2+1} unchanged. The index for \eref{mirrSO2USp2O2} would then be
\bes{ \label{mirrquivindxxxx}
\frac{1}{2}[\eref{indexmirrorquivc2+1}+\eref{indexmirrorquivc2-1xxxx}]_{\zeta_1=\zeta_2=1} =1 &+ (10a^{-2} + a^2)x + (49a^{-4} + 3a^4 - 10)x^2\\
&  + (165a^{-6} -92a^{-2}  - 5a^2 + 3a^6)x^3+\ldots~.
}
Remarkably, the Coulomb branch limit and the Higgs branch limit of this modified index \eref{mirrquivindxxxx} now match the Higgs branch and the Coulomb branch limits, given by $\overline{\text{n.min}\,\mathfrak{so}(5)}$ and $\BC^2/\BZ_4$ respectively, of the original theory index \eref{unrefindexO1USp2SO5} with $\delta=1$. Despite this agreement in the limits, the full indices themselves are {\it not} related by the simple mirror map $a \leftrightarrow a^{-1}$. A detailed comparison of the expansions of \eref{mirrquivindxxxx} and \eref{unrefindexO1USp2SO5} (with $\delta=1$) reveals the presence of two additional marginal operators that are neither Higgs nor Coulomb branch operators.

\paragraph{What about a mirror theory of the $\BZ^{[0]}_{2, \chi_1} \times \BZ^{[0]}_{2, \chi_2}$ gauging of \eqref{USp2SO6}?} Naively, one would propose the candidate mirror theory to be as follows:
\bes{\label{mirrO2USp2O2A}
\begin{tikzpicture}[baseline, font = \small]
\node[gauge,label=below:{$\USp(2)$}] (C1) at (0,0) {};
\node[gauge,label=below:{${\O(2)}$}] (D1r) at (2,0) {};
\node[gauge,label=below:{${\O(2)}$}] (D1l) at (-2,0) {};
\node[flavour,label=right:{$\SO(2)$}] (D1u) at (0,1) {};
\draw (D1l)--(C1)--(D1r);
\draw (C1)--(D1u);
\end{tikzpicture}
}
As before, if one were to perform a calculation omitting the crucial phase factors discussed earlier, we would obtain
\bes{ \label{indexnaiveA}
&\sum_{m_u \in\BZ_{\BZ \geq 0}}\, \sum_{m_z \in \BZ_{\geq 0}} \, \sum_{m_v \in\BZ_{\geq 0}} \,\, \zeta_1^{m_u} \zeta_2^{m_v}\,\, t^{\sum_{s = \pm 1} \left( |s m_z|+|m_u+sm_z|+|m_v+sm_z| \right)-2|2m_z|} \\
& \qquad \qquad \times  P_{\USp(2)}(t; m_u) P_{\USp(2)}(t; m_z) P_{\USp(2)}(t; m_v)  \\
&= 1 + 6 t^2 + 29 t^4 + 89 t^6 + 236 t^8 + 521 t^{10}+\ldots~, \,\, \text{for $\zeta_1=\zeta_2=1$}~.
}
This is in agreement with the Coulomb branch limits of \eqref{SO1USp2SO5delta0} and \eqref{SO1USp2SO5delta1} upon gauging $\BZ^{[0]}_{2, \chi_1} \times \BZ^{[0]}_{2, \chi_2}$. Nevertheless, let us again emphasise that omitting such a crucial phase factor leads to a result that is inconsistent with the index and known dualities. As before, if we {\it ad hoc} modify the index computation by inserting a phase factor in a similar way to \eqref{indexmirrorquivc2-1xxxx} as follows:\footnote{In the last term contributing to the sum \eref{indexZ2Z2}, we also have to replace $\CZ_{\text{chir}}^{1}(x;a^{-2};0)$ appearing in \eqref{indexmirrorquivc2-1} with $\CZ_{\text{chir}}^{1}(x;-a^{-2};0)$.}
\bes{ \label{indexZ2Z2}
\frac{1}{4} \left\{ \eqref{indexmirrorquivc2+1} + \eqref{indexmirrorquivc2-1xxxx} + [\eqref{indexmirrorquivc2-1xxxx}]_{u \leftrightarrow v, \, m_u \leftrightarrow m_v} + [\eqref{indexmirrorquivc2-1}]_{u=1, u^{-1} =-1, m_u=0} \right\}~,
}
we would obtain
\bes{
1 &+ (6 a^{-2} + a^2) x + (29 a^{-4} + 3 a^4 -6) x^2\\& + (89 a^{-6} - 48 a^{-2} - 5 a^2 + 3 a^6) x^3 +\ldots~.
}
Although the Coulomb branch limit of this index is in agreement with the Hilbert series \eqref{indexnaiveA}, the entire index is not equal to \eqref{SO1USp2SO5delta0} and \eqref{SO1USp2SO5delta1} upon exchanging $a$ and $a^{-1}$. This is in contradiction with what should be expected from mirror symmetry.

\subsubsection{Orthosymplectic quiver with an $\so(10)$ global symmetry} \label{sec:orthosymso10}
In this section, we consider the $\USp(2)$ gauge theory with five flavours of fundamental hypermultiplets, namely
\bes{ \label{USp2w5flv}
\begin{tikzpicture}[baseline, font=\footnotesize]
\node[gauge,label=below:{$\USp(2)$}] (C1) at (-4,0) {};
\node[flavour,label=below:{$\SO(10)$}] (D5) at (-2,0) {};
\draw[-,bend left=0] (C1)--(D5);
\end{tikzpicture}
}
and its mirror theory described by the following orthosymplectic quiver (see \eg~ \cite[(3.6)]{Bourget:2020xdz}):
\bes{ \label{OSpSO10}
\begin{tikzpicture}[baseline, font=\footnotesize]
\node[gauge,label=below:{$\SO(2)$}] (D1l) at (-4,0) {};
\node[gauge,label=below:{$\USp(2)$}] (C1l) at (-2,0) {};
\node[gauge,label=below:{$\SO(4)$}] (D2) at (0,0) {};
\node[gauge,label=below:{$\USp(2)$}] (C1r) at (2,0) {};
\node[gauge,label=below:{$\SO(2)$}] (D1r) at (4,0) {};
\node[gauge,label=right:{$\U(1)$}] (D1u) at (0,1) {};
\draw[-,bend left=0] (D1l)--(C1l)--(D2)--(C1r)--(D1r);
\draw[-,bend right=0] (D2) to (D1u);
\end{tikzpicture} \quad /\BZ_2
}
where $/\BZ_2$ denotes the gauging of the $\BZ_2$ one-form symmetry that is present in the quiver diagram. This gauging gives rise to a dual $\BZ_2$ zero-form symmetry, which we denote by $\BZ^{[0]}_{2, \, B}$. It is worth noting that theory \eref{USp2w5flv} is also known to be mirror dual to a theory based on the affine $D_5$ Dynkin diagram involving unitary gauge groups \cite{Intriligator:1996ex}:
\bes{ \label{affineD5}
\begin{tikzpicture}[baseline, font=\footnotesize]
\node[gauge,label=left:{$1$}] (U1lu) at (-2,1) {};
\node[gauge,label=left:{$1$}] (U1ld) at (-2,-1) {};
\node[gauge,label=below:{$2$}] (U2l) at (-1,0) {};
\node[gauge,label=below:{$2$}] (U2r) at (1,0) {};
\node[gauge,label=right:{$1$}] (U1ru) at (2,1) {};
\node[gauge,label=right:{$1$}] (U1rd) at (2,-1) {};
\draw[-,bend left=0] (U1lu)--(U2l)--(U2r)--(U1ru);
\draw[-,bend right=0] (U1ld)--(U2l);
\draw[-,bend right=0] (U1rd)--(U2r);
\end{tikzpicture} \quad /\U(1)
}
Here, we will focus primarily on the relationship between \eref{USp2w5flv} and the orthosymplectic quiver \eref{OSpSO10}, deferring discussion of the affine $D_5$ mirror \eref{affineD5} to Appendix \ref{app:affineD5}.

The index of theory \eref{OSpSO10} is given by
\bes{ \label{indOSpSO10}
&\CI_{\eref{OSpSO10}}(x; a ; g; \zeta; \chi) \\
& = 1 + \left[7+ 10\zeta + 6\chi + 6\zeta\chi + g \left(4  + 4\zeta + 4\chi  + 4\zeta\chi\right) \right] a^{-2} x \\
& \quad \,\,\,\ + \left\{ \left[120 + 102\zeta + 106\chi + 106\zeta\chi + g \left(84 + 84\zeta + 84\chi  + 84\zeta\chi \right) \right] a^{-4} \right. \\
& \quad \,\,\,\ \left. + a^{4} -\left[8 + 10\zeta + 6\chi + 6\zeta\chi + g \left(4  + 4\zeta + 4\chi  + 4\zeta\chi \right)\right] \right\} x^2+\ldots~,
}
where $g$ is the fugacity associated with $\BZ_{2, \, B}^{[0]}$, whereas $\zeta$ and $\chi$ are the fugacities for the $\BZ_{2,\, \CM}^{[0]}$ magnetic symmetry and the $\BZ_{2,\, \CC}^{[0]}$ charge conjugation symmetry associated with the $\SO(4)$ gauge group in quiver \eref{OSpSO10}. Note that the index is invariant under the transformation $(\zeta, \chi) \to (\zeta, \zeta\chi)$, consistent with the self-duality property \eref{dualityN4} applied to the $\SO(4)$ gauge node.

For simplicity in the following analysis, let us initially focus on the roles of the $\BZ_{2,\, \CM}^{[0]}$ and $\BZ_{2,\, B}^{[0]}$ symmetries (associated with fugacities $\zeta$ and $g$, respectively). The theories resulting from gauging various combinations of these discrete symmetries are summarised in Table \ref{tab:discrete_gaugings}. The table also reflects the duality \eref{dualityN4} applied to the $\SO(4)$ node, indicating that the global form involving $\Spin(4)$ is dual to the one involving $\O(4)^-$. Setting fugacities other than $a$ to unity, the unrefined index for each theory listed in the table takes the following schematic form up to order $x^2$:
\bes{ \label{unrefindexso10}
1+ \alpha a^{\pm 2} x + \left[\beta a^{\pm 4} +a^{\mp 4}-\left(\alpha+1\right)\right] x^2 +\ldots~.
}
where $\alpha$ and $\beta$ are given in the last column of Table \ref{tab:discrete_gaugings}, and the sign choice $\pm$ depends on whether the quantity is computed from \eref{USp2w5flv} or its mirror  \eref{OSpSO10}. From the mirror symmetry perspective, the discrete symmetries $\BZ_{2,\, \CM}^{[0]}$ and $\BZ^{[0]}_{2,\, B}$ of the quiver theory \eref{OSpSO10} correspond to specific actions on the hypermultiplets of the mirror $\USp(2)$ theory \eref{USp2w5flv}. This correspondence can be visualised as
\bes{ \label{actionzetagonUSp2w5}
\scalebox{0.9}{%
\begin{tikzpicture}
\node[gauge,label=below:{$\USp(2)$}] (C1) at (0,0) {};
\node[flavour,label=below:{$\SO(2)$}] (D1r) at (2,0) {};
\node[flavour,label=right:{$\SO(4)$}] (D4) at (0,1) {}; 
\node[flavour,label=below:{$\SO(4)$}] (D2l) at (-2,0) {};
\draw[-] (C1)--(D4);
\draw[red, very thick, -] (D2l) to node[above,midway]{\textcolor{red}{$\zeta$}} (C1);
\draw[new-green, very thick, -] (C1) to node[above,midway]{\textcolor{new-green}{$g$}} (D1r);
\end{tikzpicture}%
}
}
where {\red $\zeta$} and {\green $g$} denote the fugacities corresponding to $\BZ_{2,\, \CM}^{[0]}$ and $\BZ_{2,\, B}^{[0]}$ respectively, acting on specific hypermultiplets of the $\USp(2)$ gauge theory with five flavours.

\begin{longtable}{ccc} 

\caption{Variants of \eqref{OSpSO10}, the corresponding mirror theories, index coefficients, and continuous global symmetries.}
\label{tab:discrete_gaugings} \\

\toprule
Variants of \eref{OSpSO10} & Mirror theory & $\overset{\textstyle (\alpha, \beta)\,\, \text{in}\,\, \eqref{unrefindexso10}}{\text{Continuous symmetry}}$ \\
\midrule
\endfirsthead

 \toprule
Variants of \eref{OSpSO10} & Mirror theory & $\overset{\textstyle (\alpha, \beta)}{\text{Continuous symmetry}}$ \\
 \midrule
 \endhead


 \bottomrule
 \endlastfoot


\scalebox{0.65}{%
\begin{tikzpicture}
\node[gauge,label=below:{$\SO(2)$}] (D1l) at (-4,0) {};
\node[gauge,label=below:{$\USp(2)$}] (C1l) at (-2,0) {};
\node[gauge,label=below:{$\SO(4)$}] (D2) at (0,0) {};
\node[gauge,label=below:{$\USp(2)$}] (C1r) at (2,0) {};
\node[gauge,label=below:{$\SO(2)$}] (D1r) at (4,0) {};
\node[gauge,label=right:{$\U(1)$}] (D1u) at (0,1) {};
\draw[-] (D1l)--(C1l)--(D2)--(C1r)--(D1r);
\draw[-] (D2) to (D1u);
\node at (5,0) {$/\BZ_2$};
\end{tikzpicture}%
} &
\scalebox{0.65}{%
\begin{tikzpicture}
\node[gauge,label=below:{$\USp(2)$}] (C1) at (-1,0) {};
\node[flavour,label=below:{$\SO(10)$}] (D5) at (1,0) {};
\draw[-] (C1)--(D5);
\end{tikzpicture}%
} &
$\overset{\textstyle (45, 770)}{\so(10)}$ \\
\midrule 

\scalebox{0.65}{%
\begin{tikzpicture}
\node[gauge,label=below:{$\SO(2)$}] (D1l) at (-4,0) {};
\node[gauge,label=below:{$\USp(2)$}] (C1l) at (-2,0) {};
\node[gauge,label=below:{$\overset{\textstyle \Spin(4)}{\text{or} \,\, \O(4)^-}$}] (D2) at (0,0) {};
\node[gauge,label=below:{$\USp(2)$}] (C1r) at (2,0) {};
\node[gauge,label=below:{$\SO(2)$}] (D1r) at (4,0) {};
\node[gauge,label=right:{$\U(1)$}] (D1u) at (0,1) {};
\draw[-] (D1l)--(C1l)--(D2)--(C1r)--(D1r);
\draw[-] (D2) to (D1u);
\node at (5,0) {$/\BZ_2$};
\end{tikzpicture}%
} &
\scalebox{0.65}{%
\begin{tikzpicture}
\node[gauge,label=below:{$\USp(2)$}] (C1) at (0,0) {};
\node[flavour,label=below:{$\SO(4)$}] (D2) at (2,0) {};
\node[flavour,label=left:{$\SO(6)$}] (D3) at (0,1) {};
\draw[-] (C1)--(D3);
\draw[red, very thick, -] (C1) to node[above,midway]{\textcolor{red}{$/\BZ_2$}} (D2);
\end{tikzpicture}%
} &
$\overset{\textstyle (21,394)}{\so(6) \oplus \so(4)}$ \\
\midrule 

\scalebox{0.65}{%
\begin{tikzpicture}
\node[gauge,label=below:{$\SO(2)$}] (D1l) at (-4,0) {};
\node[gauge,label=below:{$\USp(2)$}] (C1l) at (-2,0) {};
\node[gauge,label=below:{$\SO(4)$}] (D2) at (0,0) {};
\node[gauge,label=below:{$\USp(2)$}] (C1r) at (2,0) {};
\node[gauge,label=below:{$\SO(2)$}] (D1r) at (4,0) {};
\node[gauge,label=right:{$\U(1)$}] (D1u) at (0,1) {};
\draw[-] (D1l)--(C1l)--(D2)--(C1r)--(D1r);
\draw[-] (D2) to (D1u);
\end{tikzpicture}%
} &
\scalebox{0.65}{%
\begin{tikzpicture}
\node[gauge,label=below:{$\USp(2)$}] (C1) at (0,0) {};
\node[flavour,label=below:{$\SO(2)$}] (D1) at (2,0) {};
\node[flavour,label=left:{$\SO(8)$}] (D4) at (0,1) {};
\draw[-] (C1)--(D4);
\draw[new-green, very thick, -] (C1) to node[above,midway]{\textcolor{new-green}{$/\BZ_2$}} (D1);
\end{tikzpicture}%
} &
$\overset{\textstyle (29, 434)}{\so(8) \oplus \so(2)}$ \\
\midrule 

\scalebox{0.65}{%
\begin{tikzpicture}
\node[gauge,label=below:{$\SO(2)$}] (D1l) at (-4,0) {};
\node[gauge,label=below:{$\USp(2)$}] (C1l) at (-2,0) {};
\node[gauge,label=below:{$\overset{\textstyle \Spin(4)}{\text{or} \,\, \O(4)^-}$}] (D2) at (0,0) {};
\node[gauge,label=below:{$\USp(2)$}] (C1r) at (2,0) {};
\node[gauge,label=below:{$\SO(2)$}] (D1r) at (4,0) {};
\node[gauge,label=right:{$\U(1)$}] (D1u) at (0,1) {};
\draw[-] (D1l)--(C1l)--(D2)--(C1r)--(D1r);
\draw[-] (D2) to (D1u);
\end{tikzpicture}%
} &
\scalebox{0.65}{%
\begin{tikzpicture}
\node[gauge,label=below:{$\USp(2)$}] (C1) at (0,0) {};
\node[flavour,label=below:{$\SO(2)$}] (D1r) at (2,0) {};
\node[flavour,label=left:{$\SO(4)$}] (D4) at (0,1) {}; 
\node[flavour,label=below:{$\SO(4)$}] (D2l) at (-2,0) {};
\draw[-] (C1)--(D4);
\draw[red, very thick, -] (D2l) to node[above,midway]{\textcolor{red}{$/\BZ_2$}} (C1);
\draw[new-green, very thick, -] (C1) to node[above,midway]{\textcolor{new-green}{$/\BZ_2$}} (D1r);
\end{tikzpicture}%
} &
$\overset{\textstyle (13, 226)}{\so(4) \oplus \so(4) \oplus \so(2)}$ \\

\end{longtable}
\noindent  Note that the quivers in the second column, representing actions on the $\USp(2)$ theory, can also be redrawn equivalently by replacing the ``/$\BZ_2$'' action on the hypermultiplets connected to the $\SO(2k)$ flavour node with $k$ bifundamental half-hypermultiplets connecting to an $\O(1)$ gauge node:
\bes{
&\scalebox{0.7}{%
\begin{tikzpicture}[baseline, font=\footnotesize]
\node[gauge,label=below:{$\USp(2)$}] (C1) at (0,0) {};
\node[flavour,label=below:{$\SO(4)$}] (D2) at (2,0) {};
\node[flavour,label=left:{$\SO(6)$}] (D3) at (0,1) {};
\draw[-] (C1)--(D3);
\draw[red, very thick, -] (C1) to node[above,midway]{\textcolor{red}{$/\BZ_2$}} (D2);
\end{tikzpicture}%
}
= 
\scalebox{0.7}{%
\begin{tikzpicture}[baseline, font=\footnotesize]
\node[gauge,label=below:{$\USp(2)$}] (C1) at (0,0) {};
\node[gauge,label=below:{$\O(1)$}] (D2) at (2,0) {};
\node[flavour,label=left:{$\SO(6)$}] (D3) at (0,1) {};
\draw[-] (C1)--(D3);
\draw[red, very thick, -, bend left =30] (C1) to (D2);
\draw[red, very thick, -, bend left =10] (C1) to (D2);
\draw[red, very thick, -, bend left =-10] (C1) to (D2);
\draw[red, very thick, -, bend left =-30] (C1) to (D2);
\end{tikzpicture}%
}
\\ &
\scalebox{0.7}{%
\begin{tikzpicture}[baseline, font=\footnotesize]
\node[gauge,label=below:{$\USp(2)$}] (C1) at (0,0) {};
\node[flavour,label=below:{$\SO(2)$}] (D2) at (2,0) {};
\node[flavour,label=left:{$\SO(8)$}] (D3) at (0,1) {};
\draw[-] (C1)--(D3);
\draw[new-green, very thick, -] (C1) to node[above,midway]{\textcolor{new-green}{$/\BZ_2$}} (D2);
\end{tikzpicture}%
}
= 
\scalebox{0.7}{%
\begin{tikzpicture}[baseline, font=\footnotesize]
\node[gauge,label=below:{$\USp(2)$}] (C1) at (0,0) {};
\node[gauge,label=below:{$\O(1)$}] (D2) at (2,0) {};
\node[flavour,label=left:{$\SO(8)$}] (D3) at (0,1) {};
\draw[-] (C1)--(D3);
\draw[new-green, very thick, -, bend left =20] (C1) to (D2);
\draw[new-green, very thick, -, bend left =-20] (C1) to (D2);
\end{tikzpicture}%
} \\
&\scalebox{0.7}{%
\begin{tikzpicture}[baseline, font=\footnotesize]
\node[gauge,label=below:{$\USp(2)$}] (C1) at (0,0) {};
\node[flavour,label=below:{$\SO(2)$}] (D1r) at (2,0) {};
\node[flavour,label=left:{$\SO(4)$}] (D4) at (0,1) {}; 
\node[flavour,label=below:{$\SO(4)$}] (D2l) at (-2,0) {};
\draw[-] (C1)--(D4);
\draw[red, very thick, -] (D2l) to node[above,midway]{\textcolor{red}{$/\BZ_2$}} (C1);
\draw[new-green, very thick, -] (C1) to node[above,midway]{\textcolor{new-green}{$/\BZ_2$}} (D1r);
\end{tikzpicture}%
}
\quad = \quad 
\scalebox{0.7}{%
\begin{tikzpicture}[baseline, font=\footnotesize]
\node[gauge,label=below:{$\USp(2)$}] (C1) at (0,0) {};
\node[gauge,label=below:{$\O(1)$}] (D1r) at (2,0) {};
\node[flavour,label=left:{$\SO(4)$}] (D4) at (0,1) {}; 
\node[gauge,label=below:{$\O(1)$}] (D2l) at (-2,0) {};
\draw[-] (C1)--(D4);
\draw[red, very thick, -, bend left =30] (D2l) to (C1);
\draw[red, very thick, -, bend left =10] (D2l) to (C1);
\draw[red, very thick, -, bend left =-10] (D2l) to (C1);
\draw[red, very thick, -, bend left =-30] (D2l) to (C1);
\draw[new-green, very thick, -, bend left =20] (C1) to (D1r);
\draw[new-green, very thick, -, bend left =-20] (C1) to (D1r);
\end{tikzpicture}}%
}

Let us now focus on the charge conjugation symmetry $\BZ^{[0]}_{2, \CC}$ associated with the $\SO(4)$ gauge group in quiver \eref{OSpSO10}. Across the mirror duality, this $\BZ^{[0]}_{2,\, \CC}$ symmetry maps to a combination of actions on the $\USp(2)$ theory \eref{USp2w5flv}. It acts on the hypermultiplet associated with the {\violet violet} edge (carrying $\zeta$) in diagram \eref{actionchizetagonUSp2w5} below, and it also induces an exchange (wreathing) between the flavour subgroups associated with the {\violet violet} edge and the black edge (the uncoloured $\SO(4)$ flavour node). The wreathing action, studied in \cite{Bourget:2020bxh, Grimminger:2024mks}, is denoted by the double-headed {\blue blue} arrow below. We depict the action of these symmetries ({\blue $\chi$} for $\BZ^{[0]}_{2,\, \CC}$, {\red $\zeta$} for $\BZ^{[0]}_{2,\, \CM}$, {\green $g$} for $\BZ^{[0]}_{2,\, B}$) on the flavour nodes of the $\USp(2)$ theory as follows:
\bes{ \label{actionchizetagonUSp2w5}
\scalebox{0.9}{%
\begin{tikzpicture}
\node[gauge,label=below:{$\USp(2)$}] (C1) at (0,0) {};
\node[flavour,label=below:{$\SO(2)$}] (D1r) at (2,0) {};
\node[flavour,label=right:{$\SO(4)$}] (D4) at (0,1) {}; 
\node[flavour,label=below:{$\SO(4)$}] (D2l) at (-2,0) {};
\draw[-] (C1)--(D4);
\draw[violet, very thick, -] (D2l) to node[above,midway]{\textcolor{blue}{$\chi$} \textcolor{red}{$\zeta$}} (C1);
\draw[new-green, very thick, -] (C1) to node[above,midway]{\textcolor{new-green}{$g$}} (D1r);
\draw[blue, very thick, <->] (D2l) to [out=150,in=60] (D4);
\end{tikzpicture}%
}
}
Quantitatively, the index of \eref{actionchizetagonUSp2w5}, for $\chi=+1$, is given by
\bes{ \label{chiplusactionchizetagonUSp2w5}
&\CI_{\eqref{actionchizetagonUSp2w5}}(x; a; f_1, \ldots, f_5; g; \zeta; \chi=+1)  \\
&= \frac{1}{2} \sum_{m_z \in \BZ} \oint \frac{dz}{2\pi i z} \CZ_{\text{vec}}^{\USp(2)}(x;z; m_z) \left[\prod_{\ell=-1}^1 \CZ_{\text{chir}}^{1}(x; a^{-2} z^{2 \ell};2 \ell m_z)  \right] \\
& \qquad \times \prod_{s_1, s_2 = \pm 1}\Big[ \CZ_{\text{chir}}^{1/2}(x;g a z^{s_1} f_1^{s_2}; s_1 m_z + s_2 m_{f_1}) \\
& \qquad \times \prod_{j=2}^3 \CZ_{\text{chir}}^{1/2} (x; {\gray \chi} \zeta a z^{s_1} f_j^{s_2}; s_1 m_z + s_2 m_{f_j}) \\
& \qquad \times  \prod_{p=4}^5 \CZ_{\text{chir}}^{1/2}(x;{\gray \chi} a z^{s_1} f_p^{s_2}; s_1 m_z + s_2 m_{f_p}) \Big]~.
} 
On the other hand, for $\chi=-1$, the index is given by
\bes{ \label{chiminusactionchizetagonUSp2w5}
&\CI_{\eqref{actionchizetagonUSp2w5}}(x; a; f_1, \tilde{f}_2,\tilde{f}_3;g=1; \zeta; \chi=-1)  \\
&= \frac{1}{2} \sum_{m_z \in \BZ} \oint \frac{dz}{2\pi i z} \CZ_{\text{vec}}^{\USp(2)}(x;z; m_z) \left[\prod_{\ell=-1}^1 \CZ_{\text{chir}}^{1}(x; a^{-2} z^{2 \ell};2 \ell m_z)  \right] \\
&\qquad \times \prod_{s_1, s_2 = \pm 1} \Big[ \CZ_{\text{chir}}^{1/2}(x;g a z^{s_1} f_1^{s_2}; s_1 m_z + s_2 m_{f_1}) \\
& \qquad \times \prod_{j=2}^3 \CZ_{\text{chir}}^{1/2}(x^2;\chi \zeta a^2 z^{2s_1} \tilde{f}_j^{s_2}; s_1 m_z + s_2 m_{\tilde{f}_j}) \Big] ~,
} 
where the squares of fugacities $z$, $a$ and $x$ in the last line are due to wreathing, as explained in \cite{Grimminger:2024mks}, and we can set the fugacity $g$ to unity since it acts trivially in \eref{chiminusactionchizetagonUSp2w5}.\footnote{The fact that the fugacity $g$ drops out of the index \eref{chiminusactionchizetagonUSp2w5} is consistent with the absence of the fugacity $g$ in the index corresponding to the $\chi=-1$ sector in the mirror theory \eref{OSpSO10}. Indeed, in the mirror quiver \eref{OSpSO10}, the fugacity $g$ is associated with the $\BZ^{[0]}_{2,B}$ symmetry arising from gauging the $\BZ_2$ one-form symmetry. At the level of the index, such gauging procedure is implemented by summing also over half-integral fluxes, \ie $\sum_{e = 0}^1 g^e \sum_{m_i \in \BZ +\frac{e}{2}}$. However, in the $\chi=-1$ sector, we set the last flux of the $\SO(4)$ gauge group to zero and, in order for the Dirac quantisation condition to be satisfied, there is no summation over half-integral fluxes, see the discussions below \eref{SO4minusUSp4} and below \eref{vecadjUSp2N}.} In the following, we will set $f_i=1$ and $\tilde{f}_i=1$. The index of \eqref{actionchizetagonUSp2w5} is therefore
\bes{
&\CI_{\eqref{actionchizetagonUSp2w5}}(x; a; f_i=1; g; \zeta; \chi) = \frac{1}{2} \sum_{s=\pm1} \CI_{\eqref{actionchizetagonUSp2w5}}(x; f_i=1; g; \zeta; \chi=s) (1+s \chi) \\
& = 1 + \left[7 + 10\zeta + 6\chi + 6\zeta\chi + g \left(4 + 4\zeta + 4\chi  + 4\zeta\chi \right) a^2 \right] x \\
& \quad \,\,\,\ + \left\{ a^{-4} + \left[120 + 102\zeta + 106\chi + 106\zeta\chi + g \left(84 + 84\zeta + 84\chi + 84\zeta\chi \right)\right] a^4 \right. \\
& \quad \,\,\,\ \left. - \left[8 + 10\zeta + 6\chi + 6\zeta\chi + g \left(4 + 4\zeta + 4\chi  + 4\zeta\chi \right) a^2 \right] \right\} x^2+\ldots \\
& = \CI_{\eref{OSpSO10}}(x; a^{-1} ; g; \zeta; \chi)~.
}
This is precisely equal to the index of theory \eref{OSpSO10} upon exchanging $a \leftrightarrow a^{-1}$, which confirms that \eqref{OSpSO10} is mirror dual to \eqref{actionchizetagonUSp2w5}. 

\subsubsection*{Inconsistency of the theory with the $\O(4)^+$ gauge group} 
Observe from \eref{chiplusactionchizetagonUSp2w5} that $\BZ_{2,\, \CM}^{[0]}$ acts non-trivially on the hypermultiplet with fugacities $f_{2,3}$, but acts trivially on those with $f_{4,5}$. Turning on $\zeta$ therefore obstructs wreathing, whose action identifies these two sets of the hypermultiplets. Recalling that the wreathing action corresponds to the $\BZ_{2,\, \CC}^{[0]}$ charge conjugation symmetry, this obstruction implies a mixed 't Hooft anomaly between the magnetic symmetry $\BZ_{2,\, \CM}^{[0]}$ and the charge conjugation symmetry $\BZ_{2,\, \CC}^{[0]}$ in the theory \eref{OSpSO10}. This is characterised by the anomaly theory
\bes{ \label{anomMCC}
i \pi \int_{M_4} A^\CM_2 \cup A^\CC_1 \cup A^\CC_1~,
}
where $A^\CM_2$ is the background gauge field associated with the one-form magnetic symmetry of the theory whose central gauge node is $\Spin(4)$, \ie~ those in the second and fourth rows of Table \ref{tab:discrete_gaugings}. This anomaly implies that we cannot gauge the one-form magnetic symmetry, together with the $\BZ_{2,\, \CC}^{[0]}$ charge conjugation symmetry, in the theory with the $\Spin(4)$ gauge group. Since gauging the one-form magnetic symmetry results in the theory with the $\SO(4)$ gauge group, it follows that we cannot gauge the $\BZ_{2,\, \CC}^{[0]}$ charge conjugation symmetry in the latter theory, which turns $\SO(4)$ into $\O(4)^+$. In other words, the following theory is inconsistent:\footnote{Note that, on the other hand, the theory with $\Spin(4)$ gauge group is consistent, since $\BZ_{2,\, \CM}^{[0]}$ is gauged and so we cannot turn on $\zeta$ in the index.}
\bes{ \label{O4plusmodZ2}
\scalebox{0.7}{%
\begin{tikzpicture}
\node[gauge,label=below:{$\SO(2)$}] (D1l) at (-4,0) {};
\node[gauge,label=below:{$\USp(2)$}] (C1l) at (-2,0) {};
\node[gauge,label=below:{$\O(4)^+$}] (D2) at (0,0) {};
\node[gauge,label=below:{$\USp(2)$}] (C1r) at (2,0) {};
\node[gauge,label=below:{$\SO(2)$}] (D1r) at (4,0) {};
\node[gauge,label=right:{$\U(1)$}] (D1u) at (0,1) {};
\draw[-] (D1l)--(C1l)--(D2)--(C1r)--(D1r);
\draw[-] (D2) to (D1u);
\node at (5,0) {$/\BZ_2$};
\end{tikzpicture}%
}}
The unrefined index of this theory takes the form \eref{unrefindexso10} with $(\alpha, \beta) = (25, 390)$, where the continuous global symmetry is $\su(5) \oplus \mathfrak{u}(1)$.  Since \eref{O4plusmodZ2} arises from gauging $\BZ_{2, \, \CC}^{[0]}$ associated with $\SO(4)$ in \eref{OSpSO10}, the operators of the former theory transform under representations of the subalgebra $\su(5) \oplus \u(1)$ embedded within the original $\so(10)$ global symmetry. In particular, the moment maps operators of \eref{O4plusmodZ2} can be derived from those of \eref{OSpSO10} using the branching rule
\bes{
\mathbf{45} \,\, &\longrightarrow \,\, 
{\orangered (\bar{{\bf 10}})(4)} \oplus ({\bf 24})(0) \oplus ({\bf 1})(0) \oplus {\orangered ({\bf 10})(-4)}~,
}
where we denote in {\orangered orange} the representations projected out by gauging $\BZ_{2, \, \CC}^{[0]}$, identified here as those carrying $\u(1)$ charges $\pm 4$. Next, we consider the Coulomb branch marginal operator in the representation $\mathbf{770}$ of \eref{OSpSO10}:
\bes{
\mathbf{770} \,\, & \longrightarrow \,\,(\mathbf{50})(8) \oplus {\orangered (\bar{\mathbf{175}})(4)} \oplus {\orangered (\bar{\mathbf{10}})(4)} \oplus (\mathbf{200})(0) \oplus (\mathbf{75})(0) \oplus (\mathbf{24})(0) \\
& \qquad \quad \oplus (\mathbf{1})(0) \oplus {\orangered (\mathbf{175})(-4)} \oplus {\orangered (\mathbf{10})(-4)} \oplus (\bar{\mathbf{50}})(-8)~.
}
The total dimension of the representations with $\u(1)$ charge $0$ is $300$, and that of those with $\u(1)$ charges $\pm 8$ is $100$. We see that there is no way, using the above representations with direct sums, we can form a representation of dimension $390$ in which the Coulomb branch marginal operators of \eref{O4plusmodZ2} transform. This is a manifestation of the inconsistency of the theory with the $\O(4)^+$ gauge group.\footnote{We can perform a similar analysis for the case of $\Spin(4)$ gauge group, say the second row of Table \ref{tab:discrete_gaugings}. We consider the branching rules between $\so(10)$ and $\su(2) \oplus \su(2) \oplus \so(6)$ as follows:
\bes{
{\bf 45} &\rightarrow ({\bf 3},{\bf 1},{\bf 1}) \oplus ({\bf 1},{\bf 3},{\bf 1}) \oplus ({\bf 1},{\bf 1},{\bf 15}) \oplus {\orangered ({\bf 2},{\bf 2},{\bf 6})}  \\
{\bf 770} &\rightarrow ({\bf 5},{\bf 1},{\bf 1}) \oplus ({\bf 3},{\bf 3},{\bf 20}') \oplus ({\bf 3},{\bf 3},{\bf 1}) \oplus ({\bf 3},{\bf 1},{\bf 15}) \oplus ({\bf 1},{\bf 5},{\bf 1}) \oplus ({\bf 1},{\bf 3},{\bf 15}) \oplus ({\bf 1},{\bf 1},{\bf 84}) \\
& \qquad \oplus ({\bf 1},{\bf 1},{\bf 20}') \oplus ({\bf 1},{\bf 1},{\bf 1}) \oplus {\orangered ({\bf 4},{\bf 2},{\bf 6})} \oplus {\orangered ({\bf 2},{\bf 4},{\bf 6})} \oplus {\orangered ({\bf 2},{\bf 2},{\bf 64})} \oplus {\orangered ({\bf 2},{\bf 2},{\bf 6})}~,
}
where gauging $\BZ_{2, \CM}^{[0]}$ projects out the terms containing even dimensional representations of $\su(2)$. Observe that the total dimension of the representations denoted in black in the second branching rule is $394$, in perfect agreement with the value of $\beta$ in the second row of Table \ref{tab:discrete_gaugings}. 
} A similar inconsistency argument applies to the parent theory without the overall $\BZ_2$ gauging:
\bes{ \label{O4plus}
\scalebox{0.7}{%
\begin{tikzpicture}
\node[gauge,label=below:{$\SO(2)$}] (D1l) at (-4,0) {};
\node[gauge,label=below:{$\USp(2)$}] (C1l) at (-2,0) {};
\node[gauge,label=below:{$\O(4)^+$}] (D2) at (0,0) {};
\node[gauge,label=below:{$\USp(2)$}] (C1r) at (2,0) {};
\node[gauge,label=below:{$\SO(2)$}] (D1r) at (4,0) {};
\node[gauge,label=right:{$\U(1)$}] (D1u) at (0,1) {};
\draw[-] (D1l)--(C1l)--(D2)--(C1r)--(D1r);
\draw[-] (D2) to (D1u);
\end{tikzpicture}%
}}
whose unrefined index of this theory takes the form \eref{unrefindexso10} with $(\alpha, \beta) = (17, 222)$, where the continuous global symmetry is $\so(6) \oplus \mathfrak{so}(2) \oplus \mathfrak{so}(2)$.

Let us briefly comment further on the inconsistency related to gauging charge conjugation, as discussed previously. We concur with the observation related to \cite[(3.6)]{Bourget:2020xdz}, suggesting that the $\U(1)$ gauge node in quiver \eqref{OSpSO10} should not be simply regarded as $\SO(2)$, but rather viewed as a $\U(1)$ subgroup of a larger $\USp(2)$ symmetry. Following this line of reasoning, we observe that if one were to formally treat such a $\U(1)$ node as $\SO(2)$ and attempt to gauge its charge conjugation symmetry, formally transitioning to $\O(2)$, the resulting index calculation leads to inconsistencies. Specifically, the coefficients of low-order terms violate expected constraints. For instance, the inferred contribution from conserved currents at order $x^2$ is inconsistent with the flavour symmetry currents contributing at order $x$.

\subsubsection*{Symmetry extensions} 
An implication of the anomaly \eqref{anomMCC} is as follows. Gauging the $\BZ_{2, \, \CC}^{[0]}$ symmetry in the theory with the $\Spin(4)$ gauge group results in a dual $\BZ_2$ one-form symmetry, and turns $\Spin(4)$ into $\Pin(4)$. The anomaly indicates that this dual symmetry and the $\BZ_2$ one-form magnetic symmetry participate in a non-trivial extension, leading to an enhanced $\BZ_4^{[1]}$ one-form symmetry in the theory with the $\Pin(4)$ gauge group. By the same token, the $\BZ_{2, \, \CC}^{[0]}$ and $\BZ_{2, \, \CM}^{[0]}$ symmetries in the theory with the $\SO(4)$ gauge group form a non-trivial extension to a $\BZ_4^{[0]}$ zero-form symmetry. To see the latter from the index, we write $d = \zeta \chi $ and treat $d$ as $\BZ^{[0]}_4$ fugacity. The following analysis is completely analogous to \cite[(3.63), (3.64)]{Grimminger:2024mks}. We write $\chi = d \zeta = q^2 \zeta$, and view $\BZ_{2, \, \CM}^{[0]}$ (whose fugacity is $\zeta$) as a normal subgroup of $\BZ^{[0]}_4$. Here $q$ is the fugacity associated with the quotient group $\BZ^{[0]}_4/\BZ_{2, \, \CM}^{[0]}$. The index of the following theory 
\bes{ \label{Pin4modZ2}
\scalebox{0.7}{%
\begin{tikzpicture}
\node[gauge,label=below:{$\SO(2)$}] (D1l) at (-4,0) {};
\node[gauge,label=below:{$\USp(2)$}] (C1l) at (-2,0) {};
\node[gauge,label=below:{$\Pin(4)$}] (D2) at (0,0) {};
\node[gauge,label=below:{$\USp(2)$}] (C1r) at (2,0) {};
\node[gauge,label=below:{$\SO(2)$}] (D1r) at (4,0) {};
\node[gauge,label=right:{$\U(1)$}] (D1u) at (0,1) {};
\draw[-] (D1l)--(C1l)--(D2)--(C1r)--(D1r);
\draw[-] (D2) to (D1u);
\node at (5,0) {$/\BZ_2$};
\end{tikzpicture}%
}}
can be computed as in \cite[(3.63), (3.64)]{Grimminger:2024mks}:
\bes{
&\frac{1}{4} \sum_{q = 1, i} \,\, \sum_{\zeta = \pm 1}\, \CI_{\eref{OSpSO10}}(x; a ; g; \zeta; \chi = q^2 \zeta) = \frac{1}{4} \sum_{\chi = \pm 1} \,\, \sum_{\zeta = \pm 1} \CI_{\eref{OSpSO10}}(x; a ; g; \zeta; \chi) \\
&= 1 +  \left(7+4g \right) a^{-2}  x + \left[ \left(120+84g \right)a^{-4} + a^{4} - \left(8+4g\right)\right] x^2 \\
& \quad \,\,\,\,\, + \left[ \left(1004+896g \right)a^{-6}- \left(236+196g \right)a^{-2} +  a^{6} \right] x^3+\ldots ~.
}
The unrefined index of this theory takes the form \eref{unrefindexso10} with $(\alpha, \beta) = (11, 204)$, where the continuous global symmetry is $\so(5) \oplus \mathfrak{so}(2)$.  Gauging $\BZ_{2,\, B}^{[0]}$, we obtain the theory
\bes{ \label{Pin4modZ2}
\scalebox{0.7}{%
\begin{tikzpicture}
\node[gauge,label=below:{$\SO(2)$}] (D1l) at (-4,0) {};
\node[gauge,label=below:{$\USp(2)$}] (C1l) at (-2,0) {};
\node[gauge,label=below:{$\Pin(4)$}] (D2) at (0,0) {};
\node[gauge,label=below:{$\USp(2)$}] (C1r) at (2,0) {};
\node[gauge,label=below:{$\SO(2)$}] (D1r) at (4,0) {};
\node[gauge,label=right:{$\U(1)$}] (D1u) at (0,1) {};
\draw[-] (D1l)--(C1l)--(D2)--(C1r)--(D1r);
\draw[-] (D2) to (D1u);
\end{tikzpicture}%
}}
whose unrefined index takes the form \eref{unrefindexso10} with $(\alpha, \beta) = (7, 120)$, where the continuous global symmetry is $\so(4) \oplus \mathfrak{so}(2)$.

\subsection{Prescription for the $\so(2N+1)$ gauge algebra}
We now turn our attention to the $\SO(2N+1)$ gauge theory with $N_f$ vector hypermultiplets.
The computation of the Coulomb branch Hilbert series for this theory, in the case of vanishing background flavour magnetic fluxes and without refinement by the charge conjugation fugacity, was detailed in \cite[Section 5.3]{Cremonesi:2013lqa}. Methods for incorporating non-zero background flavour magnetic fluxes were subsequently described in \cite{Cremonesi:2014kwa, Cremonesi:2014uva}.

Here, we propose a method to incorporate the charge conjugation fugacity $\chi$ into the Hilbert series calculation, including the effects of background flavour fluxes. Motivated by the structure observed in related index calculations, particularly the phase factor involving $\delta$ in \eqref{indexUSp2w3flv}, we propose the following formula for the $\chi$-refined Coulomb branch Hilbert series of the $\SO(2N+1)$ theory with $N_f$ vector hypermultiplets:
\bes{ \label{HS_SO2N+1}
&H_{\SO(2N+1), N_f} (t; \zeta; \chi; \vec n; \delta)\\
&= \sum_{m_1 \geq m_2 \geq \cdots \geq m_N \geq 0} \chi^{(1-\delta)\sum_{j=1}^{N_f} n_j} \zeta^{\sum_{i=1}^N m_i} t^{2\Delta(\vec m; \vec n)} P_{\USp(2N)}(t; \vec m)~.
}
Here, $\Delta(\vec m; \vec n)$ is the conformal dimension of the monopole operator with gauge magnetic fluxes $\vec m = (m_1, \dots, m_N)$ and background flavour fluxes $\vec n = (n_1, \dots, n_{N_f})$:
\bes{ \label{Delta_SO2N+1}
\Delta(\vec m; \vec n) &= \frac{1}{2} \left[  \sum_{j=1}^{N_f} \left( |n_j| + \sum_{s=\pm 1} \sum_{a=1}^N |m_a - s n_j| \right) \right] - \sum_{a=1}^N |m_a| \\
& \,\,\,\, - \sum_{1\leq a<b \leq N} \left(|m_a -m_b| +|m_a+m_b|\right)~.
}
The dressing factor ($P$-factor) of the $\SO(2N+1)$ gauge group is equal to that of the $\USp(2N)$ gauge group, namely $P_{\USp(2N)}(t; \vec m)$ given previously in \eqref{PUSp2N}. As usual, $\zeta$ is the fugacity for the zero-form magnetic (topological) symmetry, and $\chi$ is the fugacity for the zero-form charge conjugation symmetry. Note the crucial phase factor $\chi^{(1-\delta)\sum_{j=1}^{N_f} n_j}$. The parameter $\delta \in \{0, 1\}$ functions similarly to its appearance in the index \eqref{indexUSp2w3flv}, controlling the presence of this $\chi$-dependent phase, although here it appears with $(1-\delta)$ in the exponent.
Our proposal implies a specific interplay between the phase factor in the Hilbert series and a similar phase in the superconformal index. When $\delta=0$, the phase $\chi^{\sum n_j}$ appears in the Hilbert series formula \eqref{HS_SO2N+1}, but is absent in the corresponding index. Conversely, when $\delta=1$, the phase is absent in the Hilbert series, but present in the index. This inverse relationship between the presence of the phase in the Hilbert series versus the index, governed by $(1-\delta)$, is a key feature of our proposal. This was actually observed and explained around \eref{indexnaive} and \eref{indexmirrorquivc2-1xxxx}.
As a result, if the background flavour fluxes are set to zero or if $\delta=1$, the proposed Coulomb branch Hilbert series \eqref{HS_SO2N+1} becomes independent of the charge conjugation fugacity $\chi$.\footnote{The result for the case of $\delta=1$ can be obtained from that of $\delta=0$ by simply setting $\chi=1$.} In the subsequent discussion, we will primarily focus on the case $\delta=0$.

To see how the $\chi$-dependence appears in the Hilbert series according to our prescription, let us consider theory \eref{USp2SO6}, namely
\bes{ \label{SO1USp2SO5}
\begin{tikzpicture}[baseline, font=\footnotesize]
\node[flavour,label=below:{$\underset{{\chi_1}}{\SO(1)}$}] (ff1) at (-2,0) {};
\node[gauge,label=below:{$\USp(2)$}] (gg) at (0,0) {};
\node[flavour,label=below:{$\underset{{\chi_2}}{\SO(5)}$}] (ff2) at (2,0) {};
\draw[-,bend right=0] (ff1)--(gg)--(ff2);
\end{tikzpicture} 
}
and apply \eqref{HS_SO2N+1}. In this case, there is no gauge magnetic flux for $\SO(1)$. The Coulomb branch Hilbert series becomes
\bes{
&\sum_{n \geq 0} (\chi_1 \chi_2)^{(1-\delta) n} \, t^{\left(4 |n|+2 |-n|\right)-2|2n|}\, P_{\USp(2)}(t; n)\\
&= \begin{cases} 
\PE[\chi_1 \chi_2 t^2 + (1+\chi_1\chi_2)t^4 -t^8] & \\ \qquad = 1 + \chi_1\chi_2 t^2 + (2 + \chi_1 \chi_2) t^4 + (1 + 2\chi_1 \chi_2) t^6 + \ldots &\quad \delta =0 \\
\PE[ t^2 + 2t^4 -t^8] &\\
\qquad  = 1 + t^2 + 3 t^4 + 3 t^6 +\ldots &\quad \delta =1
\end{cases}~,
}
where $n$ is the $\USp(2)$ gauge magnetic flux. This Hilbert series result perfectly matches the Coulomb branch limits (coefficients of $a^{-2p} x^{p}$) extracted from the corresponding index calculations \eqref{SO1USp2SO5delta0}  (for $\delta=0$) and \eqref{SO1USp2SO5delta1} (for $\delta=1$), respectively. Averaging these results over $\chi_1=\pm 1$ yields the Hilbert series for theory \eqref{O1USp2SO5}. For $\delta=1$, the result is trivially $\PE[t^2+2t^4-t^8]$ (the Hilbert Series for $\BC^2/\BZ_4$). For $\delta=0$, the average yields $\PE[2t^4+t^6-t^{12}]$ (the Hilbert Series for $\BC^2/\hat{D}_4$). These precisely match the Coulomb branch geometries identified for theory \eqref{O1USp2SO5} in \eqref{HSO1USp2SO5} for the respective values of $\delta$. This successful matching provides support for the proposed $\chi$-dependent phase structure in our Hilbert series prescription.

Let us now examine another theory: 
\bes{ \label{SO3USp4SO7}
\begin{tikzpicture}[baseline, font=\footnotesize]
\node[gauge,label=below:{$\underset{{\chi}}{\SO(3)}$}] (ff1) at (-2,0) {};
\node[gauge,label=below:{$\USp(4)$}] (gg) at (0,0) {};
\node[flavour,label=below:{$\SO(7)$}] (ff2) at (2,0) {};
\draw[-,bend right=0] (ff1)--(gg)--(ff2);
\end{tikzpicture} 
}
where $\chi$ is the charge conjugation symmetry associated with the $\SO(3)$ gauge group. The Coulomb branch Hilbert series, where we turn off the background fluxes for the $\so(7)$ flavour symmetry and set $\delta=0$, is
\bes{
&H_{\eref{SO3USp4SO7}}(t; \zeta; \chi; \delta=0) \\
&= 
\sum_{m \geq 0} \sum_{n_1 \geq n_2 \geq 0} \chi^{n_1+n_2} \zeta^m t^{2 \Delta(m; n_1, n_2)} P_{\USp(2)}(t;m) P_{\USp(4)}(t; n_1, n_2)~, 
}
where 
\bes{
\Delta(m; n_1, n_2) &= \frac{1}{2}\sum_{i=1}^2 \left(|n_i| +\sum_{s = \pm 1} |m +s n_i|\right) + \frac{7}{2} \sum_{i=1}^2 |n_i|\\
& \,\,\,\, - |m| - \left(|n_1-n_2|+|n_1+n_2| +|2n_1|+|2n_2|\right)~.
}
The Coulomb branch of this theory is known to be a complete intersection, allowing its Hilbert series to be expressed in the following closed form:
\bes{ \label{SOUSp4SO7ref}
&H_{\eref{SO3USp4SO7}}(t; \zeta; \chi; \delta=0) \\
&= \PE\left[(\zeta + \chi + \zeta\chi) t^2 + (2 + \zeta + \chi + \zeta\chi) t^4 - t^8 - t^{12} \right]~.
}
Using this result, we can obtain the Coulomb branch Hilbert series for theories involving other global forms of the $\SO(3)$ group. For example, the Hilbert series for theory
\bes{ \label{O3pUSp4SO7}
\begin{tikzpicture}[baseline, font=\footnotesize]
\node[gauge,label=below:{$\O(3)^+$}] (ff1) at (-2,0) {};
\node[gauge,label=below:{$\USp(4)$}] (gg) at (0,0) {};
\node[flavour,label=below:{$\SO(7)$}] (ff2) at (2,0) {};
\draw[-,bend right=0] (ff1)--(gg)--(ff2);
\end{tikzpicture} 
}
is obtained by averaging \eqref{SOUSp4SO7ref} over $\chi = \pm 1$:
\bes{
&H_{\eref{O3pUSp4SO7}}(t; \zeta; \delta=0) \\
&= \frac{
\left(1+t^2\right)^2 \left(1-\zeta t^3\right) \left(1+\zeta t^3\right) \left(1-t^2+t^4\right) \left[1-2 t^2+(3+\zeta) t^4-2 t^6+t^8\right]
}{
(1-t) (1+t) \left(1-\zeta t^2\right)^4 \left(1+\zeta t^2\right)^3 \left(1-\zeta t^4\right)^2 \left(1+\zeta t^4\right)
} \\
& = 1 + \zeta t^2 + (5 + 2 \zeta) t^4 + (4 + 7 \zeta) t^6 + (20 + 13 \zeta) t^8 + \ldots~.
}
We can verify the result \eqref{SOUSp4SO7ref} by computing the Higgs branch Hilbert series of the mirror dual theory, given in the third row of \cite[Table 9]{Cremonesi:2014uva} as:
\bes{ \label{mirrorSO3USp4SO7}
\begin{tikzpicture}[baseline, font = \footnotesize]
\node[gauge,label=below:{$\SO(2)$}] (D1l) at (-6,0) {};
\node[gauge,label=below:{$\USp(2)$}] (C1l) at (-4,0) {};
\node[gauge,label=below:{$\SO(4)$}] (D2l) at (-2,0) {};
\node[gauge,label=below:{$\USp(4)$}] (C2) at (0,0) {};
\node[gauge,label=below:{$\SO(3)$}] (B1r) at (2,0) {};
\node[flavour,label=right:{$\SO(3)$}] (B1u) at (0,1) {};
\draw[-] (D1l)--(C1l)--(D2l)--(C2)--(B1r);
\draw[-] (C2)--(B1u);
\end{tikzpicture}
}
The linear chain
\bes{ 
\begin{tikzpicture}[baseline, font = \footnotesize]
\node[gauge,label=below:{$\SO(2)$}] (D1l) at (-6,0) {};
\node[gauge,label=below:{$\USp(2)$}] (C1l) at (-4,0) {};
\node[gauge,label=below:{$\SO(4)$}] (D2l) at (-2,0) {};
\node[flavour,label=below:{$\USp(4)$}] (C2) at (0,0) {};
\draw[-] (D1l)--(C1l)--(D2l)--(C2);
\end{tikzpicture}
}
within this mirror quiver corresponds to the $T[\USp(4)]$ theory. The Higgs branch Hilbert series of the latter is given by the Coulomb branch Hilbert series of the $T[\SO(5)]$ theory $H_{T[\SO(5)]}(t; x_1, x_2; \vec m =0)$ given by \cite[(4.34)]{Cremonesi:2014kwa} with the background flux $\vec m=0$, where $\vec x = (x_1, x_2)$ are the $\USp(4)$ fugacities.\footnote{Note that one needs to replace $t$ in \cite[(4.34)]{Cremonesi:2014kwa} by $t^2$ in order to conform with the convention of this paper.} The Higgs branch Hilbert series of \eqref{mirrorSO3USp4SO7} is given by
\bes{ \label{HBmirr}
&\oint \frac{d x_1}{2\pi i x_1} \frac{d x_2}{2\pi i x_2} \frac{d z}{2\pi i z} (1-x_1^2)(1-x_2^2)(1-x_1 x_2)(1-x_1 x_2^{-1}) (1-z) \\
&\quad \times \PE\left[-t^2 \chi^{C_2}_{[2,0]}(\vec x) -t^2(z+1+z^{-1})\right] \times H_{T[\SO(5)]}(t; \vec x; \vec m =0) \\
&\quad \times \PE\left[t (z+1+z^{-1}) \chi^{C_2}_{[1,0]}(\vec x) \right] \,\, \PE \left[t (\zeta+\chi+\zeta \chi) \chi^{C_2}_{[1,0]}(\vec x)\right]~,
}
where, crucially, the fugacities $\zeta, \chi$ from the original Coulomb branch Hilbert series \eqref{SOUSp4SO7ref} are mapped under mirror symmetry to flavour fugacities for the final $\SO(3)$ flavour symmetry of the mirror quiver \eqref{mirrorSO3USp4SO7}. In particular, the combination $\zeta+\chi+\zeta \chi$ that appear in \eref{SOUSp4SO7ref} is mapped to the character of the $\SO(3)$ flavour symmetry that appears in the last term in \eqref{HBmirr}. The characters of the fundamental and adjoint representations of $\USp(4)$ are as follows:
\bes{
\chi^{C_2}_{[1,0]}(\vec x) = \sum_{s = \pm 1} \sum_{i=1}^2 x_i^s~, \, \chi^{C_2}_{[2,0]}(\vec x) =2 +\sum_{s = \pm 1} [x_1^{2s}+ x_2^{2s} + (x_1x_2)^s + (x_1 x_2^{-1})^s ]~.
}
Evaluating the integrals in \eqref{HBmirr} yields a result that precisely matches the proposed closed form \eqref{SOUSp4SO7ref} for the Coulomb branch Hilbert series of the original theory, thus verifying the mirror duality and the Coulomb branch Hilbert series calculation.

\section{Charge conjugation symmetry in linear quivers}\label{sec:ccinlinearquivers}
In this section, we examine the action of the charge conjugation symmetry, as well as the topological symmetry, associated with special orthogonal gauge groups in linear quivers. In particular, we focus on the $T[\SO(2N)]$ and $T[\USp(2N)]$ quivers of \cite{Gaiotto:2008ak} and discuss how their global symmetry is affected by gauging such $\BZ_2$ symmetries. Importantly, our analysis allows us to understand how the action of charge conjugation and topological symmetries can be mapped in the mirror theories.
\subsection{The $T[\SO(2N)]$ theory and its mirror}
Let us consider the $T[\SO(2N)]$ theory introduced in \cite{Gaiotto:2008ak}. This theory arises as the infrared (IR) fixed point of a 3d $\CN=4$ linear quiver gauge theory composed of alternating $\SO(2 k)$ and $\USp(2 k)$ gauge groups (for $k = 1, \ldots, N-1$), terminated by an $\SO(2N)$ flavour symmetry group connected to the final $\USp(2N-2)$ gauge node. A key property of the $T[\SO(2N)]$ theory is that it is self-mirror.

This theory has an $\so(2N)_{F}$ flavour symmetry, an $\so(2N)_{\CM}$ zero-form magnetic (topological) symmetry, and a $\u(1)_A$ axial symmetry. The $\so(2N)_{F}$ flavour symmetry is manifest in the ultraviolet (UV) description via the quiver diagram, whereas the $\so(2N)_{\CM}$ topological symmetry is emergent in the IR. Note that the manifest topological symmetry in the quiver diagram is $\U(1)_{\CM_1} \times \prod_{k=2}^{N-1} \BZ_{2, \CM_k}^{[0]}$, where $\U(1)_{\CM_1}$ is the topological symmetry associated with the $\SO(2)$ gauge group and each $\BZ_{2, \CM_k}^{[0]}$ factor arises from the $\SO(2k)$ gauge group, with $k=2, \ldots, N-1$.

The effect of gauging a $\BZ^{[0]}_{2,\CM_k}$ topological symmetry, associated with a specific $\SO(2k)$ gauge node within the $T[\SO(2N)]$ quiver, was discussed in \cite{Nawata:2023rdx}. Gauging $\BZ^{[0]}_{2,\CM_k}$ changes the corresponding $\SO(2k)$ gauge group to $\Spin(2k)$ within the quiver description. This gauging operation affects the Coulomb branch of the theory, while leaving the Higgs branch invariant. Under the self-mirror symmetry of $T[\SO(2N)]$, this operation maps to a modification that affects the Higgs branch, but preserves the Coulomb branch of the mirror description. As shown in \cite{Nawata:2023rdx}, this mirror operation corresponds to gauging a specific discrete $\BZ_2$ subgroup of the $\so(2N)_{F}$ flavour symmetry in the mirror quiver.

Here, we extend the analysis of \cite[Section 2.5]{Nawata:2023rdx} to include the $\BZ^{[0]}_{2,\CC_k}$ charge conjugation symmetries also associated with each $\SO(2k)$ gauge node. Consider an $\SO(2k)$ gauge node within the $T[\SO(2N)]$ quiver. It possesses both a topological symmetry $\BZ^{[0]}_{2,\CM_k}$ and a charge conjugation symmetry $\BZ^{[0]}_{2,\CC_k}$, with associated fugacities denoted by $\zeta_k$ and $\chi_k$, respectively. A natural question arises: how do these discrete symmetries map onto operations or symmetries in the mirror description? Let $T[\SO(2N)^\vee]$ represent the mirror theory framework. We note that when the $\BZ^{[0]}_{2,\CM_k}$ and $\BZ^{[0]}_{2,\CC_k}$ symmetry actions are disregarded (i.e. all $\zeta_k=1, \chi_k=1$), $T[\SO(2N)^\vee]$ is described by the same quiver diagram as $T[\SO(2N)]$ due to self-mirror symmetry.

We propose that the actions of $\BZ^{[0]}_{2,\CM_k}$ and $\BZ^{[0]}_{2,\CC_k}$ under mirror symmetry can be understood via the mapping depicted in diagram \eref{TSO2Nmirror}. In the mirror description $T[\SO(2N)^\vee]$, the original $\BZ^{[0]}_{2,\CM_k}$ symmetry (with fugacity $\zeta_k$) maps to an action, here denoted by the fugacity $\textcolor{red}{\mathfrak{z}_k}$, on the hypermultiplets transforming under the $\so(2N)_F$ flavour symmetry. In particular, the ones denoted by {\red red} and {\violet violet} edges transform non-trivially under $\fz_k$. Similarly, the original $\BZ^{[0]}_{2,\CC_k}$ symmetry (with fugacity $\chi_k$) maps to an action denoted by $\textcolor{blue}{\mathfrak{c}_k}$, under which the hypermultiplets denoted by {\blue blue} and {\violet violet} edges transform non-trivially. The combined action $\textcolor{violet}{\mathfrak{z}_k \mathfrak{c}_k}$ affects the hypermultiplets denoted by the {\violet violet} edge. These actions effectively break the manifest $\so(2N)_F$ flavour symmetry of the mirror quiver $T[\SO(2N)^\vee]$ into the subgroups $\so(2k-1) \oplus \so(1) \oplus \so(1) \oplus \so(2N-2k-1)$ indicated in the diagram, with different factors transforming under $\mathfrak{z}_k$ and $\mathfrak{c}_k$. Note that a residual $\so(2N-2k-1)$ subgroup of the original $\so(2N)_F$ flavour symmetry remains uncharged under both mirror actions $\mathfrak{z}_k$ and $\mathfrak{c}_k$. The mirror duality relating the two descriptions holds when the fugacities are identified as $\zeta_k = \mathfrak{z}_k$ and $\chi_k = \mathfrak{c}_k$.
\bes
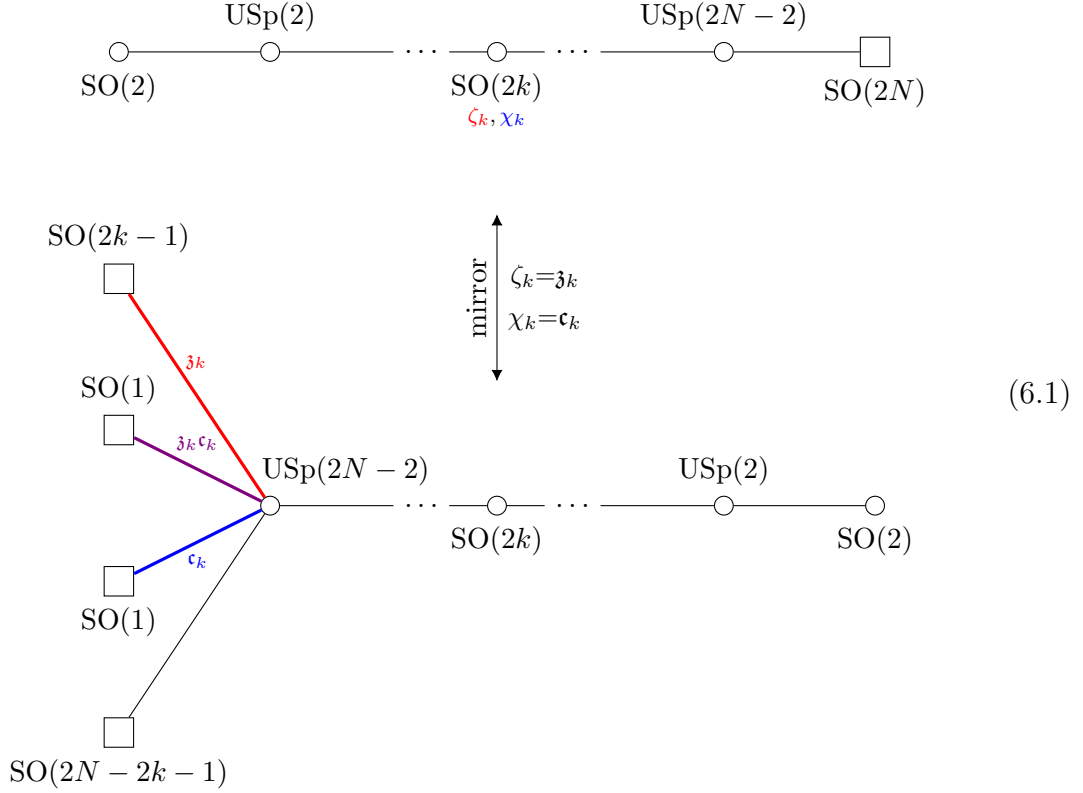
{ \label{TSO2Nmirror}
\begin{tikzpicture}[baseline, font = \small]
\node[gauge,label=below:{${\SO(2)}$}] (D1u) at (-5,6) {};
\node[gauge,label=above:{$\USp(2)$}] (C1u) at (-3,6) {};
\node[label=below:{}] (3Dlu) at (-1,6) {$\ldots$};
\node[gauge,label=below:{$\underset{\textcolor{red}{\zeta_k}, \, \textcolor{blue}{\chi_k}}{\SO(2 k)}$}] (Dku) at (0,6) {};
\node[label=below:{}] (3Dru) at (1,6) {$\ldots$};
\node[gauge,label=above:{$\USp(2N - 2)$}] (CNm1u) at (3,6) {};
\node[flavour,label=below:{$\SO(2 N)$}] (D2Nu) at (5,6) {};
\draw (D1u)--(C1u)--(3Dlu)--(Dku)--(3Dru)--(CNm1u)--(D2Nu);
\node (t) at (0,4) {};
\node (b) at (0,1.5) {};
\draw[<->] (b) to node[midway, above,sloped] {$\text{mirror}$} node[midway, right] {\large $\substack{\zeta_k = \fz_k \\ \\ \chi_k = \fc_k}$} (t);
\node[gauge,label=below:{${\SO(2)}$}] (D1) at (5,0) {};
\node[gauge,label=above:{$\USp(2)$}] (C1) at (3,0) {};
\node[label=below:{}] (3Dl) at (1,0) {$\ldots$};
\node[gauge,label=below:{${\SO(2 k)}$}] (Dk) at (0,0) {};
\node[label=below:{}] (3Dr) at (-1,0) {$\ldots$};
\node[gauge,label={above,xshift=1cm}:{$\USp(2N - 2)$}] (CNm1) at (-3,0) {};
\node[flavour,label=above:{$\SO(2 k - 1)$}] (D2Nz) at (-5,3) {};
\node[flavour,label=above:{$\SO(1)$}] (D2Nzc) at (-5,1) {};
\node[flavour,label=below:{$\SO(1)$}] (D2Nc) at (-5,-1) {};
\node[flavour,label=below:{$\SO(2 N - 2 k - 1)$}] (D2N) at (-5,-3) {};
\draw (D1)--(C1)--(3Dl)--(Dk)--(3Dr)--(CNm1)--(D2N);
\draw[very thick,red] (CNm1) to node[midway,above=0.2] {\textcolor{red}{\scriptsize $\fz_k$}} (D2Nz);
\draw[very thick,violet] (CNm1) to node[midway,above=0.1] {\textcolor{violet}{\scriptsize $\fz_k \fc_k$}} (D2Nzc);
\draw[very thick,blue] (CNm1) to node[midway,below] {\textcolor{blue}{\scriptsize $\fc_k$}} (D2Nc);
\end{tikzpicture}
}
\subsubsection*{$T[\SO(2N)]$ theory}
We now analyse the action of the discrete topological ($\BZ^{[0]}_{2,\CM_k}$) and charge conjugation ($\BZ^{[0]}_{2,\CC_k}$) symmetries, associated with the $\SO(2k)$ gauge nodes, on the Coulomb branch operators of the $T[\SO(2N)]$ theory.

We begin by examining the case $k=1$, where the relevant gauge group is $\SO(2)$. The fundamental monopole operators $X_{(\pm 1;0;\ldots)}$, carrying $\pm 1$ unit of $\SO(2)$ magnetic flux and zero flux for all other gauge groups, are odd under the topological symmetry $\BZ^{[0]}_{2,\CM_1}$, but do not have a definite parity under the charge conjugation $\BZ^{[0]}_{2,\CC_1}$. Following \cite[(2.14)]{Aharony:2013kma}, it is convenient to define the linear combinations 
\bes{ \label{Wpm}
W_{(1;0;\ldots)}^{\pm}=X_{(1;0;\ldots)} \pm X_{(-1;0;\ldots)}~, 
} 
which are even ($+$) and odd ($-$) under $\BZ^{[0]}_{2,\CC_1}$ respectively. Similar arguments apply to the $2(2N-3)$ dressed monopole operators denoted by
\bes{ \label{monopole1}
V_{(\pm 1;{\vec U}^i;{\vec 0}^{2N-3-i})}~,
}
with $\vec U = (1,0,\ldots,0 )$ and ${\vec 0} = (0,\ldots,0 )$. Here, the notation signifies a monopole operator with magnetic flux $\pm 1$ for the first gauge node (\ie the $\SO(2)$ node). For the subsequent nodes from the second (namely, the $\USp(2)$ node) up to the $(i+1)$-th gauge group (where $i=1, \ldots, 2N-3$), the magnetic flux is represented by $\vec U = (1,0,\ldots,0)$. The magnetic fluxes for the remaining gauge groups beyond node $(i+1)$ are zero, denoted by ${\vec 0}^{2N-3-i}$. Like the fundamental monopoles $X_{(\pm 1;0;\ldots)}$, the monopole operators \eref{monopole1} are odd under $\BZ^{[0]}_{2,\CM_1}$ and lack definite $\BZ^{[0]}_{2,\CC_1}$ parity. The combinations 
\bes{ \label{Whatpm}
\hat{W}_{(\pm 1;\vec{U}^i;\vec{0}^{2N-3-i})}^{\pm} = V_{(+1;\vec{U}^i;\vec{0}^{2N-3-i})} \pm V_{(-1;\vec{U}^i;\vec{0}^{2N-3-i})}
}
are even ($+$) and odd ($-$) under $\BZ^{[0]}_{2,\CC_1}$, respectively.\footnote{At the level of the index, the contribution of the monopole operators $X_{(\pm 1;0,\ldots)}$ and $V_{(\pm 1;{\vec U}^i;{\vec 0}^{2N-3-i})}$ at order $x$ is given by $\frac{1}{2} a^{-2} \zeta_1 \left(1 + \chi_1\right)$.} Finally, there exist $2N-3$ monopole operators $V_{(0;{\vec U}^i;{\vec 0}^{2N-3-i})}$ (with $i = 1, \ldots, 2N-3$) carrying zero $\SO(2)$ flux, along with the operator $\epsilon^{i_1 i_2} \varphi_{i_1 i_2}$ constructed from the $\SO(2)$ adjoint scalar $\varphi$. These operators are even under $\BZ^{[0]}_{2,\CM_1}$, but odd under $\BZ^{[0]}_{2,\CC_1}$.

The transformation properties of the operators contributing to the $\so(2N)_\CM$ Coulomb branch symmetry currents under $\BZ^{[0]}_{2,\CM_1}$ and $\BZ^{[0]}_{2,\CC_1}$ are summarised in Table \ref{tab:TSOCBmmSO2}. Based on these charges, gauging either $\BZ^{[0]}_{2,\CM_1}$ or $\BZ^{[0]}_{2,\CC_1}$ projects out exactly $4N-4$ operators from the original $\frac{1}{2} (2N)(2N-1)$ generators. This leaves $\frac{1}{2} (2N-2) (2N-3)+1$ surviving operators. Consequently, the $\so(2N)_\CM$ Coulomb branch symmetry is broken down to $\so(2N-2) \oplus \so(2)$ in either gauging scenario ($k=1$).
\begin{table}[!t]
	\begin{center}
	\begin{tabular}{c|c|c|c}
		Operators & Multiplicity & $\BZ^{[0]}_{2,\CM_1}$ & $\BZ^{[0]}_{2,\CC_1}$\\
		\hline
            $W_{(1;0;\ldots)}^{+}$ & $1$ & \ding{51} & \ding{55}\\
            $W_{(1;0;\ldots)}^{-}$ & $1$ & \ding{51} & \ding{51}\\
            $\hat{W}_{(1;{\vec U}^i;{\vec 0}^{2N-3-i})}^{+}$ & $2N-3$ & \ding{51} & \ding{55}\\
            $\hat{W}_{(1;{\vec U}^i;{\vec 0}^{2N-3-i})}^{-}$ & $2N-3$ & \ding{51} & \ding{51}\\
            $V_{(0;{\vec U}^i;{\vec 0}^{2N-3-i})}$ & $2N-3$ & \ding{55} & \ding{51}\\
            $\epsilon^{i_1 i_2} \varphi_{i_1 i_2}$ & $1$ & \ding{55} & \ding{51}
	\end{tabular}
\end{center}
	\caption{Coulomb branch moment map operators of the $T[\SO(2N)]$ theory which are charged under $\BZ^{[0]}_{2,\CM_1}$ and/or $\BZ^{[0]}_{2,\CC_1}$, where the conventions are as stated in the main text. We put either a \ding{51} or a \ding{55} to indicate whether an operator is either charged or uncharged respectively under a $\BZ_2$ symmetry.}
	\label{tab:TSOCBmmSO2}
\end{table}

For $k > 1$, a generalisation of the discussion below \eref{CBSO4w3flv} reveals that the following Coulomb branch moment map operators are odd under $\BZ^{[0]}_{2,\CM_k}$:
\begin{subequations}
\begin{align}
\begin{split} \label{TSO2Nmonzeta0}
V_{(0;{\vec 0}^{2 (k-1)-1-i};{\vec U}^i; \, {\orangered \vec U}\, ;{\vec U}^j;{\vec 0}^{2 (N-k-1) +1-j})}~,
\end{split} \\
\begin{split} \label{TSO2Nmonzetapm1}
V_{(\pm1;{\vec U}^{2 (k-1)-1}; \, {\orangered \vec U} \,;{\vec U}^j;{\vec 0}^{2 (N-k-1) +1-j})}~,
\end{split}
\end{align}
\end{subequations}
where $i=0,\ldots,2 (k-1)-1$ and $j =0, \ldots, 2 (N-k-1) +1 $. Both types of monopole operators \eref{TSO2Nmonzeta0} and \eref{TSO2Nmonzetapm1} have one unit of $\SO(2k)$ magnetic flux (denoted by ${\orangered \vec U}$ and highlighted in {\orangered orange}), while the fluxes for the other gauge groups are taken as follows. On the right-hand side of the $\SO(2k)$ node, according to the quiver on top in \eref{TSO2Nmirror}, there are $j$ gauge nodes, starting with $\USp(2k)$, with one unit of magnetic flux, while the remaining $2 (N-k-1) +1-j$ gauge nodes, ending with the $\USp(2N-2)$ node, possess zero magnetic flux. On the left-hand side of the $\SO(2k)$ gauge node, there are $i$ gauge nodes, starting from the $\USp(2k-2)$ node, possessing a unit of magnetic flux, followed by a chain of $2 (k-1)-1-i$ gauge nodes with zero flux, ending with the $\USp(2)$ node. The $\SO(2)$ gauge flux can be either zero, as in \eref{TSO2Nmonzeta0}, or $\pm1$, as in \eref{TSO2Nmonzetapm1}, that is when $i=2 (k-1)-1$. Moreover, if either $i$ and $j$ are both equal to zero or both of them are larger than zero, then the operators \eref{TSO2Nmonzeta0} and \eref{TSO2Nmonzetapm1} are even under $\BZ^{[0]}_{2,\CC_k}$. On the other hand, of either $i$ or $j$ is equal to zero, whereas the other is larger than zero, such monopole operators are odd under $\BZ^{[0]}_{2,\CC_k}$. 

Additionally, there are the Coulomb branch moment map operators with zero $\SO(2k)$ flux (denoted ${\orangered {\vec 0}}$) which are even under $\BZ^{[0]}_{2,\CM_k}$, but odd under $\BZ^{[0]}_{2,\CC_k}$:
\begin{subequations}
\begin{align}
\begin{split} \label{TSO2Nmonchi0}
V_{(0;{\vec 0}^{2 (k-1)-1-i};{\vec U}^i;\, {\orangered {\vec 0}}\, ;{\vec U}^j;{\vec 0}^{2 (N-k-1) +1-j})}~,
\end{split} \\
\begin{split} \label{TSO2Nmonchipm1}
V_{(\pm1;{\vec U}^{2 (k-1)-1};\, {\orangered {\vec 0}}\, ;{\vec U}^j;{\vec 0}^{2 (N-k-1) +1-j})}~,
\end{split}
\end{align}
\end{subequations}
where the notation for fluxes other than at the $\SO(2k)$ node is as in \eref{TSO2Nmonzeta0} and \eref{TSO2Nmonzetapm1}. The only difference is that, in both \eref{TSO2Nmonchi0} and \eref{TSO2Nmonchipm1}, one of  $i$ and $j$ is equal to zero and the other is larger than zero.
\begin{table}[!t]
	\begin{center}
        \scalebox{0.91}{
	\begin{tabular}{c|l|c|c}
		Operators & Multiplicity & $\BZ^{[0]}_{2,\CM_k}$ & $\BZ^{[0]}_{2,\CC_k}$\\
		\hline
            \eref{TSO2Nmonzeta0} and \eref{TSO2Nmonzetapm1} & $1+\left[2 (k-1)+1\right] \left[2 (N-k-1)+1\right]$ & \ding{51} & \ding{55}\\
            with $i=j=0$ or both $i, j > 0$&$=4 k (N-k)-2 N+2$&\\
            \hline
            \eref{TSO2Nmonzeta0} and \eref{TSO2Nmonzetapm1} & $2+\left[2 (k-1)-1\right]+\left[2 (N-k-1)+1\right]$ & \ding{51} & \ding{51}\\
            with $i=0, j>0$ or $i>0, j=0$&$= 2 (N-1)$&\\
            \hline
             \eref{TSO2Nmonchi0} and \eref{TSO2Nmonchipm1} & $2+\left[2 (k-1)-1\right]+\left[2 (N-k-1)+1\right]$ & \ding{55} & \ding{51}\\
             with $i=0, j>0$ or $i>0, j=0$ & $= 2 (N-1)$&
	\end{tabular}
    }
\end{center}
	\caption{Coulomb branch moment map operators of the $T[\SO(2N)]$ theory which are charged under $\BZ^{[0]}_{2,\CM_k}$ and/or $\BZ^{[0]}_{2,\CC_k}$, where the conventions regarding \ding{51} and \ding{55} are the same as in Table \ref{tab:TSOCBmmSO2}.}
	\label{tab:TSOCBmmSO2k}
\end{table}

Counting the operators listed in Table \ref{tab:TSOCBmmSO2k} based on their charges allows us to determine the effect of gauging. Among the $\frac{1}{2}(2N)(2N-1)$ operators in the adjoint representation of the $\so(2N)_\CM$ symmetry, the table indicates that $4k(N-k)$ operators (sum of multiplicities in rows 1 and 2) are projected out upon gauging $\BZ^{[0]}_{2,\CM_k}$. Similarly, upon gauging the diagonal subgroup of $\BZ^{[0]}_{2,\CM_k} \times \BZ^{[0]}_{2,\CC_k}$ (which we will denote by $\BZ^{[0]}_{2,\CM_k \CC_k}$), $4k(N-k)$ operators (sum of multiplicities in rows 1 and 3) are projected out. Gauging $\BZ^{[0]}_{2,\CM_k}$ turns the $\SO(2k)$ gauge group into $\Spin(2k)$; gauging the diagonal $\BZ^{[0]}_{2,\CM_k \CC_k}$ turns it into $\O(2k)^-$. In either of these cases, the number of surviving operators is $\frac{1}{2}(2N)(2N-1) - 4k(N-k) = \frac{1}{2}(2k)(2k-1) + \frac{1}{2}(2N-2k)(2N-2k-1)$. This corresponds precisely to the dimension of $\so(2k) \oplus \so(2N-2k)$, indicating that the $\so(2N)_\CM$ symmetry is broken to $\so(2k) \oplus \so(2N-2k)$ upon gauging.

On the other hand, gauging $\BZ^{[0]}_{2,\CC_k}$ (which transforms the $\SO(2k)$ node into $\O(2k)^+$) projects out operators which are odd under $\BZ^{[0]}_{2,\CC_k}$, which are those in rows 2 and 3, totalling $4(N-1)$. In this case, the number of remaining moment map operators is $\frac{1}{2}(2N) (2N-1) - 4 (N-1) = \frac{1}{2}(2N-2) (2N-3)+1$. This matches the dimension of $\so(2N-2) \oplus \so(2)$.

In conclusion, the $\so(2N)_{\CM}$ symmetry of the $T[\SO(2N)]$ theory breaks as follows upon modifying the $\SO(2k)$ node:
\begin{itemize}
    \item Gauging $\BZ^{[0]}_{2,\CM_k}$, \ie $\SO(2k) \to \Spin(2k)$: $\so(2N)_\CM \to \so(2k) \oplus \so(2N-2k)$.
    \item Gauging $\BZ^{[0]}_{2,\CM_k \CC_k}$, \ie $\SO(2k) \to \O(2k)^-$: $\so(2N)_\CM \to \so(2k) \oplus \so(2N-2k)$.
    \item Gauging $\BZ^{[0]}_{2,\CC_k}$, \ie $\SO(2k) \to \O(2k)^+$: $\so(2N)_\CM \to \so(2N-2) \oplus \so(2)$.
\end{itemize}
Observe that the resulting global symmetry is the same when the $\SO(2k)$ node is turned into $\Spin(2k)$ or $\O(2k)^-$, which is consistent with the duality \eref{dualityN4} applied locally at that node. Moreover, when the $\SO(2k)$ node is turned into $\O(2k)^+$, the resulting Coulomb branch global symmetry is independent of $k$.

\subsubsection*{Mirror Theory $T[\SO(2N)^\vee]$: Symmetry Actions on the Higgs Branch}
Let us now analyse the action of the discrete symmetries on the mirror theory $T[\SO(2N)^\vee]$. Recall from \eref{TSO2Nmirror} that the actions corresponding to the original $\BZ^{[0]}_{2,\CM_k}$ and $\BZ^{[0]}_{2,\CC_k}$ symmetries (represented by fugacities $\mathfrak{z}_k=\zeta_k$ and $\mathfrak{c}_k=\chi_k$ in the mirror description) break the $\so(2N)_F$ flavour symmetry. We label the fundamental half-hypermultiplets transforming under the $\USp(2N-2)$ gauge group according to their charges under these mirror actions:
\begin{itemize}
    \item $A$: Half-hypermultiplets transforming under $\USp(2N-2) \times \SO(2k-1)$, charged only under the $\mathfrak{z}_k$ action ($\leftrightarrow \BZ^{[0]}_{2,\CM_k}$).
    \item $B$: Half-hypermultiplets transforming under $\USp(2N-2) \times \SO(1)$, charged under both $\mathfrak{z}_k$ and $\mathfrak{c}_k$ actions ($\leftrightarrow \BZ^{[0]}_{2,\CM_k} \times \BZ^{[0]}_{2,\CC_k}$).
    \item $C$: Half-hypermultiplets transforming under $\USp(2N-2) \times \SO(1)$, charged only under the $\mathfrak{c}_k$ action ($\leftrightarrow \BZ^{[0]}_{2,\CC_k}$).
    \item $D$: Half-hypermultiplets transforming under $\USp(2N-2) \times \SO(2N-2k-1)$, uncharged under both $\mathfrak{z}_k$ and $\mathfrak{c}_k$ actions.
\end{itemize}

From these fundamental fields $A, B, C, D$ (which are doublets under the $\USp(2N-2)$ gauge group), we can construct the Higgs branch moment map operators, which are gauge invariant mesons of the form $Q_1 J Q_2$, where $Q_1, Q_2 \in \{A, B, C, D\}$ and $J$ is the $\USp(2N-2)$ invariant tensor ($J = \ID_{N-1} \otimes i \sigma_2$). The transformation properties of these mesons under the mirror discrete actions ($\mathfrak{z}_k, \mathfrak{c}_k$) are summarized in Table \ref{tab:TSOmirrHBmm}.

\begin{table}[!t]
	\begin{center}
	\begin{tabular}{c|c|c|c}
		Mesons & Multiplicity & $\BZ^{[0]}_{2,\CM_k}$ & $\BZ^{[0]}_{2,\CC_k}$\\
		\hline
            $A J B$ & $2k-1$ & \ding{55} & \ding{51}\\
            $A J C$ & $2k-1$ & \ding{51} & \ding{51}\\
            $A J D$ & $(2k-1)(2N-2k-1)$ & \ding{51} & \ding{55}\\
            $B J C$ & $1$ & \ding{51} & \ding{55}\\
            $B J D$ & $2N-2k-1$ & \ding{51} & \ding{51}\\
            $C J D$ & $2N-2k-1$ & \ding{55} & \ding{51}
	\end{tabular}
\end{center}
	\caption{Higgs branch moment map operators of the $T[\SO(2N)^\vee]$ theory which are charged under the mirror actions corresponding to $\BZ^{[0]}_{2,\CM_k}$ and/or $\BZ^{[0]}_{2,\CC_k}$, where the conventions regarding \ding{51} and \ding{55} are the same as in Table \ref{tab:TSOCBmmSO2}.}
	\label{tab:TSOmirrHBmm}
\end{table}

From Table \ref{tab:TSOmirrHBmm}, we can count the mesonic operators based on their charges under the mirror actions $\mathfrak{z}_k (\leftrightarrow \BZ^{[0]}_{2,\CM_k})$ and $\mathfrak{c}_k (\leftrightarrow \BZ^{[0]}_{2,\CC_k})$. The operators charged only under $\mathfrak{z}_k$ (types $AJD, BJC$) sum to multiplicity $(2k-1)(2N-2k-1)+1$. Those charged only under $\mathfrak{c}_k$ (types $AJB, CJD$) sum to $2(N-1)$. Those charged under both (types $AJC, BJD$) sum to $2(N-1)$.

Therefore, gauging the symmetry corresponding to $\mathfrak{z}_k$ (mirror to $\BZ^{[0]}_{2,\CM_k}$) projects out operators odd under this action (types $AJC, AJD, BJC, BJD$), removing a total of $4k(N-k)$ operators contributing to the $\so(2N)_F$ currents. The surviving $k (2k-1)+(N-k)(2N-2k-1)$ operators generate the residual flavour symmetry $\so(2k) \oplus \so(2N-2k)$. Conversely, gauging the symmetry corresponding to $\mathfrak{c}_k$ (mirror to $\BZ^{[0]}_{2,\CC_k}$) projects out operators odd under this action (types $AJB, AJC, BJD, CJD$), removing $4(N-1)$ operators. The remaining $(N-1)(2N-3)+1$ operators generate the residual flavour symmetry $\so(2N-2) \oplus \so(2)$.

Comparing these Higgs branch symmetry breaking patterns (under gauging $\mathfrak{z}_k$ or $\mathfrak{c}_k$) with the Coulomb branch symmetry breaking patterns found previously for the $T[\SO(2N)]$ theory (under gauging $\BZ^{[0]}_{2,\CM_k}$ or $\BZ^{[0]}_{2,\CC_k}$ respectively), we find perfect agreement. This supports the mirror duality map \eref{TSO2Nmirror} with the identification $\zeta_k = \mathfrak{z}_k$ and $\chi_k = \mathfrak{c}_k$. Furthermore, the mapping extends to the operators themselves. The Coulomb branch moment map operators of $T[\SO(2N)]$ and Higgs branch moment map operators of $T[\SO(2N)^\vee]$ transforming identically under the corresponding discrete symmetries can be matched as follows:
\bes{ \label{TSO2Nmappingmm}
\left(\begin{array}{cc}
k=1: & \{W_{(1;0;\ldots)}^{+}, \hat{W}_{(1;\vec U^i;\vec 0^{2N-3-i})}^{+}\} \\
k>1: & \{\text{\eref{TSO2Nmonzeta0}, \eref{TSO2Nmonzetapm1}}\} \\ & \text{with $i=j=0$ or both $i, j > 0$}
\end{array}\right)
\qquad &\Longleftrightarrow \qquad \{\text{$A J D$, $B J C$}\}\\
\left(\begin{array}{cc}
k=1: & \{W_{(1;0;\ldots)}^{-}, \hat{W}_{(1;\vec U^i;\vec 0^{2N-3-i})}^{-}\} \\
k>1: & \{\text{\eref{TSO2Nmonzeta0}, \eref{TSO2Nmonzetapm1}}\} \\ & \text{with $i=0, j>0$ or $i>0, j=0$}
\end{array}\right)
\qquad &\Longleftrightarrow \qquad \{\text{$A J C$, $B J D$}\}~.
\\
\left(\begin{array}{cc}
k=1: & \{V_{(0;\vec U^i;\vec 0^{2N-3-i})}, \epsilon^{i_1 i_2} \varphi_{i_1 i_2}\} \\
k>1: & \{\text{\eref{TSO2Nmonchi0}, \eref{TSO2Nmonchipm1}}\} \\ & \text{with $i=0, j>0$ or $i>0, j=0$}
\end{array}\right)
\qquad &\Longleftrightarrow \qquad \{\text{$A J B$, $C J D$}\}
}
\subsubsection*{The index}
The mirror duality \eref{TSO2Nmirror} can be checked explicitly using the index. The index of the $T[\SO(2N)]$ theory with $\vec \chi_{(N-1)} = (\chi_1, \ldots, \chi_{N-1}) = \vec{+1}_{(N-1)} = (+1, \ldots, +1)$, namely the charge conjugation fugacities associated with every $\SO(2k)$ gauge group are set to $+1$, can be written recursively as
\bes{ \label{indTSO2Nchip}
\scalebox{0.99}{$
\begin{split}
&\CI\left\{T[\SO(2N)]\right\}(x;a; \vec f,\vec l;  \vec \zeta_{(N-1)}; \vec \chi_{(N-1)}=\vec{+1}_{(N-1)}) \\ &= \frac{1}{(N-1)! 2^{N-2}} \times \frac{1}{(N-1)! 2^{N-1}} \\ & \times \sum_{\vec m_{(N-1)} \in \BZ^{N-1}} \,\, \sum_{\vec n_{(N-1)} \in \BZ^{N-1}} \zeta_{N-1}^{\sum_i m_{(N-1),i}} \oint \left( \prod_{\alpha, \beta =1}^{N-1} \frac{d z_{(N-1),\alpha}}{2\pi i z_{(N-1),\alpha}}  \frac{d u_{(N-1),\beta}}{2\pi i u_{(N-1),\beta}} \right) \\ & \times  \CZ_{\text{vec}}^{\SO(2N-2)}(x; \vec z_{(N-1)}; \vec m_{(N-1)}; \chi_{N-1} = +1) \CZ_{\text{vec}}^{\USp(2N-2)}(x; \vec u_{(N-1)}; \vec n_{(N-1)}) \\
& \times \CZ^{\SO(2N-2)}_{\mathcal{A}}(x; a; \vec z_{(N-1)}; \vec m_{(N-1)}; \chi_{N-1}=+1) \CZ^{\USp(2N-2)}_{\mathcal{A}}(x; a; \vec u_{(N-1)}; \vec n_{(N-1)}) \\
& \times \prod_{\alpha, \beta=1}^{N-1} \prod_{s_1, s_2=\pm 1} \CZ_{\text{chir}}^{1/2} (x; a z^{s_1}_{(N-1),\alpha} u^{s_2}_{(N-1),\beta};s_1 m_{(N-1),\alpha} +s_2 n_{(N-1),\beta}) \\
& \times {\claret \prod_{i=1}^{N} \prod_{\beta=1}^{N-1} \prod_{s_1, s_2=\pm 1} \CZ_{\text{chir}}^{1/2} (x; a f_i^{s_1} u^{s_2}_{(N-1),\beta};s_1 l_i +s_2 n_{(N-1),\beta})} \\ & \times \CI\left\{T[\SO(2N-2)]\right\}(x;a; \vec z_{(N-1)};\vec m_{(N-1)};  \vec \zeta_{(N-2)}; \vec \chi_{(N-2)}=\vec{+1}_{(N-2)})~,
\end{split}
$}
}
where $\vec \zeta_{(N-1)} = (\zeta_1, \ldots, \zeta_{N-1})$ are the topological fugacities associated with the $\SO(2), \ldots, \SO(2N-2)$ gauge nodes. Moreover, $\vec z_{(k)}=(z_{(k),1}, \ldots, z_{(k),k})$ and $\vec m_{(k)}=(m_{(k),1}, \ldots, m_{(k),k})$ are the fugacities and magnetic fluxes associated with the $\SO(2k)$ gauge node, $\vec u_{(k)}=(u_{(k),1}\ldots, u_{(k),k})$ and $\vec n_{(k)}=(n_{(k),1}, \ldots, n_{(k),k})$ are those associated with the $\USp(2k)$ gauge node, whereas $\vec f = (f_1, \ldots, f_N)$ and $\vec l = (l_1, \ldots, l_N)$ are the fugacities and background fluxes associated with the $\so(2N)_F$ flavour symmetry. The expression \eref{indTSO2Nchip} can be defined starting from the index of the $T[\SO(4)]$ theory, that is
\bes{ \label{indTSO4chip}
&\CI\left\{T[\SO(4)]\right\}(x;a; \vec f; \vec l;  \zeta_1; \chi_1=+1) \\
& = \frac{1}{2} \sum_{m_{(1),1} \in \BZ} \,\, \sum_{n_{(1),1} \in \BZ} \zeta^{m_{(1),1}} \oint \frac{d z_{(1),1}}{2\pi i z_{(1),1}}  \frac{d u_{(1),1}}{2\pi i u_{(1),1}} \CZ_{\text{vec}}^{\USp(2)}(x; u_{(1),1}; n_{(1),1}) \\
& \qquad \times \CZ^{\SO(2)}_{\mathcal{A}}(x; a; z_{(1),1}; m_{(1),1}; \chi_1=+1) \CZ^{\USp(2)}_{\mathcal{A}}(x; a; u_{(1),1}; n_{(1),1})  \\
& \qquad \times \prod_{s_1, s_2=\pm 1} \CZ_{\text{chir}}^{1/2} (x; a z^{s_1}_{(1),1} u^{s_2}_{(1),1};s_1 m_{(1),1} +s_2 n_{(1),1}) \\
& \qquad \times \prod_{i=1}^{2} \prod_{s_1, s_2=\pm 1} \CZ_{\text{chir}}^{1/2} (x; a f_i^{s_1} u^{s_2}_{(1),1};s_1 l_i +s_2 n_{(1),1})~.
}
The index of the $T[\SO(2N)]$ theory with the charge conjugation fugacity associated with the $\SO(2k)$ gauge group $\chi_k = -1$ can be obtained from \eref{indTSO2Nchip} by replacing $\CZ_{\text{vec}/\mathcal{A}}^{\SO(2k)}(\chi_{k} = +1)$ with $\CZ_{\text{vec}/\mathcal{A}}^{\SO(2k)}(\chi_{k} = -1)$ as defined in \eref{vecchi-1} and \eref{adjchi-1}, and by setting $z_{(k),1}=1, {z_{(k),k}}^{-1}=-1$ and $m_{(k),k} = 0$. The index of the $T[\SO(2N)]$ theory refined with respect to the fugacities $\zeta_k$ and $\chi_k$, with the other fugacities $\zeta_{j \neq k}$ and $\chi_{j \neq k}$ set to one, can then be obtained as
\bes{ \label{indTSO2Nwithchi}
&\CI\left\{T[\SO(2N)]\right\}(x;a; \zeta_k; \chi_k) \\&= 
 \frac{1}{2} \sum_{s = \pm 1} \CI\left\{T[\SO(2N)]\right\}(x;a; \zeta_k; \chi_k = s) \left(1+ s \chi_k\right)~,
}
where, for compactness, we drop the dependence of the flavour fugacities $\vec f$ and of the corresponding background fluxes $\vec l$.

The index of the mirror theory is given by \eref{indTSO2Nchip}, where the topological and charge conjugation fugacities associated with each special orthogonal gauge node are set to one, and the term in {\claret claret} is replaced by the expression $\CI_A \CI_B \CI_C \CI_D$, where
\begin{subequations}
\begin{align}
\begin{split} 
\CI_A &= \prod_{i=1}^{k-1} \prod_{\beta=1}^{N-1} \prod_{s_1, s_2=\pm 1} \CZ_{\text{chir}}^{1/2} (x; a f_i^{s_1} u^{s_2}_{(N-1),\beta} \fz_k;s_1 l_i +s_2 n_{(N-1),\beta}) \\ & \times \prod_{\beta=1}^{N-1} \prod_{s=\pm 1} \CZ_{\text{chir}}^{1/2} (x; a u^{s}_{(N-1),\beta} \fz_k;s n_{(N-1),\beta})~,
\end{split} \\
\begin{split}
\CI_B &= \prod_{\beta=1}^{N-1} \prod_{s=\pm 1} \CZ_{\text{chir}}^{1/2} (x; a u^{s}_{(N-1),\beta} \fz_k \fc_k;s n_{(N-1),\beta})~,
\end{split}\\
\begin{split}
\CI_C &= \prod_{\beta=1}^{N-1} \prod_{s=\pm 1} \CZ_{\text{chir}}^{1/2} (x; a u^{s}_{(N-1),\beta} \fc_k;s n_{(N-1),\beta})~,
\end{split}\\
\begin{split} 
\CI_D &= \prod_{i=1}^{N-k-1} \prod_{\beta=1}^{N-1} \prod_{s_1, s_2=\pm 1} \CZ_{\text{chir}}^{1/2} (x; a f_i^{s_1} u^{s_2}_{(N-1),\beta};s_1 l_i +s_2 n_{(N-1),\beta}) \\ & \,\,\,\,\, \times  \prod_{\beta=1}^{N-1} \prod_{s=\pm 1} \CZ_{\text{chir}}^{1/2} (x; a u^{s}_{(N-1),\beta};s n_{(N-1),\beta})~.
\end{split}
\end{align}
\end{subequations}
Upon performing this substitution and using the same convention as in \eref{indTSO2Nwithchi}, the following equality between the indices of the two mirror theories holds:
\bes{
\CI\left\{T[\SO(2N)]\right\}(x;a; \zeta_k; \chi_k) = \CI\left\{T[\SO(2N)^{\vee}]\right\}(x;a^{-1}; \fz_k = \zeta_k; \fc_k = \chi_k)~.
}
\subsubsection*{Example: the $T[\SO(6)]$ theory}
As an explicit example, we now examine the $T[\SO(6)]$ theory, corresponding to $N=3$. This theory involves gauge nodes $\SO(2)$ ($k=1$) and $\SO(4)$ ($k=2$).

Let us first focus on the topological ($\zeta_1$) and charge conjugation ($\chi_1$) symmetries associated with the $\SO(2)$ gauge node. In this case, the general mirror map \eref{TSO2Nmirror} specialises (for $N=3, k=1$) as follows:
\bes{ \label{TSO6mirrork1}
\scalebox{0.9}{$
\begin{tikzpicture}[baseline, font = \small]
\node[gauge,label=below:{$\underset{\textcolor{red}{\zeta_1}, \, \textcolor{blue}{\chi_1}}{\SO(2)}$}] (D1u) at (0,0) {};
\node[gauge,label=above:{$\USp(2)$}] (C1u) at (1.2,0) {};
\node[gauge,label=below:{$\SO(4)$}] (Dku) at (2.4,0) {};
\node[gauge,label=above:{$\USp(4)$}] (CNm1u) at (3.6,0) {};
\node[flavour,label=below:{$\SO(6)$}] (D2Nu) at (4.8,0) {};
\draw (D1u)--(C1u)--(Dku)--(CNm1u)--(D2Nu);
\node (l) at (5.9,0) {};
\node (r) at (7.9,0) {};
\draw[<->] (l) to node[midway, above] {$\text{mirror}$} (r);
\node[gauge,label=below:{${\SO(2)}$}] (D1) at (13.8,0) {};
\node[gauge,label=above:{$\USp(2)$}] (C1) at (12.6,0) {};
\node[gauge,label=below:{${\SO(4)}$}] (Dk) at (11.4,0) {};
\node[gauge,label={above,xshift=0.5cm}:{$\USp(4)$}] (CNm1) at (10.2,0) {};
\node[flavour,label=left:{$\SO(1)$}] (D2Nz) at (9,2) {};
\node[flavour,label=left:{$\SO(1)$}] (D2Nzc) at (9,1) {};
\node[flavour,label=left:{$\SO(1)$}] (D2Nc) at (9,-1) {};
\node[flavour,label=left:{$\SO(3)$}] (D2N) at (9,-2) {};
\draw (D1)--(C1)--(Dk)--(CNm1)--(D2N);
\draw[very thick,red] (CNm1) to node[midway,above=0.2] {\textcolor{red}{\scriptsize $\fz_1$}} (D2Nz);
\draw[very thick,violet] (CNm1) to node[midway,below=0.1] {\textcolor{violet}{\scriptsize $\fz_1 \fc_1$}} (D2Nzc);
\draw[very thick,blue] (CNm1) to node[midway,above] {\textcolor{blue}{\scriptsize $\fc_1$}} (D2Nc);
\end{tikzpicture}
$}
}
where we identify the fugacities as $\mathfrak{z}_1 = \zeta_1$ and $\mathfrak{c}_1 = \chi_1$ across the duality. The index of the $T[\SO(6)]$ theory, refined by the fugacities $\zeta_1, \chi_1$ associated with the $\SO(2)$ node (and keeping $\zeta_2=\chi_2=1$), expanded up to order $x^2$, reads:
\bes{ \label{indTSO6SO2}
\scalebox{0.9}{$
\begin{split}
\CI\left\{T[\SO(6)]\right\}(x;a; \zeta_1; \chi_1) = 1&+\left[ C\left(a^{-2} x\right) a^{-2} + \mathbf{15} a^2 \right] x \\&+\left\{C\left(a^{-4} x^2\right) a^{-4} + \mathbf{15} C\left(a^{-2} x\right) + \left(\mathbf{84}+\mathbf{20'}+\mathbf{15}\right) a^4 \right. \\ & \left. \, - \left[ C\left(a^{-2} x\right) + \mathbf{15} +1 \right]\right\} x^2 + \ldots~,
\end{split}
$}
}
with coefficients depending on the discrete fugacities:
\bes{ \label{coeffPSO6m}
&C\left(a^{-2} x\right)=3 + {\red 4 \zeta_1} + {\blue 4 \chi_1} + {\violet 4 \zeta_1 \chi_1} ~, \\ &C\left(a^{-4} x^2\right)=35+28 \zeta_1 + 28 \chi_1 + 28 \zeta_1 \chi_1~,
}
and the terms contributing to $C\left(a^{-2} x\right)$ above correspond to the following operators:
\bes{
\begin{array}{cl}
{3}&: \{V_{(0;0;1,0;0,0)}, V_{(0;0;0,0;1,0)}, V_{(0;0;1,0;1,0)}\} \\
{\red 4 \zeta_1}&: \{W_{(1;0;0,0;0,0)}^{+}, \hat{W}_{(1;1;0,0;0,0)}^{+}, \hat{W}_{(1;1;1,0;0,0)}^{+}, \hat{W}_{(1;1;1,0;1,0)}^{+}\} \\
{\blue 4 \chi_1}&: \{V_{(0;1;0,0;0,0)}, V_{(0;1;1,0;0,0)}, V_{(0;1;1,0;1,0)}, \epsilon^{i_1 i_2} \varphi_{i_1 i_2}\} \\
{\violet 4 \zeta_1 \chi_1}&: \{W_{(1;0;0,0;0,0)}^{-}, \hat{W}_{(1;1;0,0;0,0)}^{-}, \hat{W}_{(1;1;1,0;0,0)}^{-}, \hat{W}_{(1;1;1,0;1,0)}^{-}\}
\end{array}~,
}
where we use the same convention as in Table \ref{tab:TSOCBmmSO2}. The $\su(4)$ representations for the flavour symmetry appearing in \eref{indTSO6SO2} are as follows:
\bes{
\mathbf{15}=[1,0,1]~, \quad \mathbf{20'} = [0,2,0]~, \quad \mathbf{84} = [2,0,2]~.
}
Observe that such representations are uncharged under the $\BZ_4$ centre symmetry of $\su(4)$, in agreement with the global form $\PSO(6)_F$ of the flavour symmetry. Furthermore, analysing the representations contributing to the Coulomb branch limit when $\zeta_1 = \chi_1 = 1$ confirms the global form $\PSO(6)_\CM$ for the topological symmetry (as only representations invariant under the appropriate centre symmetry appear).

Let us now discuss how this $\PSO(6)_\CM$ symmetry is broken by gauging $\BZ^{[0]}_{2,\zeta_1}$ or $\BZ^{[0]}_{2,\chi_1}$. Gauging corresponds to averaging the index over the respective fugacity. The coefficients become:
\bes{
&\frac{1}{2} \sum_{\zeta_1=\pm 1}C\left(a^{-2} x\right)\overset{\chi_1 = 1}{=} \frac{1}{2} \sum_{\chi_1=\pm 1}C\left(a^{-2} x\right)\overset{\zeta_1 = 1}{=}7~, \\ &\frac{1}{2} \sum_{\zeta_1=\pm 1} C\left(a^{-4} x^2\right) \overset{\chi_1 = 1}{=}\frac{1}{2} \sum_{\chi_1=\pm 1} C\left(a^{-4} x^2\right) \overset{\zeta_1 = 1}{=}63~.
}
These coefficients can be reached starting from the $\mathbf{15}$ and the $\mathbf{84}+\mathbf{20'}+\mathbf{15}$ representation of the $\so(6)_\CM$ symmetry, by exploiting the following branching rules from $\so(6) \cong \su(4)$ to $\su(2) \oplus \su(2) \oplus \u(1)$:
\bes{ \label{brsu4tosu2su2u1}
\scalebox{0.97}{$
\begin{split}
\mathbf{15} \,\, \longrightarrow \,\, &{\orangered (\mathbf{2},\mathbf{2})(2)} \oplus  (\mathbf{3},\mathbf{1})(0) \oplus (\mathbf{1},\mathbf{3})(0) \oplus (\mathbf{1},\mathbf{1})(0) \oplus {\orangered (\mathbf{2},\mathbf{2})(-2)} \\
\mathbf{20'} \,\, \longrightarrow \,\, &(\mathbf{1},\mathbf{1})(4) \oplus {\orangered (\mathbf{2},\mathbf{2})(2)} \oplus  (\mathbf{3},\mathbf{3})(0) \oplus (\mathbf{1},\mathbf{1})(0) \oplus {\orangered (\mathbf{2},\mathbf{2})(-2)} \oplus (\mathbf{1},\mathbf{1})(-4) \\
\mathbf{84} \,\, \longrightarrow \,\, &(\mathbf{3},\mathbf{3})(4) \oplus {\orangered (\mathbf{4},\mathbf{2})(2)} \oplus {\orangered (\mathbf{2},\mathbf{4})(2)} \oplus {\orangered (\mathbf{2},\mathbf{2})(2)} \\ \oplus&  (\mathbf{5},\mathbf{1})(0) \oplus  (\mathbf{3},\mathbf{3})(0) \oplus  (\mathbf{3},\mathbf{1})(0) \oplus  (\mathbf{1},\mathbf{5})(0) \oplus  (\mathbf{1},\mathbf{3})(0) \oplus (\mathbf{1},\mathbf{1})(0) \\ \oplus & {\orangered (\mathbf{4},\mathbf{2})(-2)} \oplus {\orangered (\mathbf{2},\mathbf{4})(-2)} \oplus {\orangered (\mathbf{2},\mathbf{2})(-2)} \oplus (\mathbf{3},\mathbf{3})(-4)~,
\end{split}
$}
}
where the terms highlighted in {\orangered orange} are projected out upon gauging either $\BZ^{[0]}_{2,\zeta_1}$ or $\BZ^{[0]}_{2,\chi_1}$. Note that the surviving operators, denoted in black, are invariant under the $\BZ_2$ centre of each $\su(2)$ factor and possess $\u(1)$ charge zero (modulo four). 

We observe that, upon gauging either $\BZ^{[0]}_{2,\zeta_1}$ or $\BZ^{[0]}_{2,\chi_1}$, the topological symmetry $\so(6)_\CM$ of the $T[\SO(6)]$ theory gets reduced to $\su(2) \oplus \su(2) \oplus \u(1) \cong \so(4) \oplus \so(2)$. On the mirror side, the index perfectly matches with \eref{indTSO6SO2} as follows:
\bes{
\CI\left\{T[\SO(6)^\vee]\right\}(x;a^{-1}; \fz_1 = \zeta_1; \fc_1 = \chi_1) = \CI\left\{T[\SO(6)]\right\}(x;a; \zeta_1; \chi_1)~,
}
where, upon gauging the symmetry which, under mirror symmetry, maps to either $\BZ^{[0]}_{2,\CM_1}$ or $\BZ^{[0]}_{2,\CC_1}$, the flavour symmetry $\so(6)_F$ gets reduced to $\so(4) \oplus \so(2)$, whereas the topological symmetry remains $\so(6)_\CM$.

Let us now move on to discuss the case $k = 2$, focusing on the fugacities $\zeta_2$ and $\chi_2$ associated with the $\SO(4)$ node of the $T[\SO(6)]$ theory. In this case, the mirror map \eref{TSO2Nmirror} specialises (for $N=3, k=2$) as follows:
\bes{ \label{TSO6mirrork1}
\scalebox{0.9}{$
\begin{tikzpicture}[baseline, font = \small]
\node[gauge,label=below:{$\SO(2)$}] (D1u) at (0,0) {};
\node[gauge,label=above:{$\USp(2)$}] (C1u) at (1.2,0) {};
\node[gauge,label=below:{$\underset{\textcolor{red}{\zeta_2}, \, \textcolor{blue}{\chi_2}}{\SO(4)}$}] (Dku) at (2.4,0) {};
\node[gauge,label=above:{$\USp(4)$}] (CNm1u) at (3.6,0) {};
\node[flavour,label=below:{$\SO(6)$}] (D2Nu) at (4.8,0) {};
\draw (D1u)--(C1u)--(Dku)--(CNm1u)--(D2Nu);
\node (l) at (5.9,0) {};
\node (r) at (7.9,0) {};
\draw[<->] (l) to node[midway, above] {$\text{mirror}$} (r);
\node[gauge,label=below:{${\SO(2)}$}] (D1) at (13.8,0) {};
\node[gauge,label=above:{$\USp(2)$}] (C1) at (12.6,0) {};
\node[gauge,label=below:{${\SO(4)}$}] (Dk) at (11.4,0) {};
\node[gauge,label={above,xshift=0.5cm}:{$\USp(4)$}] (CNm1) at (10.2,0) {};
\node[flavour,label=left:{$\SO(3)$}] (D2Nz) at (9,2) {};
\node[flavour,label=left:{$\SO(1)$}] (D2Nzc) at (9,1) {};
\node[flavour,label=left:{$\SO(1)$}] (D2Nc) at (9,-1) {};
\node[flavour,label=left:{$\SO(1)$}] (D2N) at (9,-2) {};
\draw (D1)--(C1)--(Dk)--(CNm1)--(D2N);
\draw[very thick,red] (CNm1) to node[midway,above=0.2] {\textcolor{red}{\scriptsize $\fz_2$}} (D2Nz);
\draw[very thick,violet] (CNm1) to node[midway,below=0.1] {\textcolor{violet}{\scriptsize $\fz_2 \fc_2$}} (D2Nzc);
\draw[very thick,blue] (CNm1) to node[midway,above] {\textcolor{blue}{\scriptsize $\fc_2$}} (D2Nc);
\end{tikzpicture}
$}
}
where we identify $\mathfrak{z}_2 = \zeta_2$ and $\mathfrak{c}_2 = \chi_2$.  The index of the $T[\SO(6)]$ can then be refined with respect to the fugacities $\zeta_2$ and $\chi_2$, yielding the same expansion given by \eref{indTSO6SO2}, where $\zeta_1$ and $\chi_1$ are replaced by $\zeta_2$ and $\chi_2$ respectively. In particular, the following operators contribute to the term $C(a^{-2} x)$ in the index:
\bes{
\begin{array}{cl}
{3}&: \{V_{(1;0;0,0;0,0)}, V_{(-1;0;0,0;0,0)}, \epsilon^{i_1 i_2} \varphi_{i_1 i_2}\} \\
{\red 4 \zeta_2}&: \{V_{(0;0;1,0;0,0)}, V_{(0;1;1,0;1,0)}, V_{(1;1;1,0;1,0)}, V_{(-1;1;1,0;1,0)}\} \\
{\blue 4 \chi_2}&: \{V_{(0;1;0,0;0,0)}, V_{(1;1;0,0;0,0)}, V_{(-1;1;0,0;0,0)}, V_{(0;0;0,0;1,0)}\} \\
{\violet 4 \zeta_2 \chi_2}&: \{V_{(0;1;1,0;0,0)}, V_{(1;1;1,0;0,0)}, V_{(-1;1;1,0;0,0)}, V_{(0;0;1,0;1,0)}\}
\end{array}~,
}
where we use the same notation as in \eref{TSO2Nmonzeta0}--\eref{TSO2Nmonchipm1}. By the same arguments provided in case $k = 1$, it follows that, upon gauging either $\BZ^{[0]}_{2,\zeta_2}$ or $\BZ^{[0]}_{2,\chi_2}$, the topological symmetry $\so(6)_\CM$ of the $T[\SO(6)]$ theory gets reduced to $\so(4) \oplus \so(2)$, whereas the flavour symmetry $\so(6)_F$ remains unaltered. On the other hand, in the mirror theory, upon gauging the symmetry which is mapped to either $\BZ^{[0]}_{2,\CM_2}$ or $\BZ^{[0]}_{2,\CC_2}$, the topological symmetry is still $\so(6)_\CM$, while the $\so(6)_F$ flavour symmetry of the $T[\SO(6)^\vee]$ theory gets reduced to $\so(4) \oplus \so(2)$. Consistently, the indices of the two mirror theories match as follows:
\bes{
\CI\left\{T[\SO(6)^\vee]\right\}(x;a^{-1}; \fz_2 = \zeta_2; \fc_2 = \chi_2) = \CI\left\{T[\SO(6)]\right\}(x;a; \zeta_2; \chi_2)~.
}
\subsection{$T[\USp(2N)]$ and $T[\SO(2N+1)]$ theories}
Let us move on to consider the $T[\USp(2N)]$ theory introduced in \cite{Gaiotto:2008ak}. The UV description of this SCFT consists of a 3d $\CN=4$ quiver gauge theory with alternating $\SO(2k)$ and $\USp(2l)$ gauge groups (with $k = 1, \ldots, N$ and $l = 1, \ldots, N-1$), where the final $\SO(2N)$ gauge node is connected to a $\USp(2N)$ flavour node. Similarly to the $T[\SO(2N)]$ case, among the continuos global symmetries of the theory, in addition to the $\u(1)_A$ axial symmetry and to the $\usp(2N)_F$ flavour symmetry, which is manifest in the UV quiver description, there is also an enhanced $\so(2N+1)_\CM$ topological symmetry in the IR. The latter arises from the product $\U(1)_{\CM_1} \times \prod_{k=2}^{N} \BZ^{[0]}_{2,\CM_k}$ of the $\U(1)_{\CM_1}$ topological symmetry associated with the $\SO(2)$ gauge node, together with the $\BZ^{[0]}_{2,\CM_k}$ topological symmetries associated with every $\SO(2k)$ gauge node, where $k = 2, \ldots, N$. We also take into account the effect of the $\BZ^{[0]}_{2,\CC_k}$ charge conjugation symmetry associated with every $\SO(2k)$ gauge node within the $T[\USp(2N)]$ quiver. In particular, recall that gauging $\BZ^{[0]}_{2,\CM_k}$ turns an $\SO(2k)$ gauge group into $\Spin(2k)$, whereas gauging $\BZ^{[0]}_{2,\CC_k}$ turns an $\SO(2k)$ gauge group into $\O(2k)^+$. Both operations affect the Coulomb branch of the $T[\USp(2N)]$ theory, while leaving the Higgs branch invariant. 

As pointed out in \cite{Gaiotto:2008ak}, the $T[\USp(2N)]$ theory possesses a mirror description, known as the $T[\SO(2N+1)]$ theory, with $\usp(2N)^\vee_F=\so(2N+1)_F$ flavour symmetry and $\so(2N+1)^\vee_\CM = \usp(2N)_\CM$ topological symmetry. In contrast to the $T[\SO(2N)]$ and $T[\USp(2N)]$ theories, there is no ``good" UV quiver description flowing to the $T[\SO(2N+1)]$ theory in the IR. Nevertheless, the latter theory admits a ``bad" quiver description consisting of alternating $\O(2k-1)^+$ and $\USp(2k)$ gauge groups (with $k = 1, \ldots, N$), where the final $\USp(2N)$ gauge node is attached to an $\SO(2N+1)$ flavour node.\footnote{Under the proposition that the $T[\SO(3)]$ and $T[\SU(2)]$ are the same interacting superconformal field theory \cite[Section 5.2.2]{Gaiotto:2008ak}, we see that the one-form symmetry associated with the $\O(1)$ gauge group in the quiver description of $T[\SO(3)]$
\bes{ 
\begin{tikzpicture}[baseline, font=\footnotesize]
\node[gauge,label=below:{$\O(1)$}] (ff1) at (-2,0) {};
\node[gauge,label=below:{$\USp(2)$}] (gg) at (0,0) {};
\node[flavour,label=below:{$\SO(3)$}] (ff2) at (2,0) {};
\draw[-,bend right=0] (ff1)--(gg)--(ff2);
\end{tikzpicture} 
}
acts trivially on the Wilson line in the interacting SCFT.  Upon flowing to the infrared, there could be a decoupled topological field theory that supports this one-form symmetry.  We expect that this statement continues to hold for the one-form symmetry associated with $\O(2k+1)^+$ in the quiver description of the $T[\SO(2N+1)]$ theory.}

Here, we proceed with the analysis initiated in the previous subsection by examining the action of the $\BZ^{[0]}_{2,\CM_k}$ topological symmetry and the $\BZ^{[0]}_{2,\CC_k}$ charge conjugation symmetry associated with each $\SO(2k)$ gauge node in the $T[\USp(2N)]$ quiver, and map such actions under mirror symmetry. Specifically, given an $\SO(2k)$ gauge node in the $T[\USp(2N)]$ quiver, with $\BZ^{[0]}_{2,\CM_k}$ and $\BZ^{[0]}_{2,\CC_k}$ topological and charge conjugation symmetries, where, as in the previous subsection, we denote the associated fugacities by $\zeta_k$ and $\chi_k$ respectively, we aim to understand how such discrete symmetries are mapped in the mirror $T[\SO(2N+1)]$ theory. Analogously to the mirror duality depicted in \eref{TSO2Nmirror}, we propose that the action of the $\BZ^{[0]}_{2,\CM_k}$ and $\BZ^{[0]}_{2,\CC_k}$ symmetries of the $T[\USp(2N)]$ theory is mapped according to \eref{TUSp2Nmirror} in the mirror description. Here, the actions of the fugacities $\fz_k$ and $\fc_k$ in the $T[\SO(2N+1)]$ quiver is the same as explained above \eref{TSO2Nmirror}, where the mirror duality between the two quiver descriptions holds when we set $\zeta_k = \fz_k$ and $\chi_k= \fc_k$. Observe that, for $\zeta_k = \chi_k =1$, \eref{TUSp2Nmirror} reduces to the known mirror duality between the $T[\USp(2N)]$ and $T[\SO(2N+1)]$ theories.
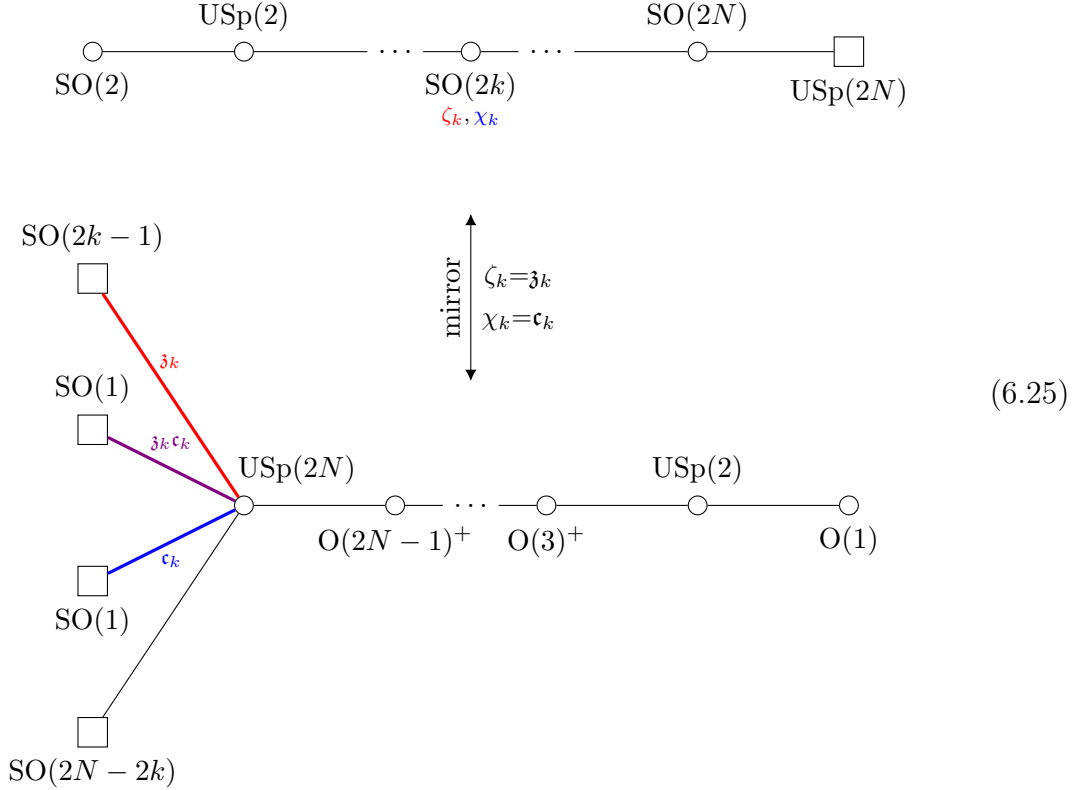
\bes{ \label{TUSp2Nmirror}
\begin{tikzpicture}[baseline, font = \small]
\node[gauge,label=below:{${\SO(2)}$}] (D1u) at (-5,6) {};
\node[gauge,label=above:{$\USp(2)$}] (C1u) at (-3,6) {};
\node[label=below:{}] (3Dlu) at (-1,6) {$\ldots$};
\node[gauge,label=below:{$\underset{\textcolor{red}{\zeta_k}, \, \textcolor{blue}{\chi_k}}{\SO(2 k)}$}] (Dku) at (0,6) {};
\node[label=below:{}] (3Dru) at (1,6) {$\ldots$};
\node[gauge,label=above:{$\SO(2N)$}] (CNm1u) at (3,6) {};
\node[flavour,label=below:{$\USp(2 N)$}] (D2Nu) at (5,6) {};
\draw (D1u)--(C1u)--(3Dlu)--(Dku)--(3Dru)--(CNm1u)--(D2Nu);
\node (t) at (0,4) {};
\node (b) at (0,1.5) {};
\draw[<->] (b) to node[midway, above,sloped] {$\text{mirror}$} node[midway, right] {\large $\substack{\zeta_k = \fz_k \\ \\ \chi_k = \fc_k}$} (t);
\node[gauge,label=below:{${\O(1)}$}] (D1) at (5,0) {};
\node[gauge,label=above:{$\USp(2)$}] (C1) at (3,0) {};
\node[gauge,label=below:{${\O(3)^+}$}] (D3) at (1,0) {};
\node[label=below:{}] (3D) at (0,0) {$\ldots$};
\node[gauge,label=below:{$\O(2N-1)^+$}] (DNm1) at (-1,0) {};
\node[gauge,label={above,xshift=0.7cm}:{$\USp(2N)$}] (CNm1) at (-3,0) {};
\node[flavour,label=above:{$\SO(2 k - 1)$}] (D2Nz) at (-5,3) {};
\node[flavour,label=above:{$\SO(1)$}] (D2Nzc) at (-5,1) {};
\node[flavour,label=below:{$\SO(1)$}] (D2Nc) at (-5,-1) {};
\node[flavour,label=below:{$\SO(2 N - 2 k)$}] (D2N) at (-5,-3) {};
\draw (D1)--(C1)--(D3)--(3D)--(DNm1)--(CNm1)--(D2N);
\draw[very thick,red] (CNm1) to node[midway,above=0.2] {\textcolor{red}{\scriptsize $\fz_k$}} (D2Nz);
\draw[very thick,violet] (CNm1) to node[midway,above=0.1] {\textcolor{violet}{\scriptsize $\fz_k \fc_k$}} (D2Nzc);
\draw[very thick,blue] (CNm1) to node[midway,below] {\textcolor{blue}{\scriptsize $\fc_k$}} (D2Nc);
\end{tikzpicture}
}
\subsubsection*{$T[\USp(2N)]$ theory}
The analysis of the action of the topological $\BZ^{[0]}_{2,\CM_k}$ and charge conjugation $\BZ^{[0]}_{2,\CC_k}$ symmetries associated with an $\SO(2k)$ gauge node inside the $T[\USp(2N)]$ quiver proceeds in the same way as in the case of the $T[\SO(2N)]$ theory, which was discussed in the previous subsection. Upon examining the Coulomb branch moment map operators and counting their multiplicities, one has to pay attention to the fact that there is one additional $\SO(2N)$ gauge node here with respect to the $T[\SO(2N)]$ quiver.

Let us start by considering the case $k = 1$, \ie we are interested in the $\BZ^{[0]}_{2,\CM_1}$ and $\BZ^{[0]}_{2,\CC_1}$ symmetries associated with the $\SO(2)$ gauge node. In addition to the $W_{(1;0;\ldots)}^{\pm}$ operators defined in \eref{Wpm}, which are even ($+$) and odd ($-$) under $\BZ^{[0]}_{2,\CC_1}$, with both of them being odd under $\BZ^{[0]}_{2,\CM_1}$, there are also $2(2N-2)$ dressed monopole operators denoted by
\bes{ \label{monopole1TUSp2N}
V_{(\pm 1;{\vec U}^i;{\vec 0}^{2N-2-i})}~,
}
with $\vec U = (1,0,\ldots,0 )$ and ${\vec 0} = (0,\ldots,0 )$. The notation is as described below \eref{monopole1}, where now the index $i$ runs from $1$ to $2N-2$, due to the presence of the extra $\SO(2N)$ gauge group with respect to the $T[\SO(2N)]$ case. Given that such monopole operators are odd under $\BZ^{[0]}_{2,\CM_1}$ and lack definite $\BZ^{[0]}_{2,\CC_1}$ parity, in analogy with \eref{Whatpm}, we can define the $2 (2N - 2)$ combinations
\bes{ \label{WhatpmTUSp2N}
\hat{W}_{(\pm 1;\vec{U}^i;\vec{0}^{2N-2-i})}^{\pm} = V_{(+1;\vec{U}^i;\vec{0}^{2N-2-i})} \pm V_{(-1;\vec{U}^i;\vec{0}^{2N-2-i})}~,
}
which are even ($+$) and odd ($-$) under $\BZ^{[0]}_{2,\CC_1}$. Finally, the $2N - 2$ monopole operators $V_{(0;{\vec U}^i;{\vec 0}^{2N-2-i})}$ and the operator $\epsilon^{i_1 i_2} \varphi_{i_1 i_2}$, with $\varphi$ being the adjoint scalar in the $\SO(2)$ vector multiplet, are even under $\BZ^{[0]}_{2,\CM_1}$ and odd under $\BZ^{[0]}_{2,\CC_1}$. The Coulomb branch moment map operators which are charged under $\BZ^{[0]}_{2,\CM_1}$ and/or $\BZ^{[0]}_{2,\CC_1}$ are summarised in Table \ref{tab:TUSpCBmmSO2}, from which it can be observed that, upon gauging either $\BZ^{[0]}_{2,\CM_1}$ or $\BZ^{[0]}_{2,\CC_1}$, there are $4N - 2$ operators that are charged under such discrete symmetry and are thus projected out. Since the number of surviving Coulomb branch moment map operators is $\frac{1}{2} (2 N +1) (2N)-(4N - 2) = \frac{1}{2} (2 N-1) (2N-2) +1$, the $\so(2N+1)_\CM$ Coulomb branch symmetry is broken down to $\so(2N-1) \oplus \u(1)$.
\begin{table}[!t]
	\begin{center}
	\begin{tabular}{c|c|c|c}
		Operators & Multiplicity & $\BZ^{[0]}_{2,\CM_1}$ & $\BZ^{[0]}_{2,\CC_1}$\\
		\hline
            $W_{(1;0;\ldots)}^{+}$ & $1$ & \ding{51} & \ding{55}\\
            $W_{(1;0;\ldots)}^{-}$ & $1$ & \ding{51} & \ding{51}\\
            $\hat{W}_{(1;{\vec U}^i;{\vec 0}^{2N-2-i})}^{+}$ & $2N-2$ & \ding{51} & \ding{55}\\
            $\hat{W}_{(1;{\vec U}^i;{\vec 0}^{2N-2-i})}^{-}$ & $2N-2$ & \ding{51} & \ding{51}\\
            $V_{(0;{\vec U}^i;{\vec 0}^{2N-2-i})}$ & $2N-2$ & \ding{55} & \ding{51}\\
            $\epsilon^{i_1 i_2} \varphi_{i_1 i_2}$ & $1$ & \ding{55} & \ding{51}
	\end{tabular}
\end{center}
	\caption{Coulomb branch moment map operators of the $T[\USp(2N)]$ theory which are charged under $\BZ^{[0]}_{2,\CM_1}$ and/or $\BZ^{[0]}_{2,\CC_1}$, where the conventions are as explained in the main text and we use the notations \ding{51} and \ding{55} as in Table \ref{tab:TSOCBmmSO2}.}
	\label{tab:TUSpCBmmSO2}
\end{table}

For $k > 1$, the operators \eref{TSO2Nmonzeta0} and \eref{TSO2Nmonzetapm1} are replaced by
\begin{subequations}
\begin{align}
\begin{split} \label{TUSp2Nmonzeta0}
V_{(0;{\vec 0}^{2 (k-1)-1-i};{\vec U}^i; \, {\orangered \vec U}\, ;{\vec U}^j;{\vec 0}^{2 (N-k) -j})}~,
\end{split} \\
\begin{split} \label{TUSp2Nmonzetapm1}
V_{(\pm1;{\vec U}^{2 (k-1)-1}; \, {\orangered \vec U} \,;{\vec U}^j;{\vec 0}^{2 (N-k) -j})}~,
\end{split}
\end{align}
\end{subequations}
where $i=0,\ldots,2 (k-1)-1$ and $j =0, \ldots, 2 (N-k) $. Such monopole operators have one unit of $\SO(2k)$ magnetic flux, denoted by {\orangered $U$}, and the notation for the other magnetic fluxes, as well as their parity under $\BZ^{[0]}_{2,\CM_k}$ and $\BZ^{[0]}_{2,\CC_k}$, is as explained below \eref{TSO2Nmonzeta0} and \eref{TSO2Nmonzetapm1}. The only difference concerns the number of nodes on the right-hand side of $\SO(2k)$. Explicitly, there are $j$ gauge nodes, starting with $\USp(2k)$, with one unit of magnetic flux, while the remaining $2 (N-k) - j$ gauge nodes, ending with $\SO(2N)$, possess zero magnetic flux.

Finally, the following monopole operators, which are even under $\BZ^{[0]}_{2,\CM_k}$ and odd under $\BZ^{[0]}_{2,\CC_k}$, play the same role as \eref{TSO2Nmonchi0} and \eref{TSO2Nmonchipm1}:
\begin{subequations}
\begin{align}
\begin{split} \label{TUSp2Nmonchi0}
V_{(0;{\vec 0}^{2 (k-1)-1-i};{\vec U}^i;\, {\orangered {\vec 0}}\, ;{\vec U}^j;{\vec 0}^{2 (N-k) -j})}~,
\end{split} \\
\begin{split} \label{TUSp2Nmonchipm1}
V_{(\pm1;{\vec U}^{2 (k-1)-1};\, {\orangered {\vec 0}}\, ;{\vec U}^j;{\vec 0}^{2 (N-k) -j})}~,
\end{split}
\end{align}
\end{subequations}
where both types of operators possess zero $\SO(2k)$ flux, denoted by {\orangered $0$}. Moreover, either $i = 0, j > 0$ or $i>0, j=0$.
\begin{table}[!t]
	\begin{center}
        \scalebox{0.91}{
	\begin{tabular}{c|l|c|c}
		Operators & Multiplicity & $\BZ^{[0]}_{2,\CM_k}$ & $\BZ^{[0]}_{2,\CC_k}$\\
		\hline
            \eref{TUSp2Nmonzeta0} and \eref{TUSp2Nmonzetapm1} & $1+\left[2 (k-1)+1\right] \left[2 (N-k)\right]$ & \ding{51} & \ding{55}\\
            with $i=j=0$ or both $i, j > 0$&$=2 (2 k-1) (N-k)+1$&\\
            \hline
            \eref{TUSp2Nmonzeta0} and \eref{TUSp2Nmonzetapm1} & $2+\left[2 (k-1)-1\right]+\left[2 (N-k)\right]$ & \ding{51} & \ding{51}\\
            with $i=0, j>0$ or $i>0, j=0$&$= 2 N-1$&\\
            \hline
             \eref{TUSp2Nmonchi0} and \eref{TUSp2Nmonchipm1} & $2+\left[2 (k-1)-1\right]+\left[2 (N-k)\right]$ & \ding{55} & \ding{51}\\
             with $i=0, j>0$ or $i>0, j=0$ & $= 2 N-1$&
	\end{tabular}
    }
\end{center}
	\caption{Coulomb branch moment map operators of the $T[\USp(2N)]$ theory which are charged under $\BZ^{[0]}_{2,\CM_k}$ and/or $\BZ^{[0]}_{2,\CC_k}$, with $k > 1$. The conventions regarding \ding{51} and \ding{55} are the same as in Table \ref{tab:TSOCBmmSO2}.}
	\label{tab:TUSpCBmmSO2k}
\end{table}
We summarise the Coulomb branch moment map operators with non-trivial charge under $\BZ^{[0]}_{2,\CM_k}$ and/or $\BZ^{[0]}_{2,\CC_k}$ in Table \ref{tab:TUSpCBmmSO2k}, along with their multiplicity. We can then study how the global symmetry of the theory is affected upon gauging the discrete topological and charge conjugation symmetries associated with the $\SO(2k)$ node. In particular, among the $\frac{1}{2} (2N+1) (2N)$ operators in the adjoint representation of the $\so(2N+1)_\CM$ symmetry, $2 k (2N-2k+1)$ of them are projected out upon gauging either $\BZ^{[0]}_{2,\CM_k}$ (sum of multiplicities in rows 1 and 2) or $\BZ^{[0]}_{2,\CM_k \CC_k}$ (sum of multiplicities in rows 1 and 3). These two operations convert the $\SO(2k)$ gauge group into $\Spin(2k)$ and $\O(2k)^-$, respectively. In both cases, $\frac{1}{2} (2N+1) (2N) - 2 k (2N-2k+1) = \frac{1}{2} [(2k) (2k-1)+(2N-2k+1) (2N-2k)]$ Coulomb branch moment map operators survive, signalling that the $\so(2N+1)_\CM$ symmetry is broken to $\so(2k) \oplus \so(2N-2k+1)$. Indeed, upon applying duality \eref{dualityN4} locally at the $\SO(2k)$ node, we expect the two $T[\USp(2N)]$ theories with the $\SO(2k)$ gauge node turned into $\Spin(2k)$ and $\O(2k)^-$ to be dual, hence their global symmetry coincides.

If the $\BZ^{[0]}_{2,\CC_k}$ symmetry is gauged instead, we observe from rows 2 and 3 of Table \ref{tab:TUSpCBmmSO2k} that $4 N - 2$ Coulomb branch moment map operators are projected out. The number of surviving operators is thus $\frac{1}{2} (2N+1) (2N) - (4N-2) = \frac{1}{2} (2N-1) (2N-2) + 1$, which coincides with the dimension of $\so(2N-1) \oplus \so(2)$. This indicates that the $\so(2N+1)_\CM$ symmetry is broken to $\so(2N-1) \oplus \so(2)$ upon gauging $\BZ^{[0]}_{2,\CC_k}$. Note that the resulting global symmetry does not depend on $k$.

In summary, upon changing the global form of the $\SO(2k)$ gauge node in the $T[\USp(2N)]$ quiver, its $\so(2N+1)_\CM$ symmetry gets modified as follows:
\begin{itemize}
    \item Gauging $\BZ^{[0]}_{2,\CM_k}$, \ie $\SO(2k) \to \Spin(2k)$: $\so(2N+1)_\CM \to \so(2k) \oplus \so(2N-2k+1)$.
    \item Gauging $\BZ^{[0]}_{2,\CM_k \CC_k}$, \ie $\SO(2k) \to \O(2k)^-$: $\so(2N+1)_\CM \to \so(2k) \oplus \so(2N-2k+1)$.
    \item Gauging $\BZ^{[0]}_{2,\CC_k}$, \ie $\SO(2k) \to \O(2k)^+$: $\so(2N+1)_\CM \to \so(2N-1) \oplus \so(2)$.
\end{itemize}
\subsubsection*{Mirror Theory $T[\SO(2N+1)]$: Symmetry Actions on the Higgs Branch}
Let us now examine how the discrete symmetries corresponding to $\BZ^{[0]}_{2,\CM_k}$ and $\BZ^{[0]}_{2,\CC_k}$ affect the Higgs branch of the mirror theory $T[\SO(2N+1)]$. As depicted in the quiver at the bottom of \eref{TUSp2Nmirror}, such $\BZ_2$ symmetries alter the $\so(2N+1)$ flavour symmetry of the mirror, where their action is represented by the fugacities $\fz_k = \zeta_k$ and $\fc_k = \chi_k$. As for the $T[\SO(2N)^\vee]$ case, let us denote the half-hypermultiplets connecting the $\USp(2N)$ gauge node of the $T[\SO(2N+1)]$ quiver with the flavour nodes corresponding to the {\red red}, {\violet violet}, {\blue blue} and black edges as $A$, $B$, $C$ and $D$, respectively. From such fields, the gauge invariant mesons $Q_1 J Q_2$, with $Q_1, Q_2 \in \{A,B,C,D\}$ can be constructed, whose multiplicities and charges under the actions of $\fz_k$ and $\fc_k$, mirror to $\BZ^{[0]}_{2,\CM_k}$ and $\BZ^{[0]}_{2,\CC_k}$ respectively, are summarised in Table \ref{tab:TSOoddHBmm}.
\begin{table}[!t]
	\begin{center}
	\begin{tabular}{c|c|c|c}
		Mesons & Multiplicity & $\BZ^{[0]}_{2,\CM_k}$ & $\BZ^{[0]}_{2,\CC_k}$\\
		\hline
            $A J B$ & $2k-1$ & \ding{55} & \ding{51}\\
            $A J C$ & $2k-1$ & \ding{51} & \ding{51}\\
            $A J D$ & $(2k-1)(2N-2k)$ & \ding{51} & \ding{55}\\
            $B J C$ & $1$ & \ding{51} & \ding{55}\\
            $B J D$ & $2N-2k$ & \ding{51} & \ding{51}\\
            $C J D$ & $2N-2k$ & \ding{55} & \ding{51}
	\end{tabular}
\end{center}
	\caption{Higgs branch moment map operators of the $T[\SO(2N+1)]$ theory which are charged under the action of $\fz_k$ ($\leftrightarrow \BZ^{[0]}_{2,\CM_k}$) and/or $\fc_k$ ($\leftrightarrow \BZ^{[0]}_{2,\CC_k}$). The conventions regarding \ding{51} and \ding{55} are the same as in Table \ref{tab:TSOCBmmSO2}.}
	\label{tab:TSOoddHBmm}
\end{table}
We see that there are $(2 k -1 )(2N-2k)+1$ operators charged only under $\fz_k$ ($AJD, BJC$) and $2N-1$ operators which are charged either under only $\fc_k$ ($AJB, CJD$) or under both $\fz_k$ and $\fc_k$ ($AJC, BJD$). Hence, upon gauging the symmetry corresponding to $\fz_k$ (mirror to $\BZ^{[0]}_{2,\CM_k}$), $2 k (2N-2k+1)$ operators are projected out and the $\so(2N+1)_F$ symmetry is broken to $\so(2k) \oplus \so(2N-2k+1)$, whose dimension is exactly $\frac{1}{2} [(2k) (2k-1)+(2N-2k+1) (2N-2k)] = \frac{1}{2} (2N+1) (2N) - 2 k (2N-2k+1)$. On the other hand, if the symmetry associated with $\fc_k$ (mirror to $\BZ^{[0]}_{2,\CC_k}$) is gauged, $4N-2$ Higgs branch moment map operators are projected out, implying that $\frac{1}{2} (2N+1) (2N) - (4N-2)= \frac{1}{2} (2N-1) (2N-2) + 1$ mesonic operators survive, indicating that the $\so(2N+1)_F$ symmetry is broken to $\so(2N-1) \oplus \so(2)$. These symmetry breaking patterns are in perfect agreement with the results previously obtained upon gauging the $\BZ^{[0]}_{2,\CM_k}$ and $\BZ^{[0]}_{2,\CC_k}$ symmetries in the $T[\USp(2N)]$ theory. Similarly to \eref{TSO2Nmappingmm}, we can establish the mapping between the Coulomb branch moment map operators of the $T[\USp(2N)]$ theory and the Higgs branch moment map operators of the mirror $T[\SO(2N+1)]$ theory as follows:
\bes{ 
\left(\begin{array}{cc}
k=1: & \{W_{(1;0;\ldots)}^{+}, \hat{W}_{(1;\vec U^i;\vec 0^{2N-2-i})}^{+}\} \\
k>1: & \{\text{\eref{TUSp2Nmonzeta0}, \eref{TUSp2Nmonzetapm1}}\} \\ & \text{with $i=j=0$ or both $i, j > 0$}
\end{array}\right)
\qquad &\Longleftrightarrow \qquad \{\text{$A J D$, $B J C$}\}\\
\left(\begin{array}{cc}
k=1: & \{W_{(1;0;\ldots)}^{-}, \hat{W}_{(1;\vec U^i;\vec 0^{2N-2-i})}^{-}\} \\
k>1: & \{\text{\eref{TUSp2Nmonzeta0}, \eref{TUSp2Nmonzetapm1}}\} \\ & \text{with $i=0, j>0$ or $i>0, j=0$}
\end{array}\right)
\qquad &\Longleftrightarrow \qquad \{\text{$A J C$, $B J D$}\}~.
\\
\left(\begin{array}{cc}
k=1: & \{V_{(0;\vec U^i;\vec 0^{2N-2-i})}, \epsilon^{i_1 i_2} \varphi_{i_1 i_2}\} \\
k>1: & \{\text{\eref{TUSp2Nmonchi0}, \eref{TUSp2Nmonchipm1}}\} \\ & \text{with $i=0, j>0$ or $i>0, j=0$}
\end{array}\right)
\qquad &\Longleftrightarrow \qquad \{\text{$A J B$, $C J D$}\}
}
\subsubsection*{The Hilbert series}
Since the quiver description of the $T[\SO(2N+1)]$ theory is ``bad", its index diverges and cannot be used to check the validity of the mirror duality \eref{TUSp2Nmirror}. Nevertheless, instead of comparing the full indices, it is sufficient to match the Coulomb branch Hilbert series of the $T[\USp(2N)]$ theory with the Higgs branch Hilbert series of the $T[\SO(2N+1)]$ theory to establish \eref{TUSp2Nmirror}. This follows from the observation that the $\BZ^{[0]}_{2,\CM_k}$ and $\BZ^{[0]}_{2,\CC_k}$ symmetries associated with an $\SO(2k)$ gauge node act trivially on the Higgs branch of the $T[\USp(2N)]$ theory, whereas, in the mirror description, the corresponding discrete symmetries act trivially on the Coulomb branch of the $T[\SO(2N+1)]$ theory.

The Coulomb branch Hilbert series of the $T[\USp(2N)]$ theory can be defined in terms of that of the $T[\SO(2N)]$ theory, namely
\bes{ \label{CBHSTUSp2N}
&H_{T[\USp(2N)]}(t; \vec \zeta_{(N)}; \vec \chi_{(N)}=\vec{+1}_{(N)}; \vec l) \\
&= \sum_{\vec m_{(N)}} \zeta^{\sum_i m_{(N),i}}_{N} \,\, t^{2 \Delta(\vec m_{(N)}; \vec l)} P_{\SO(2N)} (t;\vec m_{(N)}) \\ & \qquad \quad \times H_{T[\SO(2N)]}(t; \vec \zeta_{(N-1)}; \vec \chi_{(N-1)}=\vec{+1}_{(N-1)}; \vec m_{(N)})~,
}
where we set the charge conjugation fugacity associated with each $\SO(2k)$ gauge group to be equal to $+1$, and we use the same conventions for the fugacities and magnetic fluxes as in \eref{indTSO2Nchip}. The conformal dimension $\Delta(\vec m_{(N)}; \vec l)$ is defined as in \eref{Deltamn}, the range of summation over an $\SO(2k)$ flux is
\bes{
\sum_{\vec m_{(k)}}: \quad m_{(k),1} \ge m_{(k),2} \ge \ldots \ge |m_{(k),k}|~,
}
and the Coulomb branch Hilbert series of the $T[\SO(2N)]$ theory appearing in \eref{CBHSTUSp2N} can be expressed recursively as
\bes{ \label{CBHSTSO2N}
&H_{T[\SO(2N)]}(t; \vec \zeta_{(N-1)}; \vec \chi_{(N-1)}=\vec{+1}_{(N-1)}; \vec l) \\
&= \sum_{\vec m_{(N-1)}} \,\, \sum_{\vec n_{(N-1)}} \zeta^{\sum_i m_{(N-1),i}}_{N-1} \,\, t^{2\left[\Delta(\vec m_{(N-1)}; \vec n_{(N-1)})+\Delta(\vec l; \vec n_{(N-1)})\right]} \\ & \qquad \qquad \qquad \times P_{\SO(2N-2)} (t;\vec m_{(N-1)}) P_{\USp(2N-2)} (t; \vec n_{(N-1)}) \\ & \qquad \qquad \qquad \times H_{T[\SO(2N-2)]}(t; \vec \zeta_{(N-2)}; \vec \chi_{(N-2)}=\vec{+1}_{(N-2)}; \vec m_{(N-1)})~.
}
In the expression above, the range of summation over a $\USp(2k)$ flux is
\bes{
\sum_{\vec n_{(k)}}:& \quad n_{(k),1} \ge n_{(k),2} \ge \ldots \ge n_{(k),k} \ge 0~,
}
and the Coulomb branch Hilbert series of the $T[\SO(4)]$ theory is given by
\bes{
&H_{T[\SO(4)]}(t; \zeta; \chi=+1; \vec l) \\
&= \sum_{m_{(1),1} \in \BZ} \,\, \sum_{n_{(1),1} \ge 0} \zeta^{m_{(1),1}} \,\, t^{2\left[\Delta(m_{(1),1}; n_{(1),1})+\Delta(\vec l; n_{(1),1})\right]} \\ & \qquad \qquad \qquad \quad \times P_{\SO(2)} (t;m_{(1),1}) P_{\USp(2)} (t; n_{(1),1})~.
}
If the fugacity $\chi_k$ associated with the $\SO(2k)$ gauge group is equal to $-1$, then, in the expression \eref{CBHSTUSp2N}, following the prescription described around \eref{HSSO2NN_fchi-1}, we have to replace $\vec m_{(k)}$ with $\tilde{\vec m}_{(k)}$ (which is obtained by setting $m_{(k),k} = 0$ in $\vec m_{(k)}$) and $P_{\SO(2k)} (t;\vec m_{(k)})$ with $2 P_{\USp(2k)} (t;\tilde{\vec m}_{(k)})-P_{\SO(2k)} (t;\tilde{\vec m}_{(k)})$, and finally we have to add a phase factor $\chi_k^{{\sum_i n_{(k-1),i}}+{\sum_j n_{(k),j}}}$. The Coulomb branch Hilbert series of the $T[\USp(2N)]$ theory can then be refined with the fugacities $\zeta_k$ and $\chi_k$ as follows:
\bes{ \label{CBHSTUSp2nrefzc}
H_{T[\USp(2N)]}(t; \zeta_k; \chi_k;) = \frac{1}{2} \sum_{s = \pm 1}H_{T[\USp(2N)]}(t; \zeta_k; \chi_k=s) \left(1+ s \chi_k \right)~,
}
where the other discrete fugacities $\zeta_{j \neq k}$ and $\chi_{j \neq k}$ are set to one and, for a compactness, we drop the dependence of the background fluxes $\vec l$ for the flavour symmetry.

The expression \eref{CBHSTUSp2nrefzc} can be matched with the Higgs branch Hilbert series of the mirror theory, which takes the form
\bes{ \label{HBHSTSOodd}
&\mathrm{HS}\left\{\text{HB of $T[\SO(2N+1)]$}\right\}(t; \fz_k; \fc_k; \vec{f})\\&= \frac{1}{2^N} \sum_{i=0}^{N-1} \sum_{\epsilon_i=\pm1} \mathrm{HS}\left\{\text{HB of $T[\SO(2N+1)]_{\epsilon}$}\right\}(t; \fz_k; \fc_k; \vec{f}; \vec{\epsilon})~,
}
where we denote with $\vec{\epsilon} = (\epsilon_0, \epsilon_1, \ldots, \epsilon_{N-1})$ the charge conjugation symmetries associated with the $\SO(2 k +1)$ gauge groups, with $k=0, \ldots, N-1$. In \eref{HBHSTSOodd}, we define
\bes{ 
&\mathrm{HS}\left\{\text{HB of $T[\SO(2N+1)]_{\epsilon}$}\right\}(t; \fz_k; \fc_k; \vec{f}; \vec{\epsilon}) \\&= \int \prod_{m=1}^{N-1} d\mu_{\SO(2m+1)} (\vec v_{(m)}; \epsilon_{m}) \times \prod_{l=1}^{N} d\mu_{\USp(2l)}(\vec u_{(l)}) \\ & \,\,\,\,\,\,\ \times \prod_{m=1}^{N-1} \PE\left[-t^2 \chi^{\SO(2m+1)}_{\mathcal{A}}(\vec v_{(m)}; \epsilon_{m})\right] \times \prod_{l=1}^{N} \PE\left[-t^2 \chi^{\USp(2l)}_{\mathcal{A}}(\vec u_{(l)}) \right] \\ & \,\,\,\,\,\,\ \times \prod_{l=1}^{N} \PE \left[t \left(\epsilon_{l-1} + \prod_{\alpha=1}^{l-1} \prod_{s_1= \pm 1} v^{s_1}_{(l-1),\alpha} \right) \prod_{\beta=1}^{l} \prod_{s_2 = \pm 1} u^{s_2}_{(l),\beta}\right] \\ & \,\,\,\,\,\,\ \times \prod_{l=1}^{N-1} \PE \left[t \left(\epsilon_l + \prod_{\alpha=1}^{l} \prod_{s_1= \pm 1} v^{s_1}_{(l),\alpha} \right) \prod_{\beta=1}^{l} \prod_{s_2 = \pm 1} u^{s_2}_{(l),\beta}\right]  \\ & \,\,\,\,\,\,\ \times \mathrm{HS}_A \times \mathrm{HS}_B \times \mathrm{HS}_C \times \mathrm{HS}_D ~,
}
where
\begin{subequations}
\begin{align}
\begin{split} 
\mathrm{HS}_A &=\PE \left[t \fz_k \left(1 + \prod_{i=1}^{k-1} \prod_{s_1= \pm 1} f^{s_1}_i \right) \prod_{\beta=1}^{N} \prod_{s_2 = \pm 1} u^{s_2}_{(N),\beta}\right]~,
\end{split} \\
\begin{split}
\mathrm{HS}_B &= \PE \left[ t \fz_k \fc_k \prod_{\beta=1}^{N} \prod_{s = \pm 1} u^{s}_{(N),\beta} \right]~,
\end{split}\\
\begin{split}
\mathrm{HS}_C &= \PE \left[ t \fc_k \prod_{\beta=1}^{N} \prod_{s = \pm 1} u^{s}_{(N),\beta}  \right]~,
\end{split}\\
\begin{split}
\mathrm{HS}_D &=\PE \left[t \prod_{i=1}^{N-k} \prod_{\beta=1}^N \prod_{s_1, s_2= \pm 1} f^{s_1}_i u^{s_2}_{(N),\beta}\right]~,
\end{split}
\end{align}
\end{subequations}
the Haar measures are given by
\bes{
\scalebox{0.91}{$
\begin{split}
\int d\mu_{\SO(2m+1)}(\vec v; \epsilon) =& \oint \prod_{i=1}^{m} \frac{d v_{i}}{2 \pi i v_{i}} \left(1-\epsilon v_{i}\right) \times \left(\prod_{1 \le i < j \le m} \,\ \prod_{s=\pm 1} 1- v_{i} v^{s}_{j}\right)~, \\ \int d\mu_{\USp(2l)}(\vec u) =& \oint \prod_{i=1}^l \frac{d u_{i}}{2 \pi i u_{i}} \left(1-u^2_{i}\right) \times \left(\prod_{1 \le i < j \le l} \,\ \prod_{s=\pm 1} 1- u_{i} u^{s}_{j}\right)~,
\end{split}
$}
}
and the characters of the adjoint representations are as follows:
\bes{
\chi^{\SO(2m+1)}_{\mathcal{A}}(\vec v; \epsilon) =& m+\sum_{i=1}^{m} \sum_{s=\pm 1} \epsilon v^s_{i}+\sum_{1 \le i < j \le m} \,\, \sum_{s_1,s_2=\pm 1} v^{s_1}_{i} v^{s_2}_{j}~, \\ \chi^{\USp(2l)}_{\mathcal{A}}(\vec u) =& l+\sum_{i=1}^l \sum_{s=\pm1} u^{2 s}_{i}+\sum_{1 \le i < j \le l} \sum_{s_1,s_2=\pm 1} u^{s_1}_{i} u^{s_2}_{j}~.
}
Upon dropping the dependence of the flavour fugacities $\vec f$, which do not appear manifestly in the Coulomb branch Hilbert series \eref{CBHSTUSp2N}, the mirror duality \eref{TUSp2Nmirror} is reflected in the following equality:
\bes{
H_{T[\USp(2N)]}(t; \zeta_k; \chi_k) = \mathrm{HS}\left\{\text{HB of $T[\SO(2N+1)]$}\right\}(t; \fz_k = \zeta_k; \fc_k = \chi_k)~.
}
For reference, let us report the explicit Hilbert series computation in the case $N=3$, namely we consider the mirror pair $T[\USp(6)] \leftrightarrow T[\SO(7)]$. We consider the cases $k=1,2,3$, \ie we turn on the topological and charge conjugation fugacities associated with the $\SO(2)$, $\SO(4)$ and $\SO(6)$ gauge nodes inside the $T[\USp(6)]$ quiver. The Hilbert series results for these three cases are listed below, where we also tabulate how the $\so(7)_\CM$ symmetry of the $T[\USp(6)]$ theory, or equivalently the $\so(7)_F$ symmetry of the $T[\SO(7)]$ theory, gets broken after gauging $\BZ^{[0]}_{2,\CM_k}$, $\BZ^{[0]}_{2,\CC_k}$ or $\BZ^{[0]}_{2,\CM_k \CC_k}$.
\bes{ \label{tabTUSp6}
\scalebox{0.91}{
\begin{tabular}{c|l|c}
\hline
$k$ & \qquad \qquad \quad Hilbert series & Global symmetry  \\
\hline
$1$ & $1+\left(6+5 \zeta_1 + 5 \chi_1 + 5 \zeta_1 \chi_1 \right) t^2$ & gauge $\BZ^{[0]}_{2,\CM_1}$: $\so(5) \oplus \so(2)$\\ & $\, \, \, \,+\left(65+55 \zeta_1 + 55 \chi_1 + 55 \zeta_1 \chi_1 \right) t^4$ &gauge $\BZ^{[0]}_{2,\CC_1}$: $\so(5) \oplus \so(2)$ \\ & $\, \, \, \,+\left(445+435 \zeta_1 + 435 \chi_1 + 435 \zeta_1 \chi_1 \right) t^6+\ldots$ & gauge $\BZ^{[0]}_{2,\CM_1 \CC_1}$: $\so(5) \oplus \so(2)$ \\ 
\hline
$2$ & $1+\left(4+7 \zeta_2 + 5 \chi_2 + 5 \zeta_2 \chi_2 \right) t^2$ & gauge $\BZ^{[0]}_{2,\CM_2}$: $\so(4) \oplus \so(3)$\\  & $\, \, \, \,+\left(67+53 \zeta_2 + 55 \chi_2 + 55 \zeta_2 \chi_2 \right) t^4$ & gauge $\BZ^{[0]}_{2,\CM_2}$: $\so(5) \oplus \so(2)$\\ & $\, \, \, \,+\left(423+457 \zeta_2 + 435 \chi_2 + 435 \zeta_2 \chi_2 \right) t^6+\ldots$ & gauge $\BZ^{[0]}_{2,\CM_2 \CC_2}$: $\so(4) \oplus \so(3)$ \\ 
\hline
$3$ & $1+\left(10+\zeta_3 + 5 \chi_3 + 5 \zeta_3 \chi_3 \right) t^2$ & gauge $\BZ^{[0]}_{2,\CM_3}$: $\so(6)$\\ & $\, \, \, \,+\left(85+35 \zeta_3 + 55 \chi_3 + 55 \zeta_3 \chi_3 \right) t^4$ & gauge $\BZ^{[0]}_{2,\CM_3}$: $\so(5) \oplus \so(2)$\\ & $\, \, \, \,+\left(545+335 \zeta_3 + 435 \chi_3 + 435 \zeta_3 \chi_3 \right) t^6+\ldots$ & gauge $\BZ^{[0]}_{2,\CM_3 \CC_3}$: $\so(6)$ \\ 
\end{tabular}
}
}

\acknowledgments
We thank Riccardo Comi, Sebastiano Garavaglia,  Julius Grimminger and Yi-Nan Wang for valuable discussions and collaborations on related projects. W.H. and N.M. are partially supported by the MUR-PRIN grant No. 2022NY2MXY (Finanziato dall'Unione europea -- Next Generation EU, Missione 4 Componente 1 CUP H53D23001080006, I53D23001330006). Z.Z. is supported by the ERC Consolidator Grant \# 864828 ``Algebraic Foundations of Supersymmetric Quantum Field Theory'' (SCFTAlg).

\appendix 

\section{Indices of the theories in Section \ref{sec:noninvertible}} \label{app:souspABJlike}
In this Appendix, we discuss the indices of variants of the $\so(2N) \times \usp(2N)$ gauge theory with $n$ bifundamental half-hypermultiplets.  For simplicity, we first consider the $[\SO(2N) \times \USp(2N)]/\BZ_2$ gauge theory with $\chi =+1$, whose index is
\bes{ \label{indABJlikemodZ2}
&\CI\left\{\left[\SO(2N)\times \USp(2N)\right]/\BZ_2\right\}(x;a; \vec f;g;  \zeta; \chi=+1) \\
& = \frac{1}{N! 2^{N-1}} \times \frac{1}{N! 2^N} \,\, \sum_{e=0}^1 g^e \,\, \sum_{(m_1, \ldots, m_N) \in \left(\BZ+\frac{e}{2} \right)^N} \sum_{(n_1, \ldots, n_N) \in \left(\BZ+\frac{e}{2} \right)^N} \zeta^{\sum_i m_i} \\
& \quad \times \oint \left( \prod_{\alpha, \beta =1}^N \frac{d z_\alpha}{2\pi i z_\alpha}  \frac{d u_\beta}{2\pi i u_\beta} \right) \CZ_{\text{vec}}^{\SO(2N)}(x; \vec z; \vec m; \chi = +1) \CZ_{\text{vec}}^{\USp(2N)}(x; \vec u; \vec n)  \\
& \quad \times \CZ^{\SO(2N)}_{\mathcal{A}}(x; a; \vec z; \vec m; \chi=+1) \CZ^{\USp(2N)}_{\mathcal{A}}(x; a; \vec u; \vec n) \\
& \quad \times \prod_{i=1}^{n} \prod_{\alpha, \beta=1}^N \prod_{s_1, s_2=\pm 1} \CZ_{\text{chir}}^{1/2} (x; a f_i z_\alpha^{s_1} u_\beta^{s_2};s_1 m_\alpha +s_2 n_\beta)~,
}
where the contributions of the $\USp(2N)$ vector multiplet and the adjoint chiral field in such a multiplet are 
\bes{ \label{vecadjUSp2N}
\scalebox{0.98}{$
\begin{split}
&\CZ_{\text{vec}}^{\USp(2N)}(x; \vec u; \vec n) =  \prod_{\ell=1}^N x^{-|2n_\ell|} \prod_{s=\pm 1}  (1-(-1)^{2 s n_\ell} u_\ell^{2s} x^{|2n_\ell|} ) \\
&\qquad \times \prod_{1 \leq a< b\leq N} \,\, \prod_{s_1, s_2 = \pm 1} x^{-|s_1 n_a +s_2 n_b|/2}  \Big(1-(-1)^{|s_1 n_a +s_2 n_b|} u_a^{s_1} u_b^{s_2} x^{|s_1 n_a +s_2 n_b|} \Big)~, \\
&\CZ_{\CA}^{\USp(2N)}(x; \vec u; \vec n) = \prod_{\ell =1}^N \prod_{s = \pm 1} \CZ_{\text{chir}}^{R=1} (x; a^{-2} u_\ell^{2s}; 2s n_\ell ) \\
& \qquad \times \Big[\CZ_{\text{chir}}^{R=1} (x; a^{-2}; 0)\Big]^N  \times \prod_{1\leq i<j \leq N} \, \prod_{s_1, s_2 = \pm 1} \CZ_{\text{chir}}^{R=1} (x; a^{-2} u_i^{s_1} u_j^{s_2}; s_1 n_i +s_2 n_j )~.
\end{split}
$}
}
For $\chi = -1$, we replace $\CZ^{\SO(2N)}_{\text{vec}/\mathcal{A}}(x; a; \vec z; \vec m; \chi=+1)$ in \eqref{indABJlikemodZ2} by $\CZ^{\SO(2N)}_{\text{vec}/\mathcal{A}}(x; a; \vec z; \vec m; \chi=-1)$, and set $z_N=1$, $z_N^{-1}=-1$ and $m_N=0$. As a consequence of the latter, there is no summation over half-integral fluxes, since otherwise these would violate Dirac quantisation condition, which imposes $m_\alpha + n_\beta \in \BZ$ for all $\alpha, \beta =1, \ldots, N$. (The fact that $m_N=0$ requires $n_\beta$ to be integers, which in turn requires $m_\alpha$ to be integers, for all $\alpha=1, \ldots, N-1$.) We therefore have
\bes{
&\CI\left\{\left[\SO(2N)\times \USp(2N)\right]/\BZ_2\right\}(x;a; \vec f; \zeta; \chi=-1) \\
&= \CI\left[\SO(2N)\times \USp(2N)\right](x;a; \vec f;  \zeta; \chi=-1) \\
& = \frac{1}{(N-1)! 2^{N-1}} \times \frac{1}{N! 2^N} \,\, \sum_{(m_1, \ldots, m_N) \in \BZ^N} \sum_{(n_1, \ldots, n_N) \in \BZ^N} \zeta^{\sum_i m_i} \\
& \quad \times \oint \left( \prod_{\alpha, \beta =1}^N \frac{d z_\alpha}{2\pi i z_\alpha}  \frac{d u_\beta}{2\pi i u_\beta} \right) \CZ_{\text{vec}}^{\SO(2N)}(x; \vec z; \vec m; \chi = -1) \CZ_{\text{vec}}^{\USp(2N)}(x; \vec u; \vec n)  \\
& \quad \times \CZ^{\SO(2N)}_{\mathcal{A}}(x; a; \vec z; \vec m; \chi=-1) \CZ^{\USp(2N)}_{\mathcal{A}}(x; a; \vec u; \vec n) \\
& \quad \times \prod_{i=1}^{n} \prod_{\alpha, \beta=1}^N \prod_{s_1, s_2=\pm 1} \CZ_{\text{chir}}^{1/2} (x; a f_i z_\alpha^{s_1} u_\beta^{s_2};s_1 m_\alpha +s_2 n_\beta) \Bigg|_{z_N=1, \, z_N^{-1}=-1, \, m_N=0}~.
}
Finally, we note that the indices for the other forms of the gauge group can be computed using \eqref{indvariants}.

\section{Nilpotent orbits} \label{app:nilpotent}
Explicit examples of 3d $\mathcal{N}=4$ quiver gauge theories where the specific global form of $\so(N)$-type gauge groups is determined are relatively scarce in the literature. However, detailed examples arise in the context of quiver theories whose moduli spaces correspond to nilpotent orbit closures. In \cite{Cremonesi:2014uva, Cabrera:2017ucb}, the distinction between using an $\O(N)^+$ or $\SO(N)$ gauge group for certain nodes in these quivers was made based on the prescription outlined in \cite{Cremonesi:2014uva}. In particular, Reference \cite{Cabrera:2017ucb} identified specific quivers, labelled by partitions associated with conjectured nilpotent orbit closures on their Coulomb branch, which feature an $\O(N)^+$ gauge group at certain nodes instead of $\SO(N)$.

In this section, we revisit these specific examples related to nilpotent orbits. Using the refined Hilbert series prescription developed in Section \ref{sec:CBHS} of this paper, which distinguishes between different global forms via the refinement with respect to $\zeta$ and $\chi$ fugacities, we determine the precise global structure expected at the relevant node, assigning it to $\SO(N)$, $\O(N)^\pm$, $\Spin(N)$, or $\Pin(N)$. Our results for quivers associated with nilpotent orbits of Lie algebras of type $B_3, D_4, B_4,$ and $D_5$ are presented in Tables \ref{orbit1}, \ref{orbit2}, and \ref{orbit3}.

A consistent observation across these examples is that the Hilbert series computed for the $\O(N)^-$ case always coincides with that computed for the $\Spin(N)$ case. This finding aligns with expectations based on the duality \eqref{dualityN4}, which relates these two global forms, particularly given that the relevant orthogonal ($\so(N)$-type) gauge nodes in these quiver examples are always balanced. 

Another interesting observation arises for quivers associated with specific nilpotent orbits of $\SO(N)$, for example, those labelled by partitions $[3^2,1^2]$ ($D_4$, $N=8$), $[5,3,1^2]$ ($B_4$, $N=9$), and $[3^2,1^4]$ ($D_5$, $N=10$). In these instances, our calculations reveal that the Hilbert series for the $\O(N)^\pm$ and $\Spin(N)$ global forms are identical. A common structural feature of the quiver diagrams corresponding to these specific partitions is that the orthogonal gauge node under consideration is situated directly between two adjacent symplectic gauge nodes.  Whenever the quiver exhibits this ``sandwiched'' structure, we observe that the coefficients of $\zeta$, $\chi$ and $\zeta \chi$ in the Coulomb branch Hilbert series are equal. As a result, using \eqref{variants}, we see that the Hilbert series for the cases of $\O(N)^\pm$ and $\Spin(N)$ are all equal.  From the mirror symmetry perspective, this equality implies a specific pattern of the $\so(N)$ flavour symmetry breaking in the mirror theory. When the orthogonal node is ``sandwiched'' between two symplectic nodes, the mirror actions of charge conjugation and magnetic symmetries leave at most one non-Abelian $\so(m)$ factor (with $m \geq 3$) within the original $\so(N)$ flavor group, with the remaining factors being $\SO(1)$ nodes, including the one that is uncharged under such actions. We illustrate this phenomenon with the examples below.

\subsection*{Example: the $[3^2,1^2]$ orbit of $D_4$}
The relevant mirror dual pair is depicted below. The moduli space corresponding to the closure of the $[3^2,1^2]$ nilpotent orbit of $D_4$ arises as the Higgs branch of the left quiver and, by mirror symmetry, as the Coulomb branch of the right quiver.
\bes{ \label{mirr3212D4}
\scalebox{0.7}{
\begin{tikzpicture}[baseline]
\node[gauge,label={above,xshift=0.3cm}:{$\USp(4)$}] (C2) at (0,0) {}; 
\node[gauge,label=below:{$\SO(2)$}] (D1) at (2,0) {}; 
\node[flavour,label=left:{$\SO(5)$}] (B2) at (-2,1.5) {}; 
\node[flavour,label=left:{$\SO(1)$}] (SO1u) at (-2,0.5) {};
\node[flavour,label=left:{$\SO(1)$}] (SO1d) at (-2,-0.5) {};
\node[flavour,label=left:{$\SO(1)$}] (SO1dd) at (-2,-1.5) {};
\draw[red, thick] (C2) to node[above, midway]{$\fz$} (B2);
\draw[violet, thick] (C2) to node[above, near end]{$\fz \fc$} (SO1u);
\draw[blue, thick] (C2) to node[above, near end]{$\fc$} (SO1d);
\draw[black, thick] (C2) to (SO1dd);
\draw (D1) to (C2);
\end{tikzpicture}
}
\quad \underset{\fz=\zeta,\, \fc =\chi}{\overset{\text{mirror}}{\longleftrightarrow}} \quad
\scalebox{0.7}{
\begin{tikzpicture}[baseline]
\node[gauge,label=below:{$\SO(2)$}] (n1) at (0,0) {}; 
\node[gauge,label=below:{$\USp(2)$}] (n2) at (1.5,0) {}; 
\node[gauge,label=below:{$\SO(4)$}] (n3) at (3,0) {}; 
\node[gauge,label=below:{$\USp(4)$}] (n4) at (4.5,0) {};
\node[gauge,label=below:{$\underset{\textcolor{red}{\zeta}, \, \textcolor{blue}{\chi}}{\SO(4)}$}] (n5) at (6,0) {};
\node[gauge,label=below:{$\USp(2)$}] (n6) at (7.5,0) {};
\node[flavour,label=above:{$\SO(2)$}] (n7) at (4.5,1) {};
\node[flavour,label=above:{$\SO(2)$}] (n8) at (7.5,1) {};
\draw (n1)--(n2)--(n3)--(n4)--(n5)--(n6);
\draw (n4)--(n7);
\draw (n6)--(n8);
\end{tikzpicture}
}
}
where we identify the fugacities as $\mathfrak{z}=\zeta$ and $\mathfrak{c} = \chi$ across the duality. The Higgs branch Hilbert series of the left quiver, which by mirror symmetry equals the Coulomb branch Hilbert series of the right quiver (computed using our prescription from Section \ref{sec:CBHS}), is given by the expansion:
\bes{ \label{HS_D4_3212}
 1 &+  (10 + 6 \zeta + 6 \chi + 6 \zeta \chi) t^2 +  (127 + 102 \zeta + 102 \chi + 102 \zeta \chi) t^4 \\
&  +  (1206 + 1140 \zeta + 1140 \chi + 1140 \zeta \chi) t^6 +\ldots~.
}
This result is in agreement with the corresponding entry in Table \ref{orbit1} after applying the projections \eqref{variants} to obtain results for specific global forms. Observe that the $\SO(4)$ gauge node in the right quiver of \eqref{mirr3212D4} is indeed ``sandwiched'' between $\USp(4)$ and $\USp(2)$ nodes. As predicted by the structural argument, the coefficients multiplying $\zeta$, $\chi$, and $\zeta \chi$ in the Hilbert series expansion \eqref{HS_D4_3212} are equal at each order in $t$ (e.g., all are $6$ for the $t^2$ term, $102$ for $t^4$, $1140$ for $t^6$, etc.). As a consequence, the Hilbert series for the variants with $\O(4)^\pm$ and $\Spin(4)$ gauge groups at this node are identical.

The contributions of the moment map operators to the coefficient $(10 + 6 \zeta + 6 \chi + 6 \zeta \chi)$ of the $t^2$ term in the Hilbert series can be symbolically understood by examining the mesonic operators constructed from the fields 
\bes{ \label{labelfields}
A(\so(5),\mathfrak{z}), \,\, B(\so(1),\mathfrak{z}\mathfrak{c}), \,\, C(\so(1),\mathfrak{c}), \,\, D(\so(1),1) 
}
of the left quiver \eqref{mirr3212D4}. Schematically: the constant term $10$ relates to mesons involving only $A$ fields (in the adjoint representation of the $\so(5)$ flavour symmetry); the $6\zeta$ term relates to mesons combining $A$ with $D$ ($5$ terms) and $B$ with $C$ ($1$ term); the $6\chi$ term relates to mesons combining $A$ with $B$ ($5$ terms) and $C$ with $D$ ($1$ term); and the $6\zeta\chi$ term relates to mesons combining $A$ with $C$ ($5$ terms) and $B$ with $D$ ($1$ term), reflecting how the $\mathfrak{z}=\zeta$ and $\mathfrak{c}=\chi$ actions combine.


\subsection*{Example: the $[5,1^2]$ orbit of $B_3$}
The relevant mirror dual pair is depicted below. Notice that the left quiver features $\O(1)$ and $\SO(3)$ gauge groups. The moduli space corresponding to the closure of the $[5,1^2]$ nilpotent orbit of $B_3$ arises as the Higgs branch of the left quiver and as the Coulomb branch of the right quiver.
\bes{ \label{mirr511B3}
\scalebox{0.65}{
\begin{tikzpicture}[baseline]
\node[gauge,label=below:{$\O(1)$}] (r1) at (4.5,0) {}; 
\node[gauge,label=above:{$\USp(2)$}] (r2) at (3,0) {}; 
\node[gauge,label=below:{$\SO(3)$}] (r3) at (1.5,0) {}; 
\node[gauge,label={above,xshift=0.3cm}:{$\USp(4)$}] (r4) at (0,0) {}; 
\node[flavour,label=left:{$\SO(5)$}] (B2) at (-2,1.5) {}; 
\node[flavour,label=left:{$\SO(1)$}] (SO1u) at (-2,0) {};
\node[flavour,label=left:{$\SO(1)$}] (SO1d) at (-2,-1.5) {};
\draw[red, thick] (r4) to node[above, midway]{$\fz$} (B2);
\draw[violet, thick] (r4) to node[above, near end]{$\fz \fc$} (SO1u);
\draw[blue, thick] (r4) to node[above, near end]{$\fc$} (SO1d);
\draw (r1)--(r2)--(r3)--(r4);
\end{tikzpicture}
}
\quad \underset{\fz=\zeta,\, \fc =\chi}{\overset{\text{mirror}}{\longleftrightarrow}} \quad
\scalebox{0.65}{
\begin{tikzpicture}[baseline]
\node[gauge,label=below:{$\SO(2)$}] (n1) at (0,0) {}; 
\node[gauge,label=below:{$\USp(2)$}] (n2) at (1.5,0) {}; 
\node[gauge,label=below:{$\SO(4)$}] (n3) at (3,0) {}; 
\node[gauge,label=below:{$\USp(4)$}] (n4) at (4.5,0) {};
\node[gauge,label=below:{$\underset{\textcolor{red}{\zeta}, \, \textcolor{blue}{\chi}}{\SO(4)}$}] (n5) at (6,0) {};
\node[flavour,label=above:{$\SO(4)$}] (n7) at (4.5,1) {};
\node[flavour,label=above:{$\USp(2)$}] (n8) at (6,1) {};
\draw (n1)--(n2)--(n3)--(n4)--(n5);
\draw (n4)--(n7);
\draw (n5)--(n8);
\end{tikzpicture}
}
}
where, as before, we identify $\mathfrak{z}=\zeta$ and $\mathfrak{c} = \chi$ across the duality. By mirror symmetry, the Higgs branch Hilbert series of the left quiver equals the Coulomb branch Hilbert series of the right quiver. Computing the latter using our prescription (Section \ref{sec:CBHS}) yields:
\bes{ \label{HS_B3_511}
&\frac{1}{
(1 - \zeta t)^{10} (1 + \zeta t)^{10} (1 - \zeta t^2) (1 - \chi t^2)^5 (1 - \zeta \chi t^2)^5} \\
& \quad \times (1 - t^2)^3 (1 + t^2)^2 \Big[ 1 + t^2 + (2+5\zeta+\chi+\zeta\chi)t^4 + (2-5\zeta-\chi-\zeta\chi)t^6 \\
& \quad + (-8+4\zeta-4\chi-4\zeta\chi)t^8 + (2-5\zeta-\chi-\zeta\chi)t^{10}  \\
& \quad + (2+5\zeta+\chi+\zeta\chi)t^{12} + t^{14} + t^{16} \Big] \\
& =
1 + (10 + \zeta + 5 \chi + 5 \zeta \chi) t^2 + (85 + 40 \zeta + 56 \chi + 56 \zeta \chi) t^4 + \ldots~.
}
This result precisely matches the corresponding entry in Table \ref{orbit1} after applying the projections \eqref{variants}. In this case, the $\SO(4)$ gauge node in the right quiver is not "sandwiched" between two symplectic groups in the same manner as the previous example. Correspondingly, we observe that the coefficients multiplying $\zeta$, $\chi$, and $\zeta\chi$ in the Hilbert series expansion \eqref{HS_B3_511} are distinct (e.g., $1, 5, 5$ for $t^2$; $40, 56, 56$ for $t^4$). As expected from the projection formulae \eqref{variants} when these coefficients differ, the Hilbert series for the $\O(4)^+$ variant is distinct from those for the $\O(4)^-$ and $\Spin(4)$ variants (though the latter two still coincide due to duality). The moment map operators contributing to the coefficient of $t^2$ can be obtained in a similar way as in the previous example, except that there is no field labelled by $D$ in the left quiver of \eqref{mirr511B3} in comparison to \eqref{labelfields}.

\subsection*{Comment on the prescription and nilpotent orbits}
A comparison between the prescription previously used \cite{Cremonesi:2014uva} and the refined prescription proposed in this work warrants discussion, particularly concerning applications to nilpotent orbits. The prescription from \cite{Cremonesi:2014uva}, without appropriately accounting for the effects of background flavour magnetic fluxes, was widely adopted partly because computations based upon it, such as those in \cite{Cabrera:2017ucb} for quiver theories related to nilpotent orbits, appeared to reproduce the expected Hilbert series for the corresponding orbit closures. Conversely, calculations using our refined prescription generally yield results inconsistent with the Hilbert series expected for the full nilpotent orbit closures. As shown in our tables (Tables \ref{orbit1}-\ref{orbit3}), applying our method to different global forms ($\O(N)^\pm$, $\Spin(N)$, $\Pin(N)$) for an $\so(N)$-type node typically results in Hilbert series exhibiting manifest global symmetries that are subgroups of the expected symmetry $G$ of the nilpotent orbit closure of type $G$. Therefore, our refined prescription suggests that for the quiver theories studied, none of the aforementioned choices for the global form of the $\so(N)$-type node results in a Coulomb branch Hilbert series manifesting the full expected symmetry $G$ of the corresponding nilpotent orbit closure.
However, we contend that the apparent agreement reported in \cite{Cabrera:2017ucb}, based on the older prescription, might also not represent the definitive Hilbert series of the nilpotent orbit closure. The reason is that those results were compared against Higgs branch Hilbert series calculations from \cite{Hanany:2016gbz}. That work conjectured that the Higgs branches of certain $T_\rho(G)$ quivers (for $G$ of BCD type) correspond to the nilpotent orbit closures associated with partition $\rho$, under the specific assumption that all relevant $\so(N)$ gauge groups within the quiver should be taken as $\O(N)^+$. We argue, however, that this conjecture involving the specific $\O(N)^+$ choice may require further scrutiny or refinement. A definitive determination of the nilpotent orbit closure's Hilbert series requires an independent calculation based on its algebraic definition (generators and relations), potentially using computational algebra software like {\tt Macaulay2}. To our knowledge, such first-principles Hilbert series computations for the specific nilpotent orbits relevant to Tables \ref{orbit1}, \ref{orbit2}, and \ref{orbit3} have not appeared in the literature. Consequently, the previous agreement between results from the older prescription and the conjectured results of \cite{Hanany:2016gbz} does not definitively establish that either calculation yields the true Hilbert series for the nilpotent orbit closure.
\pgfsetlayers{edgelayer,nodelayer,main}
\tikzset{none/.style={draw=none}}
\begin{landscape}
\begin{table}[]
    \centering
    \begin{tabular}{|c|c|c|c|}
    \hline
        Nilpotent orbit & Quiver  & HS    \\ \hline

        $[5,1^2]$ of $B_3$ &\scalebox{0.6}{\begin{tikzpicture}
	\begin{pgfonlayer}{nodelayer}
		\node [style=redgauge] (0) at (0, 0) {};
		\node [style=bluegauge] (1) at (1.25, 0) {};
		\node [style=redgauge] (2) at (2.5, 0) {};
		\node [style=bluegauge] (3) at (3.75, 0) {};
		\node [style=redgauge] (4) at (5, 0) {};
		\node [style=blueflavor] (5) at (5, 1) {};
		\node [style=redflavor] (6) at (3.75, 1) {};
		\node [style=none] (7) at (0, -0.5) {\small{SO(2)}};
		\node [style=none] (8) at (1.25, -0.5) {\small{USp(2)}};
		\node [style=none] (9) at (2.5, -0.5) {\small{SO(4)}};
		\node [style=none] (10) at (3.75, -0.5) {\small{USp(4)}};
		\node [style=none] (11) at (5, -0.5) {$\mathfrak{so}(4)$};
		\node [style=none] (12) at (5, 1.5) {USp(2)};
		\node [style=none] (13) at (3.75, 1.5) {SO(2)};
		\node [style=gauge3] (14) at (0, 0) {};
		\node [style=gauge3] (15) at (1.25, 0) {};
		\node [style=gauge3] (16) at (2.5, 0) {};
		\node [style=gauge3] (17) at (3.75, 0) {};
		\node [style=gauge3] (18) at (5, 0) {};
		\node [style=flavour2] (19) at (5, 1) {};
		\node [style=flavour2] (20) at (3.75, 1) {};
	\end{pgfonlayer}
	\begin{pgfonlayer}{edgelayer}
		\draw (0) to (4);
		\draw (4) to (5);
		\draw (6) to (3);
	\end{pgfonlayer}
\end{tikzpicture}
}
&\parbox[c]{12cm}{\textcolor{red}{$\SO(4)$}:\;\; $1+21 t^2+237 t^4+1883 t^6+11696 t^8+60095 t^{10}+264881 t^{12}+1027889 t^{14}+3581403 t^{16}+11377742 t^{18}+33366886 t^{20}+O\left(t^{22}\right)$ \\ 
\textcolor{red}{$\O(4)^+$}:\;\;$1+11 t^2+125 t^4+945 t^6+5888 t^8+30061 t^{10}+132609 t^{12}+513983 t^{14}+1791259 t^{16}+5688962 t^{18}+16684998 t^{20}+O\left(t^{22}\right)$ \\ \textcolor{red}{$\text{Spin}(4)/\O(4)^-$}:\;\; $1+15 t^2+141 t^4+1029 t^6+6128 t^8+30833 t^{10}+134417 t^{12}+518515 t^{14}+1800539 t^{16}+5708826 t^{18}+16721894 t^{20}+O\left(t^{22}\right)$\\ $\textcolor{red}{\text{Pin}(4)}$:\;\;$1+10 t^2+85 t^4+560 t^6+3224 t^8+15816 t^{10}+68281 t^{12}+261562 t^{14}+905467 t^{16}+2864436 t^{18}+8380950 t^{20}+O\left(t^{22}\right)$}  \\ \hline 
    $[3^2,1^2]$ of $D_4$ &\scalebox{0.6}{\begin{tikzpicture}
	\begin{pgfonlayer}{nodelayer}
		\node [style=redgauge] (0) at (0, 0) {};
		\node [style=bluegauge] (1) at (1.25, 0) {};
		\node [style=redgauge] (2) at (2.5, 0) {};
		\node [style=bluegauge] (3) at (3.75, 0) {};
		\node [style=redgauge] (4) at (5, 0) {};
		\node [style=redflavor] (6) at (3.75, 1) {};
		\node [style=none] (7) at (0, -0.5) {SO(2)};
		\node [style=none] (8) at (1.25, -0.5) {USp(2)};
		\node [style=none] (9) at (2.5, -0.5) {SO(4)};
		\node [style=none] (10) at (3.75, -0.5) {USp(4)};
		\node [style=none] (11) at (5, -0.5) {$\mathfrak{so}(4)$};
		\node [style=none] (13) at (3.75, 1.5) {SO(2)};
		\node [style=bluegauge] (14) at (6.25, 0) {};
		\node [style=redflavor] (15) at (6.25, 1) {};
		\node [style=none] (16) at (6.25, 1.5) {SO(2)};
		\node [style=none] (17) at (6.25, -0.5) {USp(2)};
		\node [style=flavour2] (18) at (3.75, 1) {};
		\node [style=flavour2] (19) at (6.25, 1) {};
		\node [style=gauge3] (20) at (6.25, 0) {};
		\node [style=gauge3] (21) at (5, 0) {};
		\node [style=gauge3] (22) at (3.75, 0) {};
		\node [style=gauge3] (23) at (2.5, 0) {};
		\node [style=gauge3] (24) at (1.25, 0) {};
		\node [style=gauge3] (25) at (0, 0) {};
	\end{pgfonlayer}
	\begin{pgfonlayer}{edgelayer}
		\draw (0) to (4);
		\draw (6) to (3);
		\draw (4) to (14);
		\draw (14) to (15);
	\end{pgfonlayer}
\end{tikzpicture}
    }   &\parbox[c]{12cm}{\textcolor{red}{$\SO(4)$}:\;\; $1+28 t^2+433 t^4+4626 t^6+37374 t^{8}+241737 t^{10}+1304520 t^{12}+6059281 t^{14}+24813130 t^{16}+91286004 t^{18}+306261170 t^{20}+O\left(t^{22}\right)$ \\ \textcolor{red}{$\O(4)^+/\text{Spin}(4)/\O(4)^-$}:\;\; $1+16 t^2+229 t^4+2346 t^6+18814 t^8+121141 t^{10}+653060 t^{12}+3031145 t^{14}+12410258 t^{16}+45649332 t^{18}+153144258 t^{20}+O\left(t^{22}\right)$\\ \textcolor{red}{$\text{Pin}(4)$}:\;\; $1+10 t^2+127 t^4+1206 t^6+9534 t^8+60843 t^{10}+327330 t^{12}+1517077 t^{14}+6208822 t^{16}+22830996 t^{18}+76585802 t^{20}+O\left(t^{22}\right)$}\\ \hline
  $[5,1^4]$ of $B_4$ &
  \scalebox{0.6}{\begin{tikzpicture}
	\begin{pgfonlayer}{nodelayer}
		\node [style=redgauge] (0) at (0, 0) {};
		\node [style=bluegauge] (1) at (1.25, 0) {};
		\node [style=redgauge] (2) at (2.5, 0) {};
		\node [style=bluegauge] (3) at (3.75, 0) {};
		\node [style=redgauge] (4) at (5, 0) {};
		\node [style=none] (7) at (0, -0.5) {SO(2)};
		\node [style=none] (8) at (1.25, -0.5) {USp(2)};
		\node [style=none] (9) at (2.5, -0.5) {SO(4)};
		\node [style=none] (10) at (3.75, -0.5) {USp(4)};
		\node [style=none] (11) at (7.5, -0.5) {$\mathfrak{so}(4)$};
		\node [style=bluegauge] (14) at (6.25, 0) {};
		\node [style=redflavor] (15) at (6.25, 1) {};
		\node [style=none] (16) at (6.25, 1.5) {SO(1)};
		\node [style=none] (17) at (6.25, -0.5) {USp(4)};
		\node [style=redgauge] (18) at (7.5, 0) {};
		\node [style=none] (21) at (5, -0.5) {SO(5)};
		\node [style=blueflavor] (22) at (7.5, 1) {};
		\node [style=none] (23) at (7.5, 1.5) {USp(2)};
		\node [style=redflavor] (24) at (3.75, 1) {};
		\node [style=none] (25) at (3.75, 1.5) {SO(1)};
		\node [style=gauge3] (26) at (0, 0) {};
		\node [style=gauge3] (27) at (1.25, 0) {};
		\node [style=gauge3] (28) at (2.5, 0) {};
		\node [style=gauge3] (29) at (3.75, 0) {};
		\node [style=gauge3] (30) at (5, 0) {};
		\node [style=gauge3] (31) at (6.25, 0) {};
		\node [style=gauge3] (32) at (7.5, 0) {};
		\node [style=flavour2] (33) at (7.5, 1) {};
		\node [style=flavour2] (34) at (6.25, 1) {};
		\node [style=flavour2] (35) at (3.75, 1) {};
	\end{pgfonlayer}
	\begin{pgfonlayer}{edgelayer}
		\draw (0) to (4);
		\draw (4) to (14);
		\draw (14) to (15);
		\draw (14) to (18);
		\draw (18) to (22);
		\draw (3) to (24);
	\end{pgfonlayer}
\end{tikzpicture}
}  &\parbox[c]{12cm}{\textcolor{red}{$\SO(4)$}:\;\; $1+36 t^2+674 t^4+8631 t^6+84410 t^8+669825 t^{10}+4483882 t^{12}+26024643 t^{14}+133684785 t^{16}+617619665 t^{18}+2599637184 t^{20}+O\left(t^{22}\right)$\\ 
\textcolor{red}{$\O(4)^+$}:\;\;$1+22 t^2+364 t^4+4445 t^6+42776 t^8+337035 t^{10}+2249314 t^{12}+13035213 t^{14}+66909459 t^{16}+308990973 t^{18}+1300284388 t^{20}+O\left(t^{22}\right)$ \\ \textcolor{red}{$\text{Spin}(4)/\O(4)^-$}:\;\;$1+28 t^2+442 t^4+5103 t^6+46898 t^8+358377 t^{10}+2343802 t^{12}+13405515 t^{14}+68216313 t^{16}+313216337 t^{18}+1312934992 t^{20}+O\left(t^{22}\right)$ \\  \textcolor{red}{$\text{Pin}(4)$}:\;\;$1+21 t^2+287 t^4+3010 t^6+26081 t^8+191982 t^{10}+1226518 t^{12}+6910800 t^{14}+34828650 t^{16}+158901991 t^{18}+663258594 t^{20}+O\left(t^{22}\right)$}  \\ \hline
        
    \end{tabular}
    \caption{Nilpotent orbit closures of $B_3$, $D_4$ and $B_4$ labelled by their partitions given in exponential notation. The quivers whose Coulomb branches were conjectured in \cite{Cabrera:2017ucb} to describe such closures are given. The gauge node written in the algebra $\mathfrak{so}(n)$ can have different global forms $\SO(n)$, $\O(n)^+$, $\text{Spin}(n)$, $\O(n)^-$ and $\text{Pin}(n)$, whose Hilbert series are given on the right. In some cases the Hilbert series are the same. }
    \label{orbit1}
\end{table}
\begin{table}[]
    \centering
    \begin{tabular}{|c|c|c|c|}
    \hline
        Nilpotent orbit & Quiver  & HS   \\ \hline
          $[5,3,1]$ of $B_4$ & \scalebox{0.6}{\begin{tikzpicture}
	\begin{pgfonlayer}{nodelayer}
		\node [style=redgauge] (0) at (0, 0) {};
		\node [style=bluegauge] (1) at (1.25, 0) {};
		\node [style=redgauge] (2) at (2.5, 0) {};
		\node [style=bluegauge] (3) at (3.75, 0) {};
		\node [style=redgauge] (4) at (5, 0) {};
		\node [style=none] (7) at (0, -0.5) {SO(2)};
		\node [style=none] (8) at (1.25, -0.5) {USp(2)};
		\node [style=none] (9) at (2.5, -0.5) {SO(4)};
		\node [style=none] (10) at (3.75, -0.5) {USp(4)};
		\node [style=none] (11) at (7.5, -0.5) {$\mathfrak{so}(4)$};
		\node [style=bluegauge] (14) at (6.25, 0) {};
		\node [style=redflavor] (15) at (6.25, 1) {};
		\node [style=none] (16) at (6.25, 1.5) {SO(4)};
		\node [style=none] (17) at (6.25, -0.5) {USp(6)};
		\node [style=redgauge] (18) at (7.5, 0) {};
		\node [style=none] (21) at (5, -0.5) {SO(6)};
		\node [style=flavour2] (22) at (6.25, 1) {};
		\node [style=gauge3] (23) at (7.5, 0) {};
		\node [style=gauge3] (24) at (6.25, 0) {};
		\node [style=gauge3] (25) at (5, 0) {};
		\node [style=gauge3] (26) at (3.75, 0) {};
		\node [style=gauge3] (27) at (2.5, 0) {};
		\node [style=gauge3] (28) at (1.25, 0) {};
		\node [style=gauge3] (29) at (0, 0) {};
	\end{pgfonlayer}
	\begin{pgfonlayer}{edgelayer}
		\draw (0) to (4);
		\draw (4) to (14);
		\draw (14) to (15);
		\draw (14) to (18);
	\end{pgfonlayer}
\end{tikzpicture}
}  &\parbox[c]{12cm}{\textcolor{red}{$\SO(4)$}:\;\; $1+36 t^2+674 t^4+8715 t^6+87236 t^8+718680 t^{10}+5061142 t^{12}+31260629 t^{14}+172531071 t^{16}+863004616 t^{18}+3956473971 t^{20}+O\left(t^{22}\right)$\\ 
\textcolor{red}{$\O(4)^+$}:\;\; $1+22 t^2+364 t^4+4487 t^6+44162 t^8+361288 t^{10}+2536912 t^{12}+15649017 t^{14}+86316925 t^{16}+431634092 t^{18}+1978556985 t^{20}+O\left(t^{22}\right)$ \\ \textcolor{red}{$\text{Spin}(4)/\O(4)^-$}:\;\; $1+28 t^2+442 t^4+5131 t^6+48108 t^8+381192 t^{10}+2623214 t^{12}+15981397 t^{14}+87476967 t^{16}+435361368 t^{18}+1989707963 t^{20}+O\left(t^{22}\right)$ \\  \textcolor{red}{$\text{Pin}(4)$}:\;\;$1+21 t^2+287 t^4+3017 t^6+26571 t^8+202496 t^{10}+1361099 t^{12}+8175591 t^{14}+44369894 t^{16}+219676106 t^{18}+1000749470 t^{20}+O\left(t^{22}\right)$}  \\ \hline
         
            $[7,1^2]$ of $B_4$  &\scalebox{0.6}{\begin{tikzpicture}
	\begin{pgfonlayer}{nodelayer}
		\node [style=redgauge] (0) at (0, 0) {};
		\node [style=bluegauge] (1) at (1.25, 0) {};
		\node [style=redgauge] (2) at (2.5, 0) {};
		\node [style=bluegauge] (3) at (3.75, 0) {};
		\node [style=redgauge] (4) at (5, 0) {};
		\node [style=none] (7) at (0, -0.5) {SO(2)};
		\node [style=none] (8) at (1.25, -0.5) {USp(2)};
		\node [style=none] (9) at (2.5, -0.5) {SO(4)};
		\node [style=none] (10) at (3.75, -0.5) {USp(4)};
		\node [style=none] (11) at (7.5, -0.5) {$\mathfrak{so}(6)$};
		\node [style=bluegauge] (14) at (6.25, 0) {};
		\node [style=redflavor] (15) at (6.25, 1) {};
		\node [style=none] (16) at (6.25, 1.5) {SO(2)};
		\node [style=none] (17) at (6.25, -0.5) {USp(6)};
		\node [style=redgauge] (18) at (7.5, 0) {};
		\node [style=blueflavor] (19) at (7.5, 1) {};
		\node [style=none] (20) at (7.5, 1.5) {USp(4)};
		\node [style=none] (21) at (5, -0.5) {SO(6)};
		\node [style=gauge3] (22) at (0, 0) {};
		\node [style=gauge3] (23) at (1.25, 0) {};
		\node [style=gauge3] (24) at (2.5, 0) {};
		\node [style=gauge3] (25) at (3.75, 0) {};
		\node [style=gauge3] (26) at (5, 0) {};
		\node [style=gauge3] (27) at (6.25, 0) {};
		\node [style=gauge3] (28) at (7.5, 0) {};
		\node [style=flavour2] (29) at (7.5, 1) {};
		\node [style=flavour2] (30) at (6.25, 1) {};
	\end{pgfonlayer}
	\begin{pgfonlayer}{edgelayer}
		\draw (0) to (4);
		\draw (4) to (14);
		\draw (14) to (15);
		\draw (14) to (18);
		\draw (18) to (19);
	\end{pgfonlayer}
\end{tikzpicture}}   &\parbox[c]{12cm}{\textcolor{red}{$\SO(6)$}:\;\; $1+36 t^2+665 t^4+8409 t^6+81890 t^8+654882 t^{10}+4477389 t^{12}+26899665 t^{14}+144860179 t^{16}+709780863 t^{18}+3201281653 t^{20}+O\left(t^{22}\right)$\\ 
\textcolor{red}{$\O(6)^+$}:\;\; $1+22 t^2+357 t^4+4319 t^6+41404 t^8+329062 t^{10}+2243869 t^{12}+13465055 t^{14}+72471747 t^{16}+354997831 t^{18}+1600902979 t^{20}+O\left(t^{22}\right)$ \\ \textcolor{red}{$\text{Spin}(6)/\O(6)^-$}:\;\; $1+28 t^2+441 t^4+5041 t^6+46058 t^8+353730 t^{10}+2356669 t^{12}+13923625 t^{14}+74163827 t^{16}+360749439 t^{18}+1619118093 t^{20}+O\left(t^{22}\right)$ \\  \textcolor{red}{$\text{Pin}(6)$}:\;\;$1+21 t^2+287 t^4+2996 t^6+25815 t^8+190820 t^{10}+1239909 t^{12}+7206320 t^{14}+37969611 t^{16}+183357923 t^{18}+818928756 t^{20}+O\left(t^{22}\right)$}  \\ \hline
             $[3^2,1^4]$  of $D_5$  & \scalebox{0.6}{\begin{tikzpicture}
	\begin{pgfonlayer}{nodelayer}
		\node [style=redgauge] (0) at (0, 0) {};
		\node [style=bluegauge] (1) at (1.25, 0) {};
		\node [style=redgauge] (2) at (2.5, 0) {};
		\node [style=bluegauge] (3) at (3.75, 0) {};
		\node [style=redgauge] (4) at (5, 0) {};
		\node [style=none] (7) at (0, -0.5) {SO(2)};
		\node [style=none] (8) at (1.25, -0.5) {USp(2)};
		\node [style=none] (9) at (2.5, -0.5) {SO(4)};
		\node [style=none] (10) at (3.75, -0.5) {USp(4)};
		\node [style=none] (11) at (7.5, -0.5) {$\mathfrak{so}(4)$};
		\node [style=bluegauge] (14) at (6.25, 0) {};
		\node [style=redflavor] (15) at (6.25, 1) {};
		\node [style=none] (16) at (6.25, 1.5) {SO(1)};
		\node [style=none] (17) at (6.25, -0.5) {USp(4)};
		\node [style=redgauge] (18) at (7.5, 0) {};
		\node [style=none] (21) at (5, -0.5) {SO(5)};
		\node [style=redflavor] (24) at (3.75, 1) {};
		\node [style=none] (25) at (3.75, 1.5) {SO(1)};
		\node [style=bluegauge] (26) at (8.75, 0) {};
		\node [style=redflavor] (27) at (8.75, 1) {};
		\node [style=none] (28) at (8.75, 1.5) {SO(2)};
		\node [style=none] (29) at (8.75, -0.5) {USp(2)};
		\node [style=flavour2] (30) at (8.75, 1) {};
		\node [style=flavour2] (31) at (6.25, 1) {};
		\node [style=flavour2] (32) at (3.75, 1) {};
		\node [style=gauge3] (33) at (8.75, 0) {};
		\node [style=gauge3] (34) at (7.5, 0) {};
		\node [style=gauge3] (35) at (6.25, 0) {};
		\node [style=gauge3] (36) at (5, 0) {};
		\node [style=gauge3] (37) at (3.75, 0) {};
		\node [style=gauge3] (38) at (2.5, 0) {};
		\node [style=gauge3] (39) at (1.25, 0) {};
		\node [style=gauge3] (40) at (0, 0) {};
	\end{pgfonlayer}
	\begin{pgfonlayer}{edgelayer}
		\draw (0) to (4);
		\draw (4) to (14);
		\draw (14) to (15);
		\draw (14) to (18);
		\draw (3) to (24);
		\draw (18) to (26);
		\draw (26) to (27);
	\end{pgfonlayer}
\end{tikzpicture}
}  &\parbox[c]{12cm}{\textcolor{red}{$\SO(4)$}:\;\; $1+45 t^2+1079 t^4+17630 t^6+216370 t^8+2110977 t^{10}+17028305 t^{12}+116962308 t^{14}+699968215 t^{16}+3717369030 t^{18}+17782581806 t^{20}+O\left(t^{22}\right)$
\\ \textcolor{red}{$\O(4)^+/\text{Spin}(4)/\O(4)^-$}:\;\; $1+29 t^2+599 t^4+9214 t^6+110418 t^8+1066049 t^{10}+8558209 t^{12}+58645140 t^{14}+350540615 t^{16}+1860421830 t^{18}+8896344686 t^{20}+O\left(t^{22}\right)$\\ \textcolor{red}{$\text{Pin}(4)$}:\;\; $1+21 t^2+359 t^4+5006 t^6+57442 t^8+543585 t^{10}+4323161 t^{12}+29486556 t^{14}+175826815 t^{16}+931948230 t^{18}+4453226126 t^{20}+O\left(t^{22}\right)$}  \\ \hline
         
    \end{tabular}
    \caption{Nilpotent orbit closures of $B_4$ and $D_5$ labelled by their partitions given in exponential notation. The quivers whose Coulomb branches were conjectured in \cite{Cabrera:2017ucb} to describe such closures are given. The gauge node written in the algebra $\mathfrak{so}(n)$ can have different global forms $\SO(n)$, $\O(n)^+$, $\text{Spin}(n)$, $\O(n)^-$ and $\text{Pin}(n)$, whose Hilbert series are given on the right. In some cases the Hilbert series are the same.}
    \label{orbit2}
\end{table}
\begin{table}[]
    \centering
    \begin{tabular}{|c|c|c|c|} \hline
      Nilpotent orbit & Quiver  & HS    \\ 
    \hline
            $[5,3,1^2]$  of $D_5$  &\scalebox{0.6}{\begin{tikzpicture}
	\begin{pgfonlayer}{nodelayer}
		\node [style=redgauge] (0) at (0, 0) {};
		\node [style=bluegauge] (1) at (1.25, 0) {};
		\node [style=redgauge] (2) at (2.5, 0) {};
		\node [style=bluegauge] (3) at (3.75, 0) {};
		\node [style=redgauge] (4) at (5, 0) {};
		\node [style=none] (7) at (0, -0.5) {SO(2)};
		\node [style=none] (8) at (1.25, -0.5) {USp(2)};
		\node [style=none] (9) at (2.5, -0.5) {SO(4)};
		\node [style=none] (10) at (3.75, -0.5) {USp(4)};
		\node [style=none] (11) at (7.5, -0.5) {$\mathfrak{so}(6)$};
		\node [style=bluegauge] (14) at (6.25, 0) {};
		\node [style=redflavor] (15) at (6.25, 1) {};
		\node [style=none] (16) at (6.25, 1.5) {SO(2)};
		\node [style=none] (17) at (6.25, -0.5) {USp(6)};
		\node [style=redgauge] (18) at (7.5, 0) {};
		\node [style=none] (21) at (5, -0.5) {SO(6)};
		\node [style=bluegauge] (26) at (8.75, 0) {};
		\node [style=redflavor] (27) at (8.75, 1) {};
		\node [style=none] (28) at (8.75, 1.5) {SO(4)};
		\node [style=none] (29) at (8.75, -0.5) {USp(4)};
		\node [style=gauge3] (30) at (8.75, 0) {};
		\node [style=gauge3] (31) at (7.5, 0) {};
		\node [style=gauge3] (32) at (6.25, 0) {};
		\node [style=gauge3] (33) at (5, 0) {};
		\node [style=gauge3] (34) at (3.75, 0) {};
		\node [style=gauge3] (35) at (2.5, 0) {};
		\node [style=gauge3] (36) at (1.25, 0) {};
		\node [style=gauge3] (37) at (0, 0) {};
		\node [style=flavour2] (38) at (6.25, 1) {};
		\node [style=flavour2] (39) at (8.75, 1) {};
	\end{pgfonlayer}
	\begin{pgfonlayer}{edgelayer}
		\draw (0) to (4);
		\draw (4) to (14);
		\draw (14) to (15);
		\draw (14) to (18);
		\draw (18) to (26);
		\draw (26) to (27);
	\end{pgfonlayer}
\end{tikzpicture}
}  &\parbox[c]{12cm}{\textcolor{red}{$\SO(6)$}:\;\; $1+45 t^2+1034 t^4+16215 t^6+195424 t^8+1930528 t^{10}+16268461 t^{12}+120122896 t^{14}+792104940 t^{16}+4731174753 t^{18}+25878964760 t^{20}+O\left(t^{22}\right)$ \\  \textcolor{red}{$\O(6)^+/\text{Spin}(6)/\O(6)^-$}:\;\; $1+29 t^2+570 t^4+8439 t^6+99424 t^8+972944 t^{10}+8164909 t^{12}+60172992 t^{14}+396426524 t^{16}+2366758113 t^{18}+12942930440 t^{20}+O\left(t^{22}\right)$ \\  \textcolor{red}{$\text{Pin}(6)$}:\;\; $1+21 t^2+338 t^4+4551 t^6+51424 t^8+494152 t^{10}+4113133 t^{12}+30198040 t^{14}+198587316 t^{16}+1184549793 t^{18}+6474913280 t^{20}+O\left(t^{22}\right)$}  \\ \hline
         
    \end{tabular}
    \caption{Nilpotent orbit closures of $D_5$ labelled by their partitions given in exponential notation. The quivers whose Coulomb branches were conjectured in \cite{Cabrera:2017ucb} to describe such closures are given. The gauge node written in the algebra $\mathfrak{so}(n)$ can have different global forms $\SO(n)$, $\O(n)^+$, $\text{Spin}(n)$, $\O(n)^-$ and $\text{Pin}(n)$, whose Hilbert series are given on the right. In some cases the Hilbert series are the same.}
    \label{orbit3}
\end{table}
\end{landscape}

\section{Affine $D_5$ quiver} \label{app:affineD5}
In this Appendix, we examine the following affine $D_5$ quiver with unitary gauge groups and discrete gauging thereof.
\bes{ \label{affineD5a}
\scalebox{0.8}{
\begin{tikzpicture}[baseline, font=\footnotesize]
\node[gauge,label=left:{$\U(1)$}] (U1lu) at (-2,1) {};
\node[gauge,label=left:{$\U(1)$}] (U1ld) at (-2,-1) {};
\node[gauge,label={below, xshift=0.2cm}:{$\U(2)$}] (U2l) at (-1,0) {};
\node[gauge,label={below,xshift=-0.2cm}:{$\U(2)$}] (U2r) at (1,0) {};
\node[gauge,label=right:{$\U(1)$}] (U1ru) at (2,1) {};
\node[gauge,label=right:{$\U(1)$}] (U1rd) at (2,-1) {};
\draw[-,bend left=0] (U1lu)--(U2l)--(U2r)--(U1ru);
\draw[-,bend right=0] (U1ld)--(U2l);
\draw[-,bend right=0] (U1rd)--(U2r);
\node at (4,0) {$/\U(1)$};
\end{tikzpicture} }
\quad = \quad 
\scalebox{0.8}{
\begin{tikzpicture}[baseline, font=\footnotesize]
\node[gauge,label=left:{$\U(1)$}] (U1lu) at (-2,1) {};
\node[gauge,label=left:{$\U(1)$}] (U1ld) at (-2,-1) {};
\node[gauge,label={below,xshift=0.3cm}:{$\SU(2)$}] (U2l) at (-1,0) {};
\node[label=above:{$\U(1)$}] (U1m) at (0,0.4) {};
\node[gauge,label={below,xshift=-0.3cm}:{$\SU(2)$}] (U2r) at (1,0) {};
\node[gauge,label=right:{$\U(1)$}] (U1ru) at (2,1) {};
\node[gauge,label=right:{$\U(1)$}] (U1rd) at (2,-1) {};
\draw[-,bend left=0] (U1lu)--(U2l)--(U2r)--(U1ru);
\draw[-,bend right=0] (U1ld)--(U2l);
\draw[-,bend right=0] (U1rd)--(U2r);
\draw[-] (0,0) to (0,0.5);
\draw[-] (-1.5,-0.9) to node[midway,below] {$\BZ^\zeta_2 \times \BZ^g_2$} (1.5,-0.9);
\end{tikzpicture} }
}
For convenience, we choose to perform the quotient of an overall $\U(1)$ from the two $\U(2)$ gauge groups in the middle. In this case, they can be written as
\bes{ \label{quotientD5}
\frac{\U(2) \times \U(2)}{\U(1)} \cong \frac{\SU(2) \times \SU(2) \times \U(1)}{\BZ^\zeta_2 \times \BZ^g_2}~,
}
where one of the $\BZ_2$ factors (denoted by $\BZ^\zeta_2$) in the denominator is generated by the element $(-1,-1, 1)$ of the $\BZ_2 \times \BZ_2 \times \U(1)$ centre of $\SU(2) \times \SU(2) \times \U(1)$, and the other $\BZ_2$ factor (denoted by $\BZ^g_2$) is generated by the element $(1,-1, -1)$ of the $\BZ_2 \times \BZ_2 \times \U(1)$ centre. This choice is depicted in the right diagram of \eref{affineD5a}. The index of \eref{affineD5a} can be computed as
\bes{ \label{indexaffineD5a}
&\CI_{\eqref{affineD5a}}(x; a; w_1, \ldots, w_5; g; \zeta) \\
&= \frac{1}{4} \sum_{\epsilon_1, \epsilon_2 = 0}^1\,\, \sum_{m_{z_1} \in \BZ+\frac{\epsilon_1}{2}} \,\, \sum_{m_{z_2} \in \BZ+ \frac{\epsilon_1}{2} +\frac{\epsilon_2}{2}} \,\, \sum_{m_u \in \BZ +\frac{\epsilon_2}{2}} \zeta^{\epsilon_1} g^{\epsilon_2} w_5^{m_u} \oint \frac{dz_1}{2\pi i z_1} \frac{dz_2}{2\pi i z_2} \frac{du}{2\pi i u}  \\
& \qquad \times \CZ_{\text{chir}}^{1}(x;a^{-2}; 0) \prod_{j=1}^2 \left[ \CZ_{\text{vec}}^{\USp(2)}(x;z_j; m_{z_j})  \prod_{\ell=-1}^1 \CZ_{\text{chir}}^{1}(x;z_j^{2 \ell} a^{-2};2 \ell m_{z_j}) \right] \\
& \qquad \times \prod_{s,s_1, s_2 = \pm 1}  \CZ_{\text{chir}}^{1/2}(x;a z_1^{s_1} z_2^{s_2} u^s; s_1 m_{z_1}+ s_2 m_{z_2} + s m_u) \\
& \qquad \times {\claret \prod_{\alpha =1}^2 \CI_{T[\SU(2)]}(x; a; z_1, m_{z_1}; w_\alpha) \prod_{\beta =3}^4 \CI_{T[\SU(2)]}(x; a; z_2, m_{z_2}; w_\beta)}~.
}
where $g$ and $\zeta$ are the fugacities of the two $\BZ_2$ zero-form symmetries, denoted by $\BZ^{[0]}_{2, g}$ and $\BZ^{[0]}_{2, \zeta}$, which arise from the quotient in \eref{quotientD5}, and the index for the $\U(1)$ gauge theory with two hypermultiplets of charge one is
\bes{
\CI_{T[\SU(2)]}(x; a; f, m_f; w) &= \sum_{m_u \in \BZ} w^{m_u} \oint \frac{d u}{2\pi i u}  \CZ_{\text{chir}}^{1}(x;a^{-2}; 0) \\
& \qquad \times \prod_{s_1, s_2 = \pm 1} \CZ_{\text{chir}}^{1/2}(x;a f^{s_1} u^{s_2};  s_1 m_f +s_2 m_{u})~.
}
Explicitly, if we set $w_i=1$, the above index is equal to \eref{indOSpSO10} with $\chi = +1$. 
\bes{ \label{explicitindexaffineD5a}
& \CI_{\eqref{affineD5a}}(x; a; w_i =1; g; \zeta) = \CI_{\eref{OSpSO10}}(x; a ; g, \zeta; \chi=+1) \\
&= 1 +  \left[13 + 16\zeta  + g \left(8 + 8\zeta \right)\right] a^2  x + \Big\{ a^{-4} + \left[226 + 208\zeta + g \left( 168  + 168\zeta \right)\right]a^4 \\
& \quad \,\,\,\,\, - \left[14 + 16\zeta  + g \left(8 + 8\zeta \right)\right] \Big\} x^2+\ldots~.
}
We summarise the correspondence between various discrete gaugings of the $D_5$ affine quiver \eqref{affineD5a} and the variants of the orthosymplectic quiver \eqref{OSpSO10} in Table \ref{tab:unitaryD5}.

\begin{longtable}{ccc} 

\caption{Variants of \eqref{OSpSO10}, the corresponding unitary quiver, index coefficients, and continuous global symmetries.}
\label{tab:unitaryD5} \\

\toprule
Variants of \eref{OSpSO10} & Unitary quiver & $\overset{\textstyle (\alpha, \beta)\,\, \text{in}\,\, \eqref{unrefindexso10}}{\text{Continuous symmetry}}$ \\
\midrule
\endfirsthead

 \toprule
Variants of \eref{OSpSO10} & Mirror theory & $\overset{\textstyle (\alpha, \beta)}{\text{Continuous symmetry}}$ \\
 \midrule
 \endhead


 \bottomrule
 \endlastfoot


\scalebox{0.6}{%
\begin{tikzpicture}
\node[gauge,label=below:{$\SO(2)$}] (D1l) at (-4,0) {};
\node[gauge,label=below:{$\USp(2)$}] (C1l) at (-2,0) {};
\node[gauge,label=below:{$\SO(4)$}] (D2) at (0,0) {};
\node[gauge,label=below:{$\USp(2)$}] (C1r) at (2,0) {};
\node[gauge,label=below:{$\SO(2)$}] (D1r) at (4,0) {};
\node[gauge,label=right:{$\U(1)$}] (D1u) at (0,1) {};
\draw[-] (D1l)--(C1l)--(D2)--(C1r)--(D1r);
\draw[-] (D2) to (D1u);
\node at (5,0) {$/\BZ_2$};
\end{tikzpicture}%
} &
\scalebox{0.6}{
\begin{tikzpicture}[font=\footnotesize]
\node[gauge,label=left:{$\U(1)$}] (U1lu) at (-2,1) {};
\node[gauge,label=left:{$\U(1)$}] (U1ld) at (-2,-1) {};
\node[gauge,label={below,xshift=0.3cm}:{$\SU(2)$}] (U2l) at (-1,0) {};
\node[label=above:{$\U(1)$}] (U1m) at (0,0.4) {};
\node[gauge,label={below,xshift=-0.3cm}:{$\SU(2)$}] (U2r) at (1,0) {};
\node[gauge,label=right:{$\U(1)$}] (U1ru) at (2,1) {};
\node[gauge,label=right:{$\U(1)$}] (U1rd) at (2,-1) {};
\draw[-,bend left=0] (U1lu)--(U2l)--(U2r)--(U1ru);
\draw[-,bend right=0] (U1ld)--(U2l);
\draw[-,bend right=0] (U1rd)--(U2r);
\draw[-] (0,0) to (0,0.5);
\draw[-] (-1.5,-0.9) to node[midway,below] {$\BZ^\zeta_2 \times \BZ^g_2$} (1.5,-0.9);
\end{tikzpicture} }
&
$\overset{\textstyle (45, 770)}{\so(10)}$ \\
\midrule 

\scalebox{0.6}{%
\begin{tikzpicture}
\node[gauge,label=below:{$\SO(2)$}] (D1l) at (-4,0) {};
\node[gauge,label=below:{$\USp(2)$}] (C1l) at (-2,0) {};
\node[gauge,label=below:{$\overset{\textstyle \Spin(4)}{\text{or} \,\, \O(4)^-}$}] (D2) at (0,0) {};
\node[gauge,label=below:{$\USp(2)$}] (C1r) at (2,0) {};
\node[gauge,label=below:{$\SO(2)$}] (D1r) at (4,0) {};
\node[gauge,label=right:{$\U(1)$}] (D1u) at (0,1) {};
\draw[-] (D1l)--(C1l)--(D2)--(C1r)--(D1r);
\draw[-] (D2) to (D1u);
\node at (5,0) {$/\BZ_2$};
\end{tikzpicture}%
} &
\scalebox{0.6}{
\begin{tikzpicture}[font=\footnotesize]
\node[gauge,label=left:{$\U(1)$}] (U1lu) at (-2,1) {};
\node[gauge,label=left:{$\U(1)$}] (U1ld) at (-2,-1) {};
\node[gauge,label={below,xshift=0.3cm}:{$\SU(2)$}] (U2l) at (-1,0) {};
\node[gauge,label={below,xshift=-0.3cm}:{$\U(2)$}] (U2r) at (1,0) {};
\node[gauge,label=right:{$\U(1)$}] (U1ru) at (2,1) {};
\node[gauge,label=right:{$\U(1)$}] (U1rd) at (2,-1) {};
\draw[-,bend left=0] (U1lu)--(U2l)--(U2r)--(U1ru);
\draw[-,bend right=0] (U1ld)--(U2l);
\draw[-,bend right=0] (U1rd)--(U2r);
\end{tikzpicture} }
&
$\overset{\textstyle (21,394)}{\so(6) \oplus \so(4)}$ \\
\midrule 

\scalebox{0.6}{%
\begin{tikzpicture}
\node[gauge,label=below:{$\SO(2)$}] (D1l) at (-4,0) {};
\node[gauge,label=below:{$\USp(2)$}] (C1l) at (-2,0) {};
\node[gauge,label=below:{$\SO(4)$}] (D2) at (0,0) {};
\node[gauge,label=below:{$\USp(2)$}] (C1r) at (2,0) {};
\node[gauge,label=below:{$\SO(2)$}] (D1r) at (4,0) {};
\node[gauge,label=right:{$\U(1)$}] (D1u) at (0,1) {};
\draw[-] (D1l)--(C1l)--(D2)--(C1r)--(D1r);
\draw[-] (D2) to (D1u);
\end{tikzpicture}%
} &
\scalebox{0.6}{
\begin{tikzpicture}[font=\footnotesize]
\node[gauge,label=left:{$\U(1)$}] (U1lu) at (-2,1) {};
\node[gauge,label=left:{$\U(1)$}] (U1ld) at (-2,-1) {};
\node[gauge,label={below,xshift=0.3cm}:{$\SU(2)$}] (U2l) at (-1,0) {};
\node[label=above:{$\U(1)$}] (U1m) at (0,0.4) {};
\node[gauge,label={below,xshift=-0.3cm}:{$\SU(2)$}] (U2r) at (1,0) {};
\node[gauge,label=right:{$\U(1)$}] (U1ru) at (2,1) {};
\node[gauge,label=right:{$\U(1)$}] (U1rd) at (2,-1) {};
\draw[-,bend left=0] (U1lu)--(U2l)--(U2r)--(U1ru);
\draw[-,bend right=0] (U1ld)--(U2l);
\draw[-,bend right=0] (U1rd)--(U2r);
\draw[-] (0,0) to (0,0.5);
\draw[-] (-1.5,-0.9) to node[midway,below] {$\BZ^\zeta_2$} (1.5,-0.9);
\end{tikzpicture} }
&
$\overset{\textstyle (29, 434)}{\so(8) \oplus \so(2)}$ \\
\midrule 

\scalebox{0.6}{%
\begin{tikzpicture}
\node[gauge,label=below:{$\SO(2)$}] (D1l) at (-4,0) {};
\node[gauge,label=below:{$\USp(2)$}] (C1l) at (-2,0) {};
\node[gauge,label=below:{$\overset{\textstyle \Spin(4)}{\text{or} \,\, \O(4)^-}$}] (D2) at (0,0) {};
\node[gauge,label=below:{$\USp(2)$}] (C1r) at (2,0) {};
\node[gauge,label=below:{$\SO(2)$}] (D1r) at (4,0) {};
\node[gauge,label=right:{$\U(1)$}] (D1u) at (0,1) {};
\draw[-] (D1l)--(C1l)--(D2)--(C1r)--(D1r);
\draw[-] (D2) to (D1u);
\end{tikzpicture}%
} &
\scalebox{0.6}{
\begin{tikzpicture}[font=\footnotesize]
\node[gauge,label=left:{$\U(1)$}] (U1lu) at (-2,1) {};
\node[gauge,label=left:{$\U(1)$}] (U1ld) at (-2,-1) {};
\node[gauge,label={below,xshift=0.3cm}:{$\SU(2)$}] (U2l) at (-1,0) {};
\node[gauge,label={below,xshift=-0.3cm}:{$\SU(2)$}] (U2r) at (1,0) {};
\node[gauge,label=right:{$\U(1)$}] (U1ru) at (2,1) {};
\node[gauge,label=right:{$\U(1)$}] (U1rd) at (2,-1) {};
\draw[-,bend left=0] (U1lu)--(U2l)--(U2r)--(U1ru);
\draw[-,bend right=0] (U1ld)--(U2l);
\draw[-,bend right=0] (U1rd)--(U2r);
\end{tikzpicture} }
&
$\overset{\textstyle (13, 226)}{\so(4) \oplus \so(4) \oplus \so(2)}$ \\

\end{longtable}
At present, we do not manage to turn on the analogue of the charge conjugation symmetry fugacity in \eqref{indOSpSO10} for the unitary affine $D_5$ quiver. We leave as an open question for future work to investigate whether this is possible at all. In the following, we instead examine another $\BZ_2$ symmetry that acts on the affine $D_5$ quiver by wreathing.

\subsection{Wreathing by $\BZ_2$ generated by a single transposition}
Let us consider the $\BZ_2$ wreathing that acts on \eqref{affineD5a} by identifying the upper left and lower left legs, denoted by $\BZ_{2}^{(12)}$ below:
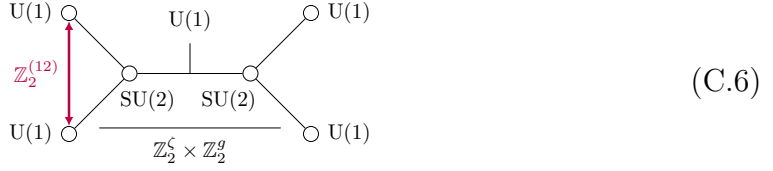
\bes{ \label{wreath12affineD5a}
\scalebox{0.8}{
\begin{tikzpicture}[baseline, font=\footnotesize]
\node[gauge,label=left:{$\U(1)$}] (U1lu) at (-2,1) {};
\node[gauge,label=left:{$\U(1)$}] (U1ld) at (-2,-1) {};
\node[gauge,label={below,xshift=0.3cm}:{$\SU(2)$}] (U2l) at (-1,0) {};
\node[label=above:{$\U(1)$}] (U1m) at (0,0.4) {};
\node[gauge,label={below,xshift=-0.3cm}:{$\SU(2)$\,\,}] (U2r) at (1,0) {};
\node[gauge,label=right:{$\U(1)$}] (U1ru) at (2,1) {};
\node[gauge,label=right:{$\U(1)$}] (U1rd) at (2,-1) {};
\draw[-,bend left=0] (U1lu)--(U2l)--(U2r)--(U1ru);
\draw[-,bend right=0] (U1ld)--(U2l);
\draw[-,bend right=0] (U1rd)--(U2r);
\draw[-] (0,0) to (0,0.5);
\draw[-] (-1.5,-0.9) to node[midway,below] {$\BZ^\zeta_2 \times \BZ^g_2$} (1.5,-0.9);
\draw[<->, purple, very thick] (U1lu) to node[midway, left]{$\BZ_{2}^{(12)}$} (U1ld);
\end{tikzpicture} }
}
The resulting index can be derived as
\bes{
&\CI_{\eqref{affineD5a}/\BZ_{2}^{(12)}}(x; a; \tilde{w}, w_3, w_4, w_5; g; \zeta) \\
&=\frac{1}{2} \Big[\CI_{\eqref{affineD5a}}(x; a; w_1=w_2=\tilde{w}, w_3 \ldots, w_5; g; \zeta) \\
& \qquad + \CI_{\eqref{affineD5a}_{(12)}}(x; a; \tilde{w}, w_3, w_4, w_5; g; \zeta) \Big]~,
}
where the second term appearing in the sum above is given by \eqref{indexaffineD5a}, in which the term in {\claret claret} is replaced by
\bes{
\CI_{T[\SU(2)]}(x^2; a^2; z_1^2, m_{z_1}; \tilde{w}^2) \prod_{\beta =3}^4 \CI_{T[\SU(2)]}(x; a; z_2, m_{z_2}; w_\beta)~.
}
Setting $\tilde{w} = w_{3,4,5} =1$, we obtain
\bes{
& \CI_{\eqref{affineD5a}/\BZ_{2}^{(12)}}(x; a; \tilde{w} = w_i=1; g; \zeta) \\ 
&= 1 +  \left[10 + 12\zeta + g \left(6 + 8\zeta \right)\right] a^{-2}  x + \Big\{ \left[161+ 140\zeta + g \left( 118 + 120\zeta \right) \right]a^{-4} \\
& \quad \,\,\,\,\,+ a^4 -\left[11 + 12\zeta + g \left(6 + 8\zeta \right)\right] \Big\} x^2 +\ldots~.
}
We can also refine the index of theory \eqref{affineD5a} with respect to the zero-form symmetry associated with the $\BZ_2$ wreathing (\ie~ associated with $\BZ_2^{(12)}$), denoted by  $\BZ_{2,\, \Xi}^{[0]}$, with the corresponding fugacity $\Xi$. The index for $\Xi=+1$ is given by 
\bes{
\CI_{\eqref{affineD5a}}(x; a; w_i=1; g; \zeta; \Xi=+1) = \text{\eref{explicitindexaffineD5a}}~,
}
whereas that for $\Xi=-1$ can be computed from 
\bes{
\scalebox{0.98}{$
\begin{split}
&\CI_{\eqref{affineD5a}}(x; a; w_i=1; g; \zeta; \Xi=-1) \\
&= 2 \times \CI_{\eqref{affineD5a}/\BZ_{2}^{(12)}}(x; a; \tilde{w} =1, w_i=1; g; \zeta) - \CI_{\eqref{affineD5a}}(x; a; w_i=1; g; \zeta; \Xi=+1)~.
\end{split}
$}
}
We can thus compute the index for \eqref{affineD5a} with refinement with respect to $\Xi$ as follows:
\bes{ \label{indexaffineD5arefXi}
&\CI_{\eqref{affineD5a}}(x; a; w_i=1; g; \zeta; \Xi)= \frac{1}{2} \sum_{s=\pm1} \CI_{\eqref{affineD5a}}(x; a; w_i=1; g; \zeta; \Xi=s) (1+s \Xi) \\
&= 1 +  \left[10 + 3\Xi + 12\zeta + 4\zeta\Xi+ g \left(6 + 2 \Xi + 8\zeta  \right) \right] a^{-2}  x \\
& \quad \,\,\,\,\, + \Big\{ \left[161 + 65\Xi + 140\zeta + 68\zeta\Xi + g \left(118  + 50 \Xi+ 120\zeta   + 48\zeta \Xi \right)\right] a^{-4}   \\
& \quad \,\,\,\,\, + a^4 - \left[11 + 3\Xi + 12\zeta + 4\zeta\Xi+ g \left(6 + 2 \Xi + 8\zeta  \right) \right]\Big\} x^2 + \ldots~.
}

In terms of the mirror theory of \eqref{affineD5a}, namely the $\USp(2)$ gauge theory with five flavours, the $\BZ_{2,\, \Xi}^{[0]}$, $\BZ_{2, \, g}^{[0]}$ and $\BZ_{2, \, \zeta}^{[0]}$ symmetries act as follows:
\bes{ \label{USp2w5withXi}
\scalebox{0.9}{%
\begin{tikzpicture}[baseline, font=\footnotesize]
\node[gauge,label=below:{\qquad \quad \,\,\, $\USp(2)$}] (C1) at (0,0) {};
\node[flavour,label=below:{$\SO(2)$}] (D1l) at (2,0) {};
\node[flavour,label=below:{$\SO(4)$}] (D2r) at (-2,0) {}; 
\node[flavour,label=right:{$\SO(1)$}] (B0) at (0,1.5) {};
\node[flavour,label=right:{$\SO(3)$}] (B1) at (0,-1.5) {};
\draw[new-green, very thick, -] (D1l) to node[above,midway]{\textcolor{new-green}{$g$}} (C1);
\draw[red, very thick, -] (C1) to node[above,midway]{\textcolor{red}{$\zeta$}} (D2r);
\draw[purple,-,very thick] (B0) to node[right, midway]{\textcolor{purple}{$\Xi$}}(C1);
\draw[black,-, thick] (C1)--(B1);
\end{tikzpicture}%
}
}
where $\Xi$ can also be identified with the charge conjugation symmetry associated with the $\SO(1)$ group.  Gauging such a symmetry turns it into the $\O(1)$ gauge group. Therefore, we see that the mirror theory of the wreathed quiver \eref{wreath12affineD5a} is
\bes{ \label{USp2w5withXiZ2}
\eqref{wreath12affineD5a} \qquad \overset{\text{mirror}}{\longleftrightarrow} \qquad
\scalebox{0.9}{%
\begin{tikzpicture}[baseline, font=\footnotesize]
\node[gauge,label=below:{\qquad \quad \,\,\, $\USp(2)$}] (C1) at (0,0) {};
\node[flavour,label=below:{$\SO(2)$}] (D1l) at (2,0) {};
\node[flavour,label=below:{$\SO(4)$}] (D2r) at (-2,0) {}; 
\node[gauge,label=right:{$\O(1)$}] (B0) at (0,1.5) {};
\node[flavour,label=right:{$\SO(3)$}] (B1) at (0,-1.5) {};
\draw[new-green, very thick, -] (D1l) to node[above,midway]{\textcolor{new-green}{$g$}} (C1);
\draw[red, very thick, -] (C1) to node[above,midway]{\textcolor{red}{$\zeta$}} (D2r);
\draw[-,thick] (B0)--(C1)--(B1);
\end{tikzpicture}%
}
}
Comparing \eref{USp2w5withXiZ2} with \eref{actionchizetagonUSp2w5}, we see that the action of $\BZ_{2, \, \Xi}^{[0]}$ in the above quiver is different from that of $\BZ_{2, \, \chi}^{[0]}$, but $\BZ_{2,\, \zeta}^{[0]}$ and $\BZ_{2,\, g}^{[0]}$ act in the same way.

\subsection{Wreathing by $\BZ_2$ generated by a double transposition}
Let us now consider the $\BZ_2$ wreathing that acts on \eqref{affineD5a} by identifying the upper left and lower left legs and simultaneously identifying the upper right and lower right legs, denoted by $\BZ_{2}^{(12)(34)}$ below:
\bes{ \label{wreath1234affineD5a}
\scalebox{0.8}{
\begin{tikzpicture}[baseline, font=\footnotesize]
\node[gauge,label=left:{$\U(1)$}] (U1lu) at (-2,1) {};
\node[gauge,label=left:{$\U(1)$}] (U1ld) at (-2,-1) {};
\node[gauge,label=below:{$\,\,\SU(2)$}] (U2l) at (-1,0) {};
\node[label=above:{$\U(1)$}] (U1m) at (0,0.4) {};
\node[gauge,label=below:{$\SU(2)$\,\,}] (U2r) at (1,0) {};
\node[gauge,label=right:{$\U(1)$}] (U1ru) at (2,1) {};
\node[gauge,label=right:{$\U(1)$}] (U1rd) at (2,-1) {};
\draw[-,bend left=0] (U1lu)--(U2l)--(U2r)--(U1ru);
\draw[-,bend right=0] (U1ld)--(U2l);
\draw[-,bend right=0] (U1rd)--(U2r);
\draw[-] (0,0) to (0,0.5);
\draw[-] (-1.5,-0.9) to node[midway,below] {$\BZ^\zeta_2 \times \BZ^g_2$} (1.5,-0.9);
\draw[<->, blue, thick] (U1lu) to node[midway, left]{$\BZ_{2}^{(12)(34)}$} (U1ld);
\draw[<->, blue, thick] (U1ru) to (U1rd);
\end{tikzpicture} }}
The resulting index is
\bes{ \label{indD5DT}
&\CI_{\eqref{affineD5a}/\BZ_{2}^{(12)(34)}}(x; a; \tilde{w}_1, \tilde{w}_3, w_5; g; \zeta)\\
&=\frac{1}{2} \Big[\CI_{\eqref{affineD5a}}(x; a; \tilde{w}_1, \tilde{w}_1, \tilde{w}_3, \tilde{w}_3, w_5; g; \zeta) \\
& \qquad + \CI_{\eqref{affineD5a}_{(12)(34)}}(x; a; \tilde{w}_1, \tilde{w}_3, w_5; g; \zeta) \Big]~,
}
where the second term appearing in the sum above can be obtained from \eqref{indexaffineD5a} by replacing the term in {\claret claret} with
\bes{
\CI_{T[\SU(2)]}(x^2; a^2; z_1^2, m_{z_1}; \tilde{w}^2_1) \,\, \CI_{T[\SU(2)]}(x^2; a^2; z_2^2, m_{z_2}; \tilde{w}^2_3)~.
}
Upon setting $\tilde{w}_1 = \tilde{w}_3 = w_5 = 1$, the index \eref{indD5DT} reads
\bes{
&\CI_{\eqref{affineD5a}/\BZ_{2}^{(12)(34)}}(x; a; \tilde{w}_1 = \tilde{w}_3 = w_5 =1; g, \zeta) \\
&= 1 + \left[7 + 10\zeta + g \left(6 + 6\zeta \right)\right] a^{-2} x + \Big\{ \left[132 + 114\zeta + g \left( 94 + 94\zeta \right) \right]a^{-4} \\
& \quad \,\,\,\,\, + a^4 - \left[8 + 10\zeta + g \left(6 + 6\zeta \right)\right] \Big\} x^2 + \ldots~.
}
Let us denote the zero-form symmetry associated with wreathing (\ie~ corresponding to $\BZ_2^{(12)(34)}$) by $\BZ_{2,\, \Gamma}^{[0]}$, with the corresponding fugacity $\Gamma$. Similarly to \eref{indexaffineD5arefXi}, the index for \eqref{affineD5a} refined with respect to $\Gamma$ is as follows:
\bes{
&\CI_{\eqref{affineD5a}}(x; a; w_i=1; g, \zeta, \Xi) \\
&= 1 + \left[7 + 6\Gamma + 10\zeta + 6\Gamma\zeta + g \left( 6  + 2\Gamma + 6\zeta   + 2\Gamma \zeta \right)\right] a^{-2}  x \\
& \quad \,\,\,\,\,  + \Big\{\left[132 + 94\Gamma + 114\zeta + 94\Gamma\zeta + g \left(94  + 74 \Gamma + 94\zeta   + 74 \Gamma\zeta \right)\right]a^{-4}  \\
& \quad \,\,\,\,\,   + a^4 - \left[8 + 6\Gamma + 10\zeta + 6\Gamma\zeta + g \left( 6  + 2\Gamma + 6\zeta   + 2\Gamma \zeta \right)\right] \Big\} x^2+\ldots~.
}
In terms of the mirror theory of \eqref{affineD5a}, namely the $\USp(2)$ gauge theory with five flavours, the $\BZ_{2,\, \Gamma}^{[0]}$, $\BZ_{2, \, g}^{[0]}$ and $\BZ_{2, \, \zeta}^{[0]}$ symmetries act as follows:
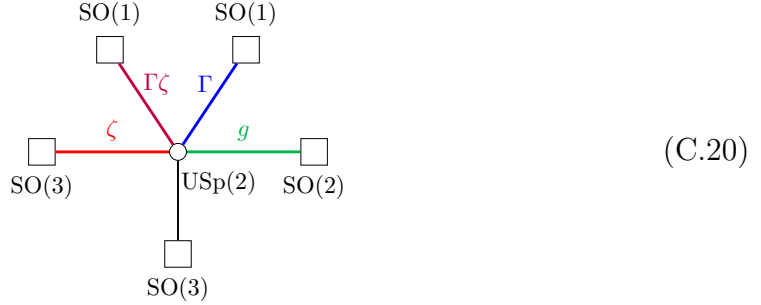
\bes{ \label{USp2w5withXi}
\scalebox{0.9}{%
\begin{tikzpicture}[baseline, font=\footnotesize]
\node[gauge,label={below,xshift=0.6cm}:{$\USp(2)$}] (C1) at (0,0) {};
\node[flavour,label=below:{$\SO(2)$}] (D1l) at (2,0) {};
\node[flavour,label=below:{$\SO(3)$}] (D2r) at (-2,0) {}; 
\node[flavour,label=above:{$\SO(1)$}] (B0a) at (-1,1.5) {};
\node[flavour,label=above:{$\SO(1)$}] (B0b) at (1,1.5) {};
\node[flavour,label=below:{$\SO(3)$}] (Bd) at (0,-1.5) {};
\draw[new-green, very thick, -] (D1l) to node[above,midway]{\textcolor{new-green}{$g$}} (C1);
\draw[red, very thick, -] (C1) to node[above,midway]{\textcolor{red}{$\zeta$}} (D2r);
\draw[purple,-,very thick] (B0a) to node[right, near start]{\textcolor{purple}{$\Gamma \zeta$}}(C1);
\draw[blue,-,very thick] (B0b) to node[left, near start]{\textcolor{blue}{$\Gamma$}}(C1);
\draw[black, thick] (C1)--(Bd);
\end{tikzpicture}%
}
}
For example, if we gauge $\BZ_{2,\, \Gamma}^{[0]}$ in \eref{USp2w5withXi} and set $g=\zeta=1$, we obtain the following theory which is mirror dual to the wreathed quiver \eref{wreath1234affineD5a}:
\bes{
\scalebox{0.9}
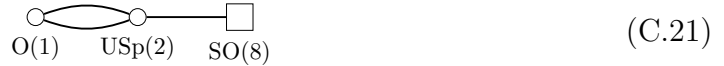
{%
\begin{tikzpicture}[baseline, font=\footnotesize]
\node[gauge,label=below:{$\USp(2)$}] (C1) at (0,0) {};
\node[gauge,label=below:{$\O(1)$}] (B0) at (-1.5,0) {};
\node[flavour,label=below:{$\SO(8)$}] (D4) at (1.5,0) {};
\draw[black, thick, -, bend left=20] (B0) to (C1);
\draw[black, thick, -, bend right=20] (B0) to (C1);
\draw[black, thick, -, bend right=0] (C1)--(D4);
\end{tikzpicture}%
}
}
On the other hand, if we gauge $\BZ_{2,\, \Gamma}^{[0]}$ and $\BZ_{2,\, g}^{[0]}$ in \eref{USp2w5withXi} and set $\zeta=1$, we obtain the following theory
\bes{
\scalebox{0.9}{%
\begin{tikzpicture}[baseline, font=\footnotesize]
\node[gauge,label=below:{$\USp(2)$}] (C1) at (0,0) {};
\node[gauge,label=below:{$\O(1)$}] (B0l) at (-1.5,0) {};
\node[gauge,label=above:{$\O(1)$}] (D1u) at (0,1.5) {};
\node[flavour,label=below:{$\SO(6)$}] (D3) at (1.5,0) {};
\draw[black, thick, -, bend left=20] (B0l) to (C1);
\draw[black, thick, -, bend right=20] (B0l) to (C1);
\draw[new-green, very thick, -, bend left=20] (D1u) to (C1);
\draw[new-green, very thick, -, bend right=20] (D1u) to (C1);
\draw[black, thick, -, bend right=0] (C1)--(D3);
\end{tikzpicture}%
}
}

\bibliographystyle{JHEP}
\bibliography{bibli.bib}

\end{document}